\documentclass{scrreprt}

\author{Thomas Wolkanowski}
\title{Poles in the second Riemann sheet}
\date{14. Januar 2013}

\usepackage[english]{babel}
\usepackage{graphics, graphicx}
\usepackage{latexsym}
\usepackage[margin=10pt,font=small,labelfont=bf]{caption}
\usepackage{amsmath}
\usepackage{cite}
\usepackage{booktabs}
\usepackage{hyperref}
\usepackage{amsfonts}
\usepackage{bbm}
\usepackage{epstopdf}
\usepackage{romannum}
\usepackage{amssymb}
\usepackage{cancel}

\DeclareMathOperator{\disc}{Disc}
\DeclareMathOperator{\sgn}{sgn}
\DeclareMathOperator{\residue}{Res}
\DeclareMathOperator{\arctanh}{arctanh}
\DeclareMathOperator{\diag}{diag}
\DeclareMathOperator{\abs}{Abs}
\DeclareMathOperator{\maxf}{max}

\addtokomafont{disposition}{\boldmath}

\begin{document}

\begin{titlepage}
	\begin{center}
		{\LARGE Johann Wolfgang Goethe--Universit\"at\\[0.15cm]
		Frankfurt am Main}\\[0.3cm]
		{\large Fachbereich Physik\\
		Institut f\"ur Theoretische Physik\\[4cm]}
		{\huge Master thesis\\[1.5cm]}
		{\Large\bf Resonances and poles in the second Riemann sheet\\[1.5cm]}
		{\large Thomas Wolkanowski}\\[0.5cm]
		14th January 2013\\[2.9cm]
		\begin{center}
			{\small\bf Supervisor}\\[0.3cm]
			{\large Prof. Dr. Dirk H. Rischke}\\[0.2cm]
			{\footnotesize Institut f\"ur Theoretische Physik\\
			Universit\"at Frankfurt am Main}\\[1cm]
		\end{center}
		\begin{center}
			{\small\bf Second reviewer}\\[0.3cm]
			{\large Dr. Francesco Giacosa}\\[0.2cm]
			{\footnotesize Institut f\"ur Theoretische Physik\\
			Universit\"at Frankfurt am Main}\\[1cm]
		\end{center}
	\end{center}
\end{titlepage}

\thispagestyle{empty}
\

\newpage
\clearpage

\thispagestyle{empty}
\

\newpage
\clearpage

\thispagestyle{empty}
\

\newpage
\clearpage

\thispagestyle{empty}
\begin{center}
\ \\[7.0cm]
''Alles Arbeit unter der Sonn, sogar Schwei{\ss} im Schlaf.\grqq\\[0.1cm]
\end{center}
\hskip 8.5cm \footnotesize{-- Georg B\"uchner}

\hskip 8.15cm \footnotesize{{\em Woyzeck}}

\newpage
\clearpage

\thispagestyle{empty}
\

\newpage
\clearpage

\thispagestyle{empty}
\section*{Preface}
In this work we study basic properties of unstable particles and scalar hadronic resonances, respectively, within simple quantum mechanical and quantum field theoretical (effective) models. The term 'particle\grq \ is usually assigned to entities, described by physical theories, that are able to propagate over sufficiently large time scales (e.g. from a source to a detector) and hence could be identified in experiments -- one especially should be able to measure some of their distinct properties like spin or charge. Nevertheless, it is well known that there exists a huge amount of {\em unstable} particles to which it seems difficult to allocate such definite values for their mass and decay width. In fact, for extremely short-lived members of that species, so called {\em resonances}, the theoretical description turns out to be highly complicated and requires some very interesting concepts of complex analysis.

In the first chapter, we start with the basic ideas of quantum field theory. In particular, we introduce the {\em Feynman propagator} for unstable scalar resonances and motivate the idea that this kind of correlation function should possess complex poles which parameterize the mass and decay width of the considered particle. We also briefly discuss the problematic scalar sector in particle physics, emphasizing that hadronic loop contributions, given by strongly coupled hadronic intermediate states, dominate its dynamics. After that, the second chapter is dedicated to the method of {\em analytic continuation} of complex functions through branch cuts. As will be seen in the upcoming sections, this method is crucial in order to describe physics of scalar resonances because the relevant functions to be investigated (namely, the Feynman propagator of interacting quantum field theories) will also have branch cuts in the complex energy plane due to the already mentioned loop contributions. As is consensus among the physical community, the understanding of the physical behaviour of resonances requires a deeper insight of what is going on beyond the branch cut. This will lead us to the idea of a {\em Riemann surface}, a one-dimensional complex manifold on which the Feynman propagator is defined.

We then apply these concepts to a simple non-relativistic Lee model in the third chapter and demonstrate the physical implications, i.e., the motion of the propagator poles and the behaviour of the spectral function. Besides that, we investigate the time evolution of a particle described by such a model. All this will serve as a detailed preparation in order to encounter the rich phenomena occuring on the Riemann surface in quantum field theory. In the last chapter, we finally concentrate on a simple quantum field theoretical model which describes the decay of a scalar state into two (pseudo)scalar ones. It is investigated how the motion of the propagator poles is influenced by loop contributions of the two (pseudo)scalar particles. We perform a numerical study for a hadronic system involving a scalar seed state (alias the $\sigma$-meson) that couples to pions. The unexpected emergence of a putative stable state below the two-pion threshold is investigated and it is clarified under which conditions such a stable state appears.
\\
\\
\\
*Note that this thesis was revised on the 24th April 2014.*

\thispagestyle{empty}
\

\newpage
\clearpage

\thispagestyle{empty}
\

\newpage
\clearpage

\pagenumbering{roman}
\setcounter{page}{1}

\renewcommand{\baselinestretch}{1.3}
\small\normalsize
\tableofcontents
\renewcommand{\baselinestretch}{1}
\small\normalsize
\clearpage

\thispagestyle{empty}
\

\newpage
\clearpage

\renewcommand{\baselinestretch}{1.3}
\small\normalsize
\addcontentsline{toc}{chapter}{List of figures}
\listoffigures
\renewcommand{\baselinestretch}{1}
\small\normalsize
\clearpage

\renewcommand{\baselinestretch}{1.3}
\small\normalsize
\addcontentsline{toc}{chapter}{List of tables}
\listoftables
\renewcommand{\baselinestretch}{1}
\small\normalsize
\clearpage

\thispagestyle{empty}
\

\newpage
\clearpage

\thispagestyle{empty}
\

\newpage
\clearpage

\thispagestyle{empty}
\

\newpage
\clearpage

\pagenumbering{arabic}
\setcounter{page}{1}

\chapter{Introduction}

\medskip

At the beginning of the 20th century most physicists around the world were faced with all the crucial changes that certainly unexpected ideas concerning the microscopic structure of nature (later on incorporated in the theory of quantum mechanics) resulted in. In retrospect, this time marks a turning point not only in the thousands-year-long history of science, but also in the mere way of how human beings look at the world surrounding them, leaving behind all coming generations with a mixture of amusement and curiosity about the universe. It is pure irony: while a huge number of our ancestors believed they were so close of obtaining a deep and conclusive understanding of the world, it seems nowadays that this kind of search for knowledge may never reach a final end. We have arranged with this. In fact, there are less people trying to reach for the answers to all things. Thinking about the universe has changed. And so we started a new venture at the beginning of the 21st centurey since it is up to us clarifying what our ancestors have left behind.

\section{Historical remarks}
Besides philosophical and fundamental challenges after finding the appropriate mathematical formalism, (non-relativistic) quantum mechanics faced a huge problem in establishing a theory of nuclear forces. In 1935, Yukawa\footnote{In 1949, H. Yukawa was honoured with the {\em Nobel Prize in Physics} for his work on nuclear forces, in which he predicted the existence of mesons.} used field theoretical methods to derive the nucleon-nucleon force as an interaction through one-pion exchange \cite{yukawa}. Although this description finally turned out to be not the correct way, it gave motivation for a lot of new approaches and theories in particle physics during the next decades. We will not try to review all those ideas, failures and milestones. This short introduction shall only sketch the basic concepts for working with one special kind of framework called {\em effective (quantum) field theory (EFT)} when dealing with {\em resonances}.

Physicists are somehow 'cursed\grq: they believe that everything should be made as simple as possible, but no simpler.\footnote{This quote is often attributed to A. Einstein.} Especially the basic first lines in books of particle physics are based on this principle, where we try to build up all matter from some very few and hopefully simple blocks of matter called elementary particles. Such efforts were first not successful after experiments discovered more and more heavy particles, named {\em hadrons}, in the early 60s whose existence was not covered by the theoretical models constructed before. It was realized soon after that most of the new particles are very short-lived states, so called {\em resonances}, during scattering processes and do not manifest themselves as propagating entities. They were usually identified from the decay products they disintegrated to, hence it became dramatically clear that they could not be taken as elementary in the above sense.

A theory to describe all the new hadrons and their interactions in a unified framework was developed from this time on, where the foundations were laid by Gell-Mann\footnote{In 1969, M. Gell-Mann was honoured with the {\em Nobel Prize in Physics} for his work on the classification of elementary particles and their interactions.} \cite{gell-mann}, Ne'eman and Zweig \cite{zweig}. One of the main questions to be answered in this context was if a classification scheme for the new and already known particles could be found. Gell-Mann and Zweig proposed a solution using group theoretical methods, namely they treated all the different hadronic states as manifestations of multiplets within the $SU(3)$ (flavour) group. As a consequence, the physical interpretation of {\em quarks}, elementary particles that are the building blocks of the observed hadrons, became the major idea to turn ones attention to. {\em Quantum chromodynamics (QCD)} is the theory of the interaction between such quarks. Without going into details, the problem with QCD is the fact that this theory is non-perturbative in the low-energy regime, for example characteristic in nuclear physics. Despite all the (technical) methods physicists and mathematicians have developed in the past to find solutions of non-pertubative problems, it was up to now not possible to solve QCD, although enormous efforts were made in recent years. By applying the fundamental theory on a discretized space-time lattice, performing all calculations by using a huge amount of computational power and finally letting the lattice spacing turn to zero, {\em lattice QCD} has become a well-established non-perturbative approach of solving QCD and though lots of poblems have not been solved yet, e.g. concerning the numerical implementation, this field of research is under continuous growth. Other approaches have been found by using holographic models and gauge/gravity correspondence, for instance to extract meson masses with good accuracy \cite{erlich,karch}.

A substantial progress in hadron physics was nevertheless made when the concept of an effective field theory was applied to the low-energy regime. Weinberg\footnote{In 1979, S. Weinberg, S. L. Glashow and A. Salam were honoured with the {\em Nobel Prize in Physics} for their work on the theory of the unified weak and electromagnetic interaction.} has pointed out the general ideas in Ref. \cite{weinberg}, i.e., the key point is to identify the appropriate degrees of freedom and to write down the most general Lagrangian consistent with the assumed symmetry principles. By this, it is not necessary anymore to solve the underlying fundamental theory due to the fact that within the new framework the set of degrees of freedom ('the basis\grq) are not quarks (and gluons) but composite particles, namely hadrons. We then only need to deal with effective fields that are obtained in the particular energy region of interest. Good introductions to EFT can be found in Refs. \cite{scherer,mosel,weise}.

Whenever studying effective field theories one needs to understand the 'replaced\grq \ theory, namely QCD, and its symmetries. The effective Lagrangian will have the same symmetries and some of them may be broken -- so clarifying the properties of the QCD-Lagrangian would give constraints on the effective Lagrangian. In this work we will not focus on specifiying an effective model for low-energy QCD. Instead we want to investigate the influence of loop contributions on the mass and decay width of a scalar particle within a simple toy model containing mesonic scalar degrees of freedom and interactions. It is therefore not necessary to summarize the properties of the QCD-Lagrangian as usually done, e.g. see Refs. \cite{phdfrancesco,phddenis}, or to go into details of {\em chiral symmetry}. We rather refer the reader to short summaries on the topic, provided for example in Ref. \cite{machleidt}.

\section{Scalar quantum field theory}
At least in modern literature it seems 'more elegant\grq \ to treat quantum field theory within the formalism of Feynman's path integrals. Yet we will choose the canonical formalism here for pedagogical reasons and follow the presentations of Refs. \cite{peskin,greiner} during the rest of this section, but use the conventions as denoted in \nameref{chapter_appendixB} within this work.

\subsection{The scalar Feynman propagator}
\begin{figure}[h]
\begin{center}
\includegraphics[scale=0.55]{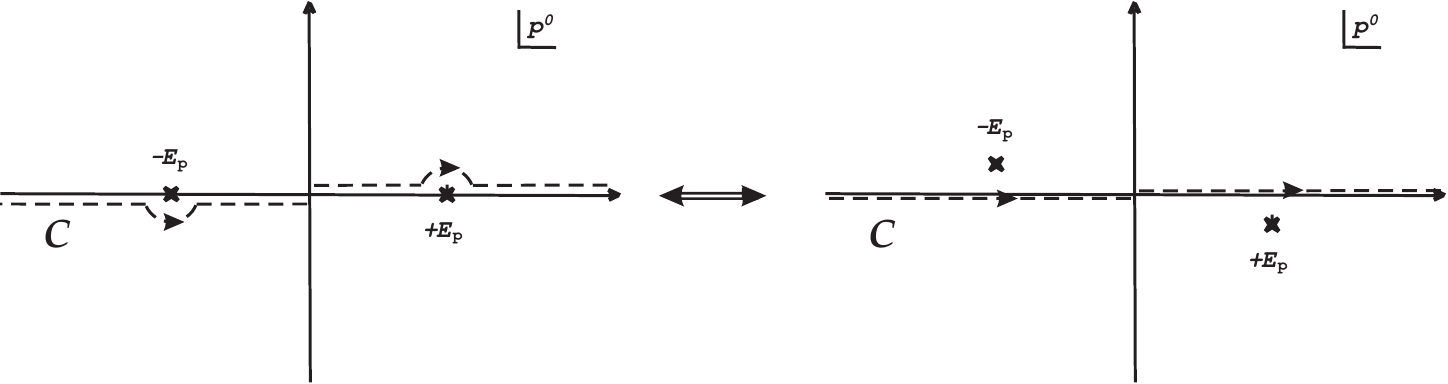}
\caption{Contour path of integration $\mathcal{C}$ (black, dashed) enclosing the two simple poles of the scalar propagator in the complex $p^{0}$-plane.}
\label{figure_proppoles}
\end{center}
\end{figure}

The free Feynman propagator for a scalar field $S$ in position space is defined as a particular expectation value of two field operators:
\begin{equation}
\Delta_{S}^{\text{free}}(x-y) = -i\langle0|\mathcal{T}\big{\{}S(x)S(y)\big{\}}|0\rangle , \label{equation_firstchapterpositionprop}
\end{equation}
where $\mathcal{T}$ is the time-ordering operator. In order to give an explicit form of this function we need to consider the action of the field operators on the vacuum state. To this end, let us first look at the case $x^{0}>y^{0}$ by writing $\Delta_{S}^{(-)\text{free}}(x-y)$. By introducing superscripts at the field $S$ denoting those parts that contain a creation or annihilation operator, i.e.,
\begin{eqnarray}
S^{(+)} & \equiv & S^{(+)}(x) \ \ = \ \ \int\frac{\text{d}^{3}p}{(2\pi)^{3}}\frac{1}{\sqrt{2E_{\textbf{p}}}} \ a_{\textbf{p}}^{\dagger}e^{ip\cdot x} \ , \\
S^{(-)} & \equiv & S^{(-)}(x) \ \ = \ \ \int\frac{\text{d}^{3}p}{(2\pi)^{3}}\frac{1}{\sqrt{2E_{\textbf{p}}}} \ a_{\textbf{p}}e^{-ip\cdot x} \ ,
\end{eqnarray}
with $p^{\mu}$ on-shell, we can immediately drop most of the emerging terms,
\begin{eqnarray}
\Delta_{S}^{(-)\text{free}}(x-y) & = & -i\langle0| \ \xcancel{S^{(+)}S^{(+)}}+\xcancel{S^{(+)}S^{(-)}}+S^{(-)}S^{(+)}+\xcancel{S^{(-)}S^{(-)}} \ |0\rangle \nonumber \\
& = & -i\int\frac{\text{d}^{3}p}{(2\pi)^{3}}\frac{1}{2E_{\textbf{p}}} \ e^{-ip\cdot(x-y)} \ ,
\end{eqnarray}
since the creation and annihilation operators combine in such a way that their scalar product vanishes. In exactly the same manner we find for the other case $x^{0}<y^{0}$:
\begin{equation}
\Delta_{S}^{(+)\text{free}}(x-y) = -i\int\frac{\text{d}^{3}p}{(2\pi)^{3}}\frac{1}{2E_{\textbf{p}}} \ e^{ip\cdot(x-y)} \ .
\end{equation}
Both results are strictly valid only for their specific time order, but can be combined when using Heaviside step functions:
\begin{eqnarray}
\Delta_{S}^{\text{free}}(x-y) & = & \Theta(x^{0}-y^{0})\Delta_{S}^{(-)\text{free}}(x-y)+\Theta(y^{0}-x^{0})\Delta_{S}^{(+)\text{free}}(x-y) \nonumber \\
& = & -i\int\frac{\text{d}^{3}p}{(2\pi)^{3}}\frac{1}{2E_{\textbf{p}}}\Big[\Theta(x^{0}-y^{0}) \ e^{-ip\cdot(x-y)}+\Theta(y^{0}-x^{0}) \ e^{ip\cdot(x-y)}\Big] \ . \nonumber \\
\end{eqnarray}
This result can be further simplified. First we split the integrand in its time and spatial coordinates and then perform the {\em analytic continuation} into the complex $p^{0}$-plane. Realizing that the remaining expression is nothing else than the summation of its two residues, we can rewrite it as a contour integral according to the residue theorem:
\begin{equation}
-\frac{1}{2\pi i}\int_{\mathcal{C}}\text{d}p^{0} \ \frac{e^{-ip^{0}(x^{0}-y^{0})}}{(p^{0}-E_{\textbf{p}})(p^{0}+E_{\textbf{p}})} \ = \ \begin{cases} \frac{1}{2E_{\textbf{p}}} \ e^{-ip^{0}(x^{0}-y^{0})}\big|_{p^{0}=E_{\textbf{p}}} & x^{0}>y^{0} \\
\frac{1}{2E_{\textbf{p}}} \ e^{ip^{0}(x^{0}-y^{0})}\big|_{p^{0}=E_{\textbf{p}}} & x^{0}<y^{0} \\
\end{cases} \ .
\end{equation}
The simple poles are located at $p^{0}=\pm E_{\textbf{p}}$. The contour path of integration $\mathcal{C}$ has to be closed either in the upper half plane $(x^{0}<y^{0})$ or in the lower half plane $(x^{0}>y^{0})$ in order to make the contribution from the semicircle tend to zero. Together with the spatial part the scalar Feynman propagator can be written as
\begin{eqnarray}
\Delta_{S}^{\text{free}}(x-y) & = & \int\frac{\text{d}^{3}p}{(2\pi)^{3}}\int_{\mathcal{C}}\frac{\text{d}p^{0}}{2\pi} \ e^{-ip\cdot(x-y)}\frac{1}{p^{2}-M_{0}^{2}} \nonumber \\
& = & \int\frac{\text{d}^{4}p}{(2\pi)^{4}} \ e^{-ip\cdot(x-y)}\frac{1}{p^{2}-M_{0}^{2}+i\epsilon} \ ,
\end{eqnarray}
where in the last step we have evaluated the $p^{0}$-integration by shifting the two poles as shown in Fig. \ref{figure_proppoles} ($p^{\mu}$ is no longer on-shell). This is the famous {\em Feynman prescription} in which the negative pole moves off the real $p^{0}$-axis into the upper half plane, while the positive one moves into the lower half plane. The propagator in momentum space now simply becomes:
\begin{equation}
\Delta_{S}^{\text{free}}(p^{2}) = \frac{1}{p^{2}-M_{0}^{2}+i\epsilon} \ . \label{equation_finalpositionprop}
\end{equation}
It should be stressed that in the rest frame the bare mass $M_{0}$ is the total energy of the particle corresponding to the field $S$. If we switch to the complex $p^{2}$-plane, then there will be a single pole just below the real axis at $p^{2}=M_{0}^{2}-i\epsilon$. Such poles describe stable states.
\\

With the result just obtained we may show that the scalar Feynman propagator as given in Eq. (\ref{equation_finalpositionprop}) is the {\em Green's function} of the Klein--Gordon equation which fully determines the evolution of the field $S$:
\begin{eqnarray}
(\Box_{x}+M_{0}^{2})\Delta_{S}^{\text{free}}(x-y) & = & \int\frac{\text{d}^{4}p}{(2\pi)^{4}} \ (-p^{2}+M_{0}^{2})e^{-ip\cdot(x-y)}\frac{1}{p^{2}-M_{0}^{2}+i\epsilon} \nonumber \\
& = & -\int\frac{\text{d}^{4}p}{(2\pi)^{4}} \ e^{-ip\cdot(x-y)}\nonumber \\
& = & -\delta^{(4)}(x-y) \ .
\end{eqnarray}

\subsection{K\"all\'{e}n--Lehmann spectral representation}
The free Feynman propagator for a scalar field $S$, defined as the two-point correlation function in position space given by Eq. (\ref{equation_firstchapterpositionprop}), plays an important role in quantum field theory, e.g. the LSZ reduction formula for the $\hat S$-matrix mainly requires these kind of correlation functions. Additionally, the propagator can be interpreted as the amplitude for a particle to propagate from the space-time point $y$ to $x$. Note that there is no problem to calculate the amplitude in a closed form when the particle is free, while this statement is not true anymore for interacting fields.

One can express the full interacting propagator as a similar product of field operators, but where the scalar product is taken between the new ground state $|\Omega\rangle$ of the interacting theory:
\begin{equation}
\Delta_{S}(x-y) = -i\langle\Omega|\mathcal{T}\big{\{}S(x)S(y)\big{\}}|\Omega\rangle \ . \label{equation_positionfullprop}
\end{equation}
Let us write down the general completeness relation as
\begin{equation}
\mathbbm{1} = \sum_{\alpha}|\alpha\rangle\langle\alpha| \ ,
\end{equation}
where we do not know the explicit form of the basis. Yet the basis will surely contain the single-particle states and all many-particle states, both types collected in $|\lambda_{\textbf{p}}\rangle$, that shall be relativistically normalized momentum eigenstates. One can choose the set of three-momentum eigenstates to be the basis, since the three-momentum operator $\hat{\textbf{P}}$ commutes with the full Hamiltonian. As a direct consequence of postulating Lorentz invariance of the theory, any eigenstate at rest $|\lambda_{0}\rangle$ of the Hamiltonian can be boosted in such a way that every possible value of the three-momentum is achieved. These new states are of course also eigenstates of the momentum operator and the Hamiltonian. By assuming the many-particle states to satisfy the energy dispersion relation with $m_{\lambda}$ as the total energy of a state at rest, this is also true for the boosted many-particle states in $|\lambda_{\textbf{p}}\rangle$. The single-particle states have invariant mass $p^{2}=m^{2}$, the total energy of an unboosted state where the particle is at rest. The above completeness relation is then given by
\begin{equation}
\mathbbm{1} = |\Omega\rangle\langle\Omega|+\sum_{\lambda}\int\frac{\text{d}^{3}p}{(2\pi)^{3}}\frac{1}{2E_{\textbf{p}}(\lambda)}|\lambda_{\textbf{p}}\rangle\langle\lambda_{\textbf{p}}| \ ,
\end{equation}
summing over all unboosted momentum states. Inserting the completeness relation between the field operators in Eq. (\ref{equation_positionfullprop}) and considering $x^{0}>y^{0}$ for the moment, we are left with
\begin{equation}
\Delta_{S}^{(-)}(x-y) = -i\sum_{\lambda}\int\frac{\text{d}^{3}p}{(2\pi)^{3}}\frac{1}{2E_{\textbf{p}}(\lambda)}\langle\Omega|S(x)|\lambda_{\textbf{p}}\rangle\langle\lambda_{\textbf{p}}|S(y)|\Omega\rangle \ .
\end{equation}
Here, we have dropped the ground state term since it usually vanishes \cite{brown}. The remaining matrix elements can be simplified using the transformation relations of scalar field operators under the translation group,
\begin{equation}
S(x) = e^{i\hat{P}\cdot x}S(0)e^{-i\hat{P}\cdot x} \ ,
\end{equation}
and the translation invariance of the ground state. Additionally, remembering that $p^{0}=E_{\textbf{p}}$ and realizing that $S(x)$ is a Lorentz scalar and that the ground state obeys Lorentz invariance, the propagator for $x^{0}>y^{0}$ reduces to
\begin{equation}
\Delta_{S}^{(-)}(x-y) = -i\sum_{\lambda}\int\frac{\text{d}^{3}p}{(2\pi)^{3}}\frac{1}{2E_{\textbf{p}}(\lambda)} \ e^{-ip\cdot(x-y)}|\langle\Omega|S(0)|\lambda_{0}\rangle|^{2} \ .
\end{equation}
An explicit calculation makes this clear:
\begin{eqnarray}
\langle\Omega|S(x)|\lambda_{\textbf{p}}\rangle & = & \langle\Omega|S(0)|\lambda_{\textbf{p}}\rangle \ e^{-ip\cdot x}\big|_{p^{0}=E_{\textbf{p}}} \nonumber \\
& = & \langle\Omega|\hat U^{-1}\hat US(0)\hat U^{-1}\hat U|\lambda_{\textbf{p}}\rangle \ e^{-ip\cdot x}\big|_{p^{0}=E_{\textbf{p}}} \nonumber \\
& = & \langle\Omega|\hat U^{-1}\hat US(0)\hat U^{-1}|\lambda_{0}\rangle \ e^{-ip\cdot x}\big|_{p^{0}=E_{\textbf{p}}} \nonumber \\
& = & \langle\Omega|S(0)|\lambda_{0}\rangle \ e^{-ip\cdot x}\big|_{p^{0}=E_{\textbf{p}}} \ .
\end{eqnarray}
Here, $\hat U$ is some unitary operator that implements a Lorentz boost from the three-momentum $\textbf{p}$ to 0. If we now define the {\em spectral density function} $\rho(s^{2})$ by the expression
\begin{equation}
\rho(s^{2}) = \pi\sum_{\lambda}\delta(s^{2}-m_{\lambda}^{2})|\langle\Omega|S(0)|\lambda_{0}\rangle|^{2} \ ,
\end{equation}
the full interacting propagator for the field $S$ can be rewritten without neglecting any terms:
\begin{eqnarray}
\Delta_{S}(x-y) & = & \Theta(x^{0}-y^{0})\Delta_{S}^{(-)}(x-y)+\Theta(y^{0}-x^{0})\Delta_{S}^{(+)}(x-y) \nonumber \\
& = & -i\sum_{\lambda}\bigg[\Theta(x^{0}-y^{0})\int\frac{\text{d}^{3}p}{(2\pi)^{3}}\frac{1}{2E_{\textbf{p}}(\lambda)} \ e^{-ip\cdot(x-y)} \nonumber \\
&  & + \ \Theta(y^{0}-x^{0})\int\frac{\text{d}^{3}p}{(2\pi)^{3}}\frac{1}{2E_{\textbf{p}}(\lambda)} \ e^{ip\cdot(x-y)}\bigg]|\langle\Omega|S(0)|\lambda_{0}\rangle|^{2} \nonumber \\
& = & \sum_{\lambda}\int\frac{\text{d}^{4}p}{(2\pi)^{4}} \ e^{-ip\cdot(x-y)}\frac{1}{p^{2}-m_{\lambda}^{2}+i\epsilon} \ |\langle\Omega|S(0)|\lambda_{0}\rangle|^{2} \nonumber \\
& = & \int\frac{\text{d}^{4}p}{(2\pi)^{4}} \ e^{-ip\cdot(x-y)} \ \frac{1}{\pi}\int_{0}^{\infty}\text{d}s^{2} \ \frac{\rho(s^{2})}{p^{2}-s^{2}+i\epsilon} \ .
\end{eqnarray}
In the last two steps an integration over $p^{0}$ was introduced. This yields a simple expression for the propagator in momentum space:
\begin{equation}
\Delta_{S}(p^{2}) = \frac{1}{\pi}\int_{0}^{\infty}\text{d}s^{2} \ \frac{\rho(s^{2})}{p^{2}-s^{2}+i\epsilon} \ . \label{equation_firstchaptermomentumprop}
\end{equation}
Obviously, the full propagator is a superposition of free single-particle propagators, each with different mass, weighted with the spectral density function $\rho(s^{2})$. This function is positive semi-definite and Lorentz invariant, it furthermore vanishes if the running argument $s^{2}$ does not lie within the future light cone. The integral representation from above is referred to as the {\em K\"all\'{e}n--Lehmann spectral representation}, independently found in the early 50s by K\"all\'{e}n \cite{kallen} and Lehmann \cite{lehmann}. If we are able to identify the spectral function of a theory, we can directly calculate its propagator. The spectral function is normalized to one, a direct consequence of the equal-time commutation relations for the interacting field operators,
\begin{eqnarray}
&  & [S(x_{0},\textbf{x}),\dot S(y_{0},\textbf{y})]\big|_{x_{0}=y_{0}} = i\delta^{(3)}(\textbf{x}-\textbf{y}) \ , \label{equation_comrelation} \\
&  & [S(x_{0},\textbf{x}),S(y_{0},\textbf{y})]\big|_{x_{0}=y_{0}} = [\dot S(x_{0},\textbf{x}),\dot S(y_{0},\textbf{y})]\big|_{x_{0}=y_{0}} = 0 \ . \label{equation_comrelation2}
\end{eqnarray}
See Ref. \cite{greiner} for a detailed derivation. It can be shown that
\begin{equation}
i\delta^{(3)}(\textbf{x}-\textbf{y}) = i\delta^{(3)}(\textbf{x}-\textbf{y}) \ \frac{1}{\pi}\int_{0}^{\infty}\text{d}s^{2} \ \rho(s^{2}) \ ,
\end{equation}
which requires the normalization condition
\begin{equation}
\frac{1}{\pi}\int_{0}^{\infty}\text{d}s^{2} \ \rho(s^{2}) = 1 \ .
\end{equation}

For typical theories all accessible intermediate many-particle states will appear in the spectral function as a continuous spectrum above some threshold (we assume a mass gap that separates single-particle states from the many-particle states). It is commonly accepted to interpret 'bumps\grq \ within this continuum as unstable particles or resonances, depending on the dimension of their broadness. The state with the lowest total energy $s^{2}=m^{2}$ is a single-particle state. If it is prepared in its rest frame the value of $m$ is the physical mass of the particle, the mass that enters the energy dispersion relation. In this work we will go into the details on how resonances can be related with analytic properties of its propagator. A first hint is given when we try to perform the integration indicated in Eq. (\ref{equation_firstchaptermomentumprop}). If there are really single-particle states in the spectrum separated from the continuous distribution, then we expect their propagators to show up since the spectral function should give zero until the threshold value $4m^{2}$:
\begin{eqnarray}
\Delta_{S}(p^{2}) & = & \frac{1}{\pi}\int_{0}^{\infty}\text{d}s^{2} \ \frac{\rho(s^{2})}{p^{2}-s^{2}+i\epsilon} \nonumber \\
& = & \frac{Z}{p^2-m^{2}+i\epsilon}+\int_{4m^{2}}^{\infty}\text{d}s^{2} \ \frac{\rho_{a.t.}(s^{2})}{p^{2}-s^{2}+i\epsilon} \ , \label{equation_propspectralwithZ}
\end{eqnarray}
where {\em a.t.} means 'above threshold\grq. The spectral function takes the form:
\begin{equation}
\rho(s^{2}) = Z\pi\delta(s^{2}-m^{2})+\rho_{a.t.}(s^{2}) \ .
\end{equation}
Here, the quantity $Z$ stands for the wave function renormalization constant, the probability for the operator $S(0)$ to create a given state from the ground state. Hence, single-particle states are manifested as simple poles in the complex $p^{2}$-plane on which the propagator $\Delta_{S}(p^{2})$ is defined, while many-particle intermediate states make the propagator to possess a {\em branch cut} starting at the threshold value. Such a singularity provides a purely imaginary discontinuity along the positive real axis, dictated by the spectral function above threshold:
\begin{eqnarray}
\frac{1}{\pi}\int_{4m^{2}}^{\infty}\text{d}s^{2} \ \frac{\rho_{a.t.}(s^{2})}{p^{2}-s^{2}+i\epsilon} & \stackrel{\mu\rightarrow s^{2}-p^{2}}{=} & -\frac{1}{\pi}\int_{4m^{2}-p^{2}}^{\infty}\text{d}\mu \ \frac{\rho_{a.t.}(\mu+p^{2})}{\mu-i\epsilon} \nonumber \\
& = & -\frac{1}{\pi} \ \mathcal{P}\int_{4m^{2}-p^{2}}^{\infty}\text{d}\mu \ \frac{\rho_{a.t.}(\mu+p^{2})}{\mu}-i\rho_{a.t.}(p^{2}) \ . \nonumber \\
\label{equaiton_propshowingdisc}
\end{eqnarray}
After we made use of the {\em Sokhotski--Plemelj theorem}, it is obvious that the additional second term only appears when $\operatorname{Re}p^{2}>4m^{2}$. We will clarify the meaning of such a behaviour in the next chapter from a mathematical point of view and apply the consequences to particle physics.

\section{Resonances}
\subsection{Decay width and Breit--Wigner parameterization} \label{subsection_breitwigner}
In the first section we have found an analytic expression for the full interacting propagator, see Eq. (\ref{equation_propspectralwithZ}), by using the K\"all\'{e}n--Lehman spectral representation. On the other hand there exists a direct connection between the propagator and the spectral function. By looking at the discontinuous behaviour shown in Eq. (\ref{equaiton_propshowingdisc}) we realize that the principal value (of the propagator) is purely real. Thus, we may take the imaginary part of the propagator right above the real axis and obtain a very important relation:
\begin{eqnarray}
\operatorname{Im}\Delta_{S}(p^{2}+i\epsilon) & = & -\frac{Z\epsilon}{(p^{2}-m^{2})^{2}+\epsilon^{2}}-\rho_{a.t.}(p^{2}) \ , \\
\nonumber \\
\Rightarrow \ \ \ -\operatorname{Im}\Delta_{S}(p^{2}+i\epsilon) & = & Z\pi\delta(p^{2}-m^{2})+\rho_{a.t.}(p^{2}) \nonumber \\
& = & \rho(p^{2}) \ ,
\end{eqnarray}
where we have applied the Lorentzian representation of the delta distribution function. The infinitesimal number $i\epsilon$ inside the argument of the propagator may look superfluous but in fact we will see later that it is crucial. So, if we are able to construct a propagator by specifying the interaction terms which lead to a continuous spectrum above some threshold value, then we can also write down the spectral function. This will be used in chapter 4 of this work to study resonances in a simple hadronic toy model.

We now have to make some remarks on resonances. In the framework of quantum field theory the term 'particle\grq \ is assigned to excitations of the fields (like the scalar field $S$) that are able to propagate over sufficiently large time scales (e.g. from a source to a detector) and hence could be identified in experiments.  Those entities would have distinct measurable properties like spin or mass, and consequently should satisfy the energy dispersion relation. It then depends on the definition how to fix terminology: a {\em stable} particle should be a particle that propagates over an infinite amount of time without decay, except when interactions with other particles occur. This leads us to {\em unstable} particles -- it is not difficult to fill dozens of pages about this topic. In order to satisfy expectations for an introductory chapter let us proceed in huge steps and simply write down results of less illuminating derivations by referring to the literature.

For example, we know that charged pions as part of the particle shower caused by cosmic radiation have a mean life time of about $10^{-8}$ s and can be described nearly as stable as long as the propagation and interaction time is much smaller than the mean life time (including relativistic time dilation effects). Nevertheless, when considering time scales of some seconds those particles decay into other particles, namely muons and neutrinos or photons. Unstable particles also build up chemical elements (and their compounds) that emit radiation in form of decay products of their nuclei -- in fact we are surrounded by usually invisible effects of decaying unstable particles.

During the past century physicists have found some 'particles\grq \ -- and constructed theories with them -- that cannot be even named as unstable, because they would be extremely short-lived unstable particles with mean life times on the order of $10^{-22}$ s. It therefore makes more sense to treat them like excitations emerging when investigating nuclear matter or when performing high-energy collision experiments. Travelling at the speed of light, these particles could only overcome a distance of about $10^{-14}$ m before decaying. Such resonant states can nevertheless be interpreted as fluctuations of a field, too, and so we may use the words 'unstable particle\grq \ and 'resonance\grq \ equivalently. A very helpful formalism to describe an unstable particle is to make connection between its mean life time $\tau$ and its spectral function (or its propagator). Treating the particle decay as a Poisson process, one usually defines
\begin{equation}
\tau = \Gamma^{-1} \ ,
\end{equation}
where $\Gamma$ is called the {\em decay width} of the resonance associated with a specific set of final states, namely its decay products. This directly yields an exponential decay law for the survival probability $p(t)$ of the particle in its rest frame \cite{brown,giacosa}:
\begin{eqnarray}
p(t) = e^{-\Gamma t} \ .
\end{eqnarray}
The particle decay width is a measurable object and theoretical determinations can surely be treated as predictions. We will calculate decay widths in the framework of non-relativistic quantum mechanics and effective quantum field theory; this is done in both cases with perturbation theory. However, it seems after all very problematic to define a propagator for a resonance if the decay width is comparable to the assigned particle mass $M$. Such a statement is sometimes made clear by a rough idea: Suppose we investigate a scattering process between two particles, then a resonant state would be the manifestation of the very short-lived unstable state created and somehow exchanged during the scattering process. In the end, of course, the resonance disappears and may decay into the former two incoming particles. During the short time of its existence, according to the uncertainty principle the unstable particle could travel for a time interval about of $1/M$. But if $\Gamma\sim M$ the resonance decays already after $1/\Gamma$ and so cannot propagate over the full interaction distance -- the main problem contains the vague definition of the unstable particle's mass.

For the moment, let us assume $\Gamma\ll M$, so that there is no problem by defining a propagator $\Delta_{S}(p^{2})$ in momentum space (where $Z=0$ in Eq. (\ref{equation_propspectralwithZ})). Since we deal with a particle that can decay, i.e., disintegrate into other particles, we are clearly faced with a problem of interacting quantum field theory. The complex pole of the free propagator situated at $p^{2}=M_{0}^{2}-i\epsilon$ can no longer be assigned to the now unstable particle, but it is also not obvious what exactly happens with this pole when the interaction is turned on. The only thing we can say for now is that there appears a branch cut above some threshold value and, respectively, the spectral function shows a continuous spectrum with maybe a narrow bump roughly at the mass $M$ of the unstable particle. The mass $M$ differs from the bare mass parameter $M_{0}$ due to mesonic quantum fluctuations which can be included in a Dyson resummation scheme. The scalar propagator for the unstable particle will then be dressed by the {\em self-energy} contribution $\Pi(p^{2})$ (see chapter 4 for details) that is specified by the interaction term(s) appearing in the Lagrangian of the theory:
\begin{equation}
\Delta_{S}(p^{2}) = \frac{1}{p^{2}-M_{0}^{2}-\Pi(p^{2})} \ . \label{equation_dressesprop}
\end{equation}
In general, the self-energy will turn out to be a complex-valued function with non-zero imaginary part, so we can drop the infinitesimal number $i\epsilon$. Whatever we may think to be 'the mass\grq \ of the unstable particle $S$, a reasonable definition is the real part of the denominator of the dressed propagator:
\begin{equation}
M^{2} = M_{0}^{2}+\operatorname{Re}\Pi(M^{2}) \ . \label{equation_firstchapterBWmass}
\end{equation}
However, if the imaginary part of $\Pi(p^{2})$ contributes to the denominator, the pole of the propagator is displaced from the real $p^{2}$-axis, determining the mass and width of the resonance through
\begin{equation}
\sqrt{p_{\text{pole}}^{2}} = M-i\frac{\Gamma}{2} \ .
\end{equation}
This follows directly from the dressed propagator (\ref{equation_dressesprop}) and the well-known {\em optical theorem},
\begin{equation}
\Gamma = -\frac{Z_{\text{pole}}}{M}\operatorname{Im}\Pi(M^{2}) \ , \label{equation_firstchapterBWwidth}
\end{equation}
which relates the decay width of the resonance to the imaginary part of the self-energy taken at the value of the particle's mass \cite{peskin}. $Z_{\text{pole}}$ is some renormalization constant. Expecting $\operatorname{Im}\Pi(M^{2})$ to be small and moreover neglecting the full energy dependence of the self-energy, i.e., approximating it by its value on the real axis at the mass of the resonance, the optical theorem can be used to write down the propagator in the vicinity of the new pole as
\begin{eqnarray}
\Delta_{S}(p^{2}) & \simeq & \frac{Z_{\text{pole}}}{p^{2}-M^{2}-iZ_{\text{pole}}\operatorname{Im}\Pi(M^{2})} \nonumber \\
& \simeq & \frac{Z_{\text{pole}}}{p^{2}-M^{2}+iM\Gamma+\frac{\Gamma^{2}}{4}} \nonumber \\
& = & \frac{Z_{\text{pole}}}{p^{2}-\left(M-i\frac{\Gamma}{2}\right)^{2}} \ ,
\end{eqnarray}
where in the second step $\Gamma^{2}/4$ was added to the denominator. It should be stressed that this is only a good approximation for our assumption $\Gamma\ll M$. In fact, the search for such poles will be the main part of this work, though we will realize soon that for broad resonances the full analytic structure and energy dependence of the self-energy has to be taken into account. This will cause the resonance pole to vanish from the complex $p^{2}$-plane.

\begin{figure}[h]
\begin{center}
\includegraphics[width=314pt]{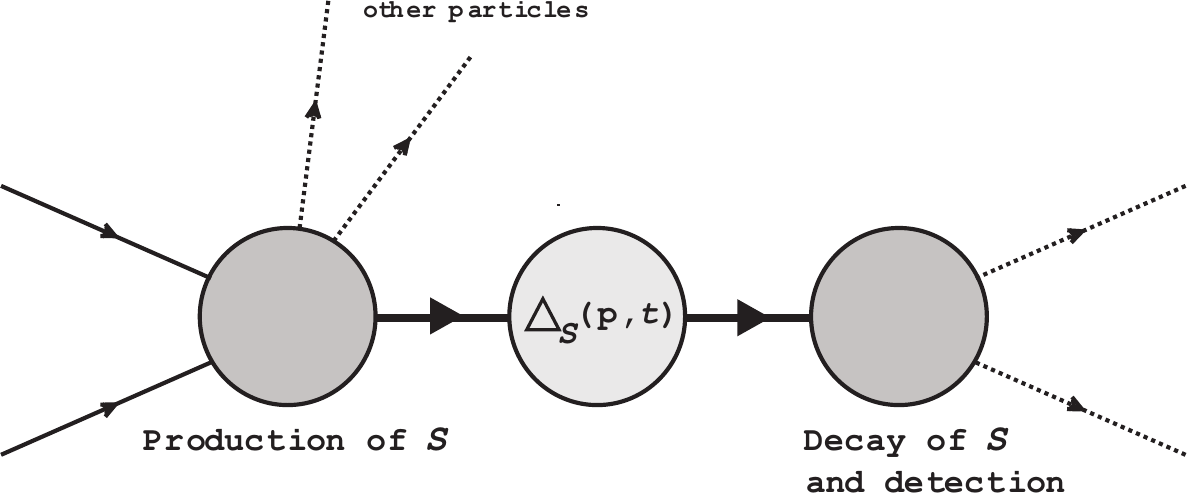}
\caption{Unstable particle production and detection: The production and decay of the particle $S$ occurs in a localized time interval such that, comparing to its mean life time, there is enough time for $S$ to propagate.}
\label{figure_generaldecay}
\end{center}
\end{figure}

The corresponding spectral function as the negative imaginary part of the propagator right above the real axis becomes
\begin{eqnarray}
\rho(p^{2}) & \simeq & \frac{Z_{\text{pole}}\operatorname{Im}\Pi(M^{2})}{(p^{2}-M^{2})^{2}+(Z_{\text{pole}}\operatorname{Im}\Pi(M^{2}))^{2}} \nonumber \\
& = & \frac{M\Gamma}{(p^{2}-M^{2})^{2}+M^{2}\Gamma^{2}} \ , \label{equation_BWspectral}
\end{eqnarray}
known as the relativistic {\em Breit--Wigner distribution} for a resonance with mass $M$ and decay width $\Gamma$. Note that the presented parameterization is strictly valid only for $\Gamma\ll M$, i.e., long-lived unstable particles and sharp resonances, respectively. For example, positively charged pions with dominant leptonic decay channel $\pi^{+}\rightarrow\mu^{+}\nu_{\mu}$ and a mass of about 140 MeV possess a mean life time of about $10^{-8}$ s, which gives a ratio $\Gamma/M\sim10^{-16}$, while for neutral pions with $\pi^{0}\rightarrow\gamma\gamma$, a mean life time of about $10^{-17}$ s and a mass of about 135 MeV we obtain $\Gamma/M\sim10^{-8}$. The decay process of such particles can be illustrated like in Fig. \ref{figure_generaldecay}.

On the contrary, very short-lived particles cannot be directly observed, yet it is possible to establish their existence from a scattering process $A+B\rightarrow C+D$ of two incoming (stable) particles $A$ and $B$, and a set of outgoing particles $C+D$, where the subset $C$ has an intermediate resonance $S$ such that $S\rightarrow C$ without detection. From literature we know that the general expression for the decay width (see chapter 4) has nearly the same formal structure as the differential cross section $\text{d}\sigma$ \cite{peskin,gross}. After performing a scattering experiment and measuring the invariant mass distribution of the particles in the subset $C$ produced during the interaction, it is possible to find a peak in the differential cross section located around a value $\sqrt{s}=M$, the mass of the resonance $S$. If all our above simplifications can be justified, namely if $\Gamma$ is not too large, the obtained curve in the region of $\sqrt{s}=M$ will be narrow and can be approximated with good accuracy by a Breit--Wigner distribution with a full width at half maximum of about $\Gamma$, the decay width of the resonance $S$ (this clarifies why one usually speaks of 'broad\grq \ and 'narrow\grq \ resonances). The differential cross section for the total process $A+B\rightarrow C+D$ is then proportional to the spectral function from Eq. (\ref{equation_BWspectral}),
\begin{equation}
\text{d}\sigma(A+B\rightarrow C+D) \sim \frac{M\Gamma}{(p^{2}-M^{2})^{2}+M^{2}\Gamma^{2}} \ ,
\end{equation}
so measurements of the cross section can look for resonances predicted by theory \cite{brown}. Note that we have only treated a non-interfering production cross section of a single resonant state with two incoming (stable) particles, while usually background reactions and other multi-channel effects distort the pure contribution from the resonance $S$, such that it is harder to observe if there really is something or not.

\subsection{The problematic scalar sector}
The original quark model based on $SU(3)$ flavour symmetry is not the full story of particle physics. A very famous sector of particles is the scalar mesonic sector. The scalar mesons have bothered the physical community for a long time because their identification and clear evidence are far from having a simple answer. Especially some of those mesons possess large decay widths, several decay channels and a huge background \cite{beringer}. In this work we will focus on the lightest scalar meson, namely the $\sigma$-meson or $f_{0}(500)$ state, which cannot be parameterized by ordinary methods, for example by a simple Breit--Wigner distribution. During the past decades a large number of theoretical approaches were invented to handle related problems and to extract particle information from experimental data, while some effective models for QCD involving chiral symmetry treated the $\sigma$-meson as existent.\footnote{The sigma model was introduced by Gell-Mann and L\'{e}vy \cite{levy} after they referred to a field corresponding to a spinless meson that was introduced by Schwinger \cite{schwinger}. In 1965, J. Schwinger, R. P. Feynman and S. Tomonaga were honoured with the {\em Nobel Prize in Physics} for their work on quantum elekctrodynamics (QED).} Various reasons for this assumption can be cited, for example it is often argued that the $\sigma$-meson is the putative chiral partner of the pion in linear sigma models, since the vacuum expectation value of this state is used to implement spontaneous chiral symmetry breaking. Furthermore, $\pi\pi$-scattering data seems to require the existence of the $\sigma$-meson in order to fit the data accurately, see Refs. \cite{phddenis,beringer} and references therein for further information. The state $f_{0}(500)$ is often interpreted as a so-called tetraquark state \cite{maiani,thooft,giacosaTetra}, a hypothetical mesonic structure first suggested by Jaffe \cite{jaffe}.

In the 70s, the Particle Data Group (PDG) removed the $\sigma$-meson from its listings only to put it back in the 90s in consequence of new theoretical efforts made by T\"ornqvist and Roos \cite{roos} when locating the pole position of $f_{0}(500)$ after fitting scattering data. It is also possible to determine the $\sigma$-pole by using {\em Roy equations} with crossing symmetry, analyticity and unitarity -- many works going in that direction were published in the last years and have found similar results. Only to mention one, Caprini et al. have shown in Ref. \cite{caprini} that the $\pi\pi$-scattering amplitude contains a pole with the quantum numbers of $f_{0}(500)$ and calculated its mass and decay width within small uncertainties:
\begin{equation}
\sqrt{s_{\text{pole}}} = \left(441_{-8}^{+16}-i272_{-12.5}^{+9}\right) \ \text{MeV} \ .
\end{equation}
The PDG recently revised its values for the mass and width estimates on the $\sigma$-meson \cite{beringer}, giving a range
\begin{equation}
\sqrt{s_{\text{pole}}} = (400\text{-}550)-i(200\text{-}350) \ \text{MeV} \ ,
\end{equation}
which is a much stronger constraint in comparison with the older window from 2010,
\begin{equation}
\sqrt{s_{\text{pole}}} = (400\text{-}1200)-i(250\text{-}500) \ \text{MeV} \ .
\end{equation}
Both specifications as well as the pole of Caprini et al. should make clear that the $f_{0}(500)$ state is a very broad resonance with a decay width $\Gamma\sim M$. The assumptions included in the Breit--Wigner parameterization derived in the last subsection are clearly invalid.

As already mentioned, the problematic scalar sector is so far not fully understood and thus it is not suprising that different interpretations and models for scalar mesons were and are in use. In 1995, T\"ornqvist studied the scalar sector by including hadronic loop contributions \cite{tornqvist}, though a similar idea was already mentioned by H\"ohler \cite{hoehler} in the late 50s. Following the pedagogical introduction of Boglione and Pennington \cite{pennington}, one can think of scalar states as belonging to ideally mixed quark multiplets. In the simplest case of a single resonance we can assign a free propagator to the mesonic state, which becomes dressed if the coupling to a decay channel is non-zero. The corresponding field $S$ cannot be considered anymore as an asymptotic state of the Lagrangian, since the loop contributions given by strongly coupled hadronic intermediate states dominate its dynamics. Because of this strong coupling to intermediate states, the scalar sector escapes from the general approach of the naive quark model. Such kind of mechanism is called {\em dynamical generation}, though this term is not used consistently by all authors working in the field, see Ref. \cite{giacosaDynamical} for an overview. The mass and width of the resonance are now determined by the position of the complex pole of the full interacting propagator, a procedure first proposed by Peierls \cite{peierls} a long time ago, making the quantum theoretical treatment of unstable particles to become an object of much interest (see Refs. \cite{hoehler,levypoles,aramaki,landshoff} for only some articles published after Peierl's work). This is usually called a {\em pole approach}. The power of such a definition of mass and width (of resonances) lies in obtaining gauge and field-redefinition invariant values in order to qualify physical observables \cite{bhattacharya}. It is well-known that the Breit--Wigner parameterization does not fulfill these properties \cite{valencia1,willenbrock1,willenbrock2,sirlin1,sirlin2}. It should also be noted that the vast majority of pole approaches first model the scattering amplitude by fitting experimental data and then extract the values of $M$ and $\Gamma$ from the propagator pole. In this work we will only focus on the pole position without looking at data, yet there is no problem of extending our work in that direction. For a short introduction on how to use the formalism regarding scattering amplitudes we highly recommend Refs. \cite{tornqvist,pennington,morgan,escribano} (and references therein).

 \label{chapter_chapter1}
\clearpage

\chapter{On multi-valued complex functions}

\medskip

\section{General problem}
In ordinary mathematical analysis, the definition of a {\em function} is not very sophisticated at first glance. It is said to be a mapping of elements $x$ from a domain $X$ to elements $y$ of another domain $Y$, where the latter is called {\em image} of the specified function $f$ \cite{koenigsberger}. One can also say that the function $f$ is a {\em relation} between $X$ and $Y$, a set of ordered pairs $(x,y)$ such that $f$ is a subset of the cartesian product $X\times Y$ \cite{weltner}. Now, the difference between a function and an arbitray mapping or relation is the fact that the first one, given domains $X$ and $Y$, is a so-called {\em unique} mapping or relation between $X$ and $Y$: by looking at all ordered pairs of elements, for every $x$ in $X$ there is a unique $y$ in $Y$ for which the pair $(x,y)$ is in $f$. Strictly speaking, if an element $x$ of $X$ is given and it is not possible to point at one and only one element $y$ of $Y$ such that the pair $(x,y)$ is in $f$, then $f$ is {\em not} a function. The same statement in the context of a mapping becomes: each input value $x$ needs to be associated with exactly one output $y$.

We already know some examples of relations of which we usually speak of functions but where we actually need modifications to be in line with the upper definition. The root function
\begin{equation}
f: \ \mathbb{R}^{+}\rightarrow\mathbb{R}^{+}, \ x\mapsto\sqrt{x}
\end{equation}
is well-defined for all positive real numbers including zero, and it is known to be {\em single-valued} because the mapping is indeed unique. Every ordinary function is single-valued (in fact, this was the key point in our definition). Nevertheless, we can expand the image of the root function onto all real numbers $\mathbb{R}$. Although the new relation is not anymore a function in the spirit of our above definition, it is commonly handled like that and simply called a {\em multi-valued} function. The main reason for such an ambivalent treatment is grounded on the possibility to make $f$ single-valued whenever it is needed (so to fix the domain $Y$ in a way that the mapping $f$ becomes unique). The full root function can be split into two single well-defined functions and the only difficulty may lie in the necessity to invent a notation for distinguishing between those two. In our case it is convenient to write
\begin{eqnarray}
&  &f_{+}:\mathbb{R}^{+}\rightarrow\mathbb{R}^{+}, \ x\mapsto+\sqrt{x} \ , \\
&  & f_{-}:\mathbb{R}^{+}\rightarrow\mathbb{R}^{-}, \ x\mapsto-\sqrt{x} \ .
\end{eqnarray}
One can now perform all calculations and mathematical manipulations with respect to the rich analytic structure of the full function and make it single-valued right at the moment when it is needed to be. In physical practice, for example when solving quadratic equations, negative results are often arising as contributions from the negative branch of the root (but are not desired, e.g. because of boundary conditions). They can be wiped out by setting $\pm\sqrt{x}\rightarrow+\sqrt{x}$. The root is then purely single-valued.

\begin{figure}[h]
\begin{center}
\includegraphics[scale=0.55]{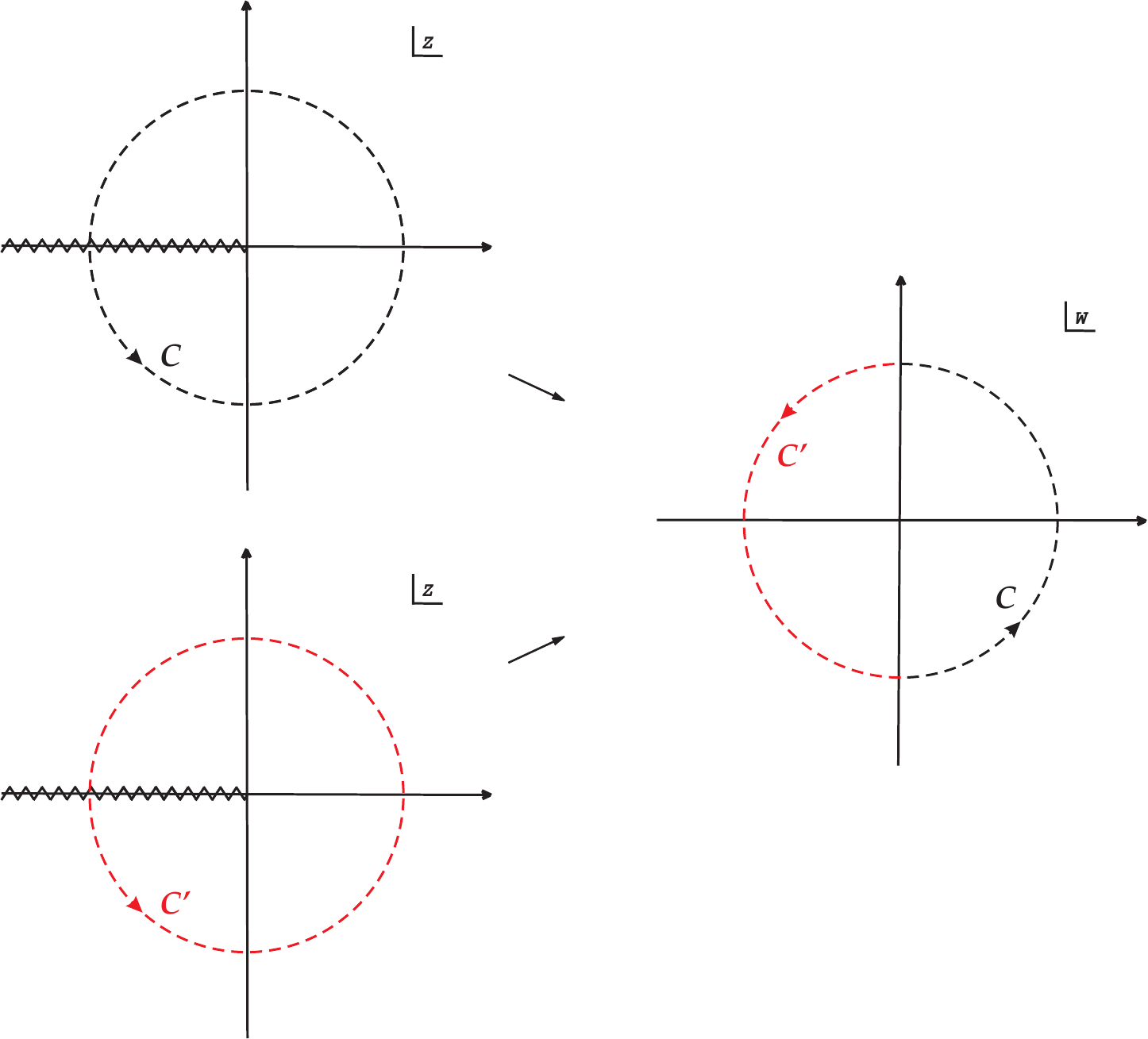}
\caption{Multi-valued character of $f(z)=\sqrt{z}$ with paths $\mathcal{C}$ (black, dashed) and $\mathcal{C}^{\prime}$ (red, dashed) in the complex $z$- and $w$-planes.}
\label{figure_rootzwplane}
\end{center}
\end{figure}

All this is quite intuitive by considering only real numbers. If we now extend our domains to complex numbers, a new kind of multi-valued character of functions comes into play. To begin with, let us study the complex root function
\begin{equation}
f:\mathbb{C}\rightarrow\mathbb{C}, \ z\mapsto+\sqrt{z}=\sqrt{z}=w \ . \label{equation_complexrootdefinition}
\end{equation}
We will encode points in the complex plane by using polar coordinates $z=\rho e^{i\phi}$. The root function then can be written as
\begin{equation}
f(z) = \sqrt{z} = \sqrt{\rho} e^{i\frac{\phi}{2}} \ , \ \ \ \phi\in(-\pi,\pi] \ .
\end{equation}
While there is no problem to give the root of a negative real number, the function $f$ as denoted in the last equation is not single-valued and therefore not well-defined for all $z\in\mathbb{C}$. We can convince ourselves of this fact by looking at the complex $z$- and $w$-planes (see Fig. \ref{figure_rootzwplane}). Imagine we take a path $\mathcal{C}$ in the $z$-plane starting from the point $z=-\rho$ and walk counterclockwise back to that point. In the $w$-plane this would correspond to a path starting from $-i\sqrt{\rho}$ and ending at $i\sqrt{\rho}$, which is a semicircle on the right side of the imaginary axis. So, coming back to the starting point in the $z$-plane does not give us the same value for $\sqrt{z}=w$. The complex root function is obviously not single-valued: we must take the same path as before in the $z$-plane, denoted as $\mathcal{C}^{\prime}$, in order to arrive at the same point in the $w$-plane.
In addition to this geometric view, we can investigate the behaviour of $f$ by approaching the negative real axis from the upper and lower side of the complex $z$-plane:
\begin{eqnarray}
\lim_{\epsilon \to 0^{+}}f(-\rho+i\epsilon) & = & \sqrt{\rho}e^{i\frac{\pi}{2}} \nonumber \\
& = & i\sqrt{\rho} \ , \label{equation_rootdisc1} \\
\lim_{\epsilon \to 0^{+}}f(-\rho-i\epsilon) & = & \sqrt{\rho}e^{-i\frac{\pi}{2}} \nonumber \\
& = & -i\sqrt{\rho} \ . \label{equation_rootdisc2}
\end{eqnarray}
By coming from different directions the limits of the root function on the negative real axis are in fact not the same. It should be clear now that there is something strange going on in this region and one may ask how we could manipulate $f$ to get rid of its multi-valued character. Usually, this is done by taking a 'mathematical scissor\grq \ and making a cut along the negative real axis of the $z$-plane, starting from complex infinity on the left and proceeding down to the origin. It is important to understand the cut only separating the upper and lower half planes on the left -- in our definition of the root function all the points on the respective axis are placed above the cut. If we now take the same paths $\mathcal{C}$ and $\mathcal{C}^{\prime}$ in this new sliced $z$-plane, there will be no problem by approaching the negative real axis from above because there is an edge that cannot be overcome. With this procedure we can rewrite our definition (\ref{equation_complexrootdefinition}) in the following way:
\begin{equation}
f:\mathbb{C}\rightarrow\mathbb{C}\backslash\{w\in\mathbb{C}:\operatorname{Re}w<0\vee(\operatorname{Re}w=0\wedge \operatorname{Im}w<0)\}, \ z\mapsto+\sqrt{z}=\sqrt{z}=w \ .
\end{equation}
This is called the {\em principal branch} of the complex root function.

\section{Branch points and branch cuts}
The example just illustrated is not as special as it might seem. There exists a lot of common functions which have a similar analytic structure as the complex root, or an even more complicated one. For another famous example let us look at the complex logarithm
\begin{equation}
f:\mathbb{C}\rightarrow\mathbb{C}\backslash\{w\in\mathbb{C}:-\pi<\operatorname{Im}w\leq\pi\}, \ z\mapsto\ln{z}=w \ . \label{equation_complexlogarithm}
\end{equation}
This function has also a cut along the negative real axis (this time including the origin). We should remark as a general statement that the cut is not a unique entity but can be {\em any} arbitrary line (or line segment) connecting two so-called {\em branch points} \cite{pucker}. In the case of the root function and the logarithm the branch points are situated at $z=0$ and $z=\infty$, while the cut is taken to be a straight line connecting them. Nevertheless, we could also have chosen the positive real axis along which to be cutted along (or any other, even curved line) and in practice one should take the one that makes calculations as simple as possible.\footnote{In the mathematical literature and computational software, branch cuts are chosen by convention to give simple analytic properties. In this work we adopt the cut structure of multi-valued functions as is used by {\em Mathematica}.} In this manner all problems of considering multi-valued functions simply reduce to the study of their branch points.

It was shown in the previous section that by circling around the branch point at the origin the value of the root function is different from the value with which we have started (see Fig. \ref{figure_rootzwplane}). This can be taken as the definition of a branch point. There would have been no difference if we took any other contour path in the $z$-plane than a simple circle; as was already stated, only the branch points are the independent objects which cannot be moved to another position. Consider the complex logarithm $f$ from Eq. (\ref{equation_complexlogarithm}) in the whole complex plane now. By taking again a circle $z=\rho e^{i\phi}$ with $\phi=-\pi...\pi$ around the branch point at $z=0$, we find:
\begin{eqnarray}
\lim_{\epsilon \to 0^{+}}f(-\rho+i\epsilon) & = & \ln\rho e^{i\pi} \nonumber \\
& = & \ln\rho+i\pi \nonumber \\
& = & \ln\rho-i\pi+i2\pi \ , \\ \label{equation_logsecond}
\lim_{\epsilon \to 0^{+}}f(-\rho-i\epsilon) & = & \ln\rho e^{-i\pi} \nonumber \\
& = & \ln\rho-i\pi \ .
\end{eqnarray}
In contrast to the root function, the multi-valued character of the logarithm appears as an additional imaginary part of $2\pi$. It can also be shown very easily that it is not possible to come back to the value of $\ln z$ at the starting point $z=-\rho-i\epsilon$ by crossing the negative real axis, since every additional circle around the branch point adds another $2\pi$. As a consequence, the cut leads to different {\em branches} of the logarithm, exactly as in the case of the root function (where only two branches exist). So it is natural to call such objects {\em branch cuts} from now on. The case without imaginary shift is taken to be the principal branch of the logarithm and one can define the $k$th branch by
\begin{equation}
\ln(z;k) = \ln|z|+i(\arg z+2k\pi) \ , \ \ \ k\in\mathbb{Z} \ , \label{equation_logkth}
\end{equation}
where the domain for every single branch is assigned in the same way as was given in the naive definition (\ref{equation_complexlogarithm}).

\begin{figure}[h]
\begin{center}
\includegraphics[width=380pt]{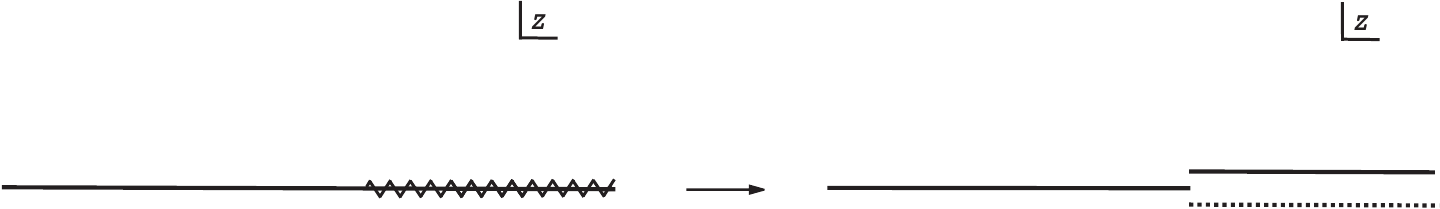}
\caption{Understanding the branch cut: In this picture the points that are on the real axis are placed above the cut (indicated by a solid line), so to be situated right above the axis. Branch cuts of a specific function can be also defined the other way around.}
\label{figure_cutregion}
\end{center}
\end{figure}

We motivated branch points and branch cuts in order to obtain a single-valued complex function out from a multi-valued one. The key point consisted in the 'exclusion\grq \ of the cut region from the domain of the function in such a way that the cut acts like an edge in the complex plane. Note that there is no exclusion at all, see also Fig. \ref{figure_cutregion} for a better understanding of this subtle but important detail. A huge amount of literature about mathematics for physicists is dealing with this kind of description of branch cuts.
As will be seen in the next section, this is not enough to describe physics of scalar resonances, because the upcoming relevant functions to be analyzed in this field (especially the Feynman propagator of interacting quantum field theories) will also have branch cuts in the complex energy plane. In order to understand the physical behaviour of resonances we need to get a deeper insight of what is going on at the cut. In fact, we need to go {\em beyond} it.

\section{Introducing Riemann sheets}
\subsection{Motivation and definition}
\begin{figure}[h]
\begin{center}
\includegraphics[scale=0.9]{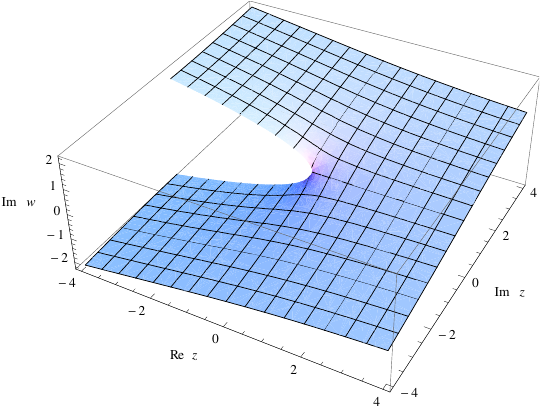}
\caption{Discontinuity of the root function $f(z)=\sqrt{z}=w$.}
\label{figure_rootwall}
\end{center}
\end{figure}

We have seen how a multi-valued complex function can be made well-defined by cutting the complex $z$-plane along a line (or line segment) that effectively behaves like an edge for every path taken from one side to the other. Within this framework it is not possible to cross the branch cut -- so we could in principle grab both separate layers of the complex plane and move them in the direction of the imaginary axis to visualize the {\em discontinuity} of the function $f$ in the $w$-plane. In case of the root function, for someone coming from below the cut this deformation would look like a wall, while it would look like an abyss for someone standing on the other side (see Fig. \ref{figure_rootwall}). By approaching the branch cut the difference of $w$ between both sides is only in the imaginary part.

We now want to abandon our above handling of branch cuts and introduce not only another, but also a more general interpretation of multi-valued complex functions. In fact, we already have worked out all necessary aspects of this new way of thinking on the last pages, we just need to merge them differently from how we did before. Remembering Fig. \ref{figure_rootzwplane}, we notice that the two sketched $z$-planes are actually the same, although the value of $w$ in both planes is not. It seems that the problem of obtaining a single-valued function is only a question of choosing one of those $z$-planes. This was also suggested when we defined the $k$th branch of the complex logarithm: every single value of $k$ corresponds to one specific $z$-plane where the branch lives in. The same could have been done in the case of the root function, yielding only two distinct complex planes. So, let us exploit the idea of having really two $z$-planes. Our complex function $f$ is then a mapping from those two planes onto a single $w$-plane and the distinction between the two planes is made by giving the $k$-value.
It is fully natural to take this value as an additional coordinate -- by giving all three coordinates, namely the real- and imaginary part of a given point in the complex $z$-plane and the $k$-value of that plane, we can assign exactly one value $w$ to every given point $z$. The function appears to be single-valued.
\begin{figure}[h]
\begin{center}
\includegraphics[width=360pt]{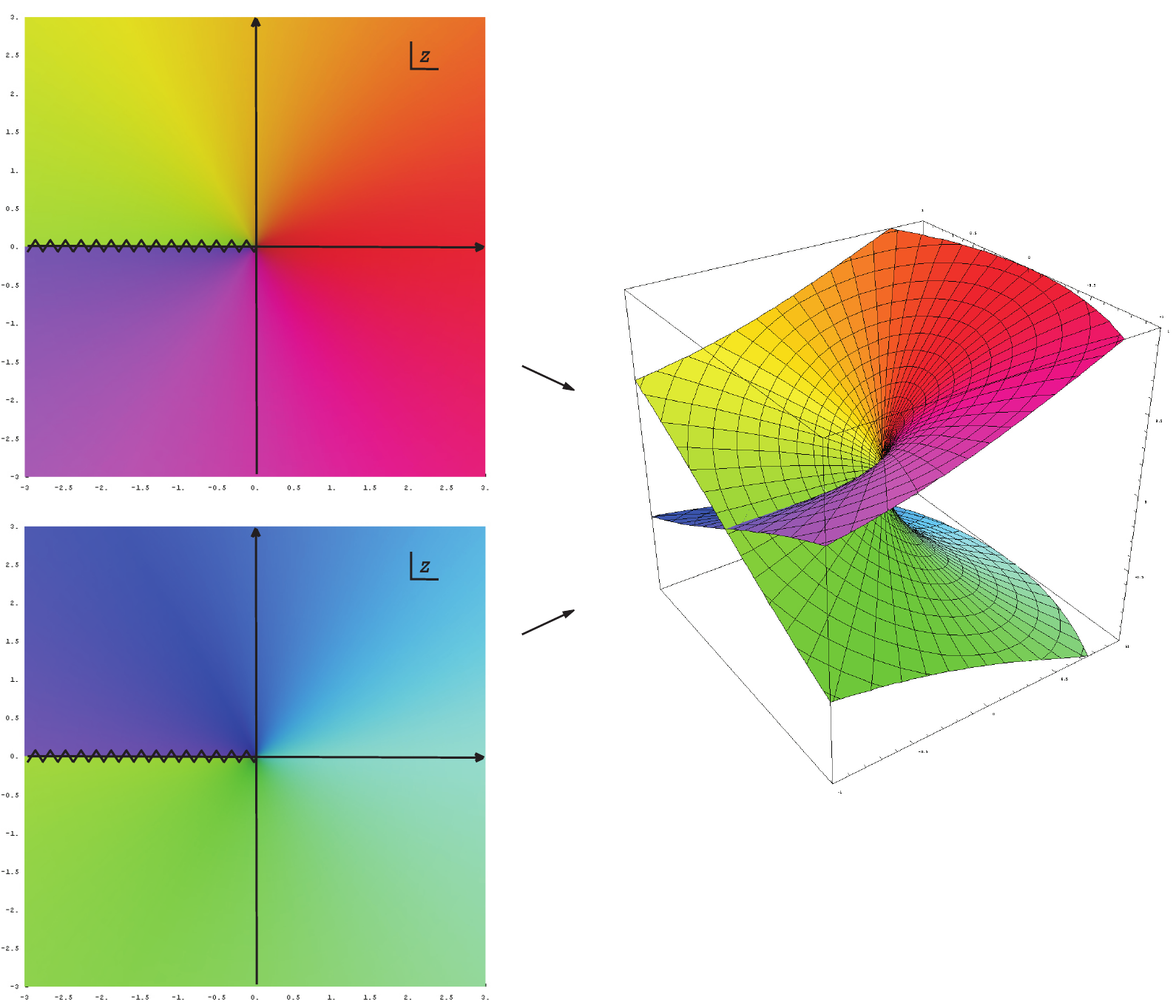}
\caption{Riemann surface of the complex root function. Each complex value $w$ is represented as a particular color: the $\arg$ of the complex number is encoded as the hue of the color, the modulus as its saturation (the colored background graphics on the left as well as the figure on the right were created by \href{http://livingmatter.physics.upenn.edu/node/38}{Jan Homann} from the University of Pennsylvania).}
\label{figure_rootsurface}
\end{center}
\end{figure}

But what has happened to the branch cut? In our new interpretation the cut can be understood as the connection between the two different $z$-planes because now there is no discontinuity at all when slipping from one plane to the other. We can attach both planes along the cut such that circling around the origin will make us leave the first plane after a full polar angle of $2\pi$, walking on the second plane and letting us arrive at the starting point in the first plane after a (global) turn of $4\pi$. The structure of the two combined $z$-planes appears as a very simple closed spiral stairway. To make this aspect clearer, we visualize the structure in Fig. \ref{figure_rootsurface} by using a color encoding scheme for complex numbers.

In this new framework any crossing of the branch cut is not special anymore since the function considered passes continuously from the first plane to the second one or vice versa from the second to the first one, depending on the path taken. We call every individual $z$-plane a {\em Riemann sheet} of the {\em Riemann surface}. For the root function the latter is the mentioned closed spiral starway structure shown in Fig. \ref{figure_rootsurface}. As a result, a new definition of a single-valued complex function $f$ is needed here:
\begin{equation}
f:X\rightarrow\mathbb{C}, \ f(z)=w \ ,
\end{equation}
where $X$ is a Riemann surface, a one-dimensional complex manifold. We do not want to get into the mathematical details of such manifolds, but shall mention the most important theorems required in this work. For a deeper mathematical introduction see for example the standard textbook by Forster \cite{forster}, whereas a very good presentation can be found in Ref.  \cite{lepage} covering a wide range of the practical aspects for calculations. For just a quick but adequate look we highly recommend Ref. \cite{nearing}. First of all, it can be proven that Stokes' theorem is still valid on a Riemann surface and this leads directly to a generalization of the residue theorem. Contour integration along a path, however, needs to be defined by using differential forms, but this will not affect us here. Apart from that, the very important {\em identity theorem} also holds true and therefore we can apply analytic continuation not only from the real to the complex, but also from one Riemann sheet to the other. Since this technique will be crucial, let us review the basic ideas and apply them to Riemann surfaces.

\subsection{Analytic continuation}
Suppose we have two holomorphic functions $f_{1}$ and $f_{2}$, defined each on domains $\Omega_{1},\Omega_{2}\subset\mathbb{C}$ such that both domains have a non-empty intersection $\Omega=\Omega_{1}\cap\Omega_{2}$ with $f_{1}(z)=f_{2}(z)$ for every $z\in\Omega$. Then both functions are analytic in their domains and can be expressed in terms of a power series,
\begin{equation}
f_{1}(z) = \sum_{n=0}^{\infty}a_{n}(z-z_{1})^{n} \ , \ \ \ \ \ f_{2}(z) = \sum_{n=0}^{\infty}b_{n}(z-z_{2})^{n} \ ,
\end{equation}
where $z_{1}\in\Omega_{1}$ and $z_{2}\in\Omega_{2}$. Either of the two expressions is certainly valid in the intersection region $\Omega$ and is consequently the series representation of one and the same analytic function $f$ around the points $z_{1}$ and $z_{2}$, respectively \cite{bronstein}. Thus, the function $f_{2}$ is called the analytic continuation of $f_{1}$ onto $\Omega_{2}$ and vice versa.

\begin{figure}[h]
\begin{center}
\includegraphics[scale=0.55]{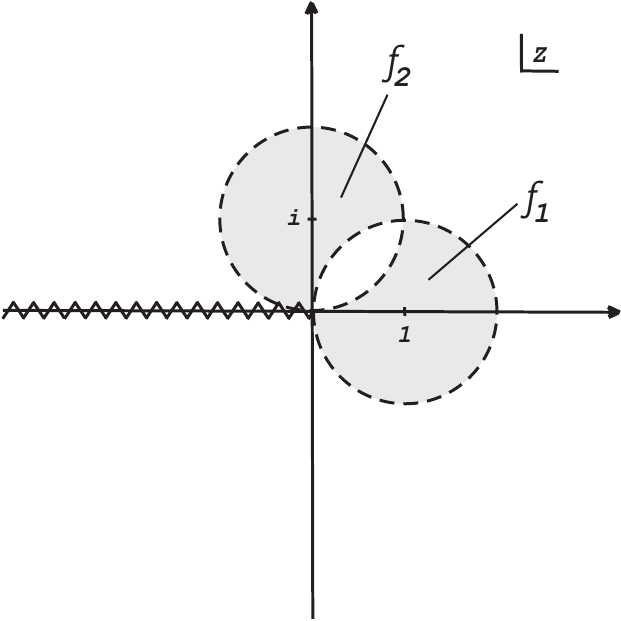}
\caption{Analytic continuation of the complex root function by expanding it in a power series in two different discs (gray) and realizing that both representations equal each other in the intersection region (white).}
\label{figure_rootanalytic}
\end{center}
\end{figure}

We can immediately apply this procedure to the complex root function $f(z)=\sqrt{z}$ around the point $z_{1}=1$ on an open disc with radius one. The latter limitation reflects the radius of convergence of the power series of $f$, so the root function seems to be only analytic on the disc. Analytic continuation now provides us with a powerful feature. We expand $f$ around the point $z_{2}=i$ in another open disc, but with the same radius as before, and hence there arises an overlapping region, see Fig. \ref{figure_rootanalytic}. The expanded function is again holomorphic in the second disc and, moreover, it is still the root function! In fact, we could take the point $z_{2}$ in such a way that it lies in the first disc and where a point $z=i$ is included in the new disc resulting from the series expansion around that $z_{2}$. Either way, we realize that both resulting functions agree in the intersection region and are consequently the same in both discs.

It is possible to extend the natural domain (i.e., a disc with finite radius of convergence) of a single series representation of a function without passing through any singularity by adopting the outlined method. For instance, if we add a third disc with center at $z_{3}=-1$ the root function is extended from the second disc to negative real numbers. Note that there is no branch cut preventing us to go to and beyond the negative real axis. Instead we would clearly rediscover its necessity when continuing further back to our starting point at $z=1$, because we would end up with the negative root function. This is a very non-trivial statement, though. Remember, the most important result of the previous subsection was the introduction of a general procedure to make a multi-valued complex function well-defined by introducing a Riemann surface. Usually, physical and mathematical literature provides us only with complex functions or (in)definite integrals on the first Riemann sheet. We are finally able to pass through the branch cut by applying analytic continuation down into the second sheet where the analytic structure can be very different than in the first one. This will be shown in the next chapters when we will consider decay processes in quantum mechanics and quantum field theory. In order to study these processes, we need to know the explicit form of the propagator on the second Riemann sheet. However, we have already seen two examples, the root function and the logarithm, where the function values are known on the complete Riemann surface. For the first one it is
\begin{eqnarray}
f(z) & = & \sqrt{z} \ , \\
f_{\text{\Romannum{2}}}(z) & = & -f(z) \ .
\end{eqnarray}
We have denoted the second Riemann sheet with a roman number. The function on the first sheet will never have such a number in this work. We got the above result by approaching the branch cut on the negative real axis from two different directions, compare Eq. (\ref{equation_rootdisc1}) and (\ref{equation_rootdisc2}). The observed discontinuity is of course only the difference between two sheets and not an intrinsic property of the root function. To gain an appropriate continuation of $f$ across the cut into the second Riemann sheet, we first need to calculate that mentioned discontinuity at $z=-\rho$:
\begin{eqnarray}
\disc f(-\rho) & = & \lim_{\epsilon \to 0^{+}}\Big[f(-\rho+i\epsilon)-f(-\rho-i\epsilon)\Big] \nonumber \\
& = & i\sqrt{\rho}-(-i\sqrt{\rho}) \nonumber \\
& = & 2i\sqrt{\rho} \ .
\end{eqnarray}
The analytic continuation of the root function down into the second sheet is then performed by accepting the requirement that the value $f_{\text{\Romannum{2}}}(-\rho-i\epsilon)$ (i.e., the function just below the cut in the second sheet) equals the value $f(-\rho+i\epsilon)$ (i.e., the function just above the cut in the first sheet) along the whole negative real axis:
\begin{eqnarray}
\lim_{\epsilon \to 0^{+}}f_{\text{\Romannum{2}}}(-\rho-i\epsilon) & = & \lim_{\epsilon \to 0^{+}}f(-\rho+i\epsilon) \nonumber \\
& = & \lim_{\epsilon \to 0^{+}}f(-\rho-i\epsilon)+2i\sqrt{\rho} \nonumber \\
& = & -i\sqrt{\rho}+2i\sqrt{\rho} \nonumber \\
& = & i\sqrt{\rho} \ . \label{equation_rootsecondinfinitesimal}
\end{eqnarray}
The analytic extension of this result into the lower half plane is after all what we were looking for,
\begin{equation}
f_{\text{\Romannum{2}}}(z) = -\sqrt{z} \ , \label{equation_rootsecondsheet}
\end{equation}
and exactly the second branch of the complex root in the context of our first interpretation of multi-valued functions. We should also notice the very useful identity between the imaginary part of $f$, taken right above the real axis, and its discontinuity:
\begin{equation}
\disc f(-\rho) = 2i\lim_{\epsilon \to 0^{+}}\operatorname{Im}f(-\rho+i\epsilon) \ . \label{equation_usefullidentity}
\end{equation}
One may ask immediately why the extension (\ref{equation_rootsecondsheet}) of the purely imaginary result from Eq. (\ref{equation_rootsecondinfinitesimal}) is so simple. In fact, we should be more precise in this point: The {\em identity theorem} states that if two given functions $f$ and $g$, both holomorphic on a domain $\Omega$, equal each other on some line segment $U$ lying in $\Omega$, then they equal each other on the whole domain $\Omega$\cite{fischerkaul}. Here, we see at first glance what is really the difference between real and complex analysis: a holomorphic function defined on $\Omega$ is completely determined by its values on a line segment $U\subset\Omega$. In this manner, there is no problem of continuing Eq. (\ref{equation_rootsecondinfinitesimal}) into the complex plane.

For completeness, we quickly calculate the complex logarithm in the second Riemann sheet and we should of course obtain the first branch as was given in Eq. (\ref{equation_logsecond}) and in (\ref{equation_logkth}) for $k=1$, respectively:\footnote{The Riemann surface of the complex logarithm is constructed by attaching the single complex $z$-planes in a slightly different way than in the case of the root function. It can be visualized by an infinite spiral stairway.}
\begin{eqnarray}
\disc f(-\rho) & = & \lim_{\epsilon \to 0^{+}}\Big[f(-\rho+i\epsilon)-f(-\rho-i\epsilon)\Big] \nonumber \\
& = & \ln\rho+i\pi-(\ln\rho-i\pi) \nonumber \\
& = & 2\pi i \ , \\
\Rightarrow \ \ \ \lim_{\epsilon \to 0^{+}}f_{\text{\Romannum{2}}}(-\rho-i\epsilon) & = & \lim_{\epsilon \to 0^{+}}f(-\rho+i\epsilon) \nonumber \\
& = & \lim_{\epsilon \to 0^{+}}f(-\rho-i\epsilon)+2\pi i \ ,
\end{eqnarray}
and so we indeed find for every $z\in\mathbb{C}\backslash\{z\in\mathbb{R}:z\leq0\}$:
\begin{equation}
f_{\text{\Romannum{2}}}(z) = \ln z+2\pi i \ .
\end{equation}

\section{Dispersion relations} \label{section_dispersion}
By using branch cuts instead of Riemann sheets one can perform contour integration in the complex plane without any problems. This is clear: in that case we simply restrict ourselves to a single (sliced) complex plane and the only new thing to deal with is the integration along the cut. In the following section we shall first derive the useful identity (\ref{equation_usefullidentity}) between a multi-valued complex function and its discontinuity at the branch cut, then we will make use of this relation to obtain the root function only from its cut structure. This can be done by performing a single integration with the help of Cauchy's integral formula. All this will be important in the next chapters to varify that every calculated expression for the propagator indeed is the right one.

\begin{figure}[h]
\begin{center}
\includegraphics[width=381pt]{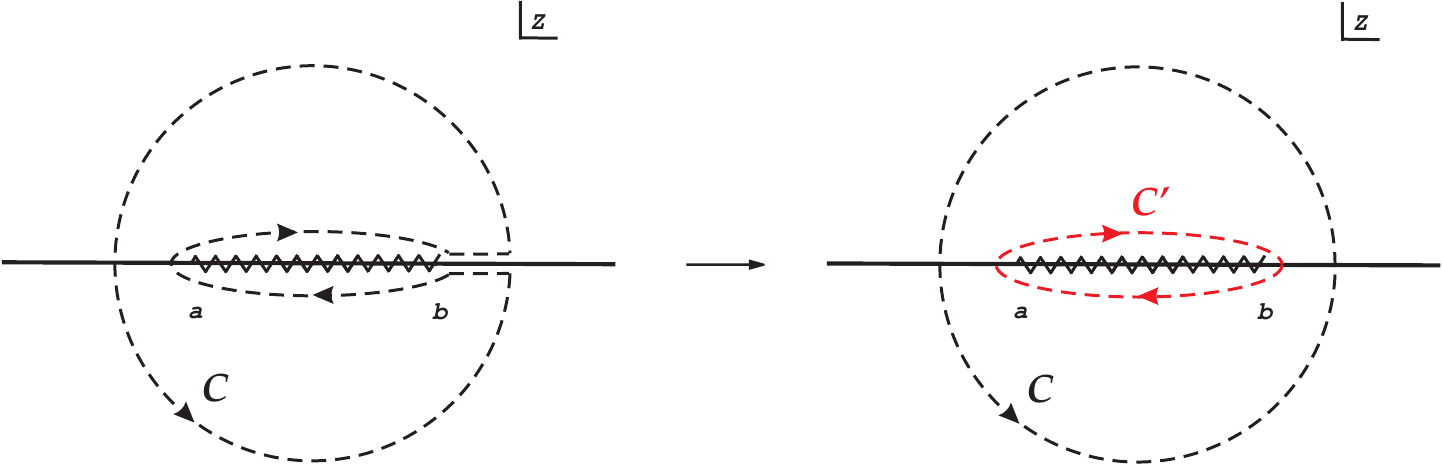}
\caption{Contour path of integration $\mathcal{C}^{\prime}$ (red, dashed) enclosing the finite branch cut when blowing up the contour $\mathcal{C}$ (black, dashed).}
\label{figure_cauchy}
\end{center}
\end{figure}
The most important complex functions in this work, like the mentioned self-energy as well as the scalar propagator, have in common the powerful property
\begin{equation}
f(z) = f^{*}(z^{*}) \ , \label{equation_hermanalytic}
\end{equation}
which leads to the useful identity from Eq. (\ref{equation_usefullidentity}), since
\begin{eqnarray}
\lim_{\epsilon \to 0^{+}}\Big[f(x+i\epsilon)-f(x-i\epsilon)\Big] & = & \lim_{\epsilon \to 0^{+}}\Big[f(x+i\epsilon)-f^{*}(x+i\epsilon)\Big] \nonumber \\
& = & 2i\lim_{\epsilon \to 0^{+}}\operatorname{Im}f(x+i\epsilon) \ .
\end{eqnarray}
From this we also obtain that the function $f$ is either purely real on the real axis or has a branch cut with the discontinuity just calculated \cite{pucker}. We now assume the function to be holomorphic except at the cut and to vanish faster than $\mathcal{O}(1/|z|^{\eta})$ (with $\eta>0$). If the cut has finite length then due to Cauchy's integral formula the function can be expressed as the sum of two integrals, namely
\begin{equation}
f(z) = \frac{1}{2\pi i}\left(\oint_{\mathcal{C}}\text{d}\xi \ \frac{f(\xi)}{\xi-z}+\oint_{\mathcal{C}^{\prime}}\text{d}\xi \ \frac{f(\xi)}{\xi-z}\right) \ ,
\end{equation}
where the contour paths of integration are taken as illustrated in Fig. \ref{figure_cauchy}. This is true when we blow up the contour, because in this case the first integral does not contribute at all. If we additionally shrink the contour $\mathcal{C}^{\prime}$ in such a way that only the paths below and above the real axis survive, the function can be represented in the limit of $\epsilon\rightarrow0^{+}$ by its imaginary part right above the branch cut:
\begin{eqnarray}
f(z) & = & \frac{1}{2\pi i}\lim_{\epsilon \to 0^{+}}\int_{a}^{b}\text{d}x\left(\frac{f(x+i\epsilon)}{x+i\epsilon-z}-\frac{f(x-i\epsilon)}{x-i\epsilon-z}\right) \nonumber \\
& = & \frac{1}{2\pi i}\int_{a}^{b}\text{d}x \ \frac{2i\lim_{\epsilon \to 0^{+}}\operatorname{Im}f(x+i\epsilon)}{x-z} \nonumber \\
& = & \frac{1}{\pi}\int_{a}^{b}\text{d}x \ \frac{\lim_{\epsilon \to 0^{+}}-\operatorname{Im}f(x+i\epsilon)}{z-x} \ , \label{equation_firstchapterdispersionintegral}
\end{eqnarray}
where $a$ and $b$ mark the branch points on the real axis. The integral representation for the Feynman propagator follows directly from this calculation. There is no difficulty at all to implement additional simple poles of $f$ on the real axis, since they can be taken into account by deforming the integration contour to make use of the residue theorem. Physically, the poles will appear as free propagators with some renormalization constants.

So, by restricting the integration to be performed only in the first Riemann sheet we need to be aware of taking the correct limit of the function $f$, depending on which side of the cut the contour path actually lies. Suppose we {\em know} that there is a branch cut on a part of the real axis and there shall exist an analytic expression for the discontinuity across this cut. We now ask for the complex function $f$ with exactly this discontinuity and which also fullfils the relation (\ref{equation_hermanalytic}). From our above considerations it is obvious that the whole function is determined only by the discontinuity and can be calculated by evaluating the dispersion integral (\ref{equation_firstchapterdispersionintegral}). As an instructive example we take again  the complex root function $f(z)=\sqrt{z}$. Since it does not even decrease for $|z|\rightarrow\infty$, we need to modify the dispersion integral by using a slightly changed function:
\begin{equation}
g(z) = \frac{\sqrt{z}}{z} \ .
\end{equation}
This new function has the same branch cut structure as $f$ (note that there is no simple pole at $z=0$). The discontinuity of the pure root function at the cut was found in the last section to be simply two times itself. For the new function $g$ this means
\begin{equation}
\disc g(-\rho) = -2i\frac{\sqrt{\rho}}{\rho} \ .
\end{equation}
The dispersion integral can then be computed by using a {\em Hankel contour} path of integration $\mathcal{C}$ with left open end:
\begin{eqnarray}
g(z) & = & \frac{1}{2\pi i}\oint_{\mathcal{C}}\text{d}\xi \ \frac{g(\xi)}{\xi-z} \nonumber \\
& = & \frac{1}{\pi}\int_{-\infty}^{0}\text{d}\rho' \ \frac{-\sqrt{-\rho'}}{\rho'(z-\rho')} \nonumber \\
& = & \frac{1}{\pi}\int_{0}^{\infty}\text{d}\rho \ \frac{\sqrt{\rho}}{\rho(z+\rho)} \ ,
\end{eqnarray}
where in the last step the variable transformation $\rho'\rightarrow-\rho$ has been introduced. Finally, the function $f$ can be denoted as
\begin{eqnarray}
f(z) & = & \frac{z}{\pi}\int_{0}^{\infty}\text{d}\rho \ \frac{\sqrt{\rho}}{\rho(z+\rho)} \nonumber \\
& \stackrel{\rho\rightarrow x^2}{=} & \frac{2z}{\pi}\int_{0}^{\infty}\text{d}x \ \frac{1}{z+x^{2}} \nonumber \\
& = & \frac{\sqrt{z}}{\pi}\int_{0}^{\infty}\text{d}x \ \left(\frac{1}{\sqrt{z}+ix}+\frac{1}{\sqrt{z}-ix}\right) \nonumber \\
& = & \frac{\sqrt{z}}{\pi}\pi \ , \ \ \text{for $\operatorname{Im}z\neq0\vee\operatorname{Re}z\geq0$} \nonumber \\
& = & \sqrt{z} \ .
\end{eqnarray}

 \label{chapter_chapter2}
\clearpage

\chapter{Quantum theory of unstable particles}

\medskip

\section{Non-relativistic Lee model}
\subsection{Lee Hamiltonian}
In 1954, Lee\footnote{In 1957, T. D. Lee and C. N. Yang were honoured with the {\em Nobel Prize in Physics} for their work on the violation of the parity conservation law in weak interaction, which C. S. Wu verified experimentally. Lee was just 31 years old.} worked out a general technique to handle, with solvable interacting quantum field theoretical models, the study of the renormalization problem \cite{lee}. In this context he introduced a so-called {\em Lee Hamiltonian} which can be used to deal with decay processes in quantum mechanics. To this end, we need to couple an initial state\footnote{As was pointed out by Fonda, Ghirardi and Rimini \cite{ghirardi}, it is a controversial problem to assign a state vector to an unstable system, so we will call it simply 'initial\grq \ state without further commenting on this point.} $|S\rangle$ to a continuum of two-particle states $|\textbf{k}\rangle$. In the following, we assume the decay process to be a transition of the form $|S\rangle\rightarrow|\textbf{k}\rangle$ where the unstable particle $S$ decays into two particles with momenta $\textbf{k}$ and $-\textbf{k}$. Following Facchi and Pascazio \cite{facchi}, we decompose the full Hilbert space $\mathcal{H}$ of all normalized states of the system in such a way that the state $|S\rangle$ lives in a subspace $\mathcal{H}_{S}$:
\begin{equation}
\mathcal{H} = \mathcal{H}_{S}\oplus\mathcal{H}_{\textbf{k}} \ . \label{equation_decomposition}
\end{equation}
In order to be more precise, we can write explicitly the set of states living in each subspace:
\begin{eqnarray}
\mathcal{H}_{S} & = & \{|S\rangle\} \ , \\
\mathcal{H}_{\textbf{k}} & = & \text{span}\big{(}\{|\textbf{k}\rangle\}\big{)} \ .
\end{eqnarray}
For the moment, the whole system is to be considered in a finite volume $V=L^{3}$ and as a consequence of applying periodic boundary conditions the momenta $\textbf{k}$ have the form $\textbf{k} = \frac{(2\pi)^{3}}{V}\textbf{n}$ with $\textbf{n}=(n_{1},n_{2},n_{3})^{T}$ and $n_{i}\in\mathbb{Z}, \ i=1,2,3$. The full Hamiltonian $\hat H$ shall be a sum of a free and an interaction part, namely
\begin{equation}
\hat H = \hat H_{0}+\hat H_{\text{int}} \ ,
\end{equation}
and we assume the states $\{|\textbf{k}\rangle\}$ to be the eigenbasis of $\hat H_{0}$ in the subspace $\mathcal{H}_{\textbf{k}}$ with eigenvalues $\omega(\textbf{k})$, while $|S\rangle$ shall be the eigenstate in $\mathcal{H}_{S}$ with eigenvalue $\omega_{S}$. It is obvious that because of our decomposition (\ref{equation_decomposition}) of the Hilbert space the states belonging to either subspace are orthonormal to each other. Hence, we can define two operators $\hat P_{S}$ and $\hat P_{\textbf{k}}$ that project arbitrary states of $\mathcal{H}$ onto $\mathcal{H}_{S}$ and $\mathcal{H}_{\textbf{k}}$:
\begin{equation}
\hat P_{S} = |S\rangle\langle S| \ , \ \ \ \ \ \hat P_{\textbf{k}} = \mathbbm{1}-|S\rangle\langle S| \ .
\end{equation}
Although we do not know the eigenbasis of the full Hamiltonian we nevertheless can work with the eigenbasis of the free part and write down a completeness relation
\begin{equation}
\mathbbm{1} = |S\rangle\langle S|+\sum_{\textbf{k}}|\textbf{k}\rangle\langle\textbf{k}| \ . \label{equation_completeness}
\end{equation}

The Lee Hamiltonian can now be constructed for the purpose to deal with our decay transition problem. The interaction part will mediate between states of the two subspaces $\mathcal{H}_{S}$ and $\mathcal{H}_{\textbf{k}}$, respectively, while the free part of course must deal with states from both:
\begin{eqnarray}
\hat H_{0} & = & \omega_{S}|S\rangle\langle S|+\sum_{\textbf{k}}\omega(\textbf{k})|\textbf{k}\rangle\langle\textbf{k}| \ , \\ \label{equation_Hfree}
\hat H_{\text{int}} & = & \sum_{\textbf{k}}g\frac{f(\textbf{k})}{\sqrt{V}}\Big(|\textbf{k}\rangle\langle S|+|S\rangle\langle\textbf{k}|\Big) \ . \label{equation_Hint}
\end{eqnarray}
These expressions arise from
\begin{eqnarray}
\hat H_{0} & = & \hat P_{S}\hat H\hat P_{S}+\hat P_{\textbf{k}}\hat H\hat P_{\textbf{k}} \ , \\
\hat H_{\text{int}} & = & \hat P_{S}\hat H\hat P_{\textbf{k}}+\hat P_{\textbf{k}}\hat H\hat P_{S} \ .
\end{eqnarray}
Thus, the interaction part operates {\em only} between the initial state $|S\rangle$ and the final states $|\textbf{k}\rangle$ (it is completely off-diagonal in the given basis), while the free part simply gives their eigenenergies and leaves a considered state untouched. Technically speaking, the main structure of our Hilbert space and the Hamiltonian is in full analogy with that of a two-state quantum system, commonly covered in an undergraduate quantum mechanics course. The only difference is that there is a set of final states $|\textbf{k}\rangle$ instead of just a single one and the interaction term to be time-independent, governed by the {\em form factor function} $f(\textbf{k})$ with explicit dependence on the momentum $\textbf{k}$ of the final state.

\subsection{Schr\"odinger propagator}
\begin{figure}[!h]
\begin{center}
\includegraphics[scale=0.55]{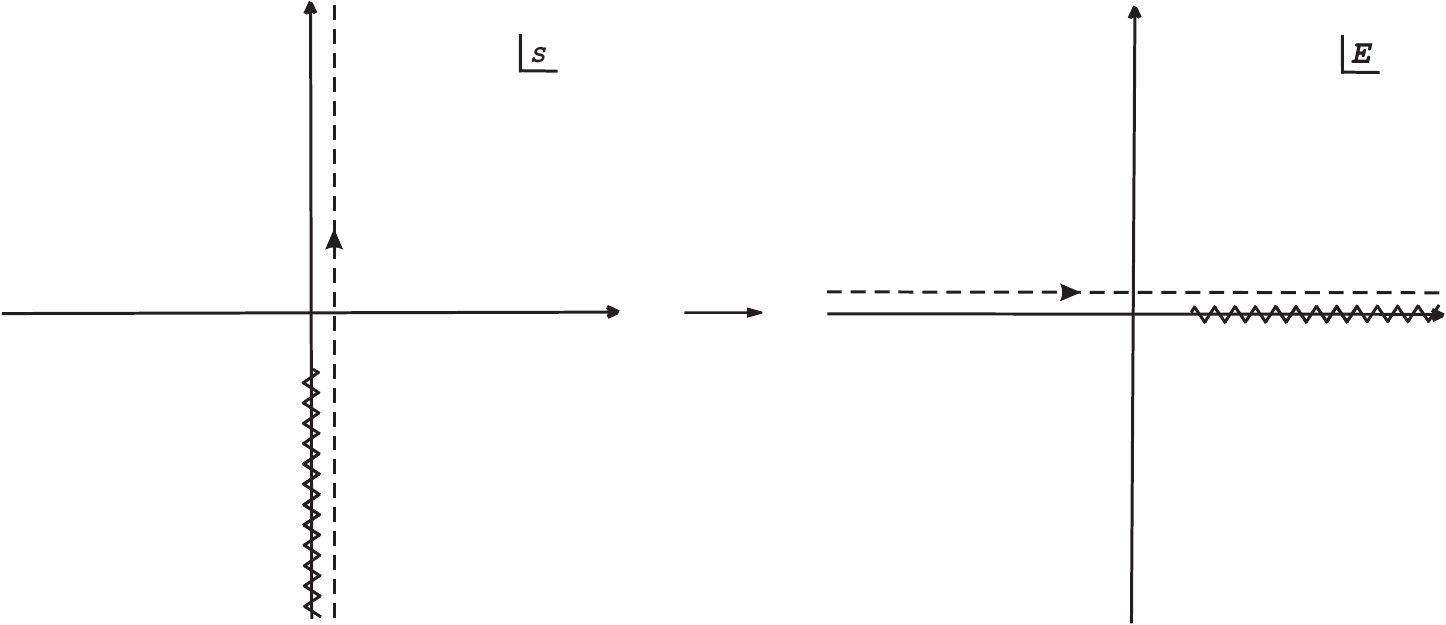}
\caption{Complex plane rotation $s\rightarrow-iE$ with branch cut and contour integration path displayed.}
\label{figure_sElaplace}
\end{center}
\end{figure}

As will be emphasized soon, the main starting point for our analysis in general is the propagator of a specific physical system, for now the non-relativistic Schr\"odinger propagator of the decaying particle $S$:
\begin{equation}
G_{S}(E) = \langle S|\frac{1}{E-\hat H+i\epsilon}|S\rangle \ . \label{equation_Spropagator}
\end{equation}
It should be mentioned that there may be some confusion by looking into traditional literature due to a subtle difference in terminology. In quantum mechanics, referring to the term 'propagator\grq \ in the context of Feynman's path integral formalism, we address the (probability) amplitude for a particle to travel from one spatial point at one time to another spatial point at a later time. One usually derives the propagator, often denoted as $K(x,t,x^{\prime},t^{\prime})$, and further the energy-dependent Green's function $G(x,x^{\prime},E)$, which is the Fourier transform of the propagator with respect to time.\footnote{A good introduction into path integrals in quantum mechanics can be found in Ref. \cite{shankar}. For a more detailed study one should consult classic literature on this topic, e.g. the textbook by Feynman and Hibbs \cite{feynmanhibbs}.} In this work, as done in most of the literature concerning (relativistic) decay processes, we will name both objects in the opposite way. In particular, $G_{S}(E)$ shall be the propagator for the particle $S$. To calculate the propagator we first take a look at the survival amplitude in the initial state $|S\rangle$,
\begin{equation}
a(t) = \langle S|\hat U(t)|S\rangle = \langle S|e^{-i\hat Ht}|S\rangle \ , \label{equation_SUS}
\end{equation}
where $\hat U(t)=e^{-i\hat Ht}$ is the time evolution operator. In Ref. \cite{moshinsky}, Moshinsky et al. demonstrated for time-independent interaction terms in one dimension that the Schr\"odinger propagator (the Green's function) in position space can be obtained from the above mentioned amplitude $K(x,t,x^{\prime},t^{\prime})$ by a Laplace transformation. In fact, this makes sense because the continuous Fourier transform is equivalent to evaluating the bilateral Laplace transform with imaginary arguments. Applying a similiar procedure to our problem, we take the Laplace transform of Eq. (\ref{equation_SUS}) and simply get the resolvent of $\hat U(t)$ by applying the geometric series:
\begin{eqnarray}
F(s) & = & \langle S|\int_{0}^{\infty}\text{d}t \ e^{-st}e^{-i\hat Ht}|S\rangle \nonumber \\
& = & \langle S|\int_{0}^{\infty}\text{d}t \ e^{-st}\sum_{n=0}^{\infty}\frac{(-i)^{n}}{n!}\hat H^{n}t^{n}|S\rangle \nonumber \\
& = & \langle S|\sum_{n=0}^{\infty}(-i)^{n}\hat H^{n}\int_{0}^{\infty}\text{d}t \ e^{-st}\frac{t^{n}}{n!}|S\rangle \nonumber \\
& = & \langle S|\sum_{n=0}^{\infty}(-i)^{n}\hat H^{n}\frac{1}{s^{n+1}}|S\rangle \nonumber \\
& = & \langle S|\frac{1}{s+i\hat H}|S\rangle \ \ \stackrel{s\rightarrow-iE}{=} \ \ \langle S|\frac{i}{E-\hat H}|S\rangle \ .
\end{eqnarray}
This in principle gives the expression in Eq. (\ref{equation_Spropagator}), yet we do not stop here, especially without commenting on the variable transformation in the last step. Let us first recover our survival amplitude by applying the inverse Laplace transformation:
\begin{eqnarray}
a(t) & = & \langle S|\frac{1}{2\pi i}\int_{\gamma-i\infty}^{\gamma+i\infty}\text{d}s \ e^{st}\frac{1}{s+i\hat H}|S\rangle \nonumber \\
& \stackrel{s\rightarrow-iE}{=} & \langle S|\frac{1}{2\pi i}\int_{i\gamma+\infty}^{i\gamma-\infty}\text{d}E \ e^{-iEt}\frac{1}{E-\hat H}|S\rangle \nonumber \\
& \stackrel{\gamma\rightarrow\epsilon}{=} & \frac{i}{2\pi}\int_{-\infty}^{+\infty}\text{d}E \ e^{-iEt}\langle S|\frac{1}{E-\hat H+i\epsilon}|S\rangle \ . \label{equation_Samplitude}
\end{eqnarray}
The integration in the complex $s$-plane has to be done along a vertical line with $\operatorname{Re}s=\gamma$ such that $\gamma$ is greater than the real part of all singularities of $F(s)$. More general, $\gamma$ is chosen in a way that the contour path of integration lies in the region of convergence of $F(s)$. In order to guarantee real values for the energy variable $E$, the propagator from Eq. (\ref{equation_Spropagator}) will have either simple poles or a branch cut (starting and ending at some branch points) on the imaginary axis in the $s$-plane, depending on the coupling $g$ in the interaction part (\ref{equation_Hint}) of the Hamiltonian. The integration path needs to be shifted by an infinitesimal number $\gamma=\epsilon>0$ to the right. The $s$-plane is then rotated by the transformation $s\rightarrow-iE$, see Fig. \ref{figure_sElaplace}. Thus, the propagator becomes analytic in the whole new physical $E$-plane, except at the simple pole positions and the branch cut, respectively. These last statements are however not obvious and so the full analytic structure of the propagator will be studied in the next subsection.

We are now able to determine the Schr\"odinger propagator explicitly. By following Giacosa \cite{giacosa}, one can rewrite the propagator using the decomposition $\hat H=\hat H_{0}+\hat H_{\text{int}}$ and performing a sequential insertion of completeness relations (\ref{equation_completeness}):
\begin{eqnarray}
G_{S}(E) & = & \langle S|\frac{1}{E-\hat H+i\epsilon}|S\rangle \nonumber \\
& = & \langle S|\frac{1}{E-\hat H_{0}+i\epsilon}\sum_{n=0}^{\infty}\bigg(\hat H_{\text{int}}\frac{1}{E-\hat H_{0}+i\epsilon}\bigg)^{n}|S\rangle \nonumber \\
& = & \langle S|\frac{1}{E-\hat H_{0}+i\epsilon}\bigg(1+\mathbbm{1}\hat H_{\text{\text{int}}}\mathbbm{1}\frac{1}{E-\hat H_{0}+i\epsilon}+\dots\bigg)|S\rangle \nonumber \\
& = & \langle S|\frac{1}{E-\hat H_{0}+i\epsilon}\bigg[1+\bigg(|S\rangle\langle S|+\sum_{\textbf{q}}|\textbf{q}\rangle\langle\textbf{q}|\bigg)\hat H_{\text{int}}\bigg(|S\rangle\langle S|+\sum_{\textbf{k}}|\textbf{k}\rangle\langle\textbf{k}|\bigg) \nonumber \\
& & \ \times \ \frac{1}{E-\hat H_{0}+i\epsilon}+\dots\bigg]|S\rangle \ .
\end{eqnarray}
We only have to evaluate matrix elements of the following form:
\begin{eqnarray}
\langle S|\frac{1}{E-\hat H_{0}+i\epsilon}|S\rangle & = & \frac{1}{E-\omega_{S}+i\epsilon} \ , \\
\langle \textbf{q}|\frac{1}{E-\hat H_{0}+i\epsilon}|\textbf{k}\rangle & = & \frac{\delta_{\textbf{qk}}}{E-\omega(\textbf{k})+i\epsilon} \ , \\
\langle S|\frac{1}{E-\hat H_{0}+i\epsilon}|\textbf{k}\rangle & = & 0 \ ,
\end{eqnarray}
and since the interaction part is completely off-diagonal we find
\begin{eqnarray}
\langle S|\hat H_{\text{int}}|S\rangle & = & \langle\textbf{q}|\hat H_{\text{int}}|\textbf{k}\rangle \ \ = \ \ 0 \ , \\
\langle S|\hat H_{\text{int}}|\textbf{k}\rangle & = & \langle \textbf{k}|\hat H_{\text{int}}|S\rangle^{*} \ = \ \ g\frac{f(\textbf{k})}{\sqrt{V}} \ .
\end{eqnarray}
All terms in the infinite sum over $n$ can be collected in a resummation after taking all the Kronecker deltas into account. We are left with:
\begin{eqnarray}
G_{S}(E) & = & \frac{1}{E-\omega_{S}+i\epsilon}\sum_{n=0}^{\infty}\bigg(\frac{1}{E-\omega_{S}+i\epsilon}\sum_{\textbf{k}}\frac{1}{V}\frac{g^{2}f^{2}(\textbf{k})}{E-\omega({\textbf{k}})+i\epsilon}\bigg)^{n} \nonumber \\
& = & \frac{1}{E-\omega_{S}+i\epsilon}\sum_{n=0}^{\infty}\bigg(\frac{-g^{2}\Sigma(E)}{E-\omega_{S}+i\epsilon}\bigg)^{n} \nonumber \\
& = & \frac{1}{E-\omega_{S}+g^{2}\Sigma(E)+i\epsilon} \ , \label{equation_leepropagator}
\end{eqnarray}
where in the last steps we have defined
\begin{equation}
\Sigma(E) = -\sum_{\textbf{k}}\frac{1}{V}\frac{f^{2}(\textbf{k})}{E-\omega({\textbf{k}})+i\epsilon} \ .
\end{equation}
Finally, performing the continuum limit we get
\begin{eqnarray}
\sum_{\textbf{k}} & \rightarrow & V\int\frac{\text{d}^{3}k}{(2\pi)^{3}} \ , \\
\Rightarrow \ \ \ \Sigma(E) & = & -\int\frac{\text{d}^{3}k}{(2\pi)^{3}}\frac{f^{2}(\textbf{k})}{E-\omega({\textbf{k}})+i\epsilon} \ . \label{equation_Leeselfenergy}
\end{eqnarray}
In the same way as was shown in section \ref{section_dispersion}, one immediately obtains an integral representation for the Schr\"odinger propagator by applying Cauchy's integral formula. This spectral representation can be also derived from a physical point of view in the context of many-body physics, similar as we did in the first chapter for relativistic quantum field theory (see for example Ref. \cite{simons}). In the end the propagator reads:
\begin{equation}
G_{S}(E) = \frac{1}{\pi}\int_{-\infty}^{\infty}\text{d}\omega \ \frac{\rho(\omega)}{E-\omega+i\epsilon} \ , \label{equation_GSspectral}
\end{equation}
with the spectral function
\begin{equation}
\rho(\omega) = -\operatorname{Im}G_{S}(\omega+i\epsilon) \ .
\end{equation}

\subsection{Analytic structure of the propagator}
We will now study a decay process within a specified Lee model. We restrict ourselves to final states $|\textbf{k}\rangle$ with eigenvalue $\omega(\textbf{k})\geq0$ with respect to the free part $\hat H_{0}$ of the full Hamiltonian $\hat H$. The decaying particle $S$ will be prepared in its rest frame with energy $\omega_{S}=M_{0}$ and the form factor function $f(\textbf{k})$ shall be a simple product of two  Heaviside step functions of the form
\begin{equation}
f(\textbf{k}) = \Theta(2k-E_{0})\Theta(\Lambda-2k) \ ,
\end{equation}
where $k=|\textbf{k}|$. The parameter $E_{0}$ is nothing else than the energy threshold; we thus assume $E_{0}<M_{0}<\Lambda$, where $\Lambda$ is the upper bound of the spectrum, acting as a cutoff (one could understand $\Theta(\Lambda-2k)$ as a cutoff-function). We chose the final energies as $\omega(\textbf{k})=2k$. Although this model represents an unrealistic simplification, the following section will provide us with the basic mathematical framework for our quantum field theoretical investigation.

\begin{figure}[!h]
\begin{center}
\includegraphics{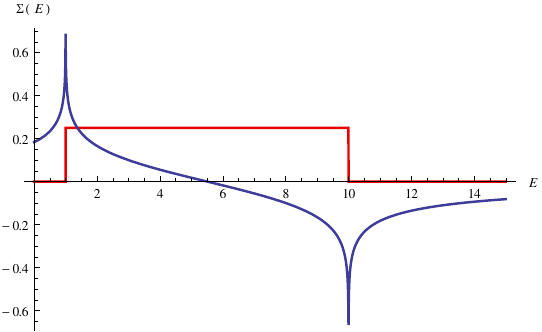}
\caption{Real (dark blue) and imaginary part (red) of the self-energy $\Sigma(E)$ on the real axix with the choice $\Lambda=10E_{0}, \ M_{0}=2E_{0}$ and $E_{0}=1$.}
\label{figure_nonrel1dSigma}
\end{center}
\end{figure}

As was discussed in section \ref{section_dispersion}, there exists a remarkable connection between the branch cut structure of some complex functions and dispersion relation integrals of the generic form
\begin{equation}
\int_{a}^{b}\text{d}x \ \frac{\rho(x)}{z-x} \ , \ \ \ z\in\mathbb{C} \ .
\end{equation}
We have also figured out in chapter 1 that whenever the (analytic) spectral function $\rho(x)$ gives support to the integral, there is a branch cut arising in this region on the real axis.\footnote{To be more precise: The function $\rho(x)$ gives us two branch points on the real axis. The branch cut can be any arbitrary line connecting those two points, but in order to keep our definition of the complex logarithm in Eq. (\ref{equation_complexlogarithm}) and (\ref{equation_logkth}) we have to draw the cut along the real axis.} Taking a look at the relation between the non-relativistic propagator $G_{S}(E)$ of the particle $S$ and the spectral function $\rho(\omega)$ in Eq. (\ref{equation_GSspectral}), we conclude that there is a branch cut if the spectral function is not a delta distribution function -- i.e., if the coupling $g$ in our Lee Hamiltonian is non-zero. Then the self-energy (\ref{equation_Leeselfenergy}) will contribute to the full propagator and influence the position of the mass-pole in the complex $E$-plane, where $E\rightarrow z=x+iy$. In one dimension, the self-energy takes the simple form
\begin{eqnarray}
\Sigma(E) & = & -\int\frac{\text{d}k}{2\pi}\frac{f^{2}(\textbf{k})}{E-\omega(\textbf{k})+i\epsilon} \nonumber \\
& = & -\frac{1}{2\pi}\int_{-\infty}^{\infty}\text{d}k \ \frac{\Theta(2k-E_{0})\Theta(\Lambda-2k)}{E-2k+i\epsilon} \nonumber \\
& \stackrel{2k\rightarrow k}{=} & \frac{1}{4\pi}\int_{E_{0}}^{\Lambda}\text{d}k \ \frac{1}{k-E-i\epsilon} \nonumber \\
& = & \frac{1}{4\pi}\Big[\ln(\Lambda-E-i\epsilon)-\ln(E_{0}-E-i\epsilon)\Big] \ . \label{equation_sigmaln}
\end{eqnarray}
There is a very important aspect here to point out: In the last step we actually performed the limit $\epsilon\rightarrow0^{+}$. From a conservative point of view we should make use of the Sokhotski--Plemelj theorem, according to which an additional discontinuity of $i\pi$ would come into play. However, this term does not appear in our expression. In the complex $E$-plane the self-energy has a branch cut on the positive real axis starting from $z=E_{0}$ and ending at $z=\Lambda$, so as long as $\operatorname{Im}z$ is finite, $\Sigma(z)$ is well-defined. Whereas, if we try to interpret the limit $\epsilon\rightarrow0^{+}$ as a falling of the pole $z=E+i\epsilon$ onto the real $k$-axis, we will have a problem with our formalism. But this problem is actually not a real one: except for the spectral function, we will {\em never} evaluate the self-energy at such points, since for $g\ne0$ the pole will be a complex number with non-zero imaginary part and for $g=0$ the self-energy will vanish in the propagator, see Eq. (\ref{equation_leepropagator}). The infinitesimal number $i\epsilon$ is only necessary if one strikes real $z=E$. All this makes the expression (\ref{equation_sigmaln}) to live on the first Riemann sheet. The $k$-integration should be taken with complex energies $z=x+iy$ right from the beginning:
\begin{equation}
\int_{E_{0}}^{\Lambda}\text{d}k \ \frac{1}{k-E-i\epsilon} \ \ \rightarrow \ \ \int_{E_{0}}^{\Lambda}\text{d}k \ \frac{1}{k-z} = \ln(\Lambda-z)-\ln(E_{0}-z) \ .
\end{equation}
The last result is the final one because common multiplication rules for the logarithm are in general lost for complex arguments. A plot of the self-energy for the parameters (\ref{equation_nonrelparameters}) can be found in Fig. \ref{figure_nonrel1dSigma}. In order to obtain the correct plot one needs to evaluate the function at $z=E+i\epsilon$, including explicitly a small imaginary number $i\epsilon$ (or use the Sokhotski--Plemelj theorem and calculate the principal value).

We have already indicated that the self-energy has a branch cut on part of the positive real axis. This follows directly from our general discussion in the second chapter, because we are dealing with a complex logarithmic function with branch points at $z=E_{0}$ and $z=\Lambda$. The point at complex infinity is not a branch point, which can be easily seen by taking a path $z=a+\rho e^{i\phi}$ with $\phi=0...2\pi$ that encircles both points $z=E_{0}$ and $z=\Lambda$ (so $a$ needs to be in between both points on the real axis and $\rho$ must be greater than the distance to the farthest one): after a single full turn of $2\pi$ we arrive again at the starting point. One may argue that both single logarithms in $\Sigma(z)$ provide us with two branch cuts, one starting at $z=E_{0}$ and the other at $z=\Lambda$, each heading to infinity along the positive real axis. But the discontinuities collected for $z>\Lambda$ cancel each other due to the minus sign. Note that this is the reason why the situation at the remaining branch cut is inverse to what we sketched in Fig. \ref{figure_cutregion}. This is also set up by the computational software {\em Mathematica}, though it will be shown in subsection \ref{subsection_nonrelcouplings50100} that the self-energy is actually continuous with the first quadrant (i.e., the real axis is situated right above the cut). Besides that, one can also investigate the behaviour of the point at infinity when making the variable transformation $z\rightarrow1/z$ and circling around the new origin. Here, too, a full circle does not give any change.

Let us concentrate on the Schr\"odinger propagator which can be constructed with the specified self-energy. For the simplest case $g=0$ we have a free particle with mass $\omega_{S}=M_{0}$ in its rest frame:
\begin{equation}
G_{S}(E) = \frac{1}{E-M_{0}+i\epsilon} \ .
\end{equation}
We apply here our definition of the (real) mass of a free particle as the pole of the propagator in the complex $E$-plane ($E\rightarrow z=x+iy$). The pole position of course gives the mass slightly shifted into the lower half plane to ensure convergence:
\begin{equation}
E_{\text{pole}} = M_{0}-i\epsilon \ .
\end{equation}
After turning on the coupling, $g>0$, there is a contribution from the (complex-valued) self-energy function to the pole position. One could rephrase this statement in the following way: The appearence of an additional function in the denominator of the propagator forces the pole to move away from the real axis down into the lower half plane, which makes the self-energy per se to be evaluated at complex values. The new pole position as the zero of the denominator is determined by a system of two equations:
\begin{eqnarray}
x-M_{0}+g^{2}\operatorname{Re}\Sigma(x+iy) & \stackrel{!}{=} & 0 \ , \nonumber \\
y+g^{2}\operatorname{Im}\Sigma(x+iy) & \stackrel{!}{=} & 0 \label{equation_systemofEq} \ .
\end{eqnarray}
We could decompose the self-energy for $E_0<x<\Lambda$ in the following way:
\begin{eqnarray}
\Sigma(z=x+iy) & = & \frac{1}{4\pi}\Big[\ln(\Lambda-x-iy)-\ln(E_{0}-x-iy)\Big] \nonumber \\
\nonumber \\
& = & \frac{1}{4\pi}\ln\Bigg(\frac{\sqrt{(\Lambda-x)^{2}+y^{2}}}{\sqrt{(E_{0}-x)^{2}+y^{2}}}\Bigg)+\frac{i}{4\pi}\bigg[\arctan\left(\frac{-y}{\Lambda-x}\right) \nonumber \\
&  & - \arctan\left(\frac{-y}{E_{0}-x}\right)\bigg] \ , \label{equation_sigmalncomplex}
\end{eqnarray}
where we have used general conversion rules for complex numbers with $\arg z=\phi\in(-\pi,\pi]$. But the system of equations, combined with the decomposition of the self-energy, cannot be solved with analytic methods. We have to find the solution numerically by performing a direct search for the pole of the propagator in the complex $E$-plane. The parameters are chosen to be
\begin{equation}
\Lambda = 10E_{0} \ , \ \ \ \ \ M_{0} = 2E_{0} \ , \ \ \ \ \ E_{0} = 1 \ . \label{equation_nonrelparameters}
\end{equation}
One should not be surprised to find the pole at $z=M_{0}-i\epsilon$ for vanishing coupling, yet for small $g$ there is no pole in the whole complex plane. We may get some numerical results near the previous pole position, but inserting those points inside the propagator reveals that they in fact are ordinary points. This is in full agreement with our discussion in the first two chapters: the complex mass-pole leaves the first Riemann sheet when we turn on the coupling! Our Lee model describes by construction the decay of the unstable particle $S$ and consequently the particle should be represented as a pole with non-vanishing imaginary part on the second Riemann sheet.\footnote{It was actually proven by Aramaki and Osawa in Ref. \cite{aramaki} that in the general Lee model the propagator of the unstable particle must have poles in the second Riemann sheet.} We have to look for the pole there.

For $g=0$ the propagator $G_{S}(z)$ is a single-valued function, while for non-vanishing coupling the multi-valued self-energy makes the propagator also not well-defined along the branch cut starting from $z=E_{0}$ and ending at $z=\Lambda$. The cut structure is 'transmitted\grq \ to $G_{S}(z)$. Although we can guess that the pure discontinuity of the logarithmic self-energy across the branch cut will be $i$, we can also calculate it directly for $E_{0}<E<\Lambda$:
\begin{eqnarray}
\disc\Sigma(E) & = & \Sigma(E+i\epsilon)-\Sigma(E-i\epsilon) \nonumber \\
& = & \frac{1}{4\pi}\Big[\ln\big(\Lambda-(E+i\epsilon)\big)-\ln\big(E_{0}-(E+i\epsilon)\big) \nonumber \\
&  & -\ln\big(\Lambda-(E-i\epsilon)\big)+\ln\big(E_{0}-(E-i\epsilon)\big)\Big] \nonumber \\
& = & \frac{1}{4\pi}\Bigg[\ln\left(\frac{\sqrt{(\Lambda-E)^{2}+\epsilon^{2}}}{\sqrt{(E_{0}-E)^{2}+\epsilon^{2}}}\right)+i\phi_{1}-i\phi_{2} \nonumber \\
&  & -\ln\left(\frac{\sqrt{(\Lambda-E)^{2}+\epsilon^{2}}}{\sqrt{(E_{0}-E)^{2}+\epsilon^{2}}}\right)-i\phi_{3}+i\phi_{4}\Bigg] \nonumber \\
& = & \frac{1}{4\pi}\Big[i(\phi_{1}-\phi_{2})-i(\phi_{3}-\phi_{4})\Big] \nonumber \\
& = & \frac{i}{4\pi}\bigg[\arctan\frac{-\epsilon}{\Lambda-E}-\left(\arctan\frac{-\epsilon}{E_{0}-E}-\pi\right)-\arctan\frac{\epsilon}{\Lambda-E} \nonumber \\
& & +\left(\arctan\frac{\epsilon}{E_{0}-E}+\pi\right)\bigg] \nonumber \\
& = & \frac{i}{4\pi}(\pi+\pi) \nonumber \\
\nonumber \\
& = & \frac{i}{2} \ ,
\end{eqnarray}
\begin{figure}[h]
\begin{center}
\includegraphics{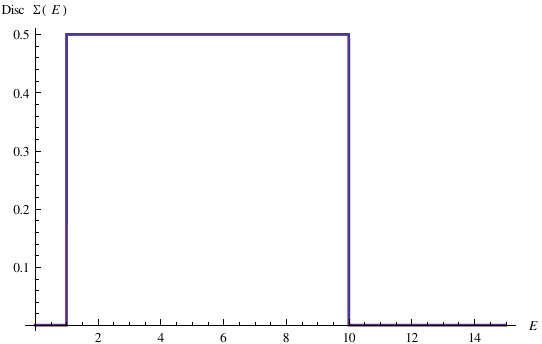}
\caption{Discontinuity of the self-energy (in units of $i$): numerical difference (dark blue) and analytic result (red).}
\label{figure_nonrel1dSigmaDisc}
\end{center}
\end{figure}
\\
where some conversion relations for the polar angles have been used. We have plotted the discontinuity and the numerical difference of $\Sigma(z)$ across the real axis in Fig. \ref{figure_nonrel1dSigmaDisc}. Surely, this result can be obtained quicker by making use of the integral representation of the self-energy and the Lorentzian form of the delta distribution function:
\begin{eqnarray}
\disc\Sigma(E) & = & \Sigma(E+i\epsilon)-\Sigma(E-i\epsilon) \nonumber \\
& = & \frac{1}{4\pi}\int_{E_{0}}^{\Lambda}\text{d}k\left(\frac{1}{k-E-i\epsilon}-\frac{1}{k-E+i\epsilon}\right) \nonumber \\
& = & \frac{1}{4\pi}\int_{E_{0}}^{\Lambda}\text{d}k \ \frac{2i\epsilon}{(k-E)^{2}+\epsilon^{2}} \nonumber \\
& = & \frac{1}{2\pi}\int_{E_{0}}^{\Lambda}\text{d}k \ i\pi\delta(k-E) \nonumber \\
& = & \frac{i}{2} \ , \ \ \ E_{0}<E<\Lambda \ ,
\end{eqnarray}
from which we can observe immediately a discontinuous behaviour only at the branch cut. The analytic continuation down into (at least the fourth quadrant of) the second Riemann sheet therefore reads:
\begin{equation}
\Sigma_{\text{\Romannum{2}}}(z) = \Sigma(z)+\frac{i}{2} \ ,
\end{equation}
and as a consequence the propagator can be written there as
\begin{eqnarray}
G_{\text{\Romannum{2}}}(z) & = & \frac{1}{z-M_{0}+g^{2}\Sigma_{\text{\Romannum{2}}}(z)} \nonumber \\
& = & \frac{1}{z-M_{0}+g^{2}\big(\Sigma(z)+\frac{i}{2}\big)} \ .
\end{eqnarray}
To make sure we have found the correct expression in the first sheet, we proceed in the same way as we did in section \ref{section_dispersion} when the complex root function was computed from its cut structure and the requirement $f(z)=f^{*}(z^{*})$. First, we define the functions
\begin{eqnarray}
f_{1}(z) & = & \frac{1}{4\pi}\ln(\Lambda-z) \ , \ \ \ \ \ g_{1}(z) \ \ = \ \ \frac{f_{1}(z)}{\Lambda-z-1} \ , \\
f_{2}(z) & = & \frac{1}{4\pi}\ln(E_{0}-z) \ , \ \ \ \ g_{2}(z) \ \ = \ \ \frac{f_{2}(z)}{E_{0}-z-1} \ ,
\end{eqnarray}
so that $\Sigma(z)=f_{1}(z)-f_{2}(z)$. Cauchy's integral formula for each of the functions $g_{1}(z)$ and $g_{2}(z)$ then leads us to the correct result, for example:
\begin{eqnarray}
f_{1}(z) & = & \frac{\Lambda-z-1}{2\pi i}\oint_{\mathcal{C}}\text{d}\xi \ \frac{g_{1}(\xi)}{\xi-z} \nonumber \\
& = & \frac{\Lambda-z-1}{4\pi i}\int_{\Lambda}^{\infty}\text{d}x \ \frac{-i}{(\Lambda-x-1)(x-z)} \nonumber \\
& = & -\frac{\Lambda-z-1}{4\pi}\int_{\Lambda}^{\infty}\text{d}x \ \frac{1}{-\Lambda+z+1}\left(\frac{1}{x+1-\Lambda}-\frac{1}{x-z}\right) \nonumber \\
& = & \frac{1}{4\pi}\Big(\ln(x+1-\Lambda)\big|_{\Lambda}^{\infty}-\ln(x-z)\big|_{\Lambda}^{\infty}\Big) \nonumber \\
& = & \frac{1}{4\pi}\ln(\Lambda-z) \ ,
\end{eqnarray}
where we have used the partial fraction decomposition method. Beware of the fact that the discontinuity of $g_{1}(z)$ is the negative imaginary unit $-i/2$, because of the minus sign in the argument of the logarithm.

\begin{figure}[!h]
\begin{center}
\includegraphics{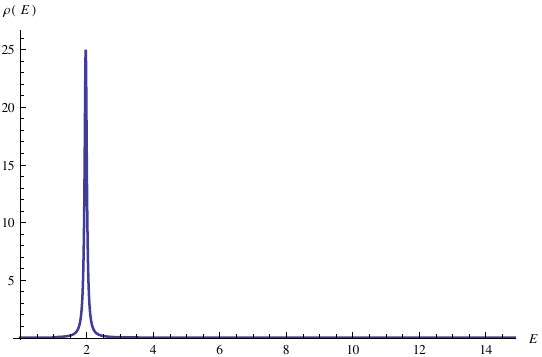}
\caption{Spectral function $\rho(E)$ of the considered Lee model for $g=0.4$.}
\label{figure_nonrel1dSpectralg04}
\end{center}
\end{figure}

Now, to find poles on the second sheet the system of equations (\ref{equation_systemofEq}) is modified to
\begin{eqnarray}
x-M_{0}+g^{2}\operatorname{Re}\Sigma(x+iy) & \stackrel{!}{=} & 0 \ , \nonumber \\
y+g^{2}\operatorname{Im}\Sigma(x+iy)+\frac{g^{2}}{2} & \stackrel{!}{=} & 0 \ .
\end{eqnarray}
In addition to solving this system numerically, we will also state results from an analytic investigation of the relevant expressions in order to support all results. Before that, we want to point out a subtle difficulty. In literature concerning the search for propagator poles in the second Riemann sheet, almost always the analytic structure of the {\em inverse} propagator is studied (e.g. in Refs. \cite{brown,higgspseudo}). A remarkably good and probably the most helpful presentation of the topic can be found in Ref. \cite{brown} (but beware of different sign conventions!). It is said that the difference of the inverse propagator at the branch cut gives the discontinuity to be used for constructing the inverse propagator on the second Riemann sheet. This is of course true since by that procedure one just gets the discontinuity of the self-energy:
\begin{eqnarray}
\disc G_{S}^{-1}(E) & = & G_{S}^{-1}(E+i\epsilon)-G_{S}^{-1}(E-i\epsilon) \nonumber \\
& = & E-M_{0}+g^{2}\Sigma(E+i\epsilon)+i\epsilon-\big(E-M_{0}+g^{2}\Sigma(E-i\epsilon)-i\epsilon\big) \nonumber \\
& = & 2i\epsilon+g^{2}\big(\Sigma(E+i\epsilon)-\Sigma(E-i\epsilon)\big) \nonumber \\
& = & g^{2}\disc\Sigma(E)
\end{eqnarray}
So instead of calculating the discontinuity of the self-energy, one can use the propagator as the multi-valued function. We know that for $g\ne0$ the propagator gains the same branch cut in the complex $E$-plane but if we want to continue it down into the second Riemann sheet we need to find its discontinuity which of course will look different:
\begin{eqnarray}
\disc G_{S}(E) & = & G_{S}(E+i\epsilon)-G_{S}(E-i\epsilon) \nonumber \\
& = & \frac{1}{\pi}\int_{-\infty}^{\infty}\text{d}\omega\left(\frac{\rho(\omega)}{E-\omega+i\epsilon}-\frac{\rho(\omega)}{E-\omega-i\epsilon}\right) \nonumber \\
& = & \frac{1}{\pi}\int_{-\infty}^{\infty}\text{d}\omega \ \rho(\omega)\frac{-2i\epsilon}{(E-\omega)^{2}+\epsilon^{2}} \nonumber \\
& = & -\frac{2i}{\pi}\int_{-\infty}^{\infty}\text{d}\omega \ \rho(\omega)\pi\delta(E-\omega) \nonumber \\
& = & -2i\rho(E) \ \ = \ \ 2i\operatorname{Im}G_{S}(E+i\epsilon) \ . \label{equation_Propdisc}
\end{eqnarray}
The last step is due to the definition of the spectral function as the negative imaginary part of the propagator. We recognize the useful identity from Eq. (\ref{equation_usefullidentity}) and remember the imaginary part to be evaluated right above the branch cut. The (preliminary) spectral function of our Lee model is shown in Fig. \ref{figure_nonrel1dSpectralg04} for an arbitrarily chosen value of the coupling $g$.
One can now carry out the analytic continuation of the spectral function into the lower half plane, i.e., by replacing it by the imaginary part of $G_{S}(E+i\epsilon)$:
\begin{equation}
G_{\text{\Romannum{2}}}(z) = G_{S}(z)-2i\rho(z) \ . \label{equation_SGsecond}
\end{equation}
But care is needed here: Computational software in principle {\em does not know} how to perform the correct extension of the spectral function into the complex plane, because we either possess only a numerical expression (and the extension of such is often non-trivial) or use the imaginary part of the propagator. Nevertheless, we cannot simply take the latter and substitute $E\rightarrow z$. If we first continue the propagator into the complex and afterwards take the imaginary part, we clearly see that Eq. (\ref{equation_SGsecond}) will not give us any pole since the resulting denominator tells us just to solve the old system of equations (\ref{equation_systemofEq}). The full analytic information never enters the final result:
\begin{eqnarray}
G_{S}(z) & = & \frac{1}{z-M_{0}+g^{2}\Sigma(z)} \nonumber \\
& = & \frac{1}{x-M_{0}+g^{2}\operatorname{Re}\Sigma(z)+i\big(y+g^{2}\operatorname{Im}\Sigma(z)\big)} \nonumber \\
\nonumber \\
& = & \frac{x-M_{0}+g^{2}\operatorname{Re}\Sigma(z)-i\big(y+g^{2}\operatorname{Im}\Sigma(z)\big)}{\big(x-M_{0}+g^{2}\operatorname{Re}\Sigma(z)\big)^{2}+\big(y+g^{2}\operatorname{Im}\Sigma(z)\big)^{2}} \ , \\
\nonumber \\
\Rightarrow \ \ \ \operatorname{Im}G_{S}(z) & = & -\frac{y+g^{2}\operatorname{Im}\Sigma(z)}{\big(x-M_{0}+g^{2}\operatorname{Re}\Sigma(z)\big)^{2}+\big(y+g^{2}\operatorname{Im}\Sigma(z)\big)^{2}} \ .
\end{eqnarray}
In fact, we first need to take the imaginary part of the propagator right above the cut and thereafter extend into the complex plane (this is the appropriate way to gain an expression for the spectral function). All this should at least make us very careful when dealing with analytic continuations at the level of the propagator, but mostly it tells us to proceed in another way. If we really want to work with the discontinuity of $G_{S}(E)$ we can make use of the self-energy:
\begin{eqnarray}
\it\Delta G(z) & = & G_{\text{\Romannum{2}}}(z)-G_{S}(z) \nonumber \\
& = & \frac{1}{z-M_{0}+g^{2}\big{(}\Sigma(z)+\frac{i}{2}\big{)}}-\frac{1}{z-M_{0}+g^{2}\Sigma(z)} \nonumber \\
& = & \frac{-ig^{2}}{\big[z-M_{0}+g^{2}\big(\Sigma(z)+\frac{i}{2}\big)\big]\big(z-M_{0}+g^{2}\Sigma(z)\big)} \ ,
\end{eqnarray}
The propagator in the second sheet then reads:
\begin{equation}
G_{\text{\Romannum{2}}}(z) = G_{S}(z)+\it\Delta G(z) \ ,
\end{equation}
which is fully consistent, because on the one hand we get the same poles (from the denominator of $\it\Delta G(z)$) and moreover we preserve a continuous transition from the first Riemann sheet to the second one,
\begin{eqnarray}
\it\Delta G(E) & = & G_{\text{\Romannum{2}}}(E-i\epsilon)-G_{S}(E+i\epsilon) \nonumber \\
& = & \frac{1}{(E-i\epsilon)-M_{0}+g^{2}\big(\Sigma(E-i\epsilon)+\frac{i}{2}\big)}-\frac{1}{(E+i\epsilon)-M_{0}+g^{2}\Sigma(E+i\epsilon)} \nonumber \\
& = & \frac{1}{E-M_{0}+g^{2}\Sigma(E+i\epsilon)-i\epsilon}-\frac{1}{E-M_{0}+g^{2}\Sigma(E+i\epsilon)+i\epsilon} \nonumber \\
\nonumber \\
& = & 0 \ ,
\end{eqnarray}
by taking $\epsilon\rightarrow0^{+}$ in the last step.

\section{Lee model poles}
\subsection{Couplings $g\in[0.1,0.4]$}
\begin{figure}[!h]
\begin{center}
\includegraphics{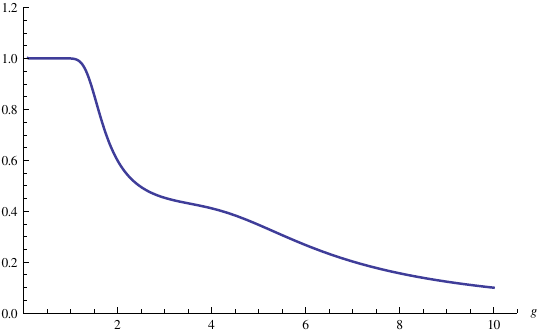}
\caption{Incomplete normalization of the spectral function $\rho(\omega)$ for different values of the coupling $g$.}
\label{figure_nonrel1dSpectralgNorm4}
\end{center}
\end{figure}

In the following subsections we search the complex $E$-plane, using the parameters (\ref{equation_nonrelparameters}), for numerical solutions of the equation
\begin{equation}
z-M_{0}+g^{2}\Sigma_{\text{\Romannum{2}}}(z) \stackrel{!}{=} 0 \label{equation_0denominator}
\end{equation}
by varying the coupling $g$ in steps of $\it\Delta g$. In our case the numerics are very sensitive to variations in $g$, so we need to give a condition for deciding whether a found solution $z_{\text{pole}}$ marks a singularity or not (on a numerical level we take it as a pole if $\abs G_{\text{\Romannum{2}}}^{-1}(z_{\text{pole}})<10^{-10}$). We may also examine the spectral function $\rho(E)$ as the negative imaginary part of the propagator. In fact, it seems normalized only for couplings in the approximate interval $[0.1,1.2]$, see Fig. \ref{figure_nonrel1dSpectralgNorm4}. (We will show in section \ref{section_spectralfunctionnormalization} that it is indeed normalized over the full range of $g$.)
Note that if $g$ tends to zero the spectral function becomes a delta distribution function due to the fact that the pole approaches the real axis:
\begin{eqnarray}
\rho(E) & = & \frac{\epsilon+g^{2}\operatorname{Im}\Sigma(E+i\epsilon)}{\big(E-M_{0}+g^{2}\operatorname{Re}\Sigma(E+i\epsilon)\big)^{2}+\big(\epsilon+g^{2}\operatorname{Im}\Sigma(E+i\epsilon)\big)^{2}} \nonumber \\
& \stackrel{g\rightarrow0}{=} & \frac{\epsilon}{(E-M_{0})^{2}+\epsilon^{2}} \nonumber \\
& = & \pi \delta(E-M_{0}) \ .
\end{eqnarray}
The numerical integration $\frac{1}{\pi}\int_{-\infty}^{\infty}\text{d}\omega \ \rho(\omega)$ is then not possible anymore (so pure poles in the vicinity of the integration contour cannot be considered). In general, $\rho(\omega)$ gets broader for larger couplings, and this effect is caused by the complex mass-pole moving deeper into the complex plane beyond the branch cut, i.e., the second Riemann sheet: the spectral function 'feels\grq \ this pole less and so shrinks very quickly the more the pole descends.
\newpage
\begin{table}[!h]\center
\scalebox{0.85}{
\begin{tabular}{l c c c c c c c c c c}
\toprule
\cmidrule(r){1-10}
$g$	& $x_{\text{(l)pole}}$	& $x_{\text{(m)pole}}$	& $x_{\text{(r)pole}}$	& $x_{\text{(l)BW}}$	& $x_{\text{(m)BW}}$	& $x_{\text{(r)BW}}$	& $x_{\text{(l)max}}$	& $x_{\text{(r)max}}$	& $x_{\text{average}}$	&Norm\\
\midrule
0.1	&-	&1.9983	&-	&-	&1.9983	&-	&-	&1.9983	&-	&-\\
0.2	&-	&1.9933	&-	&-	&1.9933	&-	&-	&1.9933	&2.0000	&0.9999\\
0.3	&-	&1.9849	&-	&-	&1.9849	&-	&-	&1.9849	&2.0000	&1.0000\\
0.4	&-	&1.9731	&-	&-	&1.9731	&-	&-	&1.9731	&2.0000	&1.0000\\
0.5	&-	&1.9577	&-	&-	&1.9576	&-	&-	&1.9576	&2.0000	&1.0000\\
0.6	&-	&1.9385	&-	&-	&1.9383	&-	&-	&1.9383	&2.0000	&1.0000\\
0.7	&-	&1.9154	&-	&-	&1.9150	&-	&-	&1.9150	&2.0000	&1.0000\\
0.8	&-	&1.8882	&-	&-	&1.8872	&-	&-	&1.8872	&2.0000	&0.9999\\
0.9	&0.9999	&1.8569	&-	&0.9999	&1.8546	&-	&-	&1.8546	&1.9999	&0.9999\\
1	&0.9999	&1.8214	&-	&0.9999	&1.8166	&-	&-	&1.8166	&1.9996	&0.9996\\
1.2	&0.9958	&1.7397	&-	&0.9985	&1.7201	&-	&1.0014	&1.7201	&1.9875	&1.0000\\
1.4	&0.9864	&1.6521	&-	&0.9864	&1.5838	&-	&1.0163	&1.5838	&1.9209	&1.0000\\
1.6	&0.9481	&1.5714	&-	&0.9481	&1.3412	&-	&1.1157	&1.3412	&1.8068	&1.0000\\
1.8	&0.8810	&1.5032	&-	&0.8810	&-	&-	&1.2656	&-	&1.7194	&1.0000\\
2	&0.7926	&1.4467	&-	&0.7926	&-	&-	&1.3304	&-	&1.6829	&1.0000\\
2.2	&0.6898	&1.4008	&-	&0.6898	&-	&-	&1.4032	&-	&1.6864	&1.0000\\
2.4	&0.5772	&1.3648	&-	&0.5772	&-	&-	&1.4844	&-	&1.7164	&0.9999\\
2.6	&0.4575	&1.3384	&-	&0.4575	&-	&-	&1.5746	&-	&1.7635	&0.9999\\
2.8	&0.3324	&1.3216	&10.0000	&0.3324	&-	&10.0000	&1.6744	&9.9999	&1.8218	&1.0000\\
3	&0.2030	&1.3140	&10.0001	&0.2030	&9.9998	&10.0001	&1.7846	&9.9998	&1.8870	&1.0000\\
3.2	&0.0701	&1.3153	&10.0005	&0.0701	&9.9995	&10.0005	&1.9061	&9.9995	&1.9548	&1.0000\\
3.4	&-0.0656	&1.3246	&10.0015	&-0.0656	&9.9984	&10.0015		&2.0401	&9.9984	&2.0207	&1.0000\\
3.6	&-0.2040	&1.3411	&10.0038	&-0.2040	&9.9961	&10.0038	&2.1881	&9.9961	&2.0791	&1.0000\\
3.8	&-0.3445	&1.3634	&10.0085	&-0.3445	&9.9914	&10.0085	&2.3522	&9.9914	&2.1243	&1.0000\\
4	&-0.4870	&1.3902	&10.0166	&-0.4870	&9.9830	&10.0166	&2.5350	&9.9830	&2.1514	&1.0000\\
4.2	&-0.6312	&1.4202	&10.0296	&-0.6312	&9.9693	&10.0296	&2.7402	&9.9693	&2.1573	&1.0000\\
4.4	&-0.7769	&1.4519	&10.0487	&-0.7769	&9.9485	&10.0487	&2.9732	&9.9485	&2.1410	&1.0000\\
4.6	&-0.9240	&1.4844	&10.0750	&-0.9240	&9.9191	&10.0750	&3.2427	&9.9191	&2.1037	&1.0000\\
4.8	&-1.0724	&1.5167	&10.1093	&-1.0724	&9.8792	&10.1093	&3.5638	&9.8792	&2.0485	&1.0000\\
5	&-1.2219	&1.5480	&10.1520	&-1.2219	&9.8273	&10.1520	&3.9686	&9.8273	&1.9791	&1.0000\\
5.5	&-1.5999	&1.6196	&10.2962	&-1.5999	&9.6382	&10.2962	&-	&9.6382	&1.7701	& 1.0000\\
6	&-1.9833	&1.6795	&10.4900	&-1.9833	&9.3595	&10.4900	&-	&9.3595	&1.5494	& 1.0000\\
6.5	&-2.3711	&1.7283	&10.7258	&-2.3711	&9.0030	&10.7258	&-	&9.0030	&1.3453	& 1.0000\\
7	&-2.7626	&1.7677	&10.9952	&-2.7626	&8.6015	&10.9952	&-	&8.6015	&1.1678 	& 1.0000\\
7.5	&-3.1571	&1.7996	&11.2912	&-3.1571	&8.1974	&11.2912	&-	&8.1974	&1.0174	& 1.0000\\
8	&-3.5544	&1.8255	&11.6081	&-3.5544	&7.8254	&11.6081	&-	&7.8254	&0.8913	& 1.0000\\
8.5	&-3.9539	&1.8468	&11.9415	&-3.9539	&7.5028	&11.9415	&-	&7.5028	&0.7858	& 1.0000\\
9	&-4.3555	&1.8645	&12.2880	&-4.3555	&7.2323	&12.2880	&-	&7.2323	&0.6971	& 1.0000\\
9.5	&-4.7588	&1.8793	&12.6450	&-4.7588	&7.0084	&12.6450	&-	&7.0084	&0.6221	& 1.0000\\
10	&-5.1637	&1.8918	&13.0106&-5.1637	&6.8235	&13.0106	&-	&6.8235	&0.5584	& 1.0000\\
10.5	&-5.5699	&1.9024	&13.3834&-5.5699	&6.6702	&13.3834	&-	&6.6702	&0.5039	& 1.0000\\
11	&-5.9774	&1.9116	&13.7620&-5.9774	&6.5421	&13.7620	&-	&6.5421	&0.4569	& 1.0000\\
11.5	&-6.3860	&1.9195	&14.1455&-6.3860	&6.4342	&14.1455	&-	&6.4342	&0.4162	& 1.0000\\
12	&-6.7955	&1.9264	&14.5332&-6.7955	&6.3426	&14.5332	&-	&6.3426	&0.3807	& 1.0000\\
\bottomrule
\end{tabular}
}
\caption{Selection of masses for the considered Lee model with parameters $\Lambda=10E_{0}, \ M_{0}=2E_{0}$ and $E_{0}=1$. All numbers after the fourth digit are dropped.}
\label{table_allpoints}
\end{table}
\newpage
\begin{table}[!h]\center
\scalebox{0.8}{
\begin{tabular}{l c c l}
\toprule
\cmidrule(r){1-4}
name	& label	& determining equation	& description\\
\midrule
resonance pole mass	&$x_{\text{(m)pole}}$	&$z-M_{0}+g^{2}\Sigma_{\text{\Romannum{2}}}(z) = 0$	&Real part of solution $z$.\\
left/right pole mass	&$x_{\text{(l/r)pole}}$	&$z-M_{0}+g^{2}\Sigma(z) = 0$	&Real part of solution $z$.\\
	&	&	&Nevertheless, $z$ is purely real.\\
Breit--Wigner resonance mass	&$x_{\text{(m)BW}}$	&$x-M_{0}+g^{2}\operatorname{Re}\Sigma(x) = 0$	&The solution $x$ is purely real.\\
Breit--Wigner left/right mass	&$x_{\text{(l/r)BW}}$	&$x-M_{0}+g^{2}\operatorname{Re}\Sigma(x) = 0$	&The solution $x$ is purely real.\\
average mass	&$x_{\text{average}}$	&$x = \frac{1}{\pi}\int_{E_{0}}^{\Lambda}\text{d}\omega \ \omega\rho(\omega)$	&Numerically evaluated.\\
maxima of spectral function	&$x_{\text{(l/r)max}}$	&$\maxf\rho(E)$	&Solution is $x=E_{\text{max}}$.\\
	&	&	&Found numerically.\\
\bottomrule
\end{tabular}
}
\caption{Description of masses for the considered Lee model. The general complex solution of an equation is denoted as $z=x+iy$.}
\label{table_allpointsexplanation}
\end{table}
\begin{figure}[!h]
\begin{center}
\includegraphics{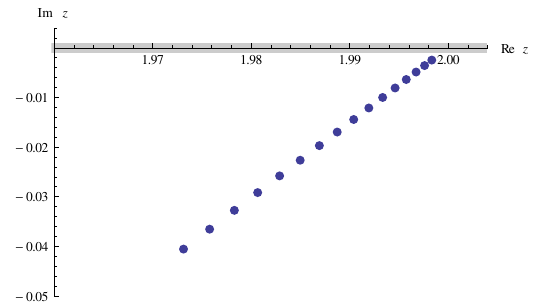}
\caption{Lee model pole (dark blue dots) for $g\in[0.1,0.4]$ in the second Riemann sheet with $\it\Delta$$g=0.02$, where the gray line marks the branch cut.}
\label{figure_nonrel1dPointsg0104gstep002}
\end{center}
\end{figure}
The pole trajectory within the regarded interval $[0.1,0.4]$ for the coupling is shown in Fig. \ref{figure_nonrel1dPointsg0104gstep002}. A summary of pole positions and additional information for different intervals can be found in Tab. \ref{table_allpoints}. To understand the adopted notation we have to define different {\em masses} of the considered Lee model (see also Tab. \ref{table_allpointsexplanation}). The pole mass $x_{\text{(m)pole}}$ of the main resonance pole is the real part of the complex solution found by solving Eq. (\ref{equation_0denominator}), while $x_{\text{(l)pole}}$ and $x_{\text{(r)pole}}$ are the real parts of the complex solutions in the first Riemann sheet, where we have replaced the self-energy $\Sigma_{\text{\Romannum{2}}}(z)$ in Eq. (\ref{equation_0denominator}) by its equivalent $\Sigma(z)$ on the first Riemann sheet. The corresponding Breit--Wigner masses are the solutions of the latter equation using just the real part of the self-energy on the first sheet (they are all real by construction). These are simply the zeros in the denominator of the propagator on the real axis. The numerical maxima of the continuous part of the spectral function are denoted as $x_{\text{(l)max}}$ and $x_{\text{(r)max}}$, whereas $x_{\text{average}}$ is the result of the numerically evaluated integral
\begin{equation}
x_{\text{average}} = \frac{1}{\pi}\int_{E_{0}}^{\Lambda}\text{d}\omega \ \omega\rho(\omega) \ . \label{equation_avmassintegral}
\end{equation}
The last column of Tab. \ref{table_allpoints} simlpy gives the normalization of the spectral function.

\subsection{Couplings $g\in[0.4,1.2]$}
For higher values of the coupling the resonance pole descends deeper into the lower half plane of the second sheet, but around $g\approx0.9$ there is a new pole arising in the first sheet left from the first branch point at $z=E_{0}$. The new pole seems to continuously enter the spectrum of found solutions.\footnote{This subtle point will be the main difference to the quantum field theoretical model in chapter 4.} Although we could clearly fix $g$ of this pole in the context of our above condition $\abs G_{S}^{-1}(z_{\text{pole}})<10^{-10}$, it is reasonable, and will be shown analytically, not to speak of the pole to suddenly 'pop up\grq \ on the real axis. Compare this situation to the behaviour of the resonance pole in the second sheet for vanishing couplings: these are obviously numerical problems we need to keep in mind.
\begin{figure}[!h]
\begin{center}
\includegraphics{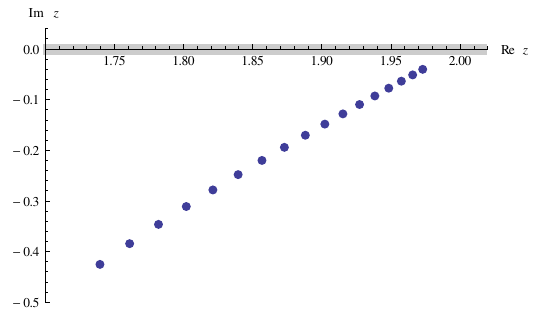}
\caption{Lee model pole for $g\in[0.4,1.2]$ in the second Riemann sheet with $\it\Delta$$g=0.05$.}
\label{figure_nonrel1dPointsg0412gstep005-1}
\end{center}
\end{figure}
\begin{figure}[!h]
\begin{center}
\includegraphics{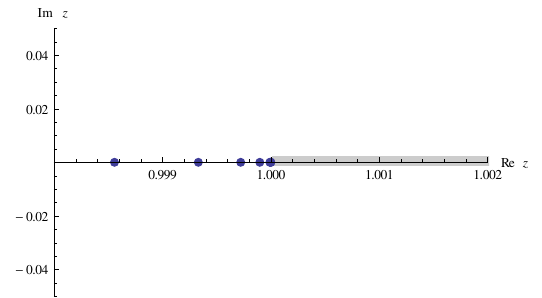}
\caption{Left Lee model pole for $g\in[0.4,1.2]$ on the real axis of the first Riemann sheet with $\it\Delta$$g=0.05$.}
\label{figure_nonrel1dPointsg0412gstep005-2}
\end{center}
\end{figure}

\subsection{Couplings $g\in[1.2,4.0]$}
The presence of a new simple pole on the first sheet can also be noticed by the abruptly falling spectral function around $g\approx1.2$, see Fig. \ref{figure_nonrel1dSpectralgNorm4}. If we further raise $g$ the pole passes the origin at $g=3.303$, while the pole on the second sheet descends deeper into the lower half plane. Here, the real part decreases more slowly and at $g=3.069$ it starts increasing.
\begin{figure}[!h]
\begin{center}
\includegraphics{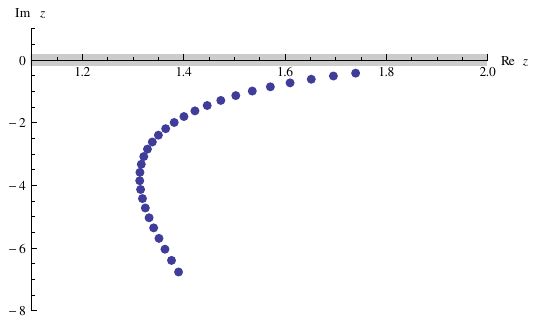}
\caption{Lee model pole for $g\in[1.2,4.0]$ in the second Riemann sheet with $\it\Delta$$g=0.1$.}
\label{figure_nonrel1dPointsg1230gstep01-4}
\end{center}
\end{figure}
\begin{figure}[!h]
\begin{center}
\includegraphics{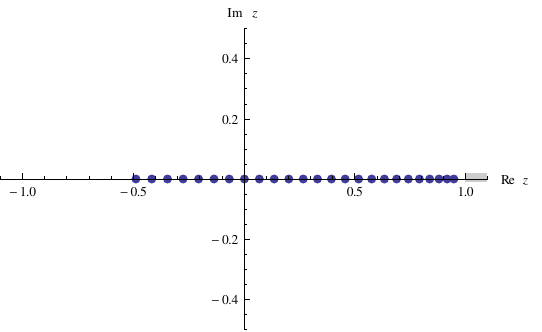}
\caption{Left Lee model pole for $g\in[1.2,4.0]$ on the real axis of the first Riemann sheet with $\it\Delta$$g=0.1$.}
\label{figure_nonrel1dPointsg1230gstep01-2}
\end{center}
\end{figure}
\begin{figure}[!h]
\begin{center}
\includegraphics{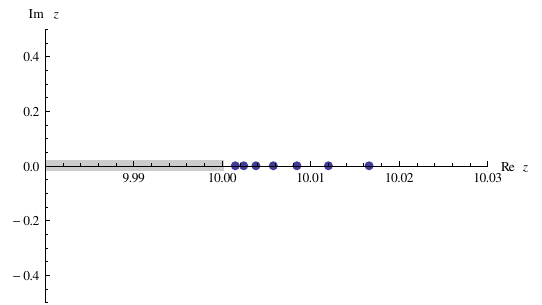}
\caption{Right Lee model pole for $g\in[1.2,4.0]$ on the real axis of the first Riemann sheet with $\it\Delta$$g=0.1$.}
\label{figure_nonrel1dPointsg1230gstep01-1}
\end{center}
\end{figure}
Additionally, there is another pole emerging around $g\approx2.7$ on the first sheet to the right of the second branch point at $z=\Lambda$. We shall call the former the left (l) one and the latter the right (r) one, respectively, as was already explained during the discussion of Tab. \ref{table_allpoints}. It should be clear that the pole masses for the left and the right poles are identical with the corresponding Breit--Wigner masses, because the imaginary part of the self-energy is only non-zero for $E_{0}<z<\Lambda$. Therefore, the purely real solutions of the equation
\begin{equation}
0 \stackrel{!}{=} x-M_{0}+g^{2}\operatorname{Re}\Sigma(x)
\end{equation}
are also solutions of
\begin{equation}
0 \stackrel{!}{=} x-M_{0}+g^{2}\Sigma(x) \ ,
\end{equation}
which is the zero of the full denominator of the propagator.

When looking at Fig \ref{figure_nonrel1dPointsg1230gstep01-2}, it seems that the left pole reaches a constant {\em g-velocity} in the parameter space spanned by $\operatorname{Re}z=x$ and the coupling $g$. Although the last equation cannot be solved for $x$ analytically, it can be solved for $g$:
\begin{equation}
g(x) = \pm\sqrt{\frac{4\pi(M_{0}-x)}{\ln\left(\frac{\Lambda-x}{E_{0}-x}\right)}} \ , \label{equation_gfunction}
\end{equation}
where we only take the positive branch from now on. The limit $x\rightarrow-\infty$ of the first derivative with respect to $x$ in fact turns out to be a constant:
\begin{eqnarray}
m_{-\infty} \ \ \equiv \ \ \lim_{x \to -\infty}\frac{\text{d}g(x)}{\text{d}x} & = & \sqrt{\pi}\lim_{x \to -\infty}\Bigg{\{}\frac{(E_{0}-\Lambda)(M_{0}-x)}{(x-E_{0})(x-\Lambda)\ln^{2}\left(\frac{\Lambda-x}{E_{0}-x}\right)\sqrt{\frac{M_{0}-x}{\ln\left(\frac{\Lambda-x}{E_{0}-x}\right)}}} \nonumber \\
& & \ + \ \frac{(E_{0}-x)(x-\Lambda)\ln\left(\frac{\Lambda-x}{E_{0}-x}\right)}{(x-E_{0})(x-\Lambda)\ln^{2}\left(\frac{\Lambda-x}{E_{0}-x}\right)\sqrt{\frac{M_{0}-x}{\ln\left(\frac{\Lambda-x}{E_{0}-x}\right)}}}\Bigg{\}} \nonumber \\
& = & \sqrt{\pi}\left(-\sqrt{\frac{1}{\Lambda-E_{0}}}-\sqrt{\frac{1}{\Lambda-E_{0}}}\right) \nonumber \\
\nonumber \\
& = & -\sqrt{\frac{4\pi}{\Lambda-E_{0}}} \ .
\end{eqnarray}
\begin{figure}[t]
\begin{center}
\includegraphics{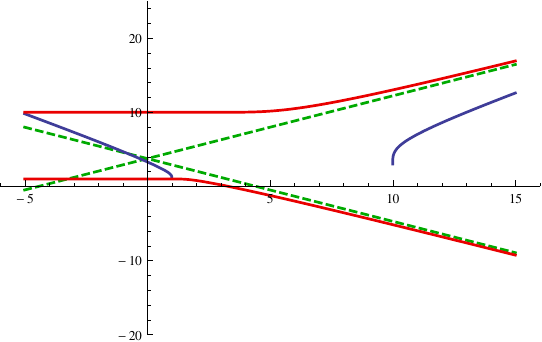}
\caption{Analytic function $g(x)$ (dark blue), numerical inverse $x(g)$ (red) and asymptotic lines from Eq. (\ref{equation_asymlines}) (green, dashed).}
\label{figure_nonrel1dInversegfunctionwithlines}
\end{center}
\end{figure}
We obtain the same expression in the limit $x\rightarrow+\infty$ but with opposite sign:
\begin{equation}
m_{+\infty} \equiv \lim_{x \to +\infty}\frac{\text{d}g(x)}{\text{d}x} = \sqrt{\frac{4\pi}{\Lambda-E_{0}}} \ .
\end{equation}
The left and right pole, respectively, can then be taken to be located on a line $g(x)=m_{\pm\infty}x+b_{\pm\infty}$ with the above coefficients $m_{\pm\infty}$. The interception points $(0|b_{\pm\infty})$ are calculated again by taking the limit $x\rightarrow\pm\infty$:
\begin{eqnarray}
\lim_{x \to \pm\infty}g(x) & \stackrel{!}{=} & m_{\pm\infty}x+b_{\pm\infty} \ , \label{equation_lineareqn} \\
\Rightarrow \ \ \ b_{\pm\infty}& = & \mp\frac{\sqrt{\pi}}{2}\left(\sqrt{\frac{1}{\Lambda-E_{0}}}(\Lambda+E_{0}+2M_{0})\right)
\end{eqnarray}
The resulting asymptotic line for the left pole gives deviations smaller than 5\% in the coupling with $x\lesssim-5.5$ for our choice of parameters. For the right pole this is the case if $x\gtrsim15.5$. By inverting the linear equation (\ref{equation_lineareqn}) we finally get an expression for the function $x(g)$ in the limit of huge couplings (here, deviations less than 5\% can be obtained for $g\gtrsim13$ and $g\gtrsim11.5$, respectively):
\begin{eqnarray}
x(g) & = & \frac{g-b_{\pm\infty}}{m_{\pm\infty}} \nonumber \\
& = & \pm\sqrt{\frac{\Lambda-E_{0}}{4\pi}}g+\frac{\Lambda+E_{0}+2M_{0}}{4} \ . \label{equation_asymlines}
\end{eqnarray}
Both of these lines intercept at the same point on the $x$-axis. The limit of constant $g$-velocity is also observed for the right pole. In Fig. \ref{figure_nonrel1dInversegfunctionwithlines} we give the numerical inverse of Eq. (\ref{equation_gfunction}) together with the asymptotic lines from Eq. (\ref{equation_asymlines}). The most important result of that plot is the fact that, although we could not find poles for some intervals of the coupling (no right pole for $g<2.7$ and neither left nor right poles for $g<1.2$), they really do exist near the branch points at $z=E_{0}$ and $z=\Lambda$. A numerical analysis cannot fully reveal the full pole structure since one needs a high starting point precision beyond $10^{-10}$. Nevertheless, we can state now that there are poles on the real axis as long as $g$ is non-zero. This is because the following limits indeed exist,
\begin{eqnarray}
\lim_{\eta \to 0^{+}}g(E_{0}-\eta) & = & \lim_{\eta \to 0^{+}}\sqrt{\frac{4\pi\big(M_{0}-(E_{0}-\eta)\big)}{\ln\left(\frac{\Lambda-(E_{0}-\eta)}{E_{0}-(E_{0}-\eta)}\right)}} \nonumber \\
& = & \lim_{\eta \to 0^{+}}\sqrt{\frac{4\pi(M_{0}-E_{0}+\eta)}{\ln\left(\frac{\Lambda-E_{0}+\eta}{\eta}\right)}} \nonumber \\
& = & 0 \ ,
\end{eqnarray}
\begin{eqnarray}
\lim_{\eta \to 0^{+}}g(\Lambda+\eta) & = & \lim_{\eta \to 0^{+}}\sqrt{\frac{4\pi\big(M_{0}-(\Lambda+\eta)\big)}{\ln\left(\frac{\Lambda-(\Lambda+\eta)}{E_{0}-(\Lambda+\eta)}\right)}} \nonumber \\
& = & \lim_{\eta \to 0^{+}}\sqrt{\frac{4\pi(M_{0}-\Lambda-\eta)}{\ln\left(\frac{-\eta}{E_{0}-\Lambda-\eta}\right)}} \nonumber \\
& = & 0 \ ,
\end{eqnarray}
and the root functions are both positive for $z<E_{0}$ and $z>\Lambda$. It is satisfying to see the few available analytic expressions not only supporting the numerical analysis but indeed completing it. We may also show that the spectral function $\rho(E)$ has delta peaks for the left $(x=E<E_{0})$ and right $(x=E>\Lambda)$ poles in the case of non-vanishing couplings:
\begin{eqnarray}
\rho(E) & = & \frac{\epsilon+g^{2}\operatorname{Im}\Sigma(E+i\epsilon)}{\big(E-M_{0}+g^{2}\operatorname{Re}\Sigma(E+i\epsilon)\big)^{2}+\big(\epsilon+g^{2}\operatorname{Im}\Sigma(E+i\epsilon)\big)^{2}} \nonumber \\
& = & \frac{\epsilon}{\big(E-M_{0}+g^{2}\operatorname{Re}\Sigma(E+i\epsilon)\big)^{2}+\epsilon^{2}} \nonumber \\
& = & \pi\delta\big(E-M_{0}+g^{2}\operatorname{Re}\Sigma(E)\big) \nonumber \\
& = & \pi\delta\big(E-M_{0}+g^{2}\Sigma(E)\big) \ .
\end{eqnarray}

\subsection{Couplings $g\in[4.0,6.0]$}
The pole in the second Riemann sheet falls deeper into the lower half plane. The real part decreases but the pole seems to approach the constant line $\operatorname{Re}z=2$. This can be investigated in a similar way to the previous cases, yet now the full complex denominator of the propagator needs to be studied. After solving Eq. (\ref{equation_0denominator}) for the positive coupling $g$ we obtain the master solution
\begin{equation}
g(z=x+iy) = \sqrt{\frac{4\pi\big(M_{0}-(x+iy)\big)}{\ln\big(\Lambda-(x+iy)\big)-\ln\big(E_{0}-(x+iy)\big)+2\pi i}} \ , \label{equation_gfunctionz}
\end{equation}
from which one immediately observes the multi-valued character of the propgator due to an imaginary shift of $2\pi i$, coming from the logarithmic structure. The numerical results already indicated that the imaginary part of the resonance pole goes to minus infinity, while the real part approaches the value $\operatorname{Re}z=2$. If we assume the latter to be true, then the coupling $g$ must grow beyond all bounds when the imaginary part $y$ goes to minus infinity. And this is in fact true since
\begin{eqnarray}
g(z=M_{0}+iy) & = & \sqrt{\frac{4\pi\big(M_{0}-(M_{0}+iy)\big)}{\ln\big(\Lambda-(M_{0}+iy)\big)-\ln\big(E_{0}-(M_{0}+iy)\big)+2\pi i}} \nonumber \\
\nonumber \\
& = & \sqrt{\frac{-4\pi i y}{\ln(\Lambda-M_{0}-iy)-\ln(E_{0}-M_{0}-iy)+2\pi i}} \ , \nonumber \\
\\
\Rightarrow \ \ \ \lim_{y \to -\infty}g(z=M_{0}+iy) & = & \infty \ .
\end{eqnarray}

The two poles in the first sheet just keep 'walking\grq \ on the real axis, the left one going to minus and the right one to plus infinity. We have just proved this behaviour in the previous subsection (the limit of constant $g$-velocity in parameter space also holds for the right pole, see lower figure).
\\
\begin{figure}[!h]
\begin{center}
\includegraphics{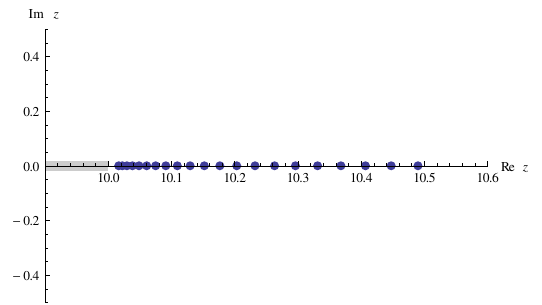}
\caption{Right Lee model pole for $g\in[4.0,6.0]$ on the real axis of the first Riemann sheet with $\it\Delta$$g=0.1$.}
\label{figure_nonrel1dPointsg3050gstep01-2}
\end{center}
\end{figure}

\newpage

\subsection{Couplings $g\in[6.0,12.0]$} \label{subsection_nonrelcouplings50100}
\begin{figure}[!h]
\begin{center}
\includegraphics{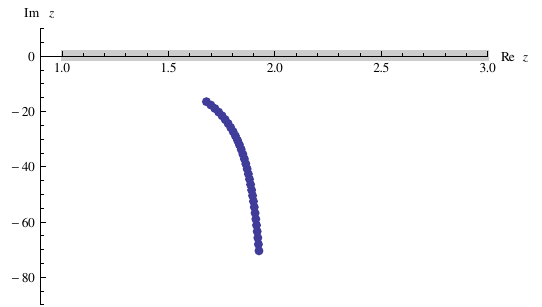}
\caption{Lee model pole for $g\in[6.0,12.0]$ in the second Riemann sheet with $\it\Delta$$g=0.2$.}
\label{figure_nonrel1dPointsg50100gstep02-1}
\end{center}
\end{figure}

The real part of the resonance pole approaches the constant line $\operatorname{Re}z=2$, see Fig. \ref{figure_nonrel1dPointsg50100gstep02-1}, while the two poles in the first sheet further walk away on the real axis. When looking back at Tab. \ref{table_allpoints}, we see that the values for the Breit--Wigner mass $x_{\text{(m)BW}}$ are very different for small and high $g$, seperated by a gap, while for $g\rightarrow\infty$ it seems to equal the value of $x_{\text{(r)max}}$. Numerically, one obtains for our choice of parameters $x_{\text{(m)BW}}\rightarrow5.5$ in the limit of very high couplings $g\gg10$. A natural suggestion therefore would be
\begin{equation}
\lim_{g \to+\infty}x_{\text{(r)max}} = \lim_{g \to+\infty}x_{\text{(m)BW}} = \frac{\Lambda+E_{0}}{M_{0}} \ ,
\end{equation}
since this involves the only free parameters of our model and gives the right numerical value.
Now, by considering the spectral function $\rho(E)$ right above the branch cut, it is possible to show explicitly that this is not the right limit. Remember that for $E_{0}<z<\Lambda$ we cannot use the self-energy as calculated at the beginning of our investigation in Eq. (\ref{equation_sigmaln}) because the $k$-integration would hit the pole on the real $k$-axis. For that reason we need to perform the integration by using the Sokhotski--Plemelj theorem:
\begin{eqnarray}
\int_{E_{0}}^{\Lambda}\text{d}k \ \frac{1}{k-E-i\epsilon} & = & \mathcal{P}\int_{E_{0}}^{\Lambda}\text{d}k \ \frac{1}{k-E}+i\pi \nonumber \\
& = & \left(\int_{E_{0}-E}^{0-\eta}\text{d}k \ \frac{1}{k}+\int_{0+\eta}^{\Lambda-E}\text{d}k \ \frac{1}{k}\right)+i\pi \nonumber \\
& = & \ln(-\eta)-\ln(E_{0}-E)+\ln(\Lambda-E)-\ln\eta+i\pi \nonumber \\
& = & \ln\eta+i\pi-\ln(E-E_{0})-i\pi+\ln(\Lambda-E)-\ln\eta+i\pi \nonumber \\
& = & \ln\left(\frac{\Lambda-E}{E-E_{0}}\right)+i\pi \ ,
\end{eqnarray}
where we have used the well-known definition of the logarithm for negative real numbers. This result can also be obtained by applying the latter to the self-energy (\ref{equation_sigmaln}) itself. The spectral function now reads:
\begin{eqnarray}
\rho(E) & = & \frac{\epsilon+g^{2}\operatorname{Im}\Sigma(E+i\epsilon)}{\big(E-M_{0}+g^{2}\operatorname{Re}\Sigma(E+i\epsilon)\big)^{2}+\big(\epsilon+g^{2}\operatorname{Im}\Sigma(E+i\epsilon)\big)^{2}} \nonumber \\
\nonumber \\
& = & \frac{1}{4}\frac{1}{\frac{1}{g^{2}}\left(E-M_{0}+\frac{g^{2}}{4\pi}\ln\left(\frac{\Lambda-E}{E-E_{0}}\right)\right)^{2}+\frac{g^{2}}{16}} \ . \label{equation_branchcutspectral}
\end{eqnarray}
A propagator pole on the real axis would correspond to a zero in the denominator of the last expression. But for the regarded interval there is no such zero except for $E=M_{0}$ and $g=0$, since the general solution for positive couplings,
\begin{equation}
g(x=E) = 2\sqrt{\pi}\sqrt{\pm\frac{(M_{0}-E)\ln\left(\frac{\Lambda-E}{E-E_{0}}\right)-\pi\sqrt{-(M_{0}-E)^{2}}}{\ln^{2}\left(\frac{\Lambda-E}{E-E_{0}}\right)+\pi^{2}}} \ ,
\end{equation}
is only real (with value $g=0$) for $E=M_{0}$ (of course it makes no sense to deal with the self-energy at all for vanishing couplings). The value $x_{\text{max}}$ is consequently the minimum of the denominator in Eq. (\ref{equation_branchcutspectral}):
\begin{eqnarray}
\frac{\text{d}}{\text{d}E}\bigg[\frac{1}{g^{2}}\left(E-M_{0}+\frac{g^{2}}{2\pi}\ln\left(\frac{\Lambda-E}{E-E_{0}}\right)\right)^{2}+\frac{g^{2}}{4}\bigg] & \stackrel{!}{=} & 0 \nonumber \\
\nonumber \\
\frac{\left[(E_{0}-\Lambda)+\frac{4\pi}{g^{2}}(E_{0}-E)(E-\Lambda)\right]\left[g^{2}\ln\left(\frac{\Lambda-E}{E-E_{0}}\right)+4\pi(E-M_{0})\right]}{8\pi^{2}(E-E_{0})(\Lambda-E)} & \stackrel{!}{=} & 0 \ . \nonumber \\
\label{equation_firstderivative}
\end{eqnarray}
The denominator $8\pi^{2}(E-E_{0})(\Lambda-E)$ vanishes for $E=E_{0}$ and $E=\Lambda$ which are independent of $g$, so there is no problem in discarding this term. On the other hand, the first part of the numerator has two solutions
\begin{equation}
E_{\pm} = \frac{E_{0}+\Lambda\pm\sqrt{\frac{g^2}{\pi}(E_{0}-\Lambda)+(E_{0}-\Lambda)^{2}}}{2} \ , \label{equation_branchcutspectralsolution}
\end{equation}
where only the negative solution belongs to a local minimum (and to a local maximum of the spectral function) for $g>0$. We cannot prove this in general, for our equations are transcendental, nevertheless the mentioned aspect becomes clear when plotting the second derivative of the denominator in the spectral function (\ref{equation_branchcutspectral}) and inserting the negative minumum solution from Eq. (\ref{equation_branchcutspectralsolution}): the graph (now with dependence of $g$) lies in the first quadrant and there is no change until both solutions become complex for a numerically computed value of $g>5.317$. This is in full agreement with the analytic condition for complex solutions coming from the root in Eq. (\ref{equation_branchcutspectralsolution}),
\begin{eqnarray}
\frac{g^{2}}{\pi}(E_{0}-\Lambda)+(E_{0}-\Lambda)^{2} \stackrel{!}{=} 0 \ , \\
\Rightarrow \ \ \ g = \sqrt{-\pi(E_{0}-\Lambda)} \approx 5.317 \ ,
\end{eqnarray}
and can be also seen in Tab. \ref{table_allpoints} where the maximum $x_{\text{(l)max}}$ vanishes for $g>5.0$. The other maximum $x_{\text{(r)max}}$ comes from the logarithmic term in Eq. (\ref{equation_firstderivative}). Its zeros are obviously the Breit--Wigner masses $x_{\text{(m)BW}}$, too:
\begin{equation}
g(x=E) = \sqrt{\frac{4\pi(M_{0}-E)}{\ln\left(\frac{\Lambda-E}{E-E_{0}}\right)}} \ .
\end{equation}

The relevant behaviour for us to study now is the limit of huge couplings. Because the logarithmic term in the first derivative from Eq. (\ref{equation_firstderivative}) also gives the remaining minimum for $g>5.317$, we only need to consider the limit of the last above expression. Our first suggestion
\begin{equation}
\lim_{g \to+\infty}x_{\text{(r)max}} = \lim_{g \to+\infty}x_{\text{(m)BW}} = \frac{\Lambda+E_{0}}{M_{0}} \nonumber
\end{equation}
yields the correct numerical value of 5.5 (this is true if we just consider the right-sided limit):
\begin{eqnarray}
\lim_{\eta \to 0^{+}}\sqrt{\frac{4\pi\big(2-(5.5+\eta)\big)}{\ln\left(\frac{10-(5.5+\eta)}{5.5+\eta-1}\right)}} & = & \lim_{\eta \to 0^{+}}\sqrt{\frac{4\pi(2-5.5-\eta)}{\ln\left(\frac{10-5.5-\eta}{5.5+\eta-1}\right)}} \nonumber \\
& = & \lim_{\eta \to 0^{+}}\sqrt{\frac{-14\pi-4\pi\eta}{\ln\left(\frac{4.5-\eta}{4.5+\eta}\right)}} \nonumber \\
& = & \infty \ .
\end{eqnarray}
Nevertheless, this does not hold in general. In fact, our first suggestion has to be modified in the following way:
\begin{equation}
\lim_{g \to +\infty}x_{\text{(r)max}} = \lim_{g \to +\infty}x_{\text{(m)BW}} = \frac{\Lambda+E_{0}}{2} \ ,
\end{equation}
which reveals the limit to be independent of the mass $M_{0}$. A real right-sided limit does not exist for arbitrary choices of parameters. We find in general:
\begin{eqnarray}
\lim_{\eta \to 0^{+}}g\big(x=(\Lambda+E_{0})/2+\eta\big) & = & \lim_{\eta \to 0^{+}}\sqrt{\frac{4\pi\left(M_{0}-\left(\frac{\Lambda+E_{0}}{2}+\eta\right)\right)}{\ln\left(\frac{\Lambda-\left(\frac{\Lambda+E_{0}}{2}+\eta\right)}{\frac{\Lambda+E_{0}}{2}+\eta-E_{0}}\right)}} \nonumber \\
& = & \lim_{\eta \to 0^{+}}\sqrt{\frac{-2\pi(\Lambda+E_{0}-2M_{0}+2\eta)}{\ln\left(\frac{\Lambda-E_{0}-2\eta}{\Lambda-E_{0}+2\eta}\right)}} \nonumber \\
& = & \infty\sqrt{-\sgn\big((E_{0}-\Lambda)(\Lambda+E_{0}-2M_{0})\big)} \ .
\end{eqnarray}

\section{Spectral function and its normalization} \label{section_spectralfunctionnormalization}
\begin{figure}[!h]
\begin{center}
\includegraphics[width=192pt]{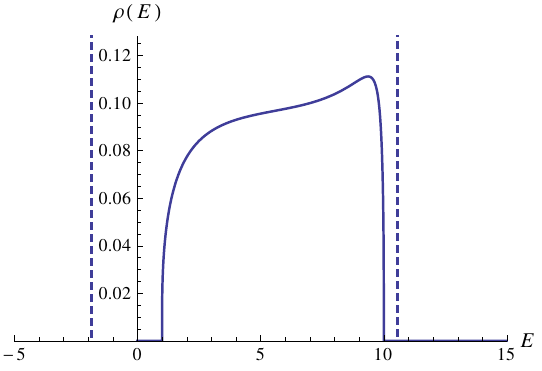}
\caption{Spectral function $\rho(E)$ of the considered Lee model for $g=6.0$.}
\label{figure_exampleleespectral}
\end{center}
\end{figure}

After having investigated the pole trajectories and the behaviour of the spectral function $\rho(E)$, let us now focus on the normalization of the latter. As was shown in Fig. \ref{figure_nonrel1dSpectralgNorm4} the spectral function is not normalized when using the numerical result of the integral
\begin{equation}
\frac{1}{\pi}\int_{-\infty}^{\infty}\text{d}\omega \ \rho(\omega) \ , \label{equation_normintegral}
\end{equation}
simply because the numerical calculation does not consider the two poles arising on the real axis in the first Riemann sheet.\footnote{The normalization integral from Eq. (\ref{equation_normintegral}), if calculated numerically, is computed only across the branch cut. Any extension to arbitrary bounds of integration beyond the cut does not change the result except for irrelevant fluctuations from the poles on the real axis. Hence, we write the bounds as plus and minus infinity.} It is possible to take them into account by splitting the Schr\"odinger propagator $G_{S}(E)$ into the contributions of two single-particle propagators and the remaining continuum part above the branch cut, similar as it was done in the first chapter:
\begin{eqnarray}
G_{S}(E) & = & \frac{1}{\pi}\int_{-\infty}^{\infty}\text{d}\omega \ \frac{\rho(\omega)}{E-\omega+i\epsilon} \nonumber \\
& = & \frac{Z_{\text{(l)pole}}}{E-x_{\text{(l)pole}}+i\epsilon}+\frac{Z_{\text{(r)pole}}}{E-x_{\text{(r)pole}}+i\epsilon}+\frac{1}{\pi}\int_{E_{0}}^{\Lambda}\text{d}\omega \ \frac{\rho(\omega)}{E-\omega+i\epsilon} \ , \label{equation_leepropagatorwithZ}
\end{eqnarray}
where $Z_{\text{(l)pole}}$ and $Z_{\text{(r)pole}}$ are some renormalization constants. We calculate them by expanding the inverse propagator in a Taylor series around the two poles on the real axis at $x=x_{\text{(l/r)pole}}$ in first order:
\begin{eqnarray}
x-M_{0}+g^{2}\Sigma(x) & \approx & \big(x-M_{0}+g^{2}\Sigma(x)\big)\big|_{x = x_{\text{(l/r)pole}}} \nonumber \\
& & \ + \ \frac{\text{d}}{\text{d}x}\Big(x-M_{0}+g^{2}\Sigma(x)\Big)\Big|_{x = x_{\text{(l/r)pole}}}\cdot(x-x_{\text{(l/r)pole}}) \nonumber \\
& = & \left(1+g^{2}\frac{\text{d}}{\text{d}x}\operatorname{Re}\Sigma(x)\right)\bigg|_{x=x_{\text{(l/r)pole}}}\cdot(x-x_{\text{(l/r)pole}}) \nonumber \\
& = & \left(1+\frac{g^{2}}{4\pi}\frac{\Lambda-E_{0}}{(\Lambda-x)(E_{0}-x)}\right)\bigg|_{x=x_{\text{(l/r)pole}}}\cdot(x-x_{\text{(l/r)pole}}) \ . \nonumber \\
\label{equation_inverseprop}
\end{eqnarray}
Since the two simple poles and the branch cut are separated on the real axis, the renormalization constants for each single-particle pole can be extracted from their residues:
\begin{eqnarray}
\lim_{x \to x_{\text{(l/r)pole}}}(x-x_{\text{(l/r)pole}})\cdot G_{S}(x) & = & \underbrace{\lim_{x \to x_{\text{(l/r)pole}}}(x-x_{\text{(l/r)pole}})\cdot\bigg[\frac{Z_{\text{(l/r)pole}}}{x-x_{\text{(l/r)pole}}+i\epsilon}}_{\residue\left(G_{S}(x), \ x=x_{\text{(l/r)pole}}\right)} \nonumber \\
& & + \ \big\{\text{other pole and branch cut}\big\}\bigg] \nonumber \\
& = & Z_{\text{(l/r)pole}} \ .
\end{eqnarray}
If we insert our expansion of the inverse propagator from Eq. (\ref{equation_inverseprop}) into the left side, we finally arrive at:
\begin{equation}
Z_{\text{(l/r)pole}} = \left[1+\frac{g^{2}}{4\pi}\frac{\Lambda-E_{0}}{(\Lambda-x_{\text{(l/r)pole}})(E_{0}-x_{\text{(l/r)pole}})}\right]^{-1} \ . \label{equation_renormconstant}
\end{equation}
Hence, the normalization condition can be written as
\begin{equation}
1 = Z_{\text{(l)pole}}+Z_{\text{(r)pole}}+\int_{E_{0}}^{\Lambda}\text{d}\omega \ \rho(\omega) \ ,
\end{equation}
which makes clear why the integral along the branch cut gives nearly the correct number for small couplings (where the renormalization constants are very small).
\begin{figure}[!h]
\begin{center}
\includegraphics{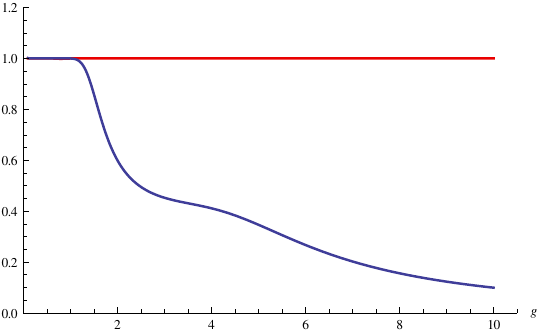}
\caption{Numerical verification that $\frac{1}{\pi}\int_{-\infty}^{\infty}\text{d}\omega \ \rho(\omega)=1$ (red) by including the single poles of the first Riemann sheet, and numerical contribution of the continuous part $\frac{1}{\pi}\int_{E_{0}}^{\Lambda}\text{d}\omega \ \rho(\omega)$ (dark blue). The necessity of the delta distribution functions is evident.}
\label{figure_nonrel1dSpectralgNorm6}
\end{center}
\end{figure}
If we perform our calculation including the renormalization constants, the spectral function is then normalized over the full range of $g$. In particular, sufficiently large numerical contributions from $Z_{\text{(l)pole}}$ are obtained for $g>1.2$ and from $Z_{\text{(r)pole}}$ for $g>2.7$.\footnote{This is also true if we perform the average mass integral in Eq. (\ref{equation_avmassintegral}) including the renormalization constants. Then, the value of $x_{\text{average}}$ is $M_{0}$ for the whole range of the coupling.} The plot in Fig. \ref{figure_nonrel1dSpectralgNorm6} shows the completely fulfilled normalization condition for our choice of parameters (the points used are those from Tab. \ref{table_allpoints}). For the sake of completeness, we provide a compilation of selected plots of the continuous part of the spectral function in \nameref{chapter_appendixE} from where one can review the behaviour described during this last section. An example for high couplings -- including the two poles in the first Riemann sheet -- can be found in Fig. \ref{figure_exampleleespectral}.

\section{Time evolution and non-exponential decay properties}
We are also interested in the survival amplitude $a(t)$ as denoted in Eq. (\ref{equation_Samplitude}) to make some statements about the decay of the unstable particle $S$ in view of the analytic structure of its propagator. We have found an explicit result for this propagator in the previous section. The survival amplitude is obtained by using a Hankel contour-like path of integration in the complex $E$-plane. We first extend the integration onto the whole complex plane and then close it counterclockwise in the lower half plane, while splitting the propagator in its single parts, see Fig. \ref{figure_hankel}. The only contributions to the amplitude come from the residues of the two simple poles and from the branch cut. In a more pedagogical way, one may also express the spectral function as the difference between the propagator across the cut (in the limit $\epsilon\rightarrow 0^{+}$) and integrate along the real axis in a first step from $-\infty$ to $+\infty$, in a second step from $+\infty$ to $-\infty$. Either way, the survival amplitude becomes
\begin{eqnarray}
a(t) & = & \frac{i}{2\pi}\int_{-\infty}^{+\infty}\text{d}E \ e^{-iEt}G_{S}(E) \ \ = \ \ \frac{i}{2\pi}\oint_{\mathcal{C}}\text{d}z \ e^{-izt}G_{S}(z) \nonumber \\
& = & \frac{i}{2\pi}\bigg[Z_{\text{(l)pole}}\oint_{\mathcal{C}_{\text{l}}}\text{d}z \ \frac{e^{-izt}}{z-x_{\text{(l)pole}}+i\epsilon}+Z_{\text{(r)pole}}\oint_{\mathcal{C}_{\text{r}}}\text{d}z \ \frac{e^{-izt}}{z-x_{\text{(r)pole}}+i\epsilon} \nonumber \\
&  & + \ \frac{1}{\pi}\int_{E_{0}}^{\Lambda}\text{d}\omega \ \rho(\omega)\oint_{\mathcal{C}^{\prime}}\text{d}z \ \frac{e^{-izt}}{z-\omega+i\epsilon}\bigg] \nonumber \\
& = & \frac{i}{2\pi}\bigg[-2\pi iZ_{\text{(l)pole}}\residue\left(\frac{e^{-izt}}{z-x_{\text{(l)pole}}+i\epsilon}, \ z=x_{\text{(l)pole}}\right) \nonumber \\
&  & - \ 2\pi iZ_{\text{(r)pole}}\residue\left(\frac{e^{-izt}}{z-x_{\text{(r)pole}}+i\epsilon}, \ z=x_{\text{(r)pole}}\right) \nonumber \\
&  & - \ 2i\int_{E_{0}}^{\Lambda}\text{d}\omega \ \rho(\omega)\residue\left(\frac{e^{-izt}}{z-\omega+i\epsilon}, \ z=\omega\right)\bigg] \nonumber \\
& = & Z_{\text{(l)pole}}e^{-ix_{\text{(l)pole}}t}+Z_{\text{(r)pole}}e^{-ix_{\text{(r)pole}}t}+\frac{1}{\pi}\int_{E_{0}}^{\Lambda}\text{d}\omega \ e^{-i\omega t}\rho(\omega) \ ,
\end{eqnarray}
and thus the survival probability $p(t)=|a(t)|^{2}$ consists of different terms:
\begin{figure}[t]
\begin{center}
\includegraphics[width=381pt]{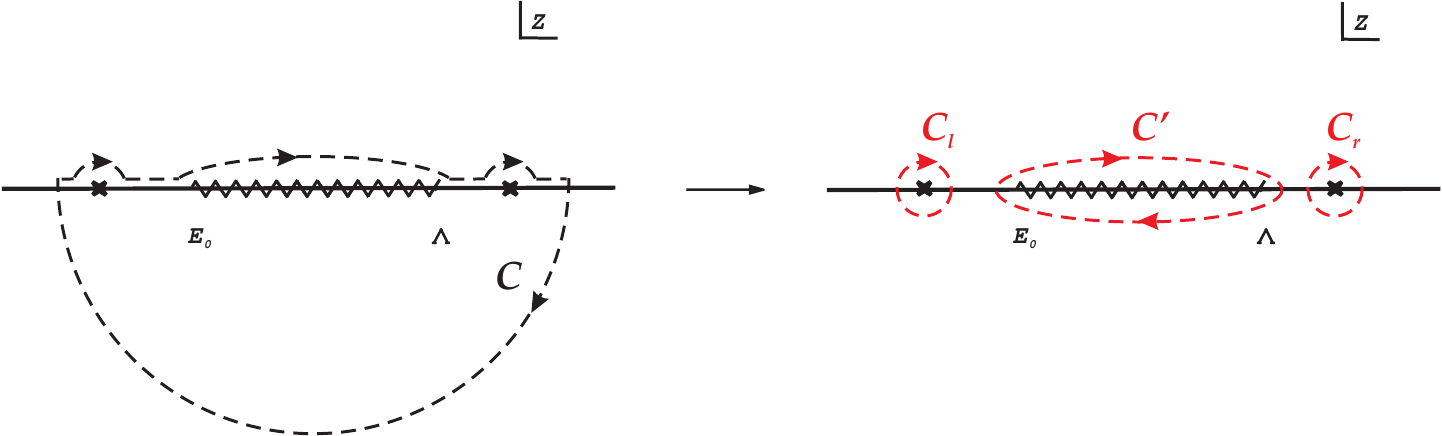}
\caption{Contour paths of integration $\mathcal{C}_{\text{l}}, \ \mathcal{C}_{\text{r}}, \ \text{and} \  \mathcal{C}^{\prime}$ (red, dashed) enclosing the two simple poles and the finite branch cut when blowing up the full contour $\mathcal{C}$ (black, dashed).}
\label{figure_hankel}
\end{center}
\end{figure}
First of all we get the squared renormalization constants with $g$-dependence belonging to stable particles, because the imaginary part of the two poles on the first Riemann sheet vanish. If we had instead just one simple pole with {\em finite} negative imaginary part at $z_{\text{pole}}=M-i\Gamma/2$ in the lower half plane, the well-known exponential decay law would have been reproduced:
\begin{eqnarray}
a(t) & = & \frac{i}{2\pi}\int_{-\infty}^{\infty}\text{d}E \ \frac{e^{-iEt}}{E-z_{\text{pole}}} \nonumber \\
& = & \frac{i}{2\pi}\oint_{\mathcal{C}}\text{d}z \ \frac{e^{-izt}}{z-z_{\text{pole}}} \nonumber \\
& = & e^{-iz_{\text{pole}}t} \ \ = \ \ e^{-iMt-\frac{\Gamma}{2}t} \ , \\
\Rightarrow \ \ \ p(t) & = & e^{-\Gamma t} \ . \label{equation_approxexp}
\end{eqnarray}
Note that in general all resonance poles lie on the second Riemann sheet. Pure exponential decays therefore cannot be realized exactly. This fact is due to the mathematical structure of the propagator: the self-energy $\Sigma(E)$ is a multi-valued complex function, creating a branch cut on the real axis whenever the spectral function gives continuous support in the integral representation of the propagator (as was already explained in the first chapter). All poles on the cut slip through it into the second sheet -- as a consequence, the survival amplitude will have a cut contribution (as well as contributions from other poles in the first sheet) which changes the familiar decay law. It is well-known that the exponential behaviour can be preserved by approximating the self-energy at its value on the real axis \cite{facchi}. In this case the resonance pole is assumed to be situated near and below the real axis, i.e., the imaginary part $y_{\text{pole}}=-\Gamma/2$ to be much smaller than the real part $x_{\text{pole}}=M$. The self-energy in the second Riemann sheet around the pole can then be taken at $z=M_{0}-\delta M+i\epsilon=M_{\text{BW}}+i\epsilon$ above the branch cut in the first sheet. This is obviously the Breit--Wigner approximation! Going over to general expressions, including a normalized form factor function $f^{2}(\textbf{k})/(2\pi)=\tilde{f}^{2}(\textbf{k})$, this means
\begin{eqnarray}
\Sigma_{\text{\Romannum{2}}}(z=M-i\Gamma/2) & \approx & \Sigma(z=M_{\text{BW}}+i\epsilon) \nonumber \\
& = & \mathcal{P}\int\text{d}k \ \frac{\tilde{f}^{2}(\textbf{k})}{k-M_{\text{BW}}}+i\pi\tilde{f}^{2}(M_{\text{BW}}) \ ,
\end{eqnarray}
and thus one can write for the Schr\"odinger propagator around the pole
\begin{eqnarray}
G_{S}(z) & \approx & \frac{1}{z-M_{0}+g^{2}\Sigma(M_{\text{BW}}+i\epsilon)} \ , \nonumber \\
& = & \frac{1}{x-M_{0}+g^{2}\operatorname{Re}\Sigma(M_{\text{BW}}+i\epsilon)+i\big(y+g^{2}\operatorname{Im}\Sigma(M_{\text{BW}}+i\epsilon)\big)} \ , \nonumber \\
\\
\Rightarrow \ \ \ \delta M & = & g^{2}\operatorname{Re}\Sigma(M_{\text{BW}}+i\epsilon) \ , \\
\Gamma_{\text{BW}} & = & 2g^{2}\operatorname{Im}\Sigma(M_{\text{BW}}+i\epsilon) \ \ = \ \ 2\pi g^{2}\tilde{f}^{2}(M_{\text{BW}})\ ,
\end{eqnarray}
where the last relation turns out to be {\em Fermi's golden rule} for the decay width (sometimes $M_{\text{BW}}$ is replaced by $M_{0}$). If we make the unphysical assumption that the energy spectrum of the full Hamiltonian has no lower bound (and no upper bound), a non-relativistic Breit--Wigner distribution is immediately obtained for the corresponding spectral function \cite{ghirardi}. On the other hand, since the spectral function displays a pole in the second sheet right below the real axis as a relatively sharp peak, the branch cut contribution can be replaced by a simple pole in the lower half plane of the first sheet, yielding the exponential decay law (\ref{equation_approxexp}) for the survival probability $p(t)$.

Besides the branch cut, we need to take the real poles on the first sheet with real renormalization constants into account. The complete survival proability reads:
\begin{eqnarray}
p(t) & = & Z^{2}_{\text{(l)pole}}+Z^{2}_{\text{(r)pole}}+Z_{\text{(l)pole}}Z_{\text{(r)pole}}\Big(e^{i(x_{\text{(r)pole}}-x_{\text{(l)pole}})t}+e^{-i(x_{\text{(r)pole}}-x_{\text{(l)pole}})t}\Big) \nonumber \\
&  & + \ \frac{1}{\pi}\Big(Z_{\text{(l)pole}}e^{ix_{\text{(l)pole}}t}+Z_{\text{(r)pole}}e^{ix_{\text{(r)pole}}t}\Big)\int_{E_{0}}^{\Lambda}\text{d}\omega \ e^{-i\omega t}\rho(\omega) \nonumber \\
&  & + \ \frac{1}{\pi}\Big(Z_{\text{(l)pole}}e^{-ix_{\text{(l)pole}}t}+Z_{\text{(r)pole}}e^{-ix_{\text{(r)pole}}t}\Big)\bigg[\int_{E_{0}}^{\Lambda}\text{d}\omega \ e^{-i\omega t}\rho(\omega)\bigg]^{*} \nonumber \\
& & + \ \frac{1}{\pi^{2}}\bigg[\int_{E_{0}}^{\Lambda}\text{d}\omega \ e^{-i\omega t}\rho(\omega)\bigg]\bigg[\int_{E_{0}}^{\Lambda}\text{d}\omega \ e^{-i\omega t}\rho(\omega)\bigg]^{*} \ .
\end{eqnarray}
After introducing
\begin{equation}
\mathcal{S} = \int_{E_{0}}^{\Lambda}\text{d}\omega \ e^{-i\omega t}\rho(\omega) \ , \ \ \ \ \ \omega_{\text{rl}} = x_{\text{(r)pole}}-x_{\text{(l)pole}} \ ,
\end{equation}
this becomes the more compact expression
\begin{eqnarray}
p(t) & = & Z^{2}_{\text{(l)pole}}+Z^{2}_{\text{(r)pole}}+2Z_{\text{(l)pole}}Z_{\text{(r)pole}}\cos(\omega_{\text{rl}}t) \nonumber \\
&  & + \ \frac{2}{\pi}\operatorname{Re}\mathcal{S}\Big(Z_{\text{(l)pole}}\cos(x_{\text{(l)pole}}t)+Z_{\text{(r)pole}}\cos(x_{\text{(r)pole}}t)\Big) \nonumber \\
&  & - \ \frac{2}{\pi}\operatorname{Im}\mathcal{S}\Big(Z_{\text{(l)pole}}\sin(x_{\text{(l)pole}}t)+Z_{\text{(r)pole}}\sin(x_{\text{(r)pole}}t)\Big)+\frac{1}{\pi^{2}}|\mathcal{S}|^{2} \ . \label{equation_survivalamp}
\end{eqnarray}
In the following Fig. \ref{figure_survivalamplitudes} we show the full survival amplitude $p(t)$ from Eq. (\ref{equation_survivalamp}) (red), the branch cut contribution $\mathcal{S}$ (dark blue) and the single pole approximation as denoted in Eq. (\ref{equation_approxexp}) (green, dashed) for some values of the coupling $g$. We observe:
\begin{itemize}
\item Small couplings $g\ll1$:

The renormalization constants and all terms containing at least one of them are negligible since the self-energy has sharp cusps at the branch points, making the expression (\ref{equation_renormconstant}) zero. The survival probability consists mainly of the branch cut contribution $\mathcal{S}$, which can be approximated as an exponential function. In the limit of very small times we realize $p(t)$ to flatten with vanishing first derivative -- a feature observable in the full range of the coupling and closely related to the famous {\em quantum Zeno effect} \cite{sudarshan,sakurai,itano,raizen,balzer}.
\item Sizable contribution from left pole, $g=1.4$:

The pure exponential law does not hold any more, and the branch cut contribution also does not give the full survival probability. This is clear because the left pole on the first Riemann sheet brings a relevant renormalization constant into play and hence the probability (though very small) stays nearly constant for large times as a direct manifestation of the first term in Eq. (\ref{equation_survivalamp}). Oscillations given by the cosinus of the fourth and the sinus of the fifth term are very weak and finally vanish due to the real part of the branch cut contribution tending to zero.
\item Sizable contributions from both poles, $g>2.8$:

For large times the survival probability $p(t)$ oscillates with the frequency $\omega_{\text{rl}}$ between the amplitude values $Z_{\text{(l)pole}}^{2}+Z_{\text{(r)pole}}^{2}+2Z_{\text{(l)pole}}Z_{\text{(r)pole}}$ and $Z_{\text{(l)pole}}^{2}+Z_{\text{(r)pole}}^{2}-2Z_{\text{(l)pole}}Z_{\text{(r)pole}}$. The intermediate oscillating behaviour with a much lower frequency is due to the interplay of the real and imaginary parts of the branch cut contribution in the last three terms. This transient overall oscillation is not important for the evolution of the system and tends to zero, because the spectral function $\rho(\omega)$ goes like $\mathcal{O}(g^{-2})$ in the highest order for huge couplings.
\end{itemize}

\newpage

\begin{figure}
\begin{minipage}[hbt]{6.2cm}
\centering
\includegraphics[width=6.1cm]{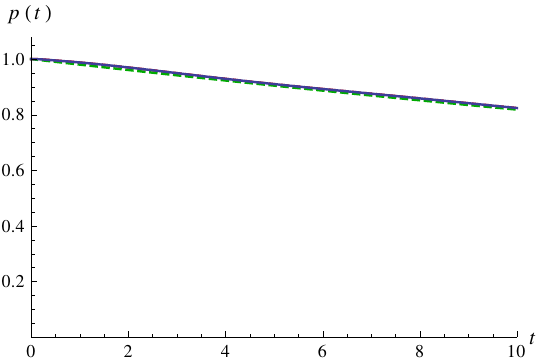}
\caption*{\footnotesize $g=0.2$}
\label{figure_nonrel1dEvolutiong015-2}
\end{minipage}
\hfill
\begin{minipage}[hbt]{6.2cm}
\centering
\includegraphics[width= 6.1cm]{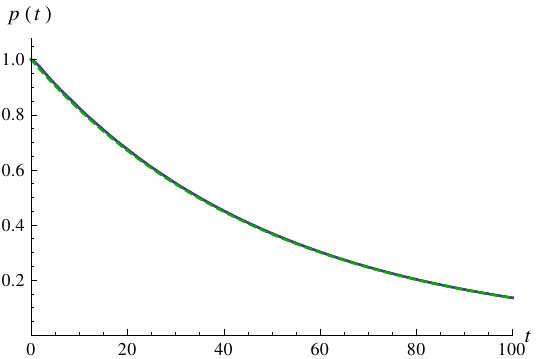}
\caption*{\footnotesize $g=0.2$}
\label{figure_nonrel1dEvolutiong015-3}
\end{minipage}
\end{figure}
\begin{figure}
\begin{minipage}[hbt]{6.2cm}
\centering
\includegraphics[width= 6.1cm]{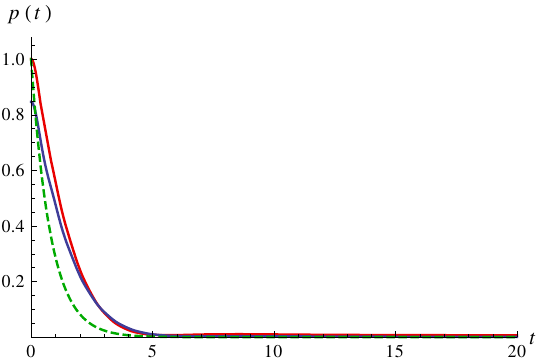}
\caption*{\footnotesize $g=1.4$}
\label{figure_nonrel1dEvolutiong10-1}
\end{minipage}
\hfill
\begin{minipage}[hbt]{6.2cm}
\centering
\includegraphics[width= 6.1cm]{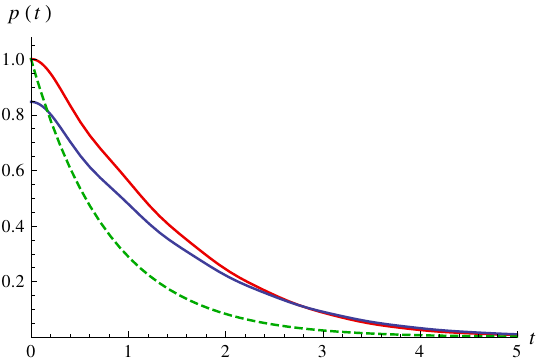}
\caption*{\footnotesize $g=1.4$}
\label{figure_nonrel1dEvolutiong10-3}
\end{minipage}
\end{figure}
\begin{figure}
\begin{minipage}[hbt]{6.2cm}
\centering
\includegraphics[width= 6.1cm]{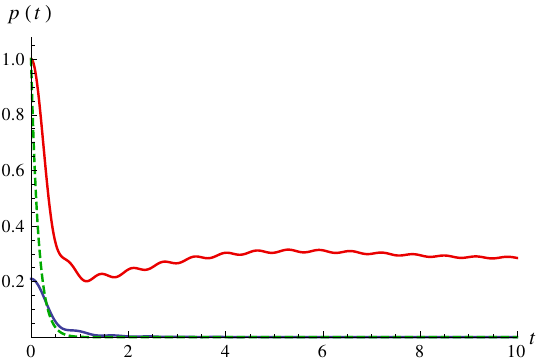}
\caption*{\footnotesize $g=2.9$}
\label{figure_nonrel1dEvolutiong25-1}
\end{minipage}
\hfill
\begin{minipage}[hbt]{6.2cm}
\centering
\includegraphics[width= 6.1cm]{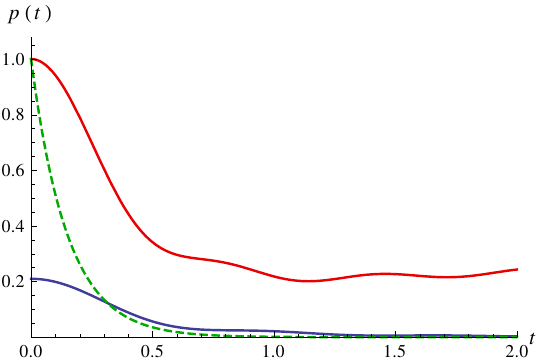}
\caption*{\footnotesize $g=2.9$}
\label{figure_nonrel1dEvolutiong25-2}
\end{minipage}
\end{figure}
\begin{figure}
\begin{minipage}[hbt]{6.2cm}
\centering
\includegraphics[width= 6.1cm]{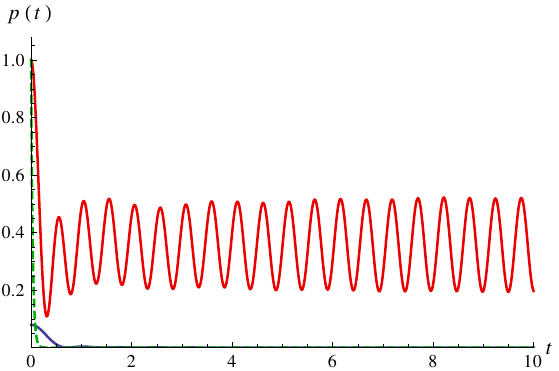}
\caption*{\footnotesize $g=5.8$}
\label{figure_nonrel1dEvolutiong45-1}
\end{minipage}
\hfill
\begin{minipage}[hbt]{6.2cm}
\centering
\includegraphics[width= 6.1cm]{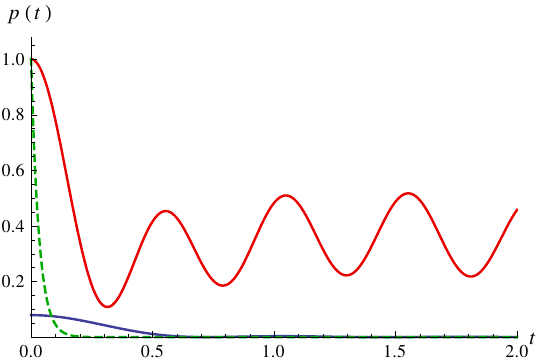}
\caption*{\footnotesize $g=5.8$}
\label{figure_nonrel1dEvolutiong45-2}
\end{minipage}
\caption{Survival amplitudes $p(t)$ for the non-relativistic Lee model  (varying scales).}
\label{figure_survivalamplitudes}
\end{figure}

\newpage
\clearpage

\section{Concluding remarks of the chapter}
The Lee model presented here provides us with a variety of interesting deviations from the general transition properties of unstable quantum systems as induced by the purely exponential decay law:

\begin{enumerate}
\item Although the Breit--Wigner parameterization reproduces very well the mass (and decay width) of the resonance pole in the regime of small couplings, there exists no such mass (and width) for intermediate values of $g$, compare Tab. \ref{table_allpoints}. Above this regime, the mass (and width) is completely detached from the pole values.
\item The spectral function describes the pole motion adequately through its maximum $x_{\text{(r)max}}=x_{\text{(m)BW}}$ and full width at half maximum as long as the two newly arising poles can be neglected. Otherwise, the maximum does not correlate with the real part of the resonance pole in the second Riemann sheet.
\item The left and right poles have their origin in the form of the self-energy. While the right pole disappears after removing the right branch point by sending $\Lambda\rightarrow\infty$ (and performing a subtraction), the other pole would remain. We interpret the latter as a bound state emerging from the strong coupling to the continuum of states. One should stress that the whole model is academic, but serves to demonstrate how to perform the analytic continuation into the second Riemann sheet. At least smooth cutoff functions have to be used since the right threshold (and consequently the right simple pole in the first sheet) is unphysical.
\item Deviations from the decay law can be obtained for small and large times due to the branch cut contribution. However, we do not find the theoretical power law \cite{ghirardi} for very large times. If present in that regime, it is covered by the dominant oscillations coming from the left and right pole.
\item The two poles in the first Riemann sheet are {\em dynamically generated} by the interaction term appearing in the full Hamiltonian. There is no evidence for other resonance poles created in the second sheet.
\end{enumerate}

 \label{chapter_chapter3}
\clearpage

\thispagestyle{empty}
\

\newpage
\clearpage

\chapter{Quantum field theory of resonances}

\medskip

\section{Construction of the $S\phi\phi$-model}
\subsection{Lagrangian}
\begin{figure}[h]
\centering
\includegraphics[scale=0.7]{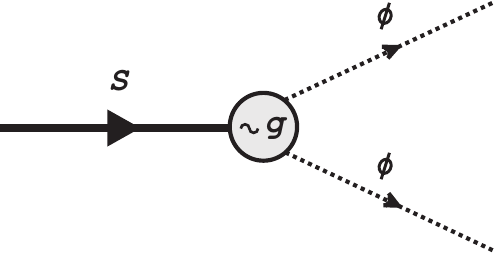}
\caption{Schematic decay process $S\rightarrow\phi\phi$.}
\label{figure_Sphiphidecay}
\end{figure}

In this chapter we study a quantum field theoretical model with two scalar fields, $S$ and $\phi$, in $d=3+1$ dimensions. Such a model can be applied to many problems in the realm of hadron physics, most notably it is possible to investigate the resonance pole of the $\sigma$-meson. The same model was already studied by Veltman \cite{veltman} as well as by Giacosa and Pagliara \cite{giacosaSpectral,giacosaHadrons} regarding its spectral function(s): it contains an interaction term for a one-channel decay process $S\rightarrow\phi\phi$ (see Fig. \ref{figure_Sphiphidecay}), described by a Lagrangian of the form
\begin{equation}
\mathcal{L} = \frac{1}{2}(\partial_{\mu}S)(\partial^{\mu}S)+\frac{1}{2}(\partial_{\mu}\phi)(\partial^{\mu}\phi)-\frac{1}{2}M_{0}^{2}S^{2}-\frac{1}{2}m^{2}\phi^{2}+gS\phi^{2} \ . \label{equation_lagrangian} \\
\end{equation}
Here, the (pseudo)scalar $\phi$-fields represent pions. The equation of motion for the $\sigma$-meson alias the field $S$,
\begin{equation}
(\Box+M_{0}^{2})S = g\phi^{2} \ ,
\end{equation}
is of course the solution of the Euler--Lagrange equation
\begin{equation}
\frac{\partial\mathcal{L}}{\partial S}-\frac{\partial}{\partial x^{\mu}}\frac{\partial\mathcal{L}}{\partial(\partial_{\mu}S)} = 0 \ .
\end{equation}
The corresponding Feynman propagator in position space for vanishing coupling $g=0$ is defined as in the first chapter,
\begin{eqnarray}
\Delta_{S}^{\text{free}}(x-y) & = & -i\langle0|\mathcal{T}\big{\{}S(x)S(y)\big{\}}|0\rangle \nonumber \\
& = & \int\frac{\text{d}^{4}p}{(2\pi)^{4}} \ e^{-ip\cdot(x-y)}\frac{1}{p^{2}-M_{0}^{2}+i\epsilon} \ , \label{equation_positionprop}
\end{eqnarray}
where $\mathcal{T}$ is the time-ordering operator. Hence, the propagator in momentum space becomes simply:
\begin{equation}
\Delta_{S}^{\text{free}}(p^{2}) = \frac{1}{p^{2}-M_{0}^{2}+i\epsilon} \ .
\end{equation}
One should mention that the theory described by the above equations is known to be {\em super-renormalizable}. In fact, the superficial divergence $D$ in terms of the number of vertices $n$ and the number of external lines is proportianal to $-n$, so the number of divergent diagrams is finite.

\subsection{Tree-level decay width} \label{subsection_decaywidth}
The decay width of the resonance $S$ can be calculated by evaluating the invariant $\hat S$-matrix element of the process $S\rightarrow\phi\phi$ in first order in the coupling. In the case of our simple interaction term $\mathcal{L}_{\text{int}}=gS\phi^{2}$ the $\hat S$-matrix can be immediately written down as
\begin{eqnarray}
\hat{S} & = & \mathbbm{1}+ig\int\text{d}^{4}x \ \mathcal{T}\big{\{}\textbf{:}S(x)\phi^{2}(x)\textbf{:}\big{\}} \nonumber \\
& = & \mathbbm{1}+\hat{S}^{(1)} \ ,
\end{eqnarray}
where the dots mark the normal ordering prescription. The crossed-out terms in the resulting transition matrix element,
\begin{eqnarray}
\langle\text{final}|\hat{S}|\text{initial}\rangle & = & \langle\textbf{p}_{1}\textbf{p}_{2}|\hat{S}^{(1)}|\textbf{p}\rangle \nonumber \\
& = & \langle\textbf{p}_{1}\textbf{p}_{2}| \ ig\int\text{d}^{4}x \ \textbf{:} \ \xcancel{S^{(+)}\phi^{(+)}\phi^{(+)}} \nonumber \\
&  & + \ \xcancel{S^{(+)}\phi^{(+)}\phi^{(-)}}+\xcancel{S^{(+)}\phi^{(-)}\phi^{(+)}} \nonumber \\
&  & + \ \xcancel{S^{(+)}\phi^{(-)}\phi^{(-)}}+S^{(-)}\phi^{(+)}\phi^{(+)} \nonumber \\
&  & + \ \xcancel{S^{(-)}\phi^{(+)}\phi^{(-)}}+\xcancel{S^{(-)}\phi^{(-)}\phi^{(+)}} \nonumber \\
&  & + \ \xcancel{S^{(-)}\phi^{(-)}\phi^{(-)}} \ \textbf{:} \ |\textbf{p}\rangle \ ,
\end{eqnarray}
give no contribution because the creation and annihilation operators combine in such a way that their scalar product vanishes. The superscript at the $S$- and $\phi$-fields denotes those parts of the field that contain a creation or annihilation operator, e.g.
\begin{eqnarray}
\phi^{(+)} & \equiv & \phi^{(+)}(x) \nonumber \\
& = &  \int\frac{\text{d}^{3}p}{(2\pi)^{3}}\frac{1}{\sqrt{2E_{\textbf{p}}}} \ b_{\textbf{p}}^{\dagger}e^{ip\cdot x} \ .
\end{eqnarray}
Writing out the Fourier expansion of the fields in full detail and using general commutation relations for the bosonic creation and annihilation operators, the remaining steps are straightforward:
\begin{eqnarray}
& = & ig\int\text{d}^{4}x\int\text{d}^{3}p_{1}^{\prime}\int\text{d}^{3}p_{2}^{\prime}\int\text{d}^{3}p^{\prime} \ \frac{\sqrt{2E_{\textbf{p}_{1}}2E_{\textbf{p}_{2}}2E_{\textbf{p}}}}{\sqrt{2E_{\textbf{p}_{1}^{\prime}}2E_{\textbf{p}_{2}^{\prime}}2E_{\textbf{p}^{\prime}}}(2\pi)^{9}} \nonumber \\
&  & \times \ e^{i(p_{1}^{\prime}+p_{2}^{\prime}-p^{\prime})\cdot x}\langle0|b_{\textbf{p}_{2}}b_{\textbf{p}_{1}}b_{\textbf{p}_{1}^{\prime}}^{\dagger}b_{\textbf{p}_{2}^{\prime}}^{\dagger}a_{\textbf{p}^{\prime}}a_{\textbf{p}}^{\dagger}|0\rangle \nonumber \\
& = & ig\int\text{d}^{4}x\int\text{d}^{3}p_{1}^{\prime}\int\text{d}^{3}p_{2}^{\prime}\int\text{d}^{3}p^{\prime} \ \frac{\sqrt{2E_{\textbf{p}_{1}}2E_{\textbf{p}_{2}}2E_{\textbf{p}}}}{\sqrt{2E_{\textbf{p}_{1}^{\prime}}2E_{\textbf{p}_{2}^{\prime}}2E_{\textbf{p}^{\prime}}}} \nonumber \\
&  & \times \ e^{i(p_{1}^{\prime}+p_{2}^{\prime}-p^{\prime})\cdot x}\Big\{\delta^{(3)}(\textbf{p}^{\prime}-\textbf{p})\delta^{(3)}(\textbf{p}_{1}-\textbf{p}_{2}^{\prime})\delta^{(3)}(\textbf{p}_{2}-\textbf{p}_{1}^{\prime}) \nonumber \\
&  & + \ \ \delta^{(3)}(\textbf{p}^{\prime}-\textbf{p})\delta^{(3)}(\textbf{p}_{1}-\textbf{p}_{1}^{\prime})\delta^{(3)}(\textbf{p}_{2}-\textbf{p}_{2}^{\prime})\Big\} \nonumber \\
& = & 2ig\int\text{d}^{4}x \ e^{i(p_{1}+p_{2}-p)\cdot x} \nonumber \\
& = & 2ig(2\pi)^{4}\delta^{(4)}(p_{1}+p_{2}-p) \nonumber \\
& \stackrel{!}{=} & -i\mathcal{M}(2\pi)^{4}\delta^{(4)}(p-p_{1}-p_{2}) \ , \\
\Rightarrow \ \ \ -i\mathcal{M} & = & 2ig \ .
\end{eqnarray}
The decay width is now obtained by performing the phase space integral over the invariant amplitude $-i\mathcal{M}$. The infinitesimal expressions read:
\begin{eqnarray}
\text{d}\Gamma & = & \frac{\mathcal{S}}{2M}|\textbf{--\hskip 0.05cm}i\mathcal{M}|^{2}\text{d}\varphi_{n} \ , \\
\text{d}\varphi_{n} & = & (2\pi)^{4}\delta^{(4)}\bigg(p-\sum_{i=1}^{n}p_{i}\bigg)\bigg(\prod_{i=1}^{n}\frac{\text{d}^{3}p_{i}}{(2\pi)^{3}2E_{i}}\bigg) \ ,
\end{eqnarray}
where $\mathcal{S}$ is a symmetry factor and $n$ is the number of particles created. For a general discussion of these formulas see for example Ref. \cite{gross}. In our case the symmetry factor is just one half -- since the directions of the outgoing momenta are determined by conservation laws, only a half sphere in position space needs to be taken into account. There is no angular dependence as a consequence of Lorentz invariance: spinless particles at rest have no preferred direction in which they decay into two other particles. We then quickly arrive at the phase space integral for the tree-level decay width,
\begin{eqnarray}
\Gamma_{\text{tree}}(\sqrt{p^{2}}=M) & = & \frac{g^{2}}{(2\pi)^{2}M}\int\text{d}^{3}p_{1}\int\text{d}^{3}p_{2} \ \frac{1}{2E_{\textbf{p}_{1}}2E_{\textbf{p}_{2}}}\underbrace{\delta^{(4)}(p-p_{1}-p_{2})}_{=\delta^{(3)}(\textbf{p}_{1}+\textbf{p}_{2})\delta(M-E_{\textbf{p}_{1}}-E_{\textbf{p}_{2}})} \nonumber \\
& = & \frac{g^{2}}{(2\pi)^{2}M}\int\text{d}^{3}p_{1} \ \frac{1}{(2E_{\textbf{p}_{1}})^{2}} \ \delta(M-2E_{\textbf{p}_{1}}) \nonumber \\
\nonumber \\
& = & \frac{4\pi g^{2}}{(2\pi)^{2}M}\int_{0}^{\infty}\text{d}u \ \frac{Mu^{2}}{16(u^{2}+m^{2})\sqrt{\frac{M^{2}}{4}-m^{2}}} \ \delta\big(u-\sqrt{M^{2}/4-m^{2}}\big) \ , \nonumber \label{equation_Gammaint} \\
\end{eqnarray}
where $u=|\textbf{p}_{1}|$. A real-valued expression requires a threshold value $2m$, so we add a Heaviside step function to the preliminary result
\begin{equation}
\Gamma_{\text{tree}}(M) = \frac{g_{S\phi\phi}^{2} \ p_{S\phi\phi}}{8\pi M^{2}} \ \Theta(M-2m) \ . \label{equation_preliminarydecaywidth}
\end{equation}
Here, we have redefined the coupling as $g_{S\phi\phi}=\sqrt{2}g$ and introduced the magnitude of the three-momentum $|\textbf{p}_{1}|=|\textbf{p}_{2}|=p_{S\phi\phi}=\sqrt{M^{2}/4-m^{2}}$ of the two outgoing $\phi$-particles. The latter is derived from ordinary kinematics, see \nameref{chapter_appendixC} for more details.

\subsection{Self-energy and interacting propagator}
In a free field theory, as was explained in the first chapter, the propagator (\ref{equation_positionprop}) has the simple interpretation of being the amplitude for a stable particle $S$ to propagate from $y$ to $x$. This is different if we turn on the interaction: the field $S$ cannot be considered anymore as an asymptotic state of the Lagrangian since loop contributions given by strongly coupled hadronic intermediate states dominate its dynamics \cite{pennington}.
\begin{figure}[h]
\centering
\includegraphics[scale=0.6]{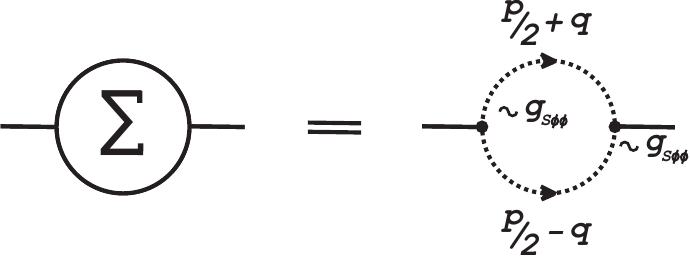}
\caption{Self-energy $\Sigma(p^{2})$ as a mesonic loop.}
\label{figure_Sigma}
\end{figure}
They appear in the full interacting propagator of the field $S$, so that the free propagator is modified:
\begin{equation}
\Delta_{S}(p^{2}) = \frac{1}{p^{2}-M_{0}^{2}+g_{S\phi\phi}^{2}\Sigma(p^{2})} \ . \label{equation_momentumfullprop}
\end{equation}
The emerging self-energy function $\Sigma(p^{2})$ in the denominator can be represented by a one-particle-irreducible (1PI) diagram in momentum space with two $\phi$-particles inside a mesonic loop where the incoming and outgoing momentum is denoted by $p$, see Fig. \ref{figure_Sigma}. The full propagator (\ref{equation_momentumfullprop}) is obtained by a resummation of the one-loop diagrams, as shown on the next page, and separating its inverse by multiplying with $\big(\Delta_{S}^{\text{free}}(p^{2})\big)^{-1}$ from the left and $\big(\Delta_{S}(p^{2})\big)^{-1}$ from the right. From general Feynman rules\footnote{We adopt the Feynman rules used by Peskin and Schroeder \cite{peskin}, yet the minus sign in front of the self-energy contribution $\Pi(p^{2})$ appearing in the denominator of the propagator (\ref{equation_dressesprop}) is absorbed into $\Sigma(p^{2})$. It is therefore $\Pi(p^{2})=-g_{S\phi\phi}^{2}\Sigma(p^{2})$.} for scalar fields we get the following expression for the self-energy:
\begin{equation}
\Sigma(p^{2}) = -i\int\frac{\text{d}^{4}q}{(2\pi)^{4}}\frac{f_{\Lambda}^{2}(q)}{\left(\frac{p}{2}+q\right)^{2}-m^{2}+i\epsilon}\frac{1}{\left(\frac{p}{2}-q\right)^{2}-m^{2}+i\epsilon} \ .
\end{equation}
\begin{figure}[h]
\centering
\includegraphics[scale=0.4]{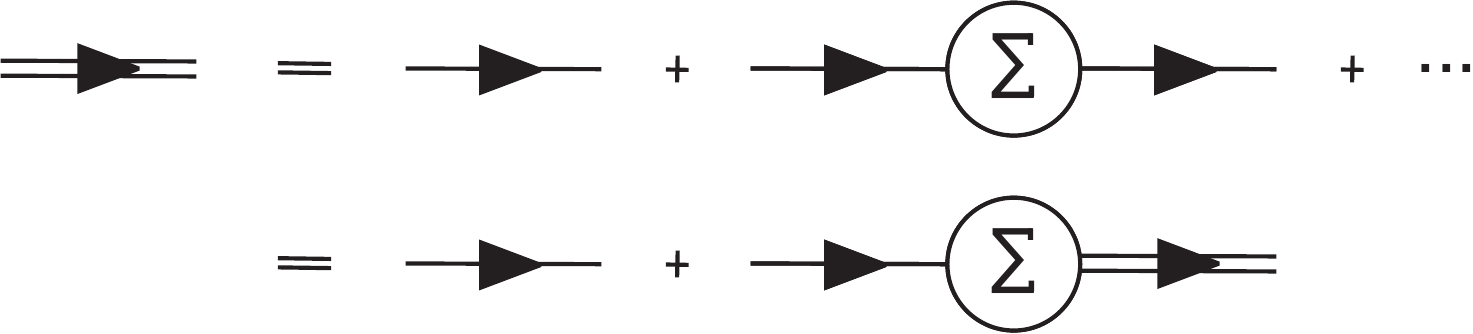}
\caption{Modification of the propagator by resummation of 1PI diagrams. The free propagator is represented by a solid line with an arrow and the interacting one by a double line with an arrow.}
\label{figure_propdiagram}
\end{figure}
\\
The regularization function $f_{\Lambda}(q)$, which depends on a UV cutoff scale $\Lambda$, was introduced to make otherwise logarithmic divergent integral finite. Since we deal with an effective Lagrangian in the low-energy regime to study light mesons, the mass scale of our model is determined: it is reasonable to set $\Lambda$ between 1 and 2 GeV, yet it should be possible to perform a full detailed study and eliminate any dependence on the cutoff. The integral is then evaluated by assuming the regularization function to depend {\em only} on the magnitude of the three-momentum $\textbf{q}$. Actually, this violates Lorentz symmetry\footnote{See also Ref. \cite{giacosaFermions} for more details on this aspect.}, but we accept this drawback in order to arrive at simple analytic expressions for $\Sigma(p^{2})$. First we split the $q$-integration into its spatial and time part, continue the latter into the complex and perform the integration in the rest frame of the particle $S$ with mass $\sqrt{p^{2}}=M$ by making use of the residue theorem:
\begin{eqnarray}
\Sigma(\sqrt{p^{2}}=M) & = & -i\int\frac{\text{d}^{3}q}{(2\pi)^{3}} \ f_{\Lambda}^{2}(|\textbf{q}|) \nonumber \\
&  & \times\int\frac{\text{d}q^{0}}{2\pi}\underbrace{\frac{1}{\left(\frac{M}{2}+q^{0}\right)^{2}-|\textbf{q}|^{2}-m^{2}+i\epsilon}}_{(1)}\underbrace{\frac{1}{\left(\frac{M}{2}-q^{0}\right)^{2}-|\textbf{q}|^{2}-m^{2}+i\epsilon}}_{(2)} \ , \nonumber \\
\end{eqnarray}
from this we have simple poles
\begin{eqnarray}
(1) \ \ \ q^{0}_{1/2} & = & -\frac{M}{2}\pm\sqrt{|\textbf{q}|^{2}+m^{2}-i\epsilon} \ \ = \ \ -\frac{M}{2}\pm\sqrt{|\textbf{q}|^{2}+m^{2}}\mp i\epsilon \ , \\
(2) \ \ \ q^{0}_{1/2} & = & \frac{M}{2}\pm\sqrt{|\textbf{q}|^{2}+m^{2}-i\epsilon} \ \ = \ \ \frac{M}{2}\pm\sqrt{|\textbf{q}|^{2}+m^{2}}\mp i\epsilon \ .
\end{eqnarray}
Because the final result does not depend on how we close the contour, only two poles need to be considered, namely those with negative root, situated above the real $q^{0}$-axis. They give a path of integration circling counterclockwise in the upper half plane. The residues
\begin{eqnarray}
\residue\big(\Sigma(M), \ z=q_{2(1)}^{0}\big) & = & -\frac{1}{2M\sqrt{|\textbf{q}|^{2}+m^{2}}\big(M+2\sqrt{|\textbf{q}|^{2}+m^{2}}\big)} \ , \\
\residue\big(\Sigma(M), \ z=q_{1(2)}^{0}\big) & = & -\frac{1}{2M\sqrt{|\textbf{q}|^{2}+m^{2}}\big(M-2\sqrt{|\textbf{q}|^{2}+m^{2}}\big)} \ ,
\end{eqnarray}
can be combined and, after performing the polar and azimuthal integration of the spatial part in spherical coordinates, the self-energy reads:
\begin{equation}
\Sigma(M) = \frac{1}{2\pi^{2}}\int_{0}^{\infty}\text{d}u \ \frac{u^{2}f_{\Lambda}^{2}(u)}{\sqrt{u^{2}+m^{2}}\big(4(u^{2}+m^{2})-M^{2}-i\epsilon\big)} \ , \label{equation_SigmaIntegral}
\end{equation}
where we have shifted the integration path slightly into the lower half plane of the complex $u$-plane ($u=|\textbf{q}|$).

Our preliminary result for the tree-level decay width in Eq. (\ref{equation_preliminarydecaywidth}) is surely correct, but it is crucial to realize that the cutoff does not exist in our model on the Lagrangian level. It can be implemented by taking a non-local interaction term into account,
\begin{equation}
\mathcal{L}_{\text{int}} = gS(x)\phi^{2}(x) \ \rightarrow \ \mathcal{L}_{\text{int}} = gS(x)\int\text{d}^{4}y \ \phi(x+y/2)\phi(x-y/2)\Phi(y) \ , \label{equation_nonlocalL}
\end{equation}
where the regularization function appears as the Fourier transform of $\Phi(y)$ in the loop integral and 'heals\grq \ the divergence in the case of $f_{\Lambda}(q)=f_{\Lambda}(|\textbf{q}|)$ \cite{giacosaSpectral}. This changes also the tree-level result for the width such that
\begin{equation}
\Gamma_{\text{tree}}(M) \ \rightarrow \ \Gamma_{\text{tree}}(M)\cdot f_{\Lambda}^{2}\big(p_{S\phi\phi}=\sqrt{M^{2}/2-m^{2}}\big) \ , \label{equation_finaldecaywidth}
\end{equation}
which is demonstrated explicitly in \nameref{chapter_appendixD}.

\section{Sharp cutoff}
\subsection{Analytic structure of the propagator}
In order to proceed we need to determine $f_{\Lambda}(|\textbf{q}|)$. Two different choices for the regularization functions will be studied in this work and for the sake of simplicity we start with a sharp cutoff,
\begin{equation}
f_{\Lambda}(|\textbf{q}|) = \Theta(\Lambda^{2}-|\textbf{q}|^{2}) \ .
\end{equation}
Inserting this function into Eq. (\ref{equation_SigmaIntegral}), we are left with a non-trivial integral, which nevertheless has an analytic solution\footnote{One should never forget that in order to look for poles in the second Riemann sheet in the same way as was done for the non-relativistic Lee model in the previous chapter, we really need an {\em analytic expression} for the self-energy. Otherwise, the continuation into the complex plane has to be performed by a (finite) series expansion, while the continuation into the second sheet would be fully non-trivial. This way can be avoided by computing the loop integral solely numerically (while numerical distortions in the vicinity of the branch cut have to be managed).} (that is found by {\em Mathematica} if we set $\epsilon=0$):
\begin{eqnarray}
\Sigma(M) = -\frac{\sqrt{4m^{2}-M^{2}}}{8\pi^{2}M}\arctan\left(\frac{\Lambda M}{\sqrt{\Lambda^{2}+m^{2}}\sqrt{4m^{2}-M^{2}}}\right)-\frac{1}{8\pi^{2}}\ln\left(\frac{m}{\Lambda+\sqrt{\Lambda^{2}+m^{2}}}\right) \ . \nonumber \\
\label{equation_relSigma}
\end{eqnarray}
Since this function has only a physical meaning for $\sqrt{p^{2}}=M>0$, we will not be concerned in what is happening for negative values of the mass.
\begin{figure}[!h]
\begin{center}
\includegraphics{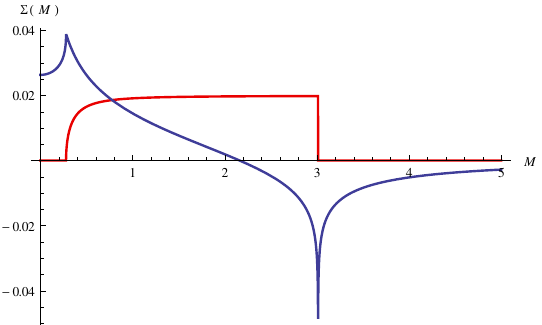}
\caption{Real (dark blue) and imaginary part (red) of the self-energy $\Sigma(M)$ on the positive real axis with $\Lambda=1.5 \ \text{GeV}$.}
\label{figure_rel4dSigma}
\end{center}
\end{figure}
This is really an important point to mention: in the literature, people usually work in the complex $p^{2}$-plane and indeed, after all calculations in this work were done, we found that it is much easier to do so as well.

A first hint of the multi-valued character of the above self-energy function is given by the root. Clearly, there should be a branch cut in the complex $M$-plane ($\sqrt{p^{2}}=M\rightarrow z=x+iy$) starting from $z=2m$ and heading to plus infinity on the real axis. Furthermore, since the complex inverse trigonometric function $\arctan$ has in general an ordinary series representation (with convergence radius $r=1$, which already indicates the presence of a singularity, e.g. a branch cut), there is no problem at all to make the entire self-energy $\Sigma(M)$ a complex multi-valued function. However, we immediately face a strange property: plotting the self-energy with $\Lambda=1.5 \ \text{GeV}$ shows no kind of discontinuous behaviour above $2\sqrt{\Lambda^{2}+m^{2}}$, see Fig. \ref{figure_rel4dSigma}. This curiosity can be clarified by investigating the quite complicated inverse tangent inside the self-energy. The common $\arctan$ has two branch points at $z=\pm i$ (excluded from its domain) and conventionally the branch cut does not connect them both by a straight line on the imaginary axis through the origin, but rather by two straight lines. One part of the cut is along the positive imaginary axis above $i$, and one along the negative imaginary axis below $-i$, both heading to complex infinity:
\begin{equation}
\text{branch cut of } z\mapsto\arctan z=w: \{(-i\infty,-i],[i,i\infty)\} \ .
\end{equation}
This cut structure determines the range of the $\arctan$ for real $z=x$ as $(-\frac{\pi}{2},\frac{\pi}{2})$ and usually the function is well-defined right from the upper and left from the lower part. Looking for branch points in our case means
\begin{eqnarray}
(1) \ \ \ \ \ \ \ \ \ \frac{\Lambda M}{\sqrt{\Lambda^{2}+m^{2}}\sqrt{4m^{2}-M^{2}}} & \stackrel{!}{=} & - i \ , \\
\Rightarrow \ \ \ M & = & \begin{cases} 2\sqrt{\Lambda^{2}+m^{2}} \\
\rightarrow\infty
\end{cases}
\ , \\
(2) \ \ \ \ \ \ \ \ \ \frac{\Lambda M}{\sqrt{\Lambda^{2}+m^{2}}\sqrt{4m^{2}-M^{2}}} & \stackrel{!}{=} & -i\infty \ , \\
\nonumber \\
\Rightarrow \ \ \ M & = & \lim_{\eta\rightarrow0^{+}}(2m+\eta) \ ,
\end{eqnarray}
so obviously the situation looks quite complicated. It is therefore much easier to find the discontinuity of $\Sigma(z)$ on the real axis in exact the same manner as we did for the non-relativistic Lee model. There, we tried two ways and we may begin with the simpler one using the integral representation of the self-energy from Eq. (\ref{equation_SigmaIntegral}). The discontinuity turns out to be:
\begin{eqnarray}
\disc\Sigma(M) & = & \Sigma(M+i\epsilon)-\Sigma(M-i\epsilon) \nonumber \\
& = & \frac{1}{2\pi^{2}}\int_{0}^{\infty}\text{d}u \ u^{2}\Theta(\Lambda^{2}-u^{2})\bigg[\frac{1}{\sqrt{u^{2}+m^{2}}\big(4(u^{2}+m^{2})-M^{2}-i\epsilon\big)} \nonumber \\
&  & - \ \frac{1}{\sqrt{u^{2}+m^{2}}\big(4(u^{2}+m^{2})-M^{2}+i\epsilon\big)}\bigg] \nonumber \\
& = & \frac{1}{2\pi^{2}}\int_{0}^{\Lambda}\text{d}u \ \frac{u^{2}}{\sqrt{u^{2}+m^{2}}}\frac{2i\epsilon}{\big(4(u^{2}+m^{2})-M^{2}\big)^{2}+\epsilon^{2}} \nonumber \\
& = & \frac{i}{\pi}\int_{0}^{\Lambda}\text{d}u \ \frac{u^{2}}{\sqrt{u^{2}+m^{2}}} \ \delta\big(4(u^{2}+m^{2})-M^{2}\big) \nonumber \\
& = & \frac{i}{\pi}\int_{0}^{\Lambda}\text{d}u \ \frac{u^{2}}{\sqrt{u^{2}+m^{2}}}\frac{1}{8\sqrt{\frac{M^{2}}{4}-m^{2}}} \ \delta\big(u-\sqrt{M^{2}/4-m^{2}}\big) \nonumber \\
& = & \frac{i\sqrt{\frac{M^{2}}{4}-m^{2}}}{4\pi M} \ , \ \ \ 2m<M\le2\sqrt{\Lambda^{2}+m^{2}} \ ,
\label{equation_relSigmaDiscontinuity}
\end{eqnarray}
where the restrictions are predefined by the cutoff and the necessity of purely real values in the argument of the delta distribution function. It is also possible to argue that in order to avoid missing a real pole in the complex $u$-plane during the integration from $0$ to $\Lambda$, the mass $M$ needs to lie inside the upper interval. We shall also verify the useful identity from Eq. (\ref{equation_usefullidentity}) between the discontinuity and the imaginary part right above the branch cut,
\begin{equation}
\disc\Sigma(M) = 2i\operatorname{Im}\Sigma(M+i\epsilon) \ ,
\end{equation}
by applying the Sokhotski--Plemelj theorem to the self-energy's integral representation:
\begin{eqnarray}
\Sigma(M) & = & \frac{1}{2\pi^{2}}\int_{0}^{\Lambda}\text{d}u \ \frac{u^{2}}{\sqrt{u^{2}+m^{2}}\big(4(u^{2}+m^{2})-M^{2}-i\epsilon\big)} \nonumber \\
& = & \frac{1}{2\pi^{2}}\int_{0}^{\Lambda}\text{d}u \ \frac{u^{2}}{4\sqrt{u^{2}+m^{2}}}\frac{1}{u^{2}-i^{2}\Big(m^{2}-\frac{M^{2}}{4}\Big)-i\epsilon} \nonumber \\
& = & \frac{1}{2\pi^{2}}\int_{0}^{\Lambda}\text{d}u \ \frac{u^{2}}{\sqrt{u^{2}+m^{2}}\Big(u+\sqrt{\frac{M^{2}}{4}-m^{2}}\Big)}\frac{1}{u-\sqrt{\frac{M^{2}}{4}-m^{2}}-i\epsilon} \ , \nonumber \\
\label{equation_relSigmaIntegralDecomposition} \\
\Rightarrow \ \ \ \operatorname{Im}\Sigma(M) & = & \frac{\sqrt{\frac{M^{2}}{4}-m^{2}}}{8\pi M} \ , \ \ \ 2m<M\le2\sqrt{\Lambda^{2}+m^{2}} \ ,
\end{eqnarray}
which indeed yields the same result as in Eq. (\ref{equation_relSigmaDiscontinuity}) after multiplying with $2i$. Note that this result is valid {\em on} the positive real axis, thus it gives us information with which quadrant the branch cut is continuous (it is continuous with the first quadrant).

\begin{figure}[!h]
\begin{center}
\includegraphics{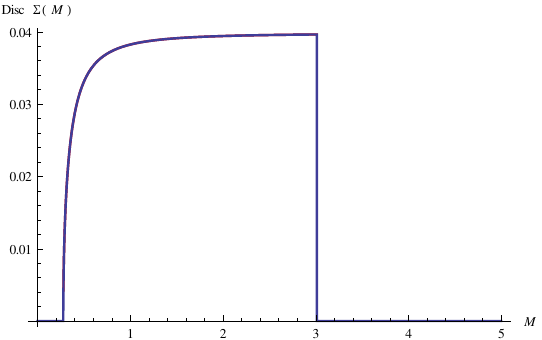}
\caption{Discontinuity of the self-energy (in units of $i$): numerical difference (dark blue) and analytic result (red).}
\label{figure_rel4dSigmaDisc}
\end{center}
\end{figure}

We now would like to use the final result of the self-energy from Eq. (\ref{equation_relSigma}) to arrive at the same result. For that purpose one has to rewrite the inverse tangent with the well-known decomposition
\begin{equation}
\arctan z = \frac{i}{2}\Big[\ln(1-iz)-\ln(1+iz)\Big] \ , \label{equation_arctandecomposition}
\end{equation}
which is especially used in computational software \cite{steele}. The two logarithmic terms determine the range and branch cut structure of the inverse tangent fully correctly.\footnote{This statement is {\em not} true if one uses
\begin{equation}
\arctan z = -i\ln\left((1+iz)\sqrt{\frac{1}{1+z^{2}}}\right) \ . \nonumber
\end{equation}
With this representation the inverse tangent would be well-defined left from the upper part of the cut and right from the lower part.} This form is actually equivalent to
\begin{equation}
\arctan z = \frac{\arctanh iz}{i} \ ,
\end{equation}
recommended by Kahan \cite{kahan}. Let us guess what will happen along the positive real axis: The ratio of negative complex root function and single variable $z$ in the expression for the self-energy will give a branch cut for $z>2m$ with an easily obtained value for the discontinuity, while we expext another conribution from the inverse tangent for values larger than $z=2\sqrt{\Lambda^{2}+m^{2}}$. If this part of the overall cut is somehow totally cancelled out it is reasonable to assume the inverse tangent to be responsible for this effect. An explicit calculation illustrates this aspect (the logarithmic constants are not important and were already subtracted):
\begin{eqnarray}
\disc\Sigma(M) & = & \Sigma(M+i\epsilon)-\Sigma(M-i\epsilon) \nonumber \\[0.3cm]
& = & -\frac{\sqrt{4m^{2}-M^{2}-i\epsilon}}{8\pi^{2}(M+i\epsilon)} \ \frac{i}{2}\bigg[\ln\left(1-i\frac{\Lambda (M+i\epsilon)}{\sqrt{\Lambda^{2}+m^{2}}\sqrt{4m^{2}-M^{2}-i\epsilon}}\right) \nonumber \\
\nonumber \\
&  & -\ln\left(1+i\frac{\Lambda (M+i\epsilon)}{\sqrt{\Lambda^{2}+m^{2}}\sqrt{4m^{2}-M^{2}-i\epsilon}}\right)\bigg] \nonumber \\
\nonumber \\
&  & +\frac{\sqrt{4m^{2}-M^{2}+i\epsilon}}{8\pi^{2}(M-i\epsilon)} \ \frac{i}{2}\bigg[\ln\left(1-i\frac{\Lambda (M-i\epsilon)}{\sqrt{\Lambda^{2}+m^{2}}\sqrt{4m^{2}-M^{2}+i\epsilon}}\right) \nonumber \\
\nonumber \\
&  & -\ln\left(1+i\frac{\Lambda (M-i\epsilon)}{\sqrt{\Lambda^{2}+m^{2}}\sqrt{4m^{2}-M^{2}+i\epsilon}}\right)\bigg] \nonumber \\[0.3cm]
& = & \frac{i\sqrt{M^{2}-4m^{2}}}{8\pi^{2}M} \ \frac{i}{2}\bigg[\ln\left(1+\frac{\Lambda M}{\sqrt{\Lambda^{2}+m^{2}}\sqrt{M^{2}-4m^{2}}}-i\epsilon\right) \nonumber \\
\nonumber \\
&  & -\ln\left(1-\frac{\Lambda M}{\sqrt{\Lambda^{2}+m^{2}}\sqrt{M^{2}-4m^{2}}}+i\epsilon\right)\bigg] \nonumber \\
\nonumber \\
&  & +\frac{i\sqrt{M^{2}-4m^{2}}}{8\pi^{2}M} \ \frac{i}{2}\bigg[\ln\left(1-\frac{\Lambda M}{\sqrt{\Lambda^{2}+m^{2}}\sqrt{M^{2}-4m^{2}}}-i\epsilon\right) \nonumber \\
\nonumber \\
&  & -\ln\left(1+\frac{\Lambda M}{\sqrt{\Lambda^{2}+m^{2}}\sqrt{M^{2}-4m^{2}}}+i\epsilon\right)\bigg] \nonumber \\[0.3cm]
& = & \frac{i\sqrt{\frac{M^{2}}{4}-m^{2}}}{4\pi M} \ , \ \ \ 2m<M\le2\sqrt{\Lambda^{2}+m^{2}} \ .
\end{eqnarray}
Here, the discontinuity vanishes for $M<2m$ because the real parts of all radicands after the second equality stay positive, so we are beyond the branch cuts of the roots and logarithms. The self-energy above and below the positive real axis consequently equal each other in the limit of $\epsilon\rightarrow0^{+}$ and subtract to zero. In order to get a discontinuity from the roots outside the arguments, we also need the logarithmic cuts to contribute -- the real part of their arguments has to be negative or zero. Only the inequality
\begin{equation}
1-\frac{\Lambda M}{\sqrt{\Lambda^{2}+m^{2}}\sqrt{M^{2}-4m^{2}}} \ \le \ 0
\end{equation}
has real positive solutions, namely $2m<M\le2\sqrt{\Lambda^{2}+m^{2}}$. We have plotted the discontinuity and the numerical difference of $\Sigma(E)$ across the real axis in Fig. \ref{figure_rel4dSigmaDisc}.

In the end, we can perform the analytic continuation of the self-energy down into the second Riemann sheet (this is true for the fourth quadrant):
\begin{equation}
\Sigma_{\text{\Romannum{2}}}(z) = \Sigma(z)+\frac{i\sqrt{\frac{z^{2}}{4}-m^{2}}}{4\pi z} \ ,
\end{equation}
and as a consequence there the propagator can be written as
\begin{eqnarray}
\Delta_{\text{\Romannum{2}}}(z) & = & \frac{1}{z^{2}-M_{0}^{2}+g_{S\phi\phi}^{2}\Sigma_{\text{\Romannum{2}}}(z)} \nonumber \\
\nonumber \\
& = & \frac{1}{z^{2}-M_{0}^{2}+g_{S\phi\phi}^{2}\bigg(\Sigma(z)+\frac{i\sqrt{\frac{z^{2}}{4}-m^{2}}}{4\pi z}\bigg)} \ ,
\end{eqnarray}
which gains the same branch cut structure in the complex $M$-plane for $g_{S\phi\phi}\ne0$ as the self-energy. Although we will work with the inverse propagator and are thus only concerned with the self-energy, one can write down, for the sake of completeness, the analytic continuation of the propagator itself by using the spectral function. Here, one should keep in mind that the starting point in the literature is the complex $p^{2}$-plane:
\begin{eqnarray}
\disc \Delta_{S}(p^{2}) & = & \Delta_{S}(p^{2}+i\epsilon)-\Delta_{S}(p^{2}-i\epsilon) \nonumber \\
& = & \frac{1}{\pi}\int_{0}^{\infty}\text{d}s^{2}\left(\frac{\rho(s^{2})}{p^{2}-s^{2}+i\epsilon}-\frac{\rho(s^{2})}{p^2-s^{2}-i\epsilon}\right) \nonumber \\
& = & \frac{1}{\pi}\int_{0}^{\infty}\text{d}s^{2} \ \rho(s^{2})\frac{-2i\epsilon}{(p^{2}-s^{2})^{2}+\epsilon^{2}} \nonumber \\
& = & -\frac{2i}{\pi}\int_{0}^{\infty}\text{d}s^{2} \ \rho(s^{2})\pi\delta(p^{2}-s^{2}) \nonumber \\
& = & -2i\rho(p^{2}) \ \ = \ \ 2i\operatorname{Im}\Delta_{S}(p^{2}+i\epsilon) \ . \label{equation_relPropdisc}
\end{eqnarray}
We of course recognize the useful identity (\ref{equation_usefullidentity}) and remember the imaginary part to be evaluated right above the branch cut. Because we work in the complex $M$-plane, the spectral function has to be changed by a variable transformation,
\begin{equation}
\rho(s^{2}) = -\operatorname{Im}\Delta_{S}(s^{2}+i\epsilon) \ \rightarrow \ d(s) = 2s\rho(s) = -2s\operatorname{Im}\Delta_{S}(s+i\epsilon) \ ,
\end{equation}
where the spectral function of the resonance $S$ is from now on denoted as $d(s)$. Finally, the propagator on the second sheet is
\begin{equation}
\Delta_{\text{\Romannum{2}}}(z) = \Delta_{S}(z)-i\frac{d(z)}{z} \ .
\end{equation}

The full interacting propagator belongs for the simplest case $g_{S\phi\phi}=0$ to a stable particle with mass $M_{0}$ in its rest frame:
\begin{equation}
\Delta_{S}(M) = \frac{1}{M^{2}-M_{0}^{2}+i\epsilon} \ .
\end{equation}
We have applied our general definition of a stable particles (real) mass as its propagator pole. The pole position of course gives a mass which is slightly shifted into the lower half plane of the complex $M$-plane ($\sqrt{p^{2}}=M\rightarrow z=x+iy$):
\begin{equation}
M_{\text{pole}} = M_{0}-i\epsilon \ .
\end{equation}
By switching on the coupling there is a contribution from the (complex-valued) self-energy to the pole position in exactly the same way as for the non-relativistic Lee model. The new pole position, as the zero of the denominator, is determined by a system of two equations
\begin{eqnarray}
x^{2}-y^{2}-M_{0}^{2}+g_{S\phi\phi}^{2}\operatorname{Re}\Sigma_{\text{\Romannum{2}}}(x+iy) & \stackrel{!}{=} & 0 \ , \nonumber \\
2ixy+g_{S\phi\phi}^{2}\operatorname{Im}\Sigma_{\text{\Romannum{2}}}(x+iy) & \stackrel{!}{=} & 0 \label{equation_relsystemofEq} \ ,
\end{eqnarray}
and we will search for numerical solutions of
\begin{equation}
z^{2}-M^{2}_{0}+g_{S\phi\phi}^{2}\Sigma_{\text{\Romannum{2}}}(z) \stackrel{!}{=} 0 \label{equation_rel0denominator}
\end{equation}
in the complex $M$-plane.

\subsection{Couplings $g_{S\phi\phi}\in[0.1,1.0]$}
We need to add some words concerning the parameters. The choice of the cutoff $\Lambda=1.5 \ \text{GeV}$ was already motivated by the energy and mass scales of the effective theory presented here. Nevertheless, there is still one free parameter in our model, namely the bare mass $M_{0}$. In a fundamental theory (where $\Lambda\rightarrow\infty$) such a quantity can neither be obtained from field theory, nor from any experiment and has to be treated in the framework of mass renormalization. This is different from our point of view: the bare mass is model dependent and was for example taken by T\"ornqvist \cite{tornqvist} to be a free parameter (that is fixed after fitting to experimental data), as it will be in this work. If we interpret this number as somehow related to the physical mass of the resonance $S$, namely its mass for a vanishing interaction, then it is modified by quantum fluctuations of the mesonic loops. The renormalized Breit--Wigner mass $M_{\text{BW}}$ of the resonance is defined like in the first chapter as the real zero of the propagator's real part:
\begin{equation}
M_{\text{BW}}^{2}-M_{0}^{2}+g_{S\phi\phi}^{2}\operatorname{Re}\Sigma(M_{\text{BW}}) = 0 \ .
\end{equation}
We fix the number by $M_{\text{BW}}=0.6 \ \text{GeV}$ and thus every resonance pole in the second Riemann sheet will describe a particle with this value of the Breit--Wigner mass (this choice was made before the PDG update on the $\sigma$-meson occurred). Such a constraint was already introduced by Achasov and Kiselev in order to take into account finite-width corrections when studying the propagators of light scalar mesons \cite{achasov}. Since $\operatorname{Re}\Sigma(0.6 \ \text{GeV})>0$, the renormalized mass is smaller than $M_{0}$. In the following subsections we search the complex $M$-plane for numerical solutions of Eq. (\ref{equation_rel0denominator}) by varying the coupling $g_{S\phi\phi}$ in steps of $\it\Delta g_{S\phi\phi}$. The condition for deciding whether a solution $z_{\text{pole}}$ marks a singularity or not shall be again $\abs\big(\Delta_{\text{\Romannum{2}}}^{-1}(z_{\text{pole}})\big)<10^{-10} \ (\text{GeV})^{2}$.

\begin{figure}[!h]
\begin{center}
\includegraphics{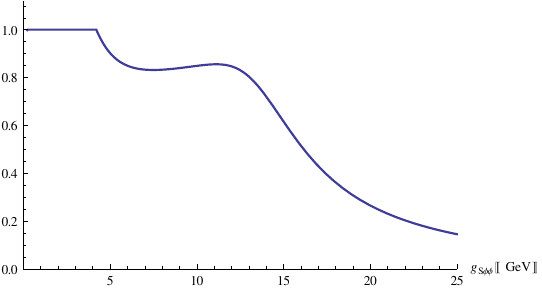}
\caption{Incomplete normalization of the spectral function $d(M)$ for different couplings.}
\label{figure_rel4dM06L15SpectralgNorm1}
\end{center}
\end{figure}

Similar as in the case of the non-relativistic Lee model, it seems the spectral function $d(M)$, defined as the negative imaginary part of the propagator right above the branch cut (see first chapter), is normalized only for couplings in the approximate interval $[0.2 \ \text{GeV},4.5 \ \text{GeV}]$, see Fig. \ref{figure_rel4dM06L15SpectralgNorm1}.
For smaller $g_{S\phi\phi}$ there would be a steep descent at the beginning which we ignore since, if the coupling tends to zero, the spectral function becomes a delta distribution function due to the fact that the pole approaches the real axis:
\begin{eqnarray}
d(M) & = & \frac{2M\big(\epsilon+g_{S\phi\phi}^{2}\operatorname{Im}\Sigma(M+i\epsilon)\big)}{\big(M^{2}-M^{2}_{0}+g_{S\phi\phi}^{2}\operatorname{Re}\Sigma(M+i\epsilon)\big)^{2}+\big(\epsilon+g_{S\phi\phi}^{2}\operatorname{Im}\Sigma(M+i\epsilon)\big)^{2}} \nonumber \\
& \stackrel{g_{S\phi\phi}\rightarrow0}{=} & \frac{2M\epsilon}{(M^{2}-M^{2}_{0})^{2}+\epsilon^{2}} \nonumber \\
& = & 2M\pi\delta(M^{2}-M^{2}_{0}) \ .
\end{eqnarray}
As we know from the previous chapter, the numerical integration $\frac{1}{\pi}\int_{0}^{\infty}\text{d}s \ d(s)$ is not possible in the vicinity of a simple pole. This was also true for poles on the real axis in the first sheet: we expect additional poles in the first Riemann sheet once the spectral function is decreasing for sufficiently large couplings. This time though, one of the new poles will {\em not} emerge right from the beginning.

\begin{figure}[t]
\begin{center}
\includegraphics{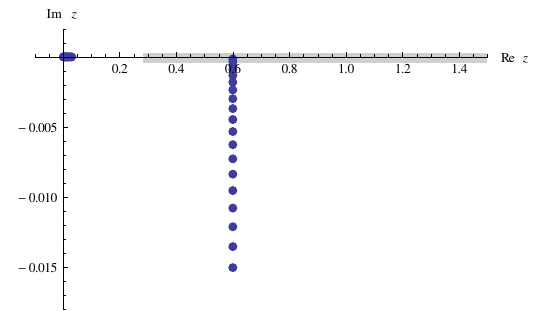}
\caption{$S\phi\phi$-model poles (dark blue dots) for $g_{S\phi\phi}\in[0.1 \ \text{GeV},1.0 \ \text{GeV}]$ in the second Riemann sheet with $\it\Delta$$g_{S\phi\phi}=0.05 \ \text{GeV}$, where the gray line marks the branch cut.}
\label{figure_rel4dM06L15Pointsg0110gstep005-1}
\end{center}
\end{figure}

In general, $d(M)$ becomes broader for larger couplings, because the complex mass-pole moves deeper down into the complex plane (i.e., the second Riemann sheet). It shrinks very quickly the more the pole descends. The pole trajectory within the regarded coupling interval [0.1 \ \text{GeV},1.0 \ \text{GeV}] is shown in Fig. \ref{figure_rel4dM06L15Pointsg0110gstep005-1}.
In contrast to the results in chapter 3, one clearly sees another pole coming from the origin and heading to the first branch point at $z=2m$.\footnote{Compared to our strategy in the non-relativistic Lee model, we have chosen this time the accuracy of the starting values on the lattice in such a way that poles on the real axis can be found the moment they really exist.} Such a phenomenon was not observed in the non-relativistic Lee model. Note that there is no possible way of finding an analytic solution of Eq. (\ref{equation_rel0denominator}), not even for purely real $z$-values. Yet, we can solve for $g_{S\phi\phi}$ and, after taking just positive solutions into account, perform the limit
\begin{eqnarray}
g_{S\phi\phi}(z=0-i\epsilon) & = & \lim_{M\rightarrow0^{+}}\sqrt{\frac{M^{2}-M_{\text{BW}}^{2}-i\epsilon}{\operatorname{Re}\Sigma(M_{\text{BW}})-\Sigma(M-i\epsilon)-\disc\Sigma(M-i\epsilon)}} \nonumber \\
& = & 0 \ . \label{equation_relLimit0Disc}
\end{eqnarray}
This result is due to the divergent behaviour of the discontinuity at the origin in the second sheet.

\begin{table}[!h]\center
\scalebox{0.8}{
\begin{tabular}{l c c l}
\toprule
\cmidrule(r){1-4}
name	& label	& determining equation	& description\\
\midrule
resonance pole mass	&$x_{\text{(m)pole}}$	&$z^{2}-M_{0}^{2}+g_{S\phi\phi}^{2}\Sigma_{\text{\Romannum{2}}}(z) = 0$	&Real part of solution $z$.\\
gap pole mass	&$x_{\text{(0)pole}}$	&$z^{2}-M_{0}^{2}+g_{S\phi\phi}^{2}\Sigma_{\text{\Romannum{2}}}(z) = 0$	&Real part of solution $z$.\\
left pole mass	&$x_{\text{(l)pole}}$	&$z^{2}-M_{0}^{2}+g_{S\phi\phi}^{2}\Sigma(z) = 0$	&Real part of solution $z$.\\
	&	&	&Nevertheless, $z$ is purely real.\\
Breit--Wigner resonance mass	&$x_{\text{(m)BW}}$	&$x^{2}-M_{0}^{2}+g_{S\phi\phi}^{2}\operatorname{Re}\Sigma(x) = 0$	&The solution $x$ is purely real.\\
Breit--Wigner left/right mass	&$x_{\text{(l)BW}}$	&$x^{2}-M_{0}^{2}+g_{S\phi\phi}^{2}\operatorname{Re}\Sigma(x) = 0$	&The solution $x$ is purely real.\\
average mass	&$x_{\text{average}}$	&$x = \frac{1}{\pi}\int_{2m}^{2\sqrt{\Lambda^{2}+m^{2}}}\text{d}s \ sd(s)$	&Numerically evaluated.\\
maxima of spectral function	&$x_{\text{(l)max}}$	&$\maxf d(s)$	&Solution is $x=M_{\text{max}}$.\\
	&	&	&Found numerically.\\
\bottomrule
\end{tabular}
}
\caption{Description of masses for the considered $S\phi\phi$-model. The general complex solution of an equation is denoted as $z=x+iy$.}
\label{table_allpointsexplanation2}
\end{table}

A summary of pole positions and other relevant informations for different intervals can be found in Tab. \ref{table_allpoints2}. The adopted notation is very similar to the one for the Lee model (for a better understanding see Tab. \ref{table_allpointsexplanation2}): The pole masses $x_{\text{(m)pole}}$ and $x_{(0)\text{pole}}$ are the real parts of the complex solutions found by solving Eq. (\ref{equation_rel0denominator}), while $x_{\text{(l)pole}}$ is the real part of the complex solution in the first Riemann sheet, where we have replaced the self-energy $\Sigma_{\text{\Romannum{2}}}(z)$ in Eq. (\ref{equation_rel0denominator}) by $\Sigma(z)$. The corresponding Breit--Wigner masses are the solutions of the latter equation using only the real part of the self-energy on the first sheet (they are all real by construction). These are of course the zeros of the propagator's denominator on the real axis. The numerical maxima of the continuous part of the spectral function are denoted as $x_{\text{(l)max}}$ and $x_{\text{(r)max}}$, whereas $x_{\text{average}}$ is the result of the numerically evaluated integral
\begin{equation}
x_{\text{average}} = \frac{1}{\pi}\int_{2m}^{2\sqrt{\Lambda^{2}+m^{2}}}\text{d}s \ sd(s) \ . \label{equation_relavmassintegral}
\end{equation}
The last column of Tab. \ref{table_allpoints2} simply gives the normalization value of the spectral function (which indeed is normalized to one, as will be shown in subsection \ref{subsection_spectralfunctionnormalization1}). All numbers after the fourth digit are dropped.

\subsection{Couplings $g_{S\phi\phi}\in[1.0,2.8]$}
The resonance pole descends deeper into the lower half plane and turns right, while the pole near the origin, from now on denoted as the gap pole, keeps approaching the first branch point with rising parameter speed. No poles are found in the first sheet.
\begin{figure}[!h]
\begin{center}
\includegraphics{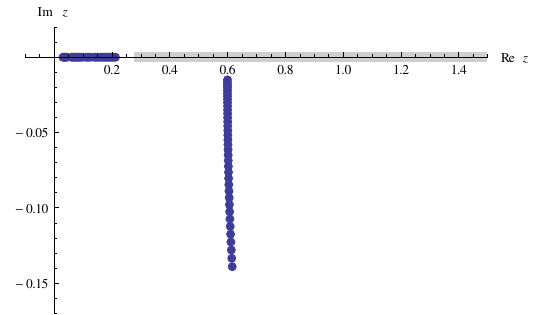}
\caption{$S\phi\phi$-model poles for $g_{S\phi\phi}\in[1.0 \ \text{GeV},2.8 \ \text{GeV}]$ in the second Riemann sheet with $\it\Delta$$g_{S\phi\phi}=0.05 \ \text{GeV}$.}
\label{figure_rel4dM06L15Pointsg1028gstep005-1}
\end{center}
\end{figure}

\begin{table}[!h]\center
\scalebox{0.85}{
\begin{tabular}{l c c c c c c c c c c}
\toprule
\cmidrule(r){1-10}
$g_{S\phi\phi}$	& $x_{(0)\text{pole}}$ & $x_{\text{(l)pole}}$	& $x_{\text{(m)pole}}$	& $x_{\text{(0)BW}}$	& $x_{\text{(l)BW}}$	& $x_{\text{(m)BW}}$	& $x_{\text{max}}$ & $x_{\text{average}}$	& $M_{0}$	&Norm\\
\midrule
0.1	&-	&-	&0.6000	&-	&-	&0.6000	&0.6000	&-	&0.6001	&-\\
0.2	&0.0012	&-	&0.6000	&-	&-	&0.6000	&0.6000	&-	&0.6007	&-\\
0.3	&0.0027	&-	&0.6000	&-	&-	&0.6000	&0.6000	&0.6009	&0.6017	&0.9999\\
0.4	&0.0049	&-	&0.6000	&-	&-	&0.6000	&0.6000	&0.6016	&0.6030	&1.0000\\
0.5	&0.0076	&-	&0.6000	&-	&-	&0.6000	&0.6000	&0.6025	&0.6047	&1.0000\\
0.6	&0.0110	&-	&0.6000	&-	&-	&0.6000	&0.6000	&0.6036	&0.6068	&1.0000\\
0.7	&0.0151	&-	&0.6000	&-	&-	&0.6000	&0.6000	&0.6050	&0.6093	&1.0000\\
0.8	&0.0197	&-	&0.6000	&-	&-	&0.6000	&0.6000	&0.6065	&0.6121	&1.0000\\
0.9	&0.0255	&-	&0.6000	&-	&-	&0.6000	&0.6000	&0.6082	&0.6153	&1.0000\\
1.0	&0.0309	&-	&0.6000	&-	&-	&0.6000	&0.6001	&0.6102	&0.6188	&1.0000\\
1.2	&0.0445	&-	&0.6002	&-	&-	&0.6000	&0.6002	&0.6147	&0.6269	&1.0000\\
1.4	&0.0605	&-	&0.6004	&-	&-	&0.6000	&0.6005	&0.6199	&0.6364	&1.0000\\
1.6	&0.0788	&-	&0.6008	&-	&-	&0.6000	&0.6009	&0.6260	&0.6471	&1.0000\\
1.8	&0.0991	&-	&0.6014	&-	&-	&0.6000	&0.6016	&0.6329	&0.6591	&1.0000\\
2.0	&0.1212	&-	&0.6025	&-	&-	&0.6000	&0.6025	&0.6405	&0.6722	&1.0000\\
2.2	&0.1444	&-	&0.6042	&-	&-	&0.6000	&0.6039	&0.6489	&0.6864	&1.0000\\
2.4	&0.1680	&-	&0.6068	&-	&-	&0.6000	&0.6058	&0.6580	&0.7016	&1.0000\\
2.6	&0.1911	&-	&0.6108	&-	&-	&0.6000	&0.6084	&0.6679	&0.7178	&1.0000\\
2.8	&0.2128	&-	&0.6166	&-	&-	&0.6000	&0.6119	&0.6785	&0.7349	&1.0000\\
3.0	&0.2318	&-	&0.6247	&-	&-	&0.6000	&0.6166	&0.6898	&0.7528	&1.0000\\
3.2	&0.2476	&-	&0.6354	&-	&-	&0.6000	&0.6228	&0.7018	&0.7715	&1.0000\\
3.4	&0.2597	&-	&0.6487	&-	&-	&0.6000	&0.6308	&0.7145	&0.7909	&1.0000\\
3.6	&0.2684	&-	&0.6646	&-	&-	&0.6000	&0.6408	&0.7279	&0.8110	&1.0000\\
3.8	&0.2740	&-	&0.6828	&-	&-	&0.6000	&0.6531	&0.7418	&0.8317	&1.0000\\
4.0	&0.2770	&-	&0.7031	&-	&-	&0.6000	&0.6678	&0.7565	&0.8529	&1.0000\\
4.2	&-	&-	&0.7251	&-	&-	&0.6000	&0.6848	&0.7717	&0.8748	&1.0000\\
4.4	&-	&0.2773	&0.7485	&0.3024	&0.2773	&0.6000	&0.7040	&0.7793	&0.8971	&1.0000\\
4.6	&-	&0.2753	&0.7732	&0.3288	&0.2753	&0.6000	&0.7252	&0.7884	&0.9198	&1.0000\\
4.8	&-	&0.2723	&0.7991	& 0.3561	&0.2723	&0.6000	&0.7481	&0.7995	&0.9430	&1.0000\\
5.0	&-	&0.2685	&0.8259	& 0.3840	&0.2685	&0.6000	&0.7725	&0.8123	&0.9666	&1.0000\\
5.5	&-	&0.2563	&0.8966	& 0.4568	&0.2563	&0.6000	&0.8388	&0.8510	&1.0271	&1.0000\\
6.0	&-	&0.2416	&0.9718	& 0.5335	&0.2416	&0.6000	&0.9109	&0.8973	&1.0895	&1.0000\\
6.5	&-	&0.2251	&1.0507	& 0.6138	&0.2251	&0.6000	&0.9875	&0.9491	&1.1535	&1.0000\\
7.0	&-	&0.2075	&1.1329	& 0.6972	&0.2075	&0.6000	&1.0679	&1.0052	&1.2189	&1.0000\\
7.5	&-	&0.1887	&1.2182	& 0.7838	&0.1887	&0.6000	&1.1518	&1.0646	&1.2854	&1.0000\\
8.0	&-	&0.1688	&1.3064	& 0.8732	&0.1688	&0.6000	&1.2390	&1.1267	&1.3529	&1.0000\\
8.5	&-	&0.1475	&1.3973	& 0.9656	&0.1475	&0.6000	&1.3297	&1.1909	&1.4212	&0.9999\\
9.0	&-	&0.1241	&1.4909	& 1.0608	&0.1241	&0.6000	&1.4238	&1.2569	&1.4902	&0.9999\\
9.5	&-	&0.0972	&1.5870	& 1.1590	&0.0972	&0.6000	&1.5218	&1.3241	&1.5599	&0.9998\\
10.0	&-	&0.0623	&1.6857	& 1.2603	&0.0623	&0.6000	&1.6239	&1.3917	&1.6301	&0.9994\\
10.2	&-	&0.0420	&1.7258	& 1.3018	&0.0420	&0.6000	&1.6660	&1.4186	&1.6584	&0.9991\\
\bottomrule
\end{tabular}
}
\caption{Selection of masses (in units of GeV) for the $S\phi\phi$-model with $\Lambda=1.5 \ \text{GeV}, \ M_{\text{BW}}=0.6 \ \text{GeV}$ and $m=0.139 \ \text{GeV}$.}
\label{table_allpoints2}
\end{table}
\clearpage

\begin{figure}[!h]
\begin{center}
\includegraphics{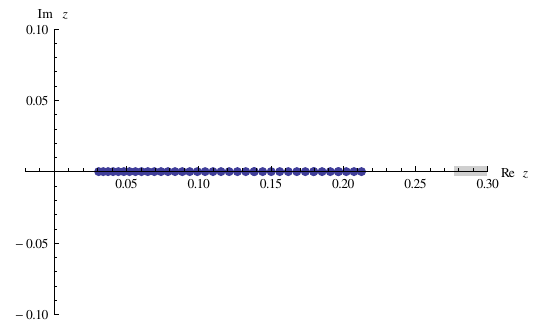}
\caption{$S\phi\phi$-model gap pole for $g_{S\phi\phi}\in[1.0 \ \text{GeV},2.8 \ \text{GeV}]$ on the real axis of the second Riemann sheet with $\it\Delta$$g_{S\phi\phi}=0.05 \ \text{GeV}$.}
\label{figure_rel4dM06L15Pointsg1028gstep005-2}
\end{center}
\end{figure}

\subsection{Couplings $g_{S\phi\phi}\in[2.8,5.0]$}
For this range of the coupling we are faced with an interesting behaviour. As we turn up $g_{S\phi\phi}$, for some value larger than $4 \ \text{GeV}$ the gap pole hits the first branch point at $z=2m$ and vanishes, in fact it slips into the first Riemann sheet. Although this strange pole has no physical interpretation, its existence below threshold lasts as long as $g_{S\phi\phi}$ is non-zero, because then the self-energy, its real part at the point $z=M_{\text{BW}}$ and the discontinuity are all three purely real and positive right below the real axis. In this case both the numerator and the denominator under the root of the master solution
\begin{equation}
g_{S\phi\phi}(M-i\epsilon) = \sqrt{\frac{M^{2}-M_{\text{BW}}^{2}-i\epsilon}{\operatorname{Re}\Sigma(M_{\text{BW}})-\Sigma(M-i\epsilon)-\disc\Sigma(M-i\epsilon)}} \label{equation_hardmastersolution}
\end{equation}
yield negative values and give a positive one for $g_{S\phi\phi}$. It is very important to realize that the pole must lie below the real axis, so the infinitesimal number $i\epsilon$ is crucial. Otherwise, the root inside the expression for the discontinuity would be evaluated in the wrong quadrant and there would be an imaginary solution for the coupling. Indeed
\begin{figure}[!h]
\begin{center}
\includegraphics{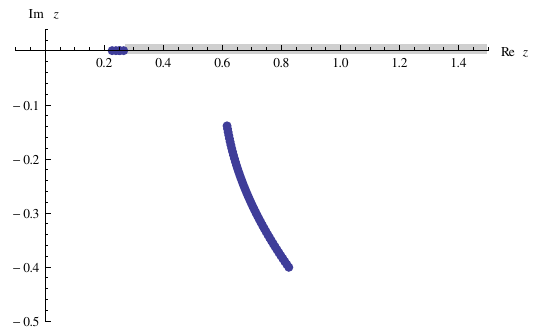}
\caption{$S\phi\phi$-model poles for $g_{S\phi\phi}\in[2.8 \ \text{GeV},5.0 \ \text{GeV}]$ in the second Riemann sheet with $\it\Delta$$g_{S\phi\phi}=0.05 \ \text{GeV}$.}
\label{figure_rel4dM06L15Pointsg2850gstep005-1}
\end{center}
\end{figure}
\begin{figure}[!h]
\begin{center}
\includegraphics{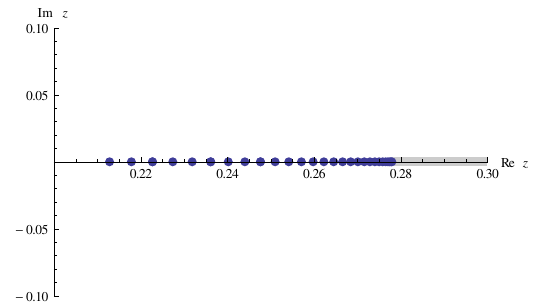}
\caption{Left $S\phi\phi$-model pole for $g_{S\phi\phi}\in[2.8 \ \text{GeV},5.0 \ \text{GeV}]$ on the real axis of the first Riemann sheet with $\it\Delta$$g_{S\phi\phi}=0.05 \ \text{GeV}$.}
\label{figure_rel4dM06L15Pointsg2850gstep005-2}
\end{center}
\end{figure}
\\
the gap pole leaves the second sheet after arriving at the first branch point for
\begin{eqnarray}
g_{S\phi\phi}(2m-i\epsilon) & = & \lim_{\eta\rightarrow0^{+}}\sqrt{\frac{(2m-\eta)^{2}-M_{\text{BW}}^{2}-i\epsilon}{\operatorname{Re}\Sigma(M_{\text{BW}})-\Sigma(2m-\eta-i\epsilon)-\disc\Sigma(2m-\eta-i\epsilon)}} \nonumber \\
\nonumber \\
& = & \sqrt{\frac{4m^{2}-M_{\text{BW}}^{2}}{\operatorname{Re}\Sigma(M_{\text{BW}})-\Sigma(2m)}} \ , \label{equation_relLimit2mDisc}
\end{eqnarray}
which numerically gives a value of $g_{S\phi\phi}=4.209 \ \text{GeV}$. In order to make this clear, one has to evaluate the limit analytically by extracting the real part of the self-energy (\ref{equation_relSigma}) by performing the principal value integral on the integral representation from Eq. (\ref{equation_SigmaIntegral}). We start with Eq. (\ref{equation_relSigmaIntegralDecomposition}):
\begin{eqnarray}
\operatorname{Re}\Sigma(M) & = & \frac{1}{2\pi^{2}} \ \mathcal{P}\int_{0}^{\Lambda}\text{d}u \ \frac{u^{2}}{\sqrt{u^{2}+m^{2}}\big(4(u^{2}+m^{2})-M^{2}\big)} \nonumber \\
& = & \frac{1}{2\pi^{2}}\bigg[\int_{0}^{\sqrt{\frac{M^{2}}{4}-m^{2}}-\eta}\text{d}u \ \frac{u^{2}}{\sqrt{u^{2}+m^{2}}\big(4(u^{2}+m^{2})-M^{2}\big)} \nonumber \\
&  & +\int_{\sqrt{\frac{M^{2}}{4}-m^{2}}+\eta}^{\Lambda}\text{d}u \ \frac{u^{2}}{\sqrt{u^{2}+m^{2}}\big(4(u^{2}+m^{2})-M^{2}\big)}\bigg] \ .
\end{eqnarray}
Only for $2m<M<2\sqrt{\Lambda^{2}+m^{2}}$ the whole integral gives another result than the already known self-energy. By using the same indefinite integral as during the calculation of the self-energy in the beginning of our investigation and performing the indicated limits, we arrive at a hardly suprising result:
\begin{eqnarray}
\operatorname{Re}\Sigma(M) & = & -\frac{\sqrt{4m^{2}-M^{2}}}{8\pi^{2}M}\arctan\left(\frac{\Lambda M}{\sqrt{\Lambda^{2}+m^{2}}\sqrt{4m^{2}-M^{2}}}\right)-\frac{1}{8\pi^{2}}\ln\left(\frac{m}{\Lambda+\sqrt{\Lambda^{2}+m^{2}}}\right) \nonumber \\
&  & -\frac{i\sqrt{\frac{M^{2}}{4}-m^{2}}}{4\pi^{2}}\underbrace{\Big[\arctan(-i+i\eta)-\arctan(-i-i\eta)\Big]}_{=\frac{\pi}{2}} \nonumber \\
& = & \Sigma(M)-i\operatorname{Im}\Sigma(M) \ .
\end{eqnarray}
Obviously, the imaginary part (which comes from the known two terms already found on the previous pages) is subtracted and leaves the real part we were looking for. We emphasize that although this last expression might look trivial, it is in fact an important result. We are now able to give explicitly the real value of the self-energy right above the real axis and in principle could also do the analytic continuation down into the second sheet with the help of the spectral function. However, with this result the denominator in the upper limit
\begin{equation}
\sqrt{\frac{4m^{2}-M_{\text{BW}}^{2}}{\operatorname{Re}\Sigma(M_{\text{BW}})-\Sigma(2m)}}
\end{equation}
can be written explicitly, but it again yields a transcendental equation, so we have to perform the numerical result for the coupling. The unphysical gap pole finally emerges on the first sheet left from the first branch point, as can be seen in Fig \ref{figure_rel4dM06L15Pointsg2850gstep005-2}. This is surely true since the master solution in the first sheet leads to
\begin{eqnarray}
g_{S\phi\phi}(2m-i\epsilon) & = & \lim_{\eta\rightarrow0^{+}}\sqrt{\frac{(2m-\eta)^{2}-M_{\text{BW}}^{2}-i\epsilon}{\operatorname{Re}\Sigma(M_{\text{BW}})-\Sigma(2m-\eta-i\epsilon)}} \nonumber \\
& = & \sqrt{\frac{4m^{2}-M_{\text{BW}}^{2}}{\operatorname{Re}\Sigma(M_{\text{BW}})-\Sigma(2m)}} \label{equation_relLimit2m} \ ,
\end{eqnarray}
thus the same result as in the second sheet. Furthermore, this equation has no real solution for $g<4.209 \ \text{GeV}$, yet they are purely real for larger values. We may show the pole appearing in the first sheet simply by considering the spectral function for $M<2m$:
\begin{eqnarray}
d(M) & = & \frac{2M\big(\epsilon+g_{S\phi\phi}^{2}\operatorname{Im}\Sigma(M+i\epsilon)\big)}{\big(M^{2}-M^{2}_{0}+g_{S\phi\phi}^{2}\operatorname{Re}\Sigma(M+i\epsilon)\big)^{2}+\big(\epsilon+g_{S\phi\phi}^{2}\operatorname{Im}\Sigma(M+i\epsilon)\big)^{2}} \nonumber \\
& = & \frac{2M\epsilon}{\big[M^{2}-M_{\text{BW}}^{2}-g_{S\phi\phi}^{2}\big(\operatorname{Re}\Sigma(M_{\text{BW}})-\operatorname{Re}\Sigma(M+i\epsilon)\big)\big]^{2}+\epsilon^{2}} \nonumber \\
& = & 2M\pi\delta\Big(M^{2}-M_{\text{BW}}^{2}-g_{S\phi\phi}^{2}\big(\operatorname{Re}\Sigma(M_{\text{BW}})-\operatorname{Re}\Sigma(M)\big)\Big) \ ,
\end{eqnarray}
where the argument of the delta distribution function naturally becomes the master solution in the first sheet (the self-energy is real below threshold). From here, we also see that the Breit--Wigner mass $x_{\text{(l)BW}}$ for the left pole equals the corresponding pole mass.

\subsection{Couplings $g_{S\phi\phi}\in[5.0,10.0]$}
The resonance pole in the second sheet is now the only one there. It descends deeper down in the lower half plane and its imaginary part becomes comparable to its real part. In the first sheet, the new pole speeds up and approaches the origin. Its existence has dramatic effects on the normalization of the spectral function. It seems we can understand the lack of normalization by finding this new pole, but for completeness reasons one should pay attention to the second branch point at $z=2\sqrt{\Lambda^{2}+m^{2}}$. We do not expect a third pole slipping from the branch cut into the complex plane, because the analytic continuation of the discontinuity into the second Riemann sheet provides us with a branch cut heading there to infinity. In spite of that, the limit
\begin{eqnarray}
g_{S\phi\phi}(2\sqrt{\Lambda^{2}+m^{2}}-i\epsilon) & = & \lim_{\eta\rightarrow0^{+}}\sqrt{\frac{(2\sqrt{\Lambda^{2}+m^{2}}+\eta)^{2}-M_{\text{BW}}^{2}-i\epsilon}{\operatorname{Re}\Sigma(M_{\text{BW}})-\Sigma(2\sqrt{\Lambda^{2}+m^{2}}+\eta-i\epsilon)}} \nonumber \\
& = & \sqrt{\frac{4\Lambda^{2}+4m^{2}-M_{\text{BW}}^{2}}{\operatorname{Re}\Sigma(M_{\text{BW}})-\Sigma(2\sqrt{\Lambda^{2}+m^{2}})}} \nonumber \\
& = & 0 \ ,
\end{eqnarray}
is obtained, where it is positive for every $z>2\sqrt{\Lambda^{2}+m^{2}}$ (since the inverse tangent in the self-energy is evaluated at a branch point, i.e., $-i\arctan(-i)=-\infty$, and stays imaginary for $M\rightarrow\infty$, while the prefactor root function is imaginary, too). So, we just made clear that there is another pole coming from the second branch point at $z=2\sqrt{\Lambda^{2}+m^{2}}$ right from the beginning, similar to the non-relativistic Lee model. This follows also from the spectral function for $M>2\sqrt{\Lambda^{2}+m^{2}}$:
\begin{eqnarray}
d(M) & = & \frac{2M\big(\epsilon+g_{S\phi\phi}^{2}\operatorname{Im}\Sigma(M+i\epsilon)\big)}{\big(M^{2}-M^{2}_{0}+g_{S\phi\phi}^{2}\operatorname{Re}\Sigma(M+i\epsilon)\big)^{2}+\big(\epsilon+g_{S\phi\phi}^{2}\operatorname{Im}\Sigma(M+i\epsilon)\big)^{2}} \nonumber \\
& = & \frac{2M\epsilon}{\big[M^{2}-M_{\text{BW}}^{2}-g_{S\phi\phi}^{2}\big(\operatorname{Re}\Sigma(M_{\text{BW}})-\operatorname{Re}\Sigma(M+i\epsilon)\big)\big]^{2}+\epsilon^{2}} \nonumber \\
& = & 2M\pi\delta\Big(M^{2}-M_{\text{BW}}^{2}-g_{S\phi\phi}^{2}\big(\operatorname{Re}\Sigma(M_{\text{BW}})-\operatorname{Re}\Sigma(M)\big)\Big) \ .
\end{eqnarray}
Again, the argument of the delta distribution function requires the master solution in the first sheet. The important point here is that this additional pole is due to the finite cutoff, thus it is not a physical effect -- one either needs to perform the limit $\Lambda\rightarrow\infty$ (and subtract the divergence), or use smooth regularization functions to make the second branch point disappear at complex infinity (where it has to be, see first chapter). That is why we completely disregard this pole\footnote{The numerical consequences are not worth mentioning: On the one hand the pole can only be found on the lattice simultaneously with tachyonic excitations (for huge couplings), on the other hand its renormalization constant is so small that it can be neglected for the presented range of the coupling.} and do not list its masses in Tab. \ref{table_allpoints2}.
\\
\begin{figure}[!h]
\begin{center}
\includegraphics{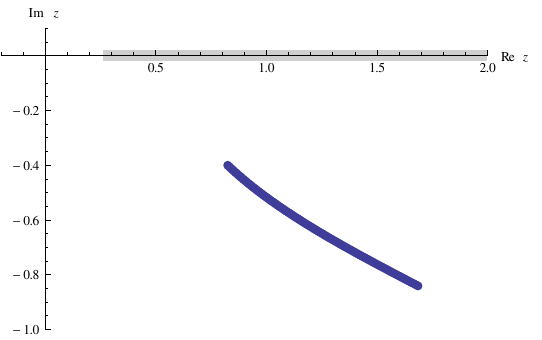}
\caption{$S\phi\phi$-model resonance pole for $g_{S\phi\phi}\in[5.0 \ \text{GeV},10.0 \ \text{GeV}]$ in the second Riemann sheet with $\it\Delta$$g_{S\phi\phi}=0.05 \ \text{GeV}$.}
\label{figure_rel4dM06L15Pointsg50100gstep005-1}
\end{center}
\end{figure}
\begin{figure}[!h]
\begin{center}
\includegraphics{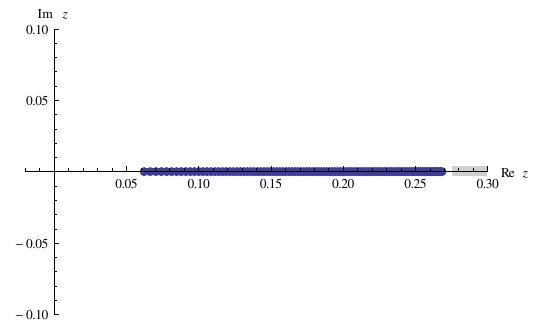}
\caption{Left $S\phi\phi$-model pole for $g_{S\phi\phi}\in[5.0 \ \text{GeV},10.0 \ \text{GeV}]$ on the real axis of the first Riemann sheet with $\it\Delta$$g_{S\phi\phi}=0.05 \ \text{GeV}$.}
\label{figure_rel4dM06L15Pointsg50100gstep005-2}
\end{center}
\end{figure}

\subsection{Couplings $g_{S\phi\phi}\in[10.0,10.5]$}
On top of that, we notice another remarkable behaviour of the left pole. It speeds up and approaches the origin but seems to vanish for some specific value of the coupling. Then, there is another pole emerging on the imaginary axis of the complex $M$-plane. We find the critical value by
\begin{eqnarray}
g_{S\phi\phi}(0-i\epsilon) & = & \lim_{M\rightarrow0^{+}}\sqrt{\frac{M^{2}-M_{\text{BW}}^{2}-i\epsilon}{\operatorname{Re}\Sigma(M_{\text{BW}})-\Sigma(M-i\epsilon)}} \nonumber \\
& = & \begin{cases} \frac{M_{\text{BW}}}{\sqrt{\Sigma(0-i\epsilon)-\operatorname{Re}\Sigma(M_{\text{BW}})}} & \operatorname{Re}\Sigma(M_{\text{BW}})<\Sigma(0-i\epsilon) \\
\frac{-iM_{\text{BW}}}{\sqrt{\operatorname{Re}\Sigma(M_{\text{BW}})-\Sigma(0-i\epsilon)}} & \operatorname{Re}\Sigma(M_{\text{BW}})>\Sigma(0-i\epsilon)
\end{cases} \ , \label{equation_relLimit0}
\end{eqnarray}
where in our case the first is true with $g_{S\phi\phi}=10.371 \ \text{GeV}$. The left pole will consequently arrive at the origin for a positive value of the coupling and the question now is whether the tachyonic pole on the imaginary axis in the lower half plane manifests at exactly the same $g_{S\phi\phi}$. The following limit clarifies this:
\begin{eqnarray}
g_{S\phi\phi}(\eta-i0) & = & \lim_{y\rightarrow0^{-}}\sqrt{\frac{(\eta+iy)^{2}-M_{\text{BW}}^{2}}{\operatorname{Re}\Sigma(M_{\text{BW}})-\Sigma(\eta+iy)}} \nonumber \\
& = & \begin{cases} \frac{M_{\text{BW}}}{\sqrt{\Sigma(\eta-i0)-\operatorname{Re}\Sigma(M_{\text{BW}})}} & \operatorname{Re}\Sigma(M_{\text{BW}})<\Sigma(\eta-i0) \\
\frac{-iM_{\text{BW}}}{\sqrt{\operatorname{Re}\Sigma(M_{\text{BW}})-\Sigma(\eta-i0)}} & \operatorname{Re}\Sigma(M_{\text{BW}})>\Sigma(\eta-i0)
\end{cases} \ . \label{equation_relLimit0i}
\end{eqnarray}
The inverse tangent inside the expression for the self-energy is purely imaginary on the negative imaginary axis and has absolute magnitude smaller than one, because the argument is smaller than one. Together with the prefactor, which is also purely imaginary, this combines to a negative real number subtracted by a positive logarithmic term. In the end, the self-energy is real and positive on the negative imaginary axis with the same limit as in Eq. (\ref{equation_relLimit0}). It declines along the negative imaginary axis and so the radicand in the first of the two above equations stays positive while increasing. This means that the purely imaginary pole keeps heading to complex infinity for increasing coupling. Hence, the tachyonic pole shows up when the left pole vanishes. Note that all these statements are strictly valid only for the first case in Eq. (\ref{equation_relLimit0i}). It clearly depends on the choice of the Breit--Wigner mass $M_{\text{BW}}$ and although we possess an expression for the real part of $\Sigma(M)$ we cannot perform further studies in that direction in virtue of the transcendental character of our equations. The numerical bound in order to obtain tachyonic excitations is $M_{\text{BW}}=0.4984 \ \text{GeV}$. Since the whole model has then become unphysical due to the tachyonic pole on the imaginary axis, it of course makes no sense to study the pole's behaviour for larger couplings.
\begin{figure}[!h]
\begin{center}
\includegraphics{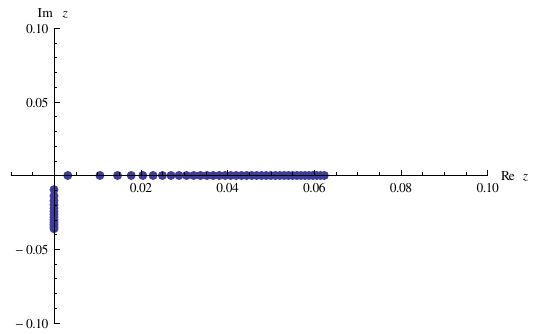}
\caption{$S\phi\phi$-model poles for $g_{S\phi\phi}\in[10.0 \ \text{GeV},10.5 \ \text{GeV}]$ in the first Riemann sheet with $\it\Delta$$g_{S\phi\phi}=0.01 \ \text{GeV}$.}
\label{figure_rel4dM06L15Pointsg100105gstep001-1}
\end{center}
\end{figure}

\subsection{Spectral function and its normalization} \label{subsection_spectralfunctionnormalization1}
\begin{figure}[!h]
\begin{center}
\includegraphics{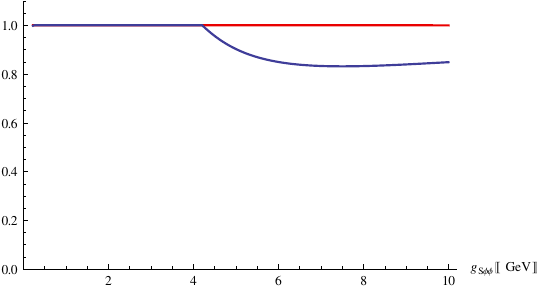}
\caption{Numerical verification that $\frac{1}{\pi}\int_{0}^{\infty}\text{d}s \ d(s)=1$ (red) by including the single poles of the first Riemann sheet, and numerical contribution of the continuous part $\frac{1}{\pi}\int_{2m}^{\sqrt{\Lambda^{2}+m^{2}}}\text{d}s \ d(s)$ (dark blue). The necessity of the delta distribution functions is evident.}
\label{figure_rel4dM06L15SpectralgNorm2}
\end{center}
\end{figure}

Let us now focus on the spectral function $d(s)$ and especially on its normalization. As was shown in Fig. \ref{figure_rel4dM06L15SpectralgNorm1} the spectral function is not normalized when using the numerical result of the integral
\begin{equation}
\frac{1}{\pi}\int_{0}^{\infty}\text{d}s \ d(s) \ , \label{equation_relnormintegral}
\end{equation}
simply because the additional simple poles are not seen by numerics. It is possible to take them into account by splitting the full propagator $\Delta_{S}(M)$ into the contributions of two single-particle propagators and the remaining continuum part above the branch cut (in particular, the right pole will be ignored completely; its renormalization constant is very small and gives no significant contribution):
\begin{eqnarray}
\Delta_{S}(M) & = & \frac{1}{\pi}\int_{0}^{\infty}\text{d}s \ \frac{d(s)}{M^{2}-s^{2}+i\epsilon} \nonumber \\
& = & \frac{Z_{\text{(l)pole}}}{M^{2}-x_{\text{(l)pole}}^{2}+i\epsilon}+\frac{1}{\pi}\int_{2m}^{2\sqrt{\Lambda^{2}+m^{2}}}\text{d}s \ \frac{d(s)}{M^{2}-s^{2}+i\epsilon} \ . \label{equation_relpropagatorwithZ}
\end{eqnarray}
The renormalization constant $Z_{\text{(l)pole}}$ is calculated by expanding the inverse of the full propagator in a Taylor series around the pole at $x=x_{\text{(l/r)pole}}$ in first order:
\begin{eqnarray}
x^{2}-M_{0}^{2}+g_{S\phi\phi}^{2}\Sigma(x) & \approx & \big(x^{2}-M_{0}^{2}+g_{S\phi\phi}^{2}\Sigma(x)\big)\big|_{x = x_{\text{(l)pole}}} \nonumber \\
& & \ + \ \frac{\text{d}}{\text{d}x}\Big(x^{2}-M_{0}^{2}+g_{S\phi\phi}^{2}\Sigma(x)\Big)\Big|_{x = x_{\text{(l)pole}}}\cdot(x-x_{\text{(l)pole}}) \nonumber \\
& = & \left(2x_{\text{(l)pole}}+g_{S\phi\phi}^{2}\frac{\text{d}\operatorname{Re}\Sigma(x)}{\text{d}x}\bigg|_{x=x_{\text{(l)pole}}}\right)\cdot(x-x_{\text{(l)pole}}) \ . \nonumber \\
\label{equation_relinverseprop}
\end{eqnarray}
The same argument as for the non-relativistic Lee model can be adopted here: The simple poles and the branch cut lie separated on the real axis at different masses, so the renormalization constants for each single-particle pole can be extracted from their residues:
\begin{eqnarray}
\lim_{x \to x_{\text{(l)pole}}}(x-x_{\text{(l)pole}})\cdot \Delta_{S}(x) & = & \underbrace{\lim_{x \to x_{\text{(l)pole}}}(x-x_{\text{(l)pole}})\cdot\bigg[\frac{Z_{\text{(l)pole}}}{x^{2}-x_{\text{(l)pole}}^{2}+i\epsilon}}_{\residue\left(\Delta_{S}(x), \ x=x_{\text{(l)pole}}\right)} \nonumber \\
& & + \ \big\{\text{other pole and branch cut}\big\}\bigg] \nonumber \\
& = & Z_{\text{(l)pole}} \ .
\end{eqnarray}
After inserting the above expansion into the left side, we finally get:
\begin{equation}
Z_{\text{(l)pole}} = \left[1+\frac{g_{S\phi\phi}^{2}}{2x_{\text{(l)pole}}}\frac{\text{d}\operatorname{Re}\Sigma(x)}{\text{d}x}\bigg|_{x = x_{\text{(l)pole}}}\right]^{-1} \ .
\end{equation}
The normalization condition thus can be written as
\begin{equation}
1 = Z_{\text{(l)pole}}+\int_{2m}^{2\sqrt{\Lambda^{2}+m^{2}}}\text{d}s \ d(s) \ .
\end{equation}
If we perform our calculation including the renormalization constants, the spectral function is normalized over the full range of $g_{S\phi\phi}$. The plot in Fig. \ref{figure_rel4dM06L15SpectralgNorm2} shows a completely fulfilled normalization condition for our choice of parameters (the points used are those from Tab. \ref{table_allpoints2}). For the sake of completeness, we provide a compilation of selected spectral function plots in \nameref{chapter_appendixF}. No analytic study is done since the expressions for the real and imaginary part of the self-energy are far too complicated to handle.
\\

\subsection{Decay width(s)}
It is reasonable to have a look at the full decay width of the resonance as two times the negative imaginary part $y_{\text{(m)pole}}=-\Gamma/2$ of the complex pole descending on the lower half plane of the second Riemann sheet. As was argued long time ago by Matthews and Salam \cite{salam}, the spectral function can be interpreted as a mass distribution of the unstable particle $S$. Remember the first chapter, where we have introduced the spectral function: given an interacting single-particle state, the spectral function $\rho(s^{2})$ describes the probability of finding a free state with invariant mass $s^{2}$. This in mind we already applied such a view when we calculated the average mass $x_{\text{average}}$ in the non-relativistic Lee model and the $S\phi\phi$-model, too. In Ref. \cite{giacosaSpectral}, Giacosa and Pagliara have given an intuitive argument for the correctness of the upper interpretation. In agreement with the authors point of view, we can define an average decay width by the expression
\begin{equation}
\Gamma_{\text{average}} = \frac{1}{\pi}\int_{2m}^{\infty}\text{d}s \ d(s)\Gamma_{\text{tree}}(s) \ . \label{equation_averagewidth}
\end{equation}
The first two panels of the next figure show the plot of the full width $\Gamma$ (red) in comparison with the tree-level result (\ref{equation_finaldecaywidth}) for the constant Breit--Wigner mass $M_{\text{BW}}$ (blue, dashed) and the average width (blue) as a function of the coupling. For small couplings $g_{S\phi\phi}<1 \ \text{GeV}$ there is actually no mentionable difference between those three, and they all vanish for $g_{S\phi\phi}\rightarrow0$. The situation changes completely for higher values: the full width becomes larger than the tree-level result, which again is larger than the average width. While this is true only for the latter two over the whole range of $g_{S\phi\phi}$, the tree-level result passes the value of the full width at $g_{S\phi\phi}=5.622 \ \text{GeV}$ and is nearly two times the full width when arriving at the end of the interval. It therefore cannot be regarded as a good approximation for large couplings, though it gives 'the best fit\grq \ for all values below the intersection point. On the other hand, the average value of the decay width is far too small in the intermediate range and large-coupling limit, but nevertheless the difference to the full width is 'only\grq \ 30\%.

\begin{figure}
\begin{minipage}[hbt]{6.9cm}
\centering
\includegraphics[width= 7.2cm]{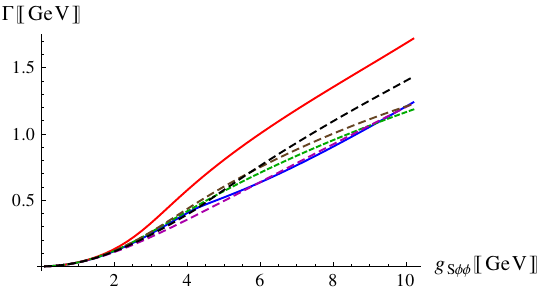}
\label{figure_rel4dM06L15DecayWidth1-2.eps}
\end{minipage}
\hfill
\begin{minipage}[hbt]{6.9cm}
\centering
\includegraphics[width= 7.2cm]{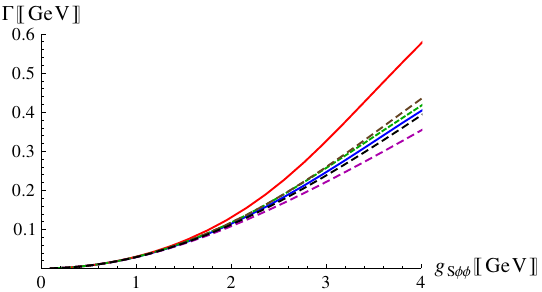}
\label{figure_rel4dM06L15DecayWidth2-2.eps}
\end{minipage}
\caption{Comparing decay widths for the $S\phi\phi$-model with sharp cutoff (and varying scales): the first two panels show the full width $\Gamma$ (red) in comparison with the tree-level result (\ref{equation_finaldecaywidth}) for the constant Breit--Wigner mass $M_{\text{BW}}$ (blue, dashed) and the average width (blue) as a function of the coupling. The lower two panels are plots of the full (red) and average width (blue), and of the tree-level results for each pole mass $x_{\text{(m)pole}}$ (green, dashed), each bare mass $M_{0}$ (magenta, dashed), each maximum value $x_{\text{max}}$ of the spectral function (brown, dashed) and each average mass $x_{\text{average}}$ (black, dashed).}
\label{figure_decaywidths}
\end{figure}

As an interesting consideration, one can study other decay widths as a combination of spectral function properties and the tree-level result. In the second part of Fig. \ref{figure_decaywidths} we provide plots of the full (red) and average width (blue), and plots of the tree-level results for each pole mass $x_{\text{(m)pole}}$ (green, dashed), each bare mass $M_{0}$ (magenta, dashed), each maximum value $x_{\text{max}}$ of the spectral function (brown, dashed) and each average mass $x_{\text{average}}$ (black, dashed). All of them may differ in several ways, but overall they are comparable and still yield too small values. The black dashed curve has the smallest error for large $g_{S\phi\phi}$, but not for intermediate ones.

\begin{figure}
\begin{minipage}[hbt]{6.9cm}
\centering
\includegraphics[width= 7.2cm]{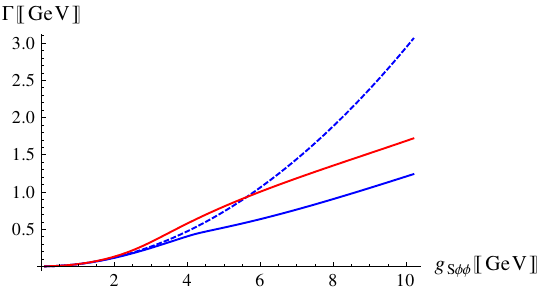}
\label{figure_rel4dM06L15DecayWidth1-1}
\end{minipage}
\hfill
\begin{minipage}[hbt]{6.9cm}
\centering
\includegraphics[width= 7.2cm]{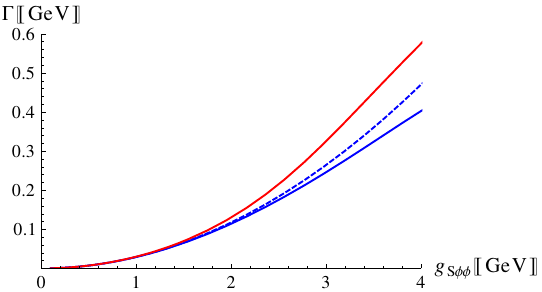}
\label{figure_rel4dM06L15DecayWidth2-1.eps}
\end{minipage}
\end{figure}
\clearpage

\subsection{Concluding remarks}
The present analysis is based on a fixed Breit--Wigner mass of $M_{\text{BW}}=0.6 \ \text{GeV}$. This is why the pole trajectory on the second Riemann sheet cannot be taken as the path of the resonance pole for different values of the coupling, but it outlines the pole position when requiring a constant Breit--Wigner mass. Some relevant points before we continue:

\begin{enumerate}
\item We have shown that a single {\em seed pole} for the $\sigma$-meson entails two other emerging poles, once the interaction term in the Lagrangian is turned on. One pole appears in the first sheet and can be discarded as non-physical, induced by the finite branch cut, while the other pole is {\em dynamically generated} in the vicinity of the origin in the second sheet. This pole moves to the branch point at threshold and slips through the cut onto the first sheet, where it has to be taken into account in the normalization condition of the spectral function. It can be interpreted as a bound state. The whole model becomes unstable due to the occurrence of tachyonic excitations for large couplings.
\item Although the Breit--Wigner parameterization reproduces very well the mass (and decay width) of the resonance pole in the small-coupling regime, this is not true anymore for intermediate values of $g_{S\phi\phi}$, compare Tab. \ref{table_allpoints2}. Above this regime, the mass is completely detached from the pole mass, while the average and bare mass, and the maximum value of the spectral function are sometimes in acceptable agreement with the pole mass.
\item The tree-level result for the decay width cannot be regarded as a good approximation for large couplings, yet it gives 'the best fit\grq \ in the intermediate regime (with deviations up to $100 \ \text{MeV}$).
\end{enumerate}

\section{Smooth cutoff}
\subsection{Analytic structure of the propagator}
A regularization function with sharp cutoff behaviour can only be motivated for reasons of simplicity. Fortunately, as long as $f_{\Lambda}(|\textbf{q}|)$ depends only on the magnitude of the three-momentum $\textbf{q}$, any smooth function modifies the self-energy integral
\begin{equation}
\Sigma(M) = \frac{1}{2\pi^{2}}\int_{0}^{\infty}\text{d}u \ \frac{u^{2}f_{\Lambda}^{2}(u)}{\sqrt{u^{2}+m^{2}}\big(4(u^{2}+m^{2})-M^{2}-i\epsilon\big)}
\end{equation}
as a multiplication factor that has to be included during the $u$-integration.
\begin{figure}[!h]
\begin{center}
\includegraphics{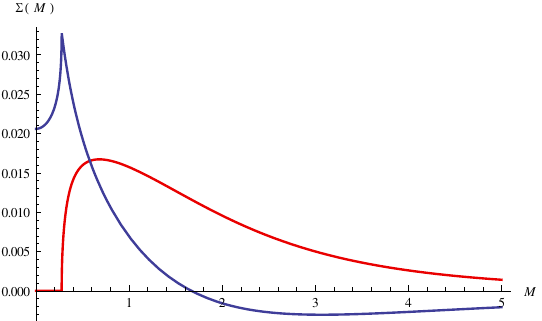}
\caption{Real (dark blue) and imaginary part (red) of the self-energy $\Sigma(M)$ on the positive real axis with $\Lambda=1.5 \ \text{GeV}$.}
\label{figure_rel4dSigma2}
\end{center}
\end{figure}
In Ref. \cite{giacosaSpectral}, a regularization function
\begin{equation}
f_{\Lambda}(|\textbf{q}|) = \frac{1}{1+\frac{|\textbf{q}|^{2}}{\Lambda^{2}}}
\end{equation}
was chosen, yielding an exponentially decreasing interaction strength for increasing distance of the two $\phi$-particles (see non-local Lagrangian (\ref{equation_nonlocalL})). Here, the analytic result for $\Sigma(M)$ reads:
\begin{eqnarray}
\Sigma(M) & = & \frac{\Lambda^{3}}{4(4\Lambda^{2}-4m^{2}+M^{2})^{2}\pi^{2}}\bigg[-\frac{\Lambda(4\Lambda^{2}-4m^{2}+M^{2})}{\Lambda^{2}-m^{2}} \nonumber \\
&  & -\frac{8\Lambda^{2}-4\Lambda^{2}m^{2}-4m^{2}+m^{2}M^{2}}{(m^{2}-\Lambda^{2})^{\frac{3}{2}}}\arctan\frac{\sqrt{m^{2}-\Lambda^{2}}}{\Lambda} \nonumber \\
&  & -\frac{8\Lambda\sqrt{4m^{2}-M^{2}}}{M}\arctan\frac{M}{\sqrt{4m^{2}-M^{2}}}\bigg] \ . \label{equation_relSigma2}
\end{eqnarray}
Considering only positive values $\sqrt{p^{2}}=M>0$, the imaginary part of $\Sigma(M)$ will be non-zero above threshold and falls rapidly for $M\rightarrow\infty$, see Fig. \ref{figure_rel4dSigma2}. This gives an infinite branch cut. In comparison to the self-energy for the sharp cutoff, all additional terms in Eq. (\ref{equation_relSigma2}) are single-valued, so it is clearly the root function and the inverse tangent in the last line that provide us with the branch cut starting at $z=2m$. The overall prefactor in the first line then makes the imaginary part of the self-energy fall for high $M$ as an effect of the incorporated regularization function. Note that the inverse tangent in the last term has two branch points in the complex $M$-plane, which are the same as for the root function:
\begin{eqnarray}
(1) \ \ \ \ \ \ \ \ \ \frac{M}{\sqrt{4m^{2}-M^{2}}} & \stackrel{!}{=} & - i \ , \\
\nonumber \\
\Rightarrow \ \ \ M & \rightarrow & \infty
\ , \\
\nonumber \\
(2) \ \ \ \ \ \ \ \ \ \frac{M}{\sqrt{4m^{2}-M^{2}}} & \stackrel{!}{=} & -i\infty \ , \\
\nonumber \\
\Rightarrow \ \ \ M & = & \lim_{\eta\rightarrow0^{+}}(2m+\eta) \ .
\end{eqnarray}
Let us look for the discontinuity by using the integral representation of the self-energy. As seen after the calculation carried out in the last section, we also here find the regularization function to be only a multiplicative factor\footnote{In particular, this would make further studies with more complicated regularization functions easier: Instead of working with loop integrals, one can use the general expression
\begin{equation}
\disc\Sigma(\sqrt{p^{2}}=M) = \frac{i\sqrt{\frac{M^{2}}{4}-m^{2}}}{4\pi M}f_{\Lambda}^{2}(p_{S\phi\phi}) \nonumber
\end{equation}
and perform a dispersion integral in the complex $p^{2}$-plane. This should simplify the computation in most cases. By adopting such a procedure, used by authors when fitting the scattering amplitude to experimental data, we automatically satisfy physically correct analytic properties, i.e., we get no spurious poles or cuts and have the correct asymptotic behaviour.} evaluated at the three-momentum $u=p_{S\phi\phi}=\sqrt{M^{2}/4-m^{2}}$, namely:
\begin{eqnarray}
\disc\Sigma(M) & = & \Sigma(M+i\epsilon)-\Sigma(M-i\epsilon) \nonumber \\
& = & \frac{1}{2\pi^{2}}\int_{0}^{\infty}\text{d}u \ \frac{u^{2}}{\big(1+\frac{u^{2}}{\Lambda^{2}}\big)^{2}}\bigg[\frac{1}{\sqrt{u^{2}+m^{2}}\big(4(u^{2}+m^{2})-M^{2}-i\epsilon\big)} \nonumber \\
&  & - \ \frac{1}{\sqrt{u^{2}+m^{2}}\big(4(u^{2}+m^{2})-M^{2}+i\epsilon\big)}\bigg] \nonumber \\
\nonumber \\
& = & \frac{i}{\pi}\int_{0}^{\infty}\text{d}u \ \frac{u^{2}}{\sqrt{u^{2}+m^{2}}\big(1+\frac{u^{2}}{\Lambda^{2}}\big)^{2}}\frac{1}{8\sqrt{\frac{M^{2}}{4}-m^{2}}} \ \delta\big(u-\sqrt{M^{2}/4-m^{2}}\big) \nonumber \\
& = & \frac{i\sqrt{\frac{M^{2}}{4}-m^{2}}}{4\pi M\Big(1+\frac{\frac{M^{2}}{4}-m^{2}}{\Lambda^{2}}\Big)^{2}} \ , \ \ \ M>2m \ .
\label{equation_relSigmaDiscontinuity2}
\end{eqnarray}
The useful identity (\ref{equation_usefullidentity}) between the discontinuity and the imaginary part right above the branch cut still holds true, since
\begin{eqnarray}
\Sigma(M) & = & \frac{1}{2\pi^{2}}\int_{0}^{\infty}\text{d}u \ \frac{u^{2}}{\sqrt{u^{2}+m^{2}}\big(1+\frac{u^{2}}{\Lambda^{2}}\big)^{2}\big(4(u^{2}+m^{2})-M^{2}-i\epsilon\big)} \nonumber \\
& = & \frac{1}{2\pi^{2}}\int_{0}^{\Lambda}\text{d}u \ \frac{u^{2}}{\sqrt{u^{2}+m^{2}}\big(1+\frac{u^{2}}{\Lambda^{2}}\big)^{2}\Big(u+\sqrt{\frac{M^{2}}{4}-m^{2}}\Big)}\frac{1}{u-\sqrt{\frac{M^{2}}{4}-m^{2}}-i\epsilon} \nonumber \\
\label{equation_relSigmaIntegralDecomposition2} \\
& \Rightarrow & \ \operatorname{Im}\Sigma(M) \ \ = \ \ \frac{\sqrt{\frac{M^{2}}{4}-m^{2}}}{8\pi M\Big(1+\frac{\frac{M^{2}}{4}-m^{2}}{\Lambda^{2}}\Big)^{2}} \ , \ \ \ M>2m \ ,
\end{eqnarray}
which indeed yields the same as in Eq. (\ref{equation_relSigmaDiscontinuity2}) after multiplying by $2i$. This result is valid {\em on} the positive real axis, so that the branch cut is continuous with the first quadrant.

\begin{figure}[t]
\begin{center}
\includegraphics{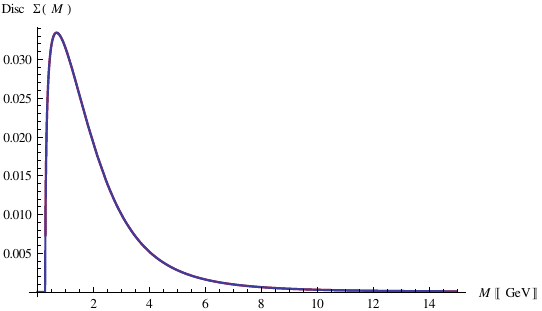}
\caption{Discontinuity of the self-energy (in units of $i$): numerical difference (dark blue) and analytic result (red).}
\label{figure_rel4dSigmaDisc2}
\end{center}
\end{figure}

The same expression can be obtained directly from the self-energy function (\ref{equation_relSigma2}) by using the logarithmic decomposition (\ref{equation_arctandecomposition}) for the inverse tangent. The ratio of negative complex root and single variable $z$ will give a branch cut for $z>2m$ along the positive real axis, while the inverse tangent once more just acts as a 'doorman\grq \ allowing the discontinuity of the root function to appear. Since this part gives support for all $M>2m$, the overall cut goes to infinity -- here lies the difference to the sharp cutoff. Explicitly this means (the single-valued terms were already subtracted):
\begin{eqnarray}
\disc\Sigma(M) & = & \Sigma(M+i\epsilon)-\Sigma(M-i\epsilon) \nonumber \\[0.3cm]
& = & -\frac{8\Lambda^{4}\sqrt{4m^{2}-M^{2}-i\epsilon}}{4(4\Lambda^{2}-4m^{2}+M^{2}+i\epsilon)^{2}\pi^{2}(M+i\epsilon)} \ \frac{i}{2}\bigg[\ln\left(1-i\frac{M+i\epsilon}{\sqrt{4m^{2}-M^{2}-i\epsilon}}\right) \nonumber \\
\nonumber \\
&  & -\ln\left(1+i\frac{M+i\epsilon}{\sqrt{4m^{2}-M^{2}-i\epsilon}}\right)\bigg] \nonumber \\
\nonumber \\
&  & +\frac{8\Lambda^{4}\sqrt{4m^{2}-M^{2}+i\epsilon}}{4(4\Lambda^{2}-4m^{2}+M^{2}-i\epsilon)^{2}\pi^{2}(M-i\epsilon)} \ \frac{i}{2}\bigg[\ln\left(1-i\frac{M-i\epsilon}{\sqrt{4m^{2}-M^{2}+i\epsilon}}\right) \nonumber \\
\nonumber \\
&  & -\ln\left(1+i\frac{\Lambda (M-i\epsilon)}{\sqrt{\Lambda^{2}+m^{2}}\sqrt{4m^{2}-M^{2}+i\epsilon}}\right)\bigg] \nonumber \\[0.3cm]
& = & \frac{8\Lambda^{4}i\sqrt{M^{2}-4m^{2}}}{4(4\Lambda^{2}-4m^{2}+M^{2})^{2}\pi^{2}M} \ \frac{i}{2}\bigg[\ln\left(1+\frac{M}{\sqrt{M^{2}-4m^{2}}}-i\epsilon\right) \nonumber \\
\nonumber \\
&  & -\ln\left(1-\frac{M}{\sqrt{M^{2}-4m^{2}}}+i\epsilon\right)\bigg] \nonumber \\
\nonumber \\
&  & +\frac{8\Lambda^{4}i\sqrt{M^{2}-4m^{2}}}{4(4\Lambda^{2}-4m^{2}+M^{2})^{2}\pi^{2}M} \ \frac{i}{2}\bigg[\ln\left(1-\frac{M}{\sqrt{M^{2}-4m^{2}}}-i\epsilon\right) \nonumber \\
\nonumber \\
&  & -\ln\left(1+\frac{M}{\sqrt{M^{2}-4m^{2}}}+i\epsilon\right)\bigg] \nonumber \\[0.3cm]
& = & \frac{i\sqrt{\frac{M^{2}}{4}-m^{2}}}{4\pi M\Big(1+\frac{\frac{M^{2}}{4}-m^{2}}{\Lambda^{2}}\Big)^{2}} \ , \ \ \ M>2m \ .
\end{eqnarray}
The discontinuity vanishes for $M<2m$ because the real parts of all radicands after the second equality stay positive, so we are beyond the branch cuts of the roots and logarithms. The self-energy above and below the positive real axis consequently equal each other in the limit of $\epsilon\rightarrow0^{+}$ and subtract to zero. In order to get a discontinuity from the roots outside the arguments, we also need the logarithmic cuts to contribute -- the real part of their arguments has to be negative or zero. Only the inequality
\begin{equation}
1-\frac{M}{\sqrt{M^{2}-4m^{2}}} \ \le \ 0
\end{equation}
has real positive solutions, namely $M>2m$. We have plotted the above discontinuity and the numerical difference of $\Sigma(E)$ across the real axis in Fig. \ref{figure_rel4dSigmaDisc2}.

The analytic continuation of the self-energy down into (at least the fourth quadrant of) the second Riemann sheet can now be written as
\begin{equation}
\Sigma_{\text{\Romannum{2}}}(z) = \Sigma(z)+\frac{i\sqrt{\frac{z^{2}}{4}-m^{2}}}{4\pi z\Big(1+\frac{\frac{z^{2}}{4}-m^{2}}{\Lambda^{2}}\Big)^{2}} \ ,
\end{equation}
and as a consequence there the propagator reads:
\begin{eqnarray}
\Delta_{\text{\Romannum{2}}}(z) & = & \frac{1}{z^{2}-M_{0}^{2}+g_{S\phi\phi}^{2}\Sigma_{\text{\Romannum{2}}}(z)} \nonumber \\
\nonumber \\
& = & \frac{1}{z^{2}-M_{0}^{2}+g_{S\phi\phi}^{2}\bigg(\Sigma(z)+\frac{i\sqrt{\frac{z^{2}}{4}-m^{2}}}{4\pi z\big(1+\frac{\frac{z^{2}}{4}-m^{2}}{\Lambda^{2}}\big)^{2}}\bigg)} \ .
\end{eqnarray}

\subsection{Couplings $g_{S\phi\phi}\in[0.1,1.0]$}
In the following, there will be one difference to the case of a sharp cutoff concerning the choice of parameters. The values for $m$ and $\Lambda$ will stay unchanged, whereas it is interesting to study the poles in the second Riemann sheet under the view of two different mass definitions. We have already introduced the (renormalized) Breit--Wigner mass $M_{\text{BW}}$ of the resonance as the real zero of the inverse propagator's real part,
\begin{equation}
M_{\text{BW}}^{2}-M_{0}^{2}+g_{S\phi\phi}^{2}\operatorname{Re}\Sigma(M_{\text{BW}}) = 0 \ ,
\end{equation}
which was fixed by setting it to $0.6 \ \text{GeV}$, a rough estimate of the $\sigma$-meson mass before the PDG update. In addition to that, we will now investigate the pole structure by fixing the bare mass $M_{0}=0.6 \ \text{GeV}$. We will solve Eq. (\ref{equation_rel0denominator}) numerically by varying the coupling $g_{S\phi\phi}$ in steps of $\it\Delta g_{S\phi\phi}$ using these parameters. The condition for deciding whether a solution $z_{\text{pole}}$ marks a singularity is still $\text{Abs}\big(\Delta_{\text{\Romannum{2}}}^{-1}(z_{\text{pole}})\big)<10^{-10} \ (\text{GeV})^{2}$.

\begin{figure}[!h]
\begin{center}
\includegraphics{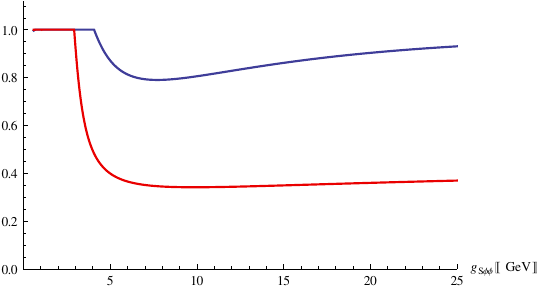}
\caption{Incomplete normalization of the spectral function $d(M)$ for different couplings. The dark blue curve belongs to the case of fixed $M_{\text{BW}}$, the red curve to the case of fixed $M_{0}$.}
\label{figure_rel4dM06L15SpectralgcombinedNorm1}
\end{center}
\end{figure}

Again, the spectral function $d(M)$ is normalized for both cases only for couplings belonging to an interval more or less $[0.2 \ \text{GeV},4.5 \ \text{GeV}]$, see Fig. \ref{figure_rel4dM06L15SpectralgcombinedNorm1}. All distortions for small $g_{S\phi\phi}$ are neglected, since if the coupling tends to zero, the spectral function becomes a delta distribution function due to the fact that the pole approaches the real axis.
Numerical integration of $\frac{1}{\pi}\int_{0}^{\infty}\text{d}s \ d(s)$ is not possible when poles on the real axis in the first sheet are present. We therefore expect extra poles in the first Riemann sheet for sufficiently large $g_{S\phi\phi}$.

\begin{figure}[!h]
\begin{center}
\includegraphics{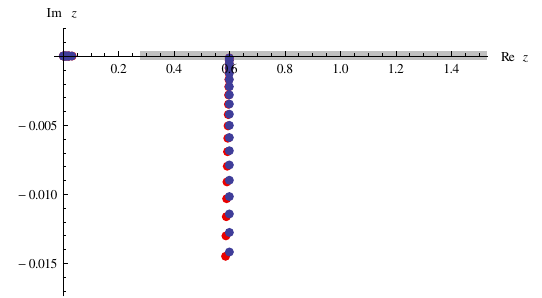}
\caption{$S\phi\phi$-model poles for $g_{S\phi\phi}\in[0.1 \ \text{GeV},1.0 \ \text{GeV}]$ in the second Riemann sheet with $\it\Delta$$g_{S\phi\phi}=0.05 \ \text{GeV}$, where the gray line marks the branch cut. The dark blue dots correspond to the case of fixed $M_{\text{BW}}$ and the red dots to the case of fixed $M_{0}$.}
\label{figure_rel4dM06L15Pointscombinedg0110gstep005-1}
\end{center}
\end{figure}

In general, both spectral functions become broader for larger couplings: the resonance poles move deeper down into the complex plane (i.e., the second Riemann sheet). The spectral functions 'shrink\grq \ very quickly the more the poles descend. The pole trajectories within the regarded coupling interval $[0.1 \ \text{GeV},1.0 \ \text{GeV}]$ are shown in Fig. \ref{figure_rel4dM06L15Pointscombinedg0110gstep005-1}. It seems that the pole in the case of fixed $M_{0}$ is faster (concerning $g_{S\phi\phi}$-velocity) and its path is twisted to the left, while the real part of the other one (with fixed $M_{\text{BW}}$) never falls below its starting value. Nevertheless, in both cases one clearly notices a gap pole heading to the first branch point at $z=2m$. The one for fixed bare mass takes precedence. To catch up with the analysis of a sharp cutoff, we have chosen the accuracy of the starting values on the lattice in such a way that poles on the real axis can be found the moment they really exist and indeed, the whole situation is very similar to the previous investigation. Equation (\ref{equation_rel0denominator}) can be solved for $g_{S\phi\phi}$ and the usual limits can be studied, whereas we will only consider the solutions for fixed bare mass $M_{0}$, since all analytic expressions for the other case were already found in the last section (see especially Eqs. (\ref{equation_relLimit0Disc}), (\ref{equation_relLimit2mDisc}), (\ref{equation_relLimit2m}), (\ref{equation_relLimit0}) and (\ref{equation_relLimit0i})). Clearly, just the numerical values will be different. We first look at
\begin{eqnarray}
g_{S\phi\phi}(0-i\epsilon) & = & \lim_{M\rightarrow0^{+}}\sqrt{\frac{M_{0}^{2}-M^{2}+i\epsilon}{\Sigma(M-i\epsilon)+\disc\Sigma(M-i\epsilon)}} \nonumber \\
& = & 0 \ .
\end{eqnarray}
This limit is due to the divergent behaviour of the discontinuity at the origin in the second sheet. It should be stressed that the infinitesimal number $i\epsilon$ is crucial. The coupling increases if the mass increases (as an issue of the interplay between discontinuity and self-energy).

A summary of pole positions and other relevant information for different intervals can be found in Tab. \ref{table_allpoints3} for fixed $M_{\text{BW}}$ and in Tab. \ref{table_allpoints4} for fixed $M_{0}$. As a reminder (see also Tab. \ref{table_allpointsexplanation2}): The pole masses $x_{\text{(m)pole}}$ and $x_{(0)\text{pole}}$ are the real parts of the complex solutions found by solving Eq. (\ref{equation_rel0denominator}), while $x_{\text{(l)pole}}$ is the real part of the complex solution in the first Riemann sheet, where we have replaced the self-energy $\Sigma_{\text{\Romannum{2}}}(z)$ in Eq. (\ref{equation_rel0denominator}) by $\Sigma(z)$. The corresponding Breit--Wigner masses are the solutions of the latter equation using only the real part of the self-energy on the first sheet (they are all real by construction). These are of course the zeros of the propagator's denominator on the real axis. The numerical maxima of the continuous part of the spectral function are denoted as $x_{\text{max}}$ and $x_{\text{average}}$ is the result of the numerically evaluated integral
\begin{equation}
x_{\text{average}} = \frac{1}{\pi}\int_{2m}^{\infty}\text{d}s \ sd(s) \ . \label{equation_relavmassintegral2}
\end{equation}
The last column of both tables gives the normalization value of the spectral function. All numbers after the fourth digit are dropped.
\\

\begin{table}[!h]\center
\scalebox{0.85}{
\begin{tabular}{l c c c c c c c c c c}
\toprule
\cmidrule(r){1-10}
$g_{S\phi\phi}$	& $x_{(0)\text{pole}}$ & $x_{\text{(l)pole}}$	& $x_{\text{(m)pole}}$	& $x_{\text{(0)BW}}$	& $x_{\text{(l)BW}}$	& $x_{\text{(m)BW}}$	& $x_{\text{max}}$ & $x_{\text{average}}$	& $M_{0}$	&Norm\\
\midrule
0.1	&0.0003	&-	&0.6000	&-	&-	&0.6000	&0.6000	&-	&0.6001	&-\\
0.2	&0.0012	&-	&0.6000	&-	&-	&0.6000	&0.6000	&-	&0.6005	&-\\
0.3	&0.0028	&-	&0.6000	&-	&-	&0.6000	&0.6000	&-	&0.6011	&-\\
0.4	&0.0050	&-	&0.6000	&-	&-	&0.6000	&0.6000	&-	&0.6021	&-\\
0.5	&0.0078	&-	&0.6000	&-	&-	&0.6000	&0.6000	&-	&0.6032	&-\\
0.6	&0.0113	&-	&0.6000	&-	&-	&0.6000	&0.6000	&-	&0.6047	&-\\
0.7	&0.0154	&-	&0.6000	&-	&-	&0.6000	&0.6000	&0.5983	&0.6064	&0.9996\\
0.8	&0.0201	&-	&0.6000	&-	&-	&0.6000	&0.6000	&0.6027	&0.6084	&0.9999\\
0.9	&0.0255	&-	&0.6000	&-	&-	&0.6000	&0.6000	&0.6051	&0.6106	&1.0000\\
1.0	&0.0315	&-	&0.6001	&-	&-	&0.6000	&0.6001	&0.6066	&0.6130	&0.9999\\
1.2	&0.0455	&-	&0.6003	&-	&-	&0.6000	&0.6003	&0.6097	&0.6187	&1.0000\\
1.4	&0.0619	&-	&0.6006	&-	&-	&0.6000	&0.6005	&0.6132	&0.6253	&1.0000\\
1.6	&0.0808	&-	&0.6011	&-	&-	&0.6000	&0.6010	&0.6172	&0.6329	&1.0000\\
1.8	&0.1017	&-	&0.6019	&-	&-	&0.6000	&0.6017	&0.6217	&0.6414	&1.0000\\
2.0	&0.1244	&-	&0.6033	&-	&-	&0.6000	&0.6027	&0.6268	&0.6507	&1.0000\\
2.2	&0.1483	&-	&0.6054	&-	&-	&0.6000	&0.6042	&0.6324	&0.6608	&1.0000\\
2.4	&0.1725	&-	&0.6086	&-	&-	&0.6000	&0.6063	&0.6384	&0.6718	&1.0000\\
2.6	&0.1960	&-	&0.6135	&-	&-	&0.6000	&0.6092	&0.6450	&0.6835	&1.0000\\
2.8	&0.2178	&-	&0.6203	&-	&-	&0.6000	&0.6132	&0.6521	&0.6959	&1.0000\\
3.0	&0.2366	&-	&0.6297	&-	&-	&0.6000	&0.6184	&0.6597	&0.7090	&1.0000\\
3.2	&0.2519	&-	&0.6419	&-	&-	&0.6000	&0.6253	&0.6677	&0.7228	&1.0000\\
3.4	&0.2634	&-	&0.6570	&-	&-	&0.6000	&0.6341	&0.6762	&0.7371	&1.0000\\
3.6	&0.2712	&-	&0.6747	&-	&-	&0.6000	&0.6452	&0.6852	&0.7520	&1.0000\\
3.8	&0.2758	&-	&0.6948	&-	&-	&0.6000	&0.6586	&0.6947	&0.7675	&1.0000\\
4.0	&0.2778	&-	&0.7169	&-	&-	&0.6000	&0.6743	&0.7046	&0.7834	&1.0000\\
4.2	&-	&0.2776	&0.7407	&0.2938	&0.2776	&0.6000	&0.6923	&0.7088	&0.7999	&1.0000\\
4.4	&-	&0.2758	&0.7658	&0.3216	&0.2758	&0.6000	&0.7123	&0.7103	&0.8167	&1.0000\\
4.6	&-	&0.2725	&0.7921	&0.3503	&0.2725	&0.6000	&0.7341	&0.7139	&0.8340	&1.0000\\
4.8	&-	&0.2682	&0.8193	&0.3798	&0.2682	&0.6000	&0.7573	&0.7195	&0.8517	&1.0000\\
5.0	&-	&0.2629	&0.8473	&0.4100	&0.2629	&0.6000	&0.7818	&0.7267	&0.8698	&1.0000\\
5.5	&-	&0.2467	&0.9198	&0.4886	&0.2467	&0.6000	&0.8468	&0.7514	&0.9164	&1.0000\\
6.0	&-	&0.2270	&0.9946	&0.5706	&0.2270	&0.6000	&0.9157	&0.7834	&0.9649	&1.0000\\
6.5	&-	&0.2044	&1.0706	&0.6552	&0.2044	&0.6000	&0.9869	&0.8208	&1.0150	&1.0000\\
7.0	&-	&0.1789	&1.1470	&0.7417	&0.1789	&0.6000	&1.0597	&0.8623	&1.0664	&1.0000\\
7.5	&-	&0.1498	&1.2233	&0.8294	&0.1498	&0.6000	&1.1333	&0.9069	&1.1190	&1.0000\\
8.0	&-	&0.1147	&1.2991	&0.9177	&0.1147	&0.6000	&1.2074	&0.9541	&1.1727	&1.0000\\
8.5	&-	&0.0653	&1.3743	&1.0062	&0.0653	&0.6000	&1.2818	&1.0034	&1.2272	&1.0000\\
\bottomrule
\end{tabular}
}
\caption{Selection of masses (in units of GeV) for the $S\phi\phi$-model with $\Lambda=1.5 \ \text{GeV}, \ M_{\text{BW}}=0.6 \ \text{GeV}$ and $m=0.139 \ \text{GeV}$. Here, $M_{\text{BW}}$ is fixed.}
\label{table_allpoints3}
\end{table}
\clearpage

\begin{table}[!h]\center
\scalebox{0.85}{
\begin{tabular}{l c c c c c c c c c c}
\toprule
\cmidrule(r){1-10}
$g_{S\phi\phi}$	& $x_{(0)\text{pole}}$ & $x_{\text{(l)pole}}$	& $x_{\text{(m)pole}}$	& $x_{\text{(0)BW}}$	& $x_{\text{(l)BW}}$	& $x_{\text{(m)BW}}$	& $x_{\text{max}}$ & $x_{\text{average}}$	& $M_{0}$	&Norm\\
\midrule
0.1	&0.0003	&-	&0.5998	&-	&-	&0.5998	&-	&-	&0.6000	&-\\
0.2	&0.0012	&-	&0.5994	&-	&-	&0.5994	&-	&-	&0.6000	&-\\
0.3	&0.0028	&-	&0.5988	&-	&-	&0.5988	&0.5988	&-	&0.6000	&-\\
0.4	&0.0050	&-	&0.5978	&-	&-	&0.5978	&0.5978	&-	&0.6000	&-\\
0.5	&0.0079	&-	&0.5966	&-	&-	&0.5966	&0.5966	&-	&0.6000	&-\\
0.6	&0.0114	&-	&0.5951	&-	&-	&0.5951	&0.5951	&-	&0.6000	&-\\
0.7	&0.0157	&-	&0.5934	&-	&-	&0.5934	&0.5934	&0.5967	&0.6000	&0.9999\\
0.8	&0.0207	&-	&0.5913	&-	&-	&0.5913	&0.5913	&0.5958	&0.6000	&0.9999\\
0.9	&0.0207	&-	&0.5913	&-	&-	&0.5913	&0.5913	&0.5958	&0.6000	&0.9999\\
1.0	&0.0330	&-	&0.5863	&-	&-	&0.5862	&0.5864	&0.5934	&0.6000	&1.0000\\
1.2	&0.0486	&-	&0.5801	&-	&-	&0.5798	&0.5801	&0.5904	&0.6000	&1.0000\\
1.4	&0.0679	&-	&0.5725	&-	&-	&0.5719	&0.5725	&0.5869	&0.6000	&1.0000\\
1.6	&0.0911	&-	&0.5635	&-	&-	&0.5623	&0.5635	&0.5827	&0.6000	&1.0000\\
1.8	&0.1187	&-	&0.5529	&-	&-	&0.5507	&0.5527	&0.5778	&0.6000	&1.0000\\
2.0	&0.1506	&-	&0.5408	&-	&-	&0.5366	&0.5400	&0.5722	&0.6000	&1.0000\\
2.2	&0.1864	&-	&0.5274	&-	&-	&0.5193	&0.5250	&0.5658	&0.6000	&1.0000\\
2.4	&0.2236	&-	&0.5138	&-	&-	&0.4976	&0.5071	&0.5585	&0.6000	&1.0000\\
2.6	&0.2558	&-	&0.5032	&-	&-	&0.4690	&0.4851	&0.5502	&0.6000	&1.0000\\
2.8	&0.2745	&-	&0.4996	&-	&-	&0.4268	&0.4561	&0.5408	&0.6000	&1.0000\\
3.0	&-	&0.2772	&0.5040	&-	&0.2772	&-	&-	&0.5122	&0.6000	&1.0000\\
3.2	&-	&0.2672	&0.5146	&-	&0.2672	&-	&-	&0.4574	&0.6000	&1.0000\\
3.4	&-	&0.2476	&0.5292	&-	&0.2476	&-	&0.4378	&0.4197	&0.6000	&1.0000\\
3.6	&-	&0.2196	&0.5465	&-	&0.2196	&-	&0.4730	&0.3948	&0.6000	&1.0000\\
3.8	&-	&0.1818	&0.5658	&-	&0.1818	&-	&0.5032	&0.3784	&0.6000	&1.0000\\
4.0	&-	&0.1279	&0.5866	&-	&0.1279	&-	&0.5317	&0.3679	&0.6000	&1.0000\\
\bottomrule
\end{tabular}
}
\caption{Selection of masses (in units of GeV) for the $S\phi\phi$-model with $\Lambda=1.5 \ \text{GeV}, \ M_{0}=0.6 \ \text{GeV}$ and $m=0.139 \ \text{GeV}$. Here, $M_{0}$ is fixed.}
\label{table_allpoints4}
\end{table}

\subsection{Couplings $g_{S\phi\phi}\in[1.0,2.8]$}
\begin{figure}[!h]
\begin{center}
\includegraphics{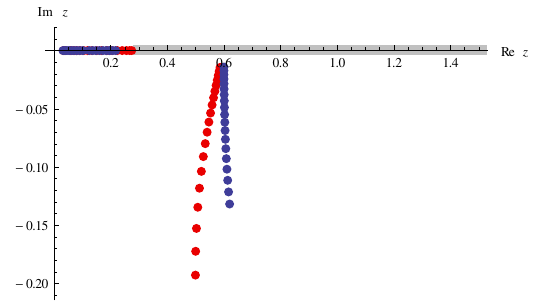}
\caption{$S\phi\phi$-model poles for $g_{S\phi\phi}\in[1.0 \ \text{GeV},2.8 \ \text{GeV}]$ in the second Riemann sheet with $\it\Delta$$g_{S\phi\phi}=0.1 \ \text{GeV}$. The dark blue dots correspond to the case of fixed $M_{\text{BW}}$ and the red dots to the case of fixed $M_{0}$.}
\label{figure_rel4dM06L15Pointscombinedg1028gstep01-1}
\end{center}
\end{figure}

No poles are found in the first sheet. The resonance poles descend deeper into the lower half plane, displaying a huge difference between the two mass definitions: the dark blue pole turns right (as in the case of a sharp cutoff), while the red one is twisted to the left (as in the non-relativistic case), yet one can see that this curve becomes flatter (see Fig. \ref{figure_rel4dM06L15Pointscombinedg1028gstep01-1}). We also realize that the red poles leave the blue one behind, e.g. the red gap pole keeps approaching the first branch point with a much higher parameter velocity. This makes sense, because the limit
\begin{eqnarray}
g_{S\phi\phi}(2m-i\epsilon) & = & \lim_{\eta\rightarrow0^{+}}\sqrt{\frac{M_{0}^{2}-(2m-\eta)^{2}+i\epsilon}{\Sigma(2m-\eta-i\epsilon)+\disc\Sigma(2m-\eta-i\epsilon)}} \nonumber \\
& = & \sqrt{\frac{M_{0}^{2}-4m^{2}}{\Sigma(2m)}}
\end{eqnarray}
is larger than the result in Eq. (\ref{equation_relLimit2mDisc}) for every point $z<2m$ on the real axis, namely for $M_{0}=M_{\text{BW}}$ the overall ratio is
\begin{equation}
\sqrt{\frac{\Sigma(M)-\operatorname{Re}\Sigma(M_{\text{BW}})}{\Sigma(M)}}
\end{equation}
as long as $\Sigma(M)>\operatorname{Re}\Sigma(M_{\text{BW}})$. The latter, though, depends on the choice of the Breit--Wigner mass. The numerical bound in order to make the upper condition hold is $M_{\text{BW}}>0.4721 \ \text{GeV}$ for the smooth cutoff (this will turn out to be the same number so as to make tachyonic excitations appear for sufficiently large values of the coupling).
\begin{figure}[!h]
\begin{center}
\includegraphics{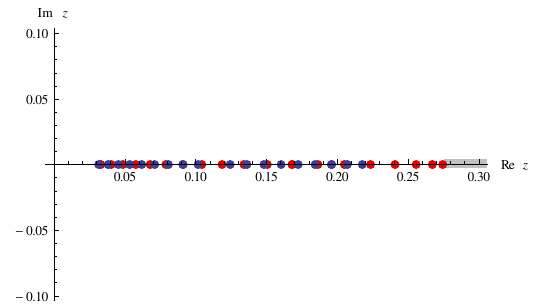}
\caption{$S\phi\phi$-model gap poles for $g_{S\phi\phi}\in[1.0 \ \text{GeV},2.8 \ \text{GeV}]$ on the real axis of the second Riemann sheet with $\it\Delta$$g_{S\phi\phi}=0.1 \ \text{GeV}$. The dark blue dots correspond to the case of fixed $M_{\text{BW}}$ and the red dots to the case of fixed $M_{0}$.}
\label{figure_rel4dM06L15Pointscombinedg1028gstep01-2}
\end{center}
\end{figure}

\subsection{Couplings $g_{S\phi\phi}\in[2.8,5.0]$}
\begin{figure}[t]
\begin{center}
\includegraphics{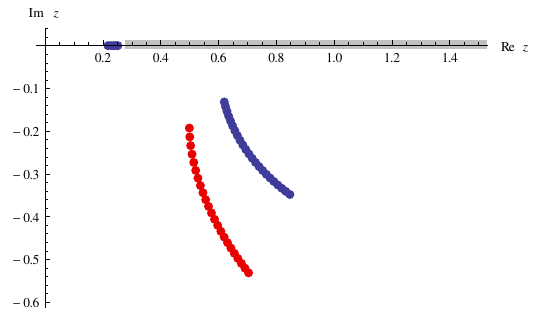}
\caption{$S\phi\phi$-model poles for $g_{S\phi\phi}\in[2.8 \ \text{GeV},5.0 \ \text{GeV}]$ in the second Riemann sheet with $\it\Delta$$g_{S\phi\phi}=0.1 \ \text{GeV}$. The dark blue dots correspond to the case of fixed $M_{\text{BW}}$ and the red dots to the case of fixed $M_{0}$.}
\label{figure_rel4dM06L15Pointscombinedg2850gstep01-1}
\end{center}
\end{figure}

For this range of the coupling we are once again faced with an interesting behaviour. As we increase $g_{S\phi\phi}$, for some values the gap poles hit the first branch point at $z=2m$ and vanish, possibly they slip into the first Riemann sheet. Although these strange poles have no physical interpretation, their existence below threshold lasts as long as $g_{S\phi\phi}$ is non-zero. For fixed $M_{0}$ the master solution
\begin{equation}
g_{S\phi\phi}(M-i\epsilon) = \sqrt{\frac{M_{0}^{2}-M^{2}+i\epsilon}{\Sigma(M-i\epsilon)+\disc\Sigma(M-i\epsilon)}}
\end{equation}
yields a positive and real-valued result for the coupling as long as the bare mass is larger than $2m$, because for $M$ lying below threshold the self-energy $\Sigma(M)$ and the discontinuity are both purely real and positive right below the real axis. It is very important to realize that the pole must lie below the real axis. Otherwise, the root inside the expression for the discontinuity would be evaluated in the wrong quadrant giving an imaginary solution for $g_{S\phi\phi}$. Indeed, we find the gap poles leaving the second sheet after they arrived at the first branch point. In the case of fixed $M_{0}$ this happens for
\begin{eqnarray}
g_{S\phi\phi}(2m-i\epsilon) & = & \lim_{\eta\rightarrow0^{+}}\sqrt{\frac{M_{0}^{2}-(2m-\eta)^{2}+i\epsilon}{\Sigma(2m-\eta-i\epsilon)+\disc\Sigma(2m-\eta-i\epsilon)}} \nonumber \\
& = & \sqrt{\frac{M_{0}^{2}-4m^{2}}{\Sigma(2m)}} \nonumber \\
& = & \sqrt{\frac{16\pi^{2}(m^{2}-\Lambda^{2})^{\frac{3}{2}}(M_{0}^{2}-4m^{2})}{\Lambda\big(\sqrt{m^{2}-\Lambda^{2}}+(m^{2}+2\Lambda^{2})\arctan\frac{\sqrt{m^{2}-\Lambda^{2}}}{\Lambda}\big)}} \ ,
\end{eqnarray}
or numerically $g_{S\phi\phi}=2.935 \ \text{GeV}$ for our choice of parameters. For fixed $M_{\text{BW}}$ we find $g_{S\phi\phi}=4.083 \ \text{GeV}$. It is evident that there consists a huge difference inside the model when we either hold the bare mass constant or the Breit--Wigner mass. As one can see from Fig. \ref{figure_rel4dM06L15Pointscombinedg2850gstep01-1}, not only the resonance pole lies deeper in the lower half plane, also the gap pole on the real axis of the second sheet slips much earlier through the branch cut. Both gap poles finally emerge in the first sheet left from the first branch point due to the outcome of the previous section and the limit
\begin{eqnarray}
g_{S\phi\phi}(2m-i\epsilon) & = & \lim_{\eta\rightarrow0^{+}}\sqrt{\frac{M_{0}^{2}-(2m-\eta)^{2}+i\epsilon}{\Sigma(2m-\eta-i\epsilon)}} \nonumber \\
& = & \sqrt{\frac{M_{0}^{2}-4m^{2}}{\Sigma(2m)}} \ .
\end{eqnarray}
This equation has no real solution for $g_{S\phi\phi}< 2.935 \ \text{GeV}$ and is purely real for larger $g_{S\phi\phi}$. Note that the new left pole for fixed bare mass tends to the origin in the first sheet and reaches it for
\begin{eqnarray}
g_{S\phi\phi}(0-i\epsilon) & = & \lim_{M\rightarrow0^{+}}\sqrt{\frac{M_{0}^{2}-M^{2}+i\epsilon}{\Sigma(M-i\epsilon)}} \nonumber \\
& = & \sqrt{\frac{M_{0}^{2}}{\Sigma(0-i\epsilon)}} \nonumber \\
& = & \sqrt{\frac{16M_{0}^{2}\pi^{2}(m^{2}-\Lambda^{2})^{\frac{5}{2}}}{\Lambda^{3}\big(-3\Lambda\sqrt{m^{2}-\Lambda^{2}}+(m^{2}+2\Lambda^{2})\arctan\frac{\sqrt{m^{2}-\Lambda^{2}}}{\Lambda}\big)}} \ ,
\end{eqnarray}
or numerically $g_{S\phi\phi}=4.184 \ \text{GeV}$.

In the end we may illustrate the new left pole showing up in the first sheet simply by considering the spectral function for $M<2m$:
\begin{eqnarray}
d(M) & = & \frac{2M\big(\epsilon+g_{S\phi\phi}^{2}\operatorname{Im}\Sigma(M+i\epsilon)\big)}{\big(M^{2}-M^{2}_{0}+g_{S\phi\phi}^{2}\operatorname{Re}\Sigma(M+i\epsilon)\big)^{2}+\big(\epsilon+g_{S\phi\phi}^{2}\operatorname{Im}\Sigma(M+i\epsilon)\big)^{2}} \nonumber \\
& = & \frac{2M\epsilon}{\big(M^{2}-M_{0}^{2}+g_{S\phi\phi}^{2}\operatorname{Re}\Sigma(M+i\epsilon)\big)^{2}+\epsilon^{2}} \nonumber \\
& = & 2M\pi\delta\big(M^{2}-M_{0}^{2}+g_{S\phi\phi}^{2}\operatorname{Re}\Sigma(M)\big) \ .
\end{eqnarray}
From here we also see that the Breit--Wigner mass $x_{\text{(l)BW}}$ equals the corresponding pole mass (the argument of the delta distribution function is the master solution in the first sheet). But, compare Tab. \ref{table_allpoints4}, once the coupling exceeds a critical value of $g_{S\phi\phi}=2.969 \ \text{GeV}$, there is actually no Breit--Wigner mass for the resonance pole in the second sheet!
\begin{figure}[!h]
\begin{center}
\includegraphics{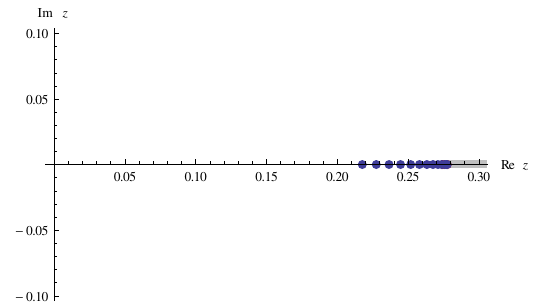}
\caption{Left $S\phi\phi$-model poles for $g_{S\phi\phi}\in[2.8 \ \text{GeV},5.0 \ \text{GeV}]$ on the real axis of the second Riemann sheet with $\it\Delta$$g_{S\phi\phi}=0.1 \ \text{GeV}$. The dark blue dots correspond to the case of fixed $M_{\text{BW}}$ and the red dots to the case of fixed $M_{0}$.}
\label{figure_rel4dM06L15Pointscombinedg2850gstep01-2}
\end{center}
\end{figure}
\begin{figure}[!h]
\begin{center}
\includegraphics{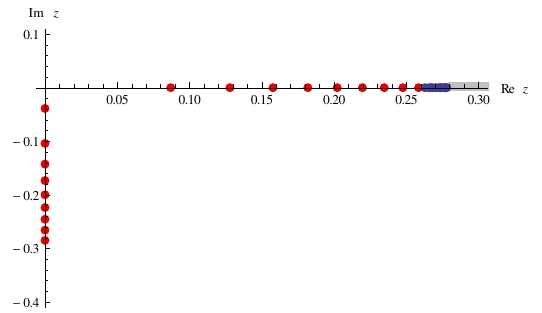}
\caption{$S\phi\phi$-model poles for $g_{S\phi\phi}\in[2.8 \ \text{GeV},5.0 \ \text{GeV}]$ in the first Riemann sheet with $\it\Delta$$g_{S\phi\phi}=0.1 \ \text{GeV}$. The dark blue dots correspond to the case of fixed $M_{\text{BW}}$ and the red dots to the case of fixed $M_{0}$.}
\label{figure_rel4dM06L15Pointscombinedg2850gstep01-3}
\end{center}
\end{figure}
This is not true in the case of fixed $M_{\text{BW}}$, both for sharp and smooth cutoff, since we forced all equations to satisfy
\begin{equation}
M_{\text{BW}}^{2}-M_{0}^{2}+g_{S\phi\phi}^{2}\operatorname{Re}\Sigma(M_{\text{BW}}) = 0 \ .
\end{equation}
Figure \ref{figure_rel4dM06L15Pointscombinedg2850gstep01-3} shows the left pole arriving at the origin for a positive value of the coupling and after that it becomes tachyonic in the regarded range of the coupling for
\begin{eqnarray}
g_{S\phi\phi}(\eta-i0) & = & \lim_{y\rightarrow0^{-}}\sqrt{\frac{M_{0}^{2}-(\eta+iy)^{2}}{\Sigma(\eta+iy)}} \nonumber \\
& = & \sqrt{\frac{M_{0}^{2}}{\Sigma(\eta-i0)}} \ .
\end{eqnarray}

\subsection{Couplings $g_{S\phi\phi}\in[5.0,9.0]$}
The only thing that is left is to find the coupling for which the left pole on the real axis becomes tachyonic in the case of fixed $M_{\text{BW}}$. We already know the limit
\begin{eqnarray}
g_{S\phi\phi}(0-i\epsilon) & = & \lim_{M\rightarrow0^{+}}\sqrt{\frac{M^{2}-M_{\text{BW}}^{2}-i\epsilon}{\operatorname{Re}\Sigma(M_{\text{BW}})-\Sigma(M-i\epsilon)}} \nonumber \\
& = & \begin{cases} \frac{M_{\text{BW}}}{\sqrt{\Sigma(0-i\epsilon)-\operatorname{Re}\Sigma(M_{\text{BW}})}} & \operatorname{Re}\Sigma(M_{\text{BW}})<\Sigma(0-i\epsilon) \\
\frac{-iM_{\text{BW}}}{\sqrt{\operatorname{Re}\Sigma(M_{\text{BW}})-\Sigma(0-i\epsilon)}} & \operatorname{Re}\Sigma(M_{\text{BW}})>\Sigma(0-i\epsilon)
\end{cases} \ ,
\end{eqnarray}
where the first is true with value $g_{S\phi\phi}=8.748 \ \text{GeV}$. Of course, this is the same as
\begin{eqnarray}
g_{S\phi\phi}(\eta-i0) & = & \lim_{y\rightarrow0^{-}}\sqrt{\frac{(\eta+iy)^{2}-M_{\text{BW}}^{2}}{\operatorname{Re}\Sigma(M_{\text{BW}})-\Sigma(\eta+iy)}} \nonumber \\
& = & \begin{cases} \frac{M_{\text{BW}}}{\sqrt{\Sigma(\eta-i0)-\operatorname{Re}\Sigma(M_{\text{BW}})}} & \operatorname{Re}\Sigma(M_{\text{BW}})<\Sigma(\eta-i0) \\
\frac{-iM_{\text{BW}}}{\sqrt{\operatorname{Re}\Sigma(M_{\text{BW}})-\Sigma(\eta-i0)}} & \operatorname{Re}\Sigma(M_{\text{BW}})>\Sigma(\eta-i0)
\end{cases} \ . \label{equation_relLimit0i2}
\end{eqnarray}
Hence, the tachyonic pole appears when the left pole vanishes. This statement is only valid for the first case in Eq. (\ref{equation_relLimit0i2}). It depends on the choice of the Breit--Wigner mass $M_{\text{BW}}$ and although we possess an expression for the real part of $\Sigma(M)$ we cannot perform further studies in that direction in virtue of the transcendental character of our equations. The numerical bound in order to obtain tachyonic excitations is $M_{\text{BW}}=0.4721 \ \text{GeV}$.
\\
\begin{figure}[!h]
\begin{center}
\includegraphics{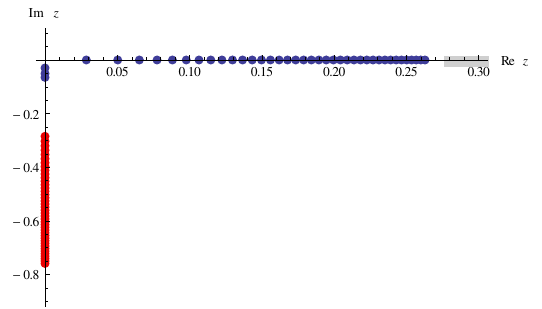}
\caption{$S\phi\phi$-model poles for $g_{S\phi\phi}\in[5.0 \ \text{GeV},9.0 \ \text{GeV}]$ in the first Riemann sheet with $\it\Delta$$g_{S\phi\phi}=0.1 \ \text{GeV}$. The dark blue dots correspond to the case of fixed $M_{\text{BW}}$ and the red dots to the case of fixed $M_{0}$.}
\label{figure_rel4dM06L15Pointscombinedg5090gstep01}
\end{center}
\end{figure}

\subsection{Spectral function and its normalization}
Let us now have a quick look at the spectral functions $d(s)$ for both cases and especially on their normalization. As was shown in Fig. \ref{figure_rel4dM06L15SpectralgcombinedNorm1}, the spectral functions are not normalized when using the numerical results for the integral
\begin{equation}
\frac{1}{\pi}\int_{0}^{\infty}\text{d}s \ d(s) \ , \label{equation_relnormintegral2}
\end{equation}
simply because the additional simple poles are not seen by numerics. As usual, we need to compute renormalization constants:
\begin{eqnarray}
\Delta_{S}(M) & = & \frac{1}{\pi}\int_{0}^{\infty}\text{d}s \ \frac{d(s)}{M^{2}-s^{2}+i\epsilon} \nonumber \\
& = & \frac{Z_{\text{(l)pole}}}{M^{2}-x_{\text{(l)pole}}^{2}+i\epsilon}+\frac{1}{\pi}\int_{2m}^{\infty}\text{d}s \ \frac{d(s)}{M^{2}-s^{2}+i\epsilon} \ . \label{equation_relpropagatorwithZ2}
\end{eqnarray}
The renormalization constant $Z_{\text{(l)pole}}$ is calculated by expanding the inverse of the full propagator in a Taylor series around the pole at $x=x_{\text{(l)pole}}$ in first order and since the simple pole and the branch cut lie separated on the real axis at different masses, the renormalization constant for the single pole can be extracted from its residue.
\begin{figure}[!h]
\begin{center}
\includegraphics{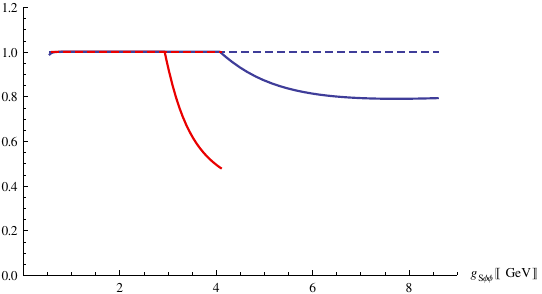}
\caption{Numerical verification that $\frac{1}{\pi}\int_{0}^{\infty}\text{d}s \ d(s)=1$ (blue/red dashed for fixed $M_{\text{BW}}$/$M_{0}$) by including the single pole of the first Riemann sheet, and numerical contribution of the continuous part $\frac{1}{\pi}\int_{2m}^{\infty}\text{d}s \ d(s)$ (dark blue and dark blue dashed). The necessity of the delta distribution function is evident.}
\label{figure_rel4dM06L15SpectralgcombinedNorm2}
\end{center}
\end{figure}
The final result equals the one for the sharp cutoff:
\begin{equation}
Z_{\text{(l)pole}} = \left[1+\frac{g_{S\phi\phi}^{2}}{2x_{\text{(l)poles}}}\frac{\text{d}\operatorname{Re}\Sigma(x)}{\text{d}x}\bigg|_{x = x_{\text{(l)pole}}}\right]^{-1} \ . \label{equation_renormconstant2}
\end{equation}
The normalization condition can thus be written as
\begin{equation}
1 = Z_{\text{(l)pole}}+\int_{2m}^{\infty}\text{d}s \ d(s) \ .
\end{equation}
If we perform our calculation including the renormalization constants, the spectral functions are then normalized over the full range of $g_{S\phi\phi}$. The plots in Fig. \ref{figure_rel4dM06L15SpectralgcombinedNorm2} show the completely fulfilled normalization condition for our choice of parameters (the used points are those in Tab. \ref{table_allpoints3} and \ref{table_allpoints4}). For the sake of completeness, we provide a compilation of selected spectral function plots in \nameref{chapter_appendixG}.

\subsection{Decay width(s)}
It is reasonable to have a look at the full decay width of the resonance in both cases, as two times the negative imaginary part $y_{\text{(m)pole}}=-\Gamma/2$ of the complex pole descending on the lower half plane of the second Riemann sheet. In the next figure the first two panels show plots of the full width (red) in comparison with the tree-level result
\begin{equation}
\Gamma_{\text{tree}}(M) = \frac{\sqrt{\frac{M^{2}}{4}-m^{2}}}{4\pi M^{2}\Big(1+\frac{\frac{M^{2}}{4}-m^{2}}{\Lambda^{2}}\Big)^{2}} \ \Theta(M-2m) \label{equation_smoothtreelevelwidth}
\end{equation}
for a constant Breit--Wigner mass (blue, dashed) and the average width (blue) from Eq. (\ref{equation_averagewidth}) as a function of the coupling. For small couplings $g_{S\phi\phi}<1 \ \text{GeV}$ there is not a large difference between those three, and they all vanish for $g_{S\phi\phi}\rightarrow0$. The situation changes completely for larger values: the full width becomes larger than the tree-level result, which again is larger than the average width. While this is true only for the latter two over the whole range of $g_{S\phi\phi}$, the tree-level result passes the value of the full width at $g_{S\phi\phi}= 5.044 \ \text{GeV}$ and is more than two times the full width when arriving at the end of the interval. It therefore cannot be regarded as a good approximation for large couplings, though it gives 'the best fit\grq \ for all values below the intersection point. On the other hand, the average value of the decay width is far too small in the intermediate range and high-coupling limit.

\begin{figure}
\begin{minipage}[hbt]{6.9cm}
\centering
\includegraphics[width= 7.2cm]{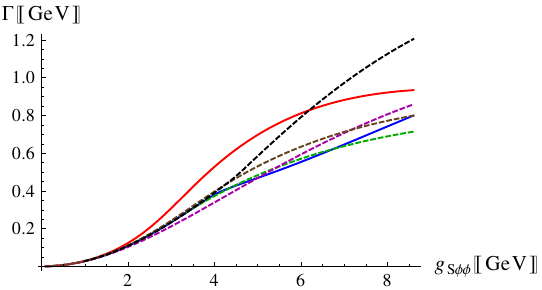}
\label{figure_rel4dM06L15DecayWidth1-2Mfixed.eps}
\end{minipage}
\hfill
\begin{minipage}[hbt]{6.9cm}
\centering
\includegraphics[width= 7.2cm]{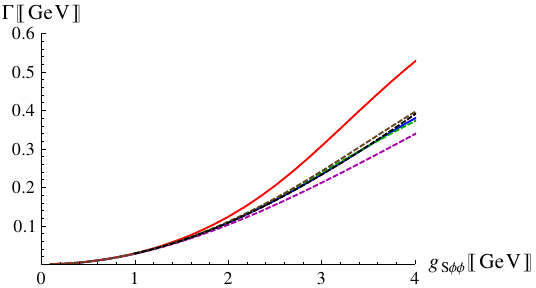}
\label{figure_rel4dM06L15DecayWidth2-2Mfixed.eps}
\end{minipage}
\caption{Comparing decay widths for the $S\phi\phi$-model with smooth cutoff and varying scales (here, $M_{\text{BW}}$ is fixed): the first two panels show plots of the full width (red) in comparison with the tree-level result (\ref{equation_smoothtreelevelwidth}) for a constant Breit--Wigner mass (blue, dashed) and the average width (blue) from Eq. (\ref{equation_averagewidth}) as a function of the coupling. The lower two panels are plots of the full (red) and average width (blue), and plots of the tree-level results for each pole mass $x_{\text{(m)pole}}$ (green, dashed), each bare mass $M_{0}$ (magenta, dashed), each maximum value $x_{\text{max}}$ of the spectral function (brown, dashed) and each average mass $x_{\text{average}}$ (black, dashed).}
\label{figure_decaywidths2}
\end{figure}

We can also study other decay widths as a combination of spectral function properties and the tree-level result. In the lower panels of Fig. \ref{figure_decaywidths2} we provide plots of the full (red) and average width (blue), and plots of the tree-level results for each pole mass $x_{\text{(m)pole}}$ (green, dashed), each bare mass $M_{0}$ (magenta, dashed), each maximum value $x_{\text{max}}$ of the spectral function (brown, dashed) and each average mass $x_{\text{average}}$ (black, dashed). All of them may differ in several ways, but overall they are comparable and still yield too small values. The black dashed curve surprisinglyhas an intersection point with the full width at $g_{S\phi\phi}=6.195 \ \text{GeV}$.

\begin{figure}
\begin{minipage}[hbt]{6.9cm}
\centering
\includegraphics[width= 7.2cm]{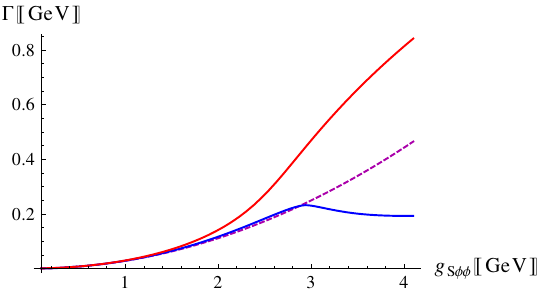}
\label{figure_rel4dM06L15DecayWidth1-1M0fixed}
\end{minipage}
\hfill
\begin{minipage}[hbt]{6.9cm}
\centering
\includegraphics[width= 7.2cm]{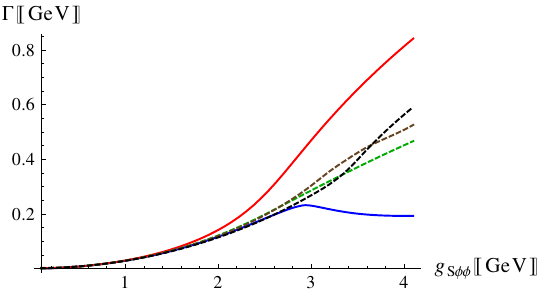}
\label{figure_rel4dM06L15DecayWidth2-1M0fixed.eps}
\end{minipage}
\caption{Comparing decay widths for the $S\phi\phi$-model with smooth cutoff (here, $M_{0}$ is fixed): On the left, full width (red) in comparison with the tree-level result for a constant bare mass (magenta, dashed) and the average width (blue) as a function of the coupling. On the right, the tree-level result with each average mass (black, dashed) and plots of the tree-level results for each pole mass $x_{\text{(m)pole}}$ (green, dashed) and each maximum value $x_{\text{max}}$ of the spectral function (brown, dashed).}
\label{figure_decaywidths3}
\end{figure}

An analogous discussion is obtained for the full width (red) in comparison with the tree-level result for a constant bare mass (magenta, dashed) and the average width (blue) in Fig. \ref{figure_decaywidths3}. It is not necessary to repeat all similar properties again -- the important point is the fact that up to intermediate values of the coupling the tree-level result is not a better approach than the average width, it does not even produce 'the best fit\grq, which is clearly given by the tree-level result with each average mass (black, dashed). The plots of the tree-level results for each pole mass $x_{\text{(m)pole}}$ (green, dashed) and each maximum value $x_{\text{max}}$ of the spectral function (brown, dashed) are also showing comparable, yet too small, values for intermediate and large couplings.
\begin{figure}
\begin{minipage}[hbt]{6.9cm}
\centering
\includegraphics[width= 7.2cm]{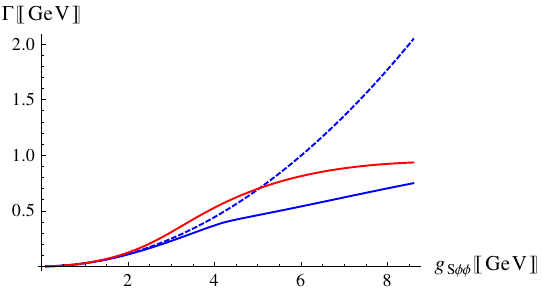}
\label{figure_rel4dM06L15DecayWidth1-1Mfixed}
\end{minipage}
\hfill
\begin{minipage}[hbt]{6.9cm}
\centering
\includegraphics[width= 7.2cm]{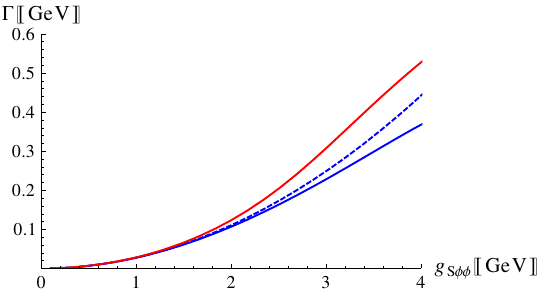}
\label{figure_rel4dM06L15DecayWidth2-1Mfixed.eps}
\end{minipage}
\end{figure}

\subsection{Concluding remarks}
The analysis presented has compared the two cases of a fixed Breit--Wigner and bare mass, in particular we specified $M_{\text{BW}}=M_{0}=0.6 \ \text{GeV}$. Some relevant points:

\begin{enumerate}
\item We have shown that in both cases a single seed pole for the $\sigma$-meson entails two other emerging poles, once the interaction term in the Lagrangian is turned on. One pole appears in the first sheet and can be discarded as non-physical, induced by the finite branch cut, while the other pole is dynamically generated in the vicinity of the origin of the second sheet. This pole moves to the branch point at threshold and slips through the cut onto the first sheet, where it has to be taken into account in the normalization condition of the spectral function. It can be interpreted as a bound state. The whole model becomes unstable due to the occurence of tachyonic excitations for large couplings.
\item Although the Breit--Wigner parameterization reproduces very well the mass (and decay width) of the resonance pole in the small-coupling regime, this is not true anymore for intermediate values of $g_{S\phi\phi}$, compare Tab. \ref{table_allpoints3} and \ref{table_allpoints4}. Above this regime, the mass is completely detached from the pole mass, while the average mass and the maximum value of the spectral function are sometimes in acceptable agreement with the pole mass.
\end{enumerate}

One can try to connect our introductory discussion about the Breit--Wigner parameterization of narrow resonances with the numerically obtained pole positions on the second Riemann sheet. If we start with a single seed pole for the $\sigma$-meson, the trajectory for fixed bare mass $M_{0}$ can be taken as the path of the resonance pole for different values of the coupling. The spectral function describes this motion due to the fact that poles slightly below the branch cut dominate its shape through the maximum and full width at half maximum -- these are the appropriate parameters to describe the unstable particle. Though we have not focussed on the full width at half maximum, we listed the maximum of the spectral function in Tab. \ref{table_allpoints4}, from which we see very good agreement with the pole mass and the Breit--Wigner mass in the case of small and up to intermediate couplings. The whole situation changes when the pole in the second sheet starts turning right: both masses become too small and the Breit--Wigner mass has even no value for $g_{S\phi\phi}> 2.969 \ \text{GeV}$. We plot the latter and the pole mass as a function of $g_{S\phi\phi}$ in the next figure.
\begin{figure}[!h]
\begin{center}
\includegraphics{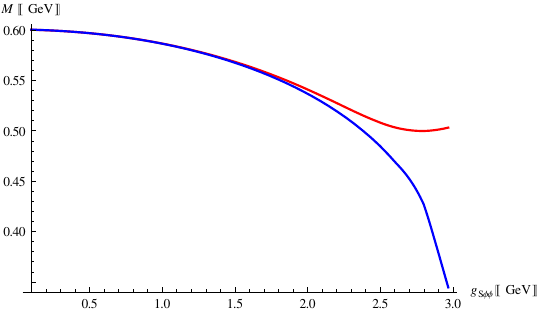}
\caption{Full resonance mass (red) as the real part $x_{\text{(m)pole}}=M$ of the complex pole descending on the lower half plane of the second Riemann sheet compared to the Breit--Wigner mass $x_{\text{(m)BW}}$ (blue).}
\label{figure_rel4dM06L15MassBW}
\end{center}
\end{figure}
The reason for the described difference lies in the real part of the self-energy. As can be seen in Fig. \ref{figure_rel4dSigma2}, the self-energy has a square root cusp near threshold that affects the solution of
\begin{equation}
x_{\text{(m)BW}}^{2}-M_{0}^{2}+g_{S\phi\phi}^{2}\operatorname{Re}\Sigma(x_{\text{(m)BW}}) = 0 \ .
\end{equation}
dramatically. Such influence is expected near every threshold and is well-known, see for example Refs. \cite{tornqvist,pennington}. The resonance is too broad to describe it with the help of the two equations (\ref{equation_firstchapterBWmass}) and (\ref{equation_firstchapterBWwidth}), while one could still try to fit the spectral function by a Breit--Wigner distribution.

\begin{figure}[!h]
\begin{center}
\includegraphics{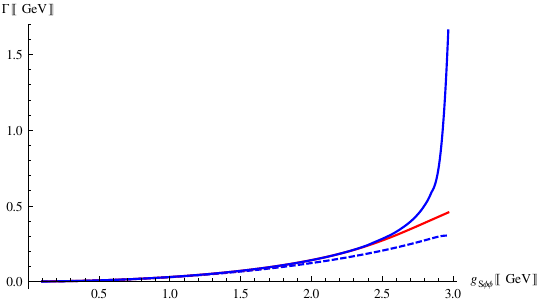}
\caption{Full decay width (red) as two times the negative imaginary part $y_{\text{(m)pole}}=-\Gamma/2$ of the complex pole descending on the lower half plane of the second Riemann sheet compared to the Breit--Wigner width (blue) and the tree-level result (blue, dashed) evaluated at the Breit--Wigner mass.}
\label{figure_rel4dM06L15DecayWidth3}
\end{center}
\end{figure}

On the other hand, the Breit--Wigner width is usually defined by Eq. (\ref{equation_firstchapterBWwidth}) if we include loop contributions, whereas in effective field theories one instead identifies the (preliminary) tree-level result (\ref{equation_preliminarydecaywidth}). Yet, since the quantum fluctuations from mesonic loops cannot be neglected in the intermediate and large-coupling regime, we may choose the modified expression (\ref{equation_smoothtreelevelwidth}), so that the Breit--Wigner width just gains an enhancement factor, namely the renormalization constant evaluated at $x_{\text{(m)BW}}$. This has a significant effect (see Fig. \ref{figure_rel4dM06L15DecayWidth3}): the Breit--Wigner width maps the full width much better as long as $x_{\text{(m)BW}}$ is sufficiently far from threshold. Otherwise, it grows beyond all bounds, while the expression without renormalization stays finite but too small.

\subsection{Determination of the $\sigma$-meson or $f_{0}(500)$ state}
It was stated in the first chapter that many works in recent years attempted to determine the $\sigma$-pole by using {\em Roy equations} with crossing symmetry, analyticity and unitarity. Similar results were found, which actually differ quite a lot from the values of T\"ornqvist and Roos \cite{roos} in 1996. We have quoted one result by Caprini et al. \cite{caprini} according to whom the $\pi\pi$-scattering amplitude would contain a pole with the quantum numbers of $f_{0}(500)$:
\begin{equation}
\sqrt{s_{\text{pole}}} = \left(441_{-8}^{+16}-i272_{-12.5}^{+9}\right) \ \text{MeV} \ .
\end{equation}
The Particle Data Group 2012 also just recently revised its values for the mass and width estimates on the $\sigma$-meson \cite{beringer}, giving a range
\begin{equation}
\sqrt{s_{\text{pole}}} = (400\text{-}550)-i(200\text{-}350) \ \text{MeV} \ .
\end{equation}
Our very simple model can indeed reproduce this pole in the case of fixed bare mass by setting
\begin{equation}
M_{0} = 0.5447 \ \text{GeV} \ , \ \ \ \ \ g_{S\phi\phi} = 2.783 \ \text{GeV} \ .
\end{equation}
The pole lies inside the PDG estimate. Besides that, we notice the left pole below threshold to exist in the first sheet with coordinates
\begin{equation}
z_{\text{pole}} = (0.2705-i\epsilon) \ \text{GeV} \ .
\end{equation}
In other words, if we take the $\sigma$-pole of Caprini et al. as a correct determination of its mass and decay width, then we additionally would get another pole, interpreted as a bound state, slightly below threshold. Both are then one and the same manifestation of the seed state included in our Lagrangian. Note that we cannot exclude the possibility of having found a spurious pole not belonging to a correct description of nature since no experiment has yet indicated a hadronic particle with a mass lower than the two-pion threshold.
\begin{figure}[t]
\begin{center}
\includegraphics{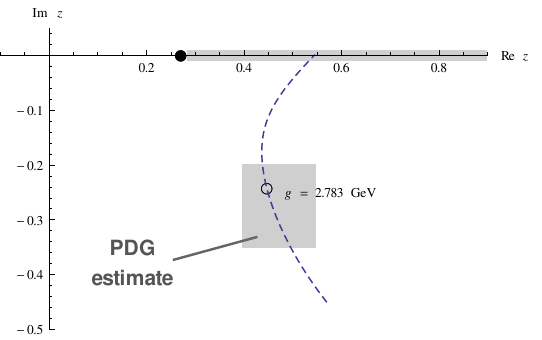}
\caption{Trajectory for the $\sigma$-pole (white dot) in the second Riemann sheet when the result of Caprini et al. is used. The black dot marks the position of the dynamically generated pole (bound state) in the first sheet.}
\label{figure_rel4dM06L15sigmapoleTrajectory}
\end{center}
\end{figure}

 \label{chapter_chapter4}
\clearpage

\thispagestyle{empty}
\

\newpage
\clearpage

\chapter{Conclusion and outlook}

\medskip

In conclusion, we studied the propagator poles of a scalar hadronic resonance within a simple quantum field theoretical model in which this scalar state, namely the $\sigma$-meson, decays into two pions. The initial state was settled as a seed pole above the corresponding two-pion threshold in the limit of vanishing coupling. We investigated how the motion of the propagator pole(s) is influenced by loop contributions of the two pions if the interaction is turned on. For this, the self-energy diagram of the two-pion loop was calculated analytically and taken into account by applying a Dyson resummation scheme -- the scalar propagator was then continued into the second sheet of the Riemann energy surface in order to numerically look for poles there. This method was also demonstrated for a non-relativistic Lee model.

The Lee model exposed some interesting deviations from the general transition properties of unstable quantum systems as induced by the purely exponential decay law. Although the Breit--Wigner parameterization was fully adequate in the small-coupling regime, it completely failed for intermediate and large values. It seems quite obvious that strongly coupled unstable quantum systems lose this feature and should be described by including the full propagator and the complete spectral function, respectively. Moreover, two poles were dynamically generated in the first sheet within the Lee model, either absent in the continuous spectrum. The one above threshold, discarded as non-physical, was assumed to be a direct consequence of a non-smooth cutoff function which therefore should be avoided in general. We nevertheless interpreted the pole below threshold as a bound state emerging from the strong coupling to the continuum of states. The pole trajectory in the second sheet revealed that the unstable particle has a constant mass for large couplings, thus, while the interaction with intermediate states first lowers its mass, it finally stabilizes it. When this complex seed pole is taken to mirror the mass and decay width by
\begin{equation}
E_{\text{pole}} = M-i\frac{\Gamma}{2} \ , \nonumber
\end{equation}
those values are completely detached from quantities like the maximum of the spectral function in the large-coupling regime. There was no evidence for other resonance poles created in the second sheet.

In the case of the $\sigma$-meson, we compared the two cases of a fixed Breit--Wigner and a fixed bare mass, as well as two different cutoff functions for the former. As a main result of the analysis based on a smooth cutoff function, a single seed pole for the $\sigma$-meson entailed another pole emerging in our model, once the interaction term in the Lagrangian was turned on. It was dynamically generated in the vicinity of the origin in the second sheet and moved to the branch point at threshold, finally slipping through the cut onto the first sheet where it had to be taken into account in the normalization condition of the spectral function. We interpreted this pole as a putative bound state belonging to a complete description of one and the same object, i.e., the state $f_{0}(500)$. Note that by increasing the coupling the whole model became unstable due to the occurrence of tachyonic excitations.

Thus, besides the expected resonance pole in the lower half plane of the second Riemann sheet, we have found -- for a sufficiently large coupling constant -- an additional pole on the first sheet below the two-pion threshold. It was clarified under which conditions such a stable state can emerge and it turned out that it factually {\em can} be generated simultaneously with the $\sigma$-pole\footnote{It should be mentioned that according to the recent claim \cite{vanBeveren1,vanBeveren2,abraamyan} of a novel scalar boson with mass 38 MeV (termed $E(38)$), this result was already put under further investigation, leading to a publication \cite{giacosaE38}.} found by Caprini et al. in Ref. \cite{caprini}. The appearence of such a pole was observed for two different kinds of cutoff functions and seems to have its origin in the (resummed) two-pion loop. Further studies should indeed target the self-energy contribution in more detail and reconsider other form factors, as well as different values for the cutoff parameter $\Lambda$. Note that we cannot exclude the possiblity of having found a spurious pole not belonging to a correct description of nature (except for the claims in the footnote, no experiment has yet indicated a hadronic particle with a mass lower than the two-pion threshold). It is additionally worth to mention that although there exist phenomenological models in which the state $f_{0}(500)$ is interpreted as a tetraquark state and where the theory reduces to our Lagrangian (\ref{equation_lagrangian}) when only the pion-pion channel is studied \cite{maiani,thooft,giacosaTetra}, more advanced investigations have to go beyond our simple model. One should especially include chiral symmetry.

Despite all that, our work provided numerical bounds for the breakdown of the tree-level values for mass and decay width of a resonance below 1 GeV characterized by our model, as well as displayed the limits of the Breit--Wigner parameterization in the case of small and intermediate couplings. When the seed pole in the second Riemann sheet moved deep down into the complex plane, especially the Breit--Wigner mass became too small and had even no value for some critical coupling due to the square root cusp of the self-energy near threshold. Such influence is indeed expected near every threshold and is well-known, see for example Refs. \cite{tornqvist,pennington}. Concerning the decay width we have found that the Breit--Wigner width maps the full width much better than the tree-level one as long as the corresponding mass is far enough from threshold.

The main topic of our future work could be the study of hadronic resonances within a specified effective quantum field theory, e.g. the extended linear sigma model (eLSM) as was established in the group of Prof. Dr. Dirk H. Rischke, see for example Refs. \cite{parganlijaBig,rischke}. One should investigate the influence of loop contributions on, for instance, the mass and decay width of scalar resonances since these observables are so far calculated only at tree-level, leading to the Breit--Wigner parameterization. This seems at least questionable for the scalar mesonic sector. A reasonable starting point would be the lightest scalar mesons, i.e., the $f_{0}(500)$ state, the $f_{0}(980)$ and $a_{0}(980)$ states, which are not yet included explicitly in the mentioned eLSM. It is important to notice that the precise value of the (relatively small) decay width of the isovector state $a_{0}(980)$ is still under debate. Moreover, it can be interpreted as a glueball or kaonic bound state \cite{locher}. Bound states actually need to be studied by solving a Bethe--Salpeter equation, so this could be another interesting project for the future.

In such a prespective work we would like to clarify if the scalar resonances not yet included in the eLSM can be found as propagator poles due to hadronic loop contributions. From such an investigation one could possibly also learn more about the general dependence of the eLSM -- and effective field theories in general -- on strongly coupled hadronic intermediate states, possibly giving new insight into the low-energy regime, scalar resonances and both its theoretical description and physical interpretation.

 \label{chapter_chapter5}
\clearpage

\thispagestyle{empty}
\

\newpage
\clearpage

\pagenumbering{roman}
\setcounter{page}{4}

\addcontentsline{toc}{chapter}{Appendix A - Some mathematical formulas}
\chapter*{Appendix A}

\medskip

\begin{itemize}
\item Fourier transformation:
\begin{eqnarray}
\text{three dimensions:} \ \ \ f(\textbf{x}) & = & \int\frac{\text{d}^{3}k}{(2\pi)^{3}} \ e^{i\textbf{k}\cdot\textbf{x}}\tilde{f}(\textbf{k}) \ , \nonumber \\
\tilde{f}(\textbf{k}) & = & \int\text{d}^{3}x \ e^{-i\textbf{k}\cdot\textbf{x}}f(\textbf{x}) \ , \nonumber \\
\Rightarrow \ \ \  \delta^{(3)}(\textbf{k}-\textbf{q}) & = & \int\frac{\text{d}^{3}x}{(2\pi)^{3}} \ e^{-i(\textbf{k}-\textbf{q})\cdot\textbf{x}} \nonumber \\
\nonumber \\
\text{four dimensions:} \ \ \ \ f(x) & = & \int\frac{\text{d}^{4}k}{(2\pi)^{4}} \ e^{-ik\cdot x}\tilde{f}(k) \ , \nonumber \\
\tilde{f}(k) & = & \int\text{d}^{4}x \ e^{ik\cdot x}f(x) \ , \nonumber \\
\Rightarrow \ \ \  \delta^{(4)}(k-q) & = & \int\frac{\text{d}^{4}x}{(2\pi)^{4}} \ e^{i(k-q)\cdot x} \nonumber
\end{eqnarray}
\item Laplace transformation
\begin{eqnarray}
f(t) & = & \frac{1}{2\pi i}\int_{\gamma-i\infty}^{\gamma+i\infty}\text{d}s \ e^{st}F(s) \ , \nonumber \\
F(s) & = & \int_{0}^{\infty}\text{d}t \ e^{-st}f(t) \ , \ \ \ s\in\mathbb{C} \nonumber
\end{eqnarray}
Here, the integration is done along a vertical line with $\operatorname{Re}s=\gamma$ such that $\gamma$ is larger than the real part of all singularities of $F(s)$. To make a more general statement, $\gamma$ is chosen in a way that the contour path of integration is in the region of convergence of $F(s)$.
\item Representation of the delta distribution function:
\begin{equation}
\delta(a-b) = \lim_{\epsilon \to 0^{+}}\frac{1}{\pi}\frac{\epsilon}{(a-b)^{2}+\epsilon^{2}} \nonumber
\end{equation}
\item Sokhotski--Plemelj theorem:
\begin{eqnarray}
\lim_{\epsilon \to 0^{+}}\int_{a}^{b}\text{d}x \ \frac{f(x)}{x\pm i\epsilon} & = & \mathcal{P}\int_{a}^{b}\text{d}x \ \frac{f(x)}{x}\mp i\pi f(0) \ , \nonumber \\
\Rightarrow \ \ \ \frac{1}{x\pm i\epsilon} & = & \mathcal{P}\left(\frac{1}{x}\right)\mp i\pi\delta(x) \ , \nonumber
\end{eqnarray}
with the Cauchy principal value
\begin{equation}
\mathcal{P}\int_{a}^{c}\text{d}x \ f(x) = \lim_{\eta \to 0^{+}}\left[\int_{a}^{b-\eta}\text{d}x \ f(x)+\int_{b+\eta}^{c}\text{d}x \ f(x)\right] \nonumber
\end{equation}
\item Polar form of a complex number $z=x+iy=\rho e^{i\phi}$ for $\phi\in(-\pi,\pi]$:
\begin{eqnarray}
\rho & = & \sqrt{x^{2}+y^{2}} \ , \nonumber \\
\phi & = & \arg z \ \ = \ \ \begin{cases} \arctan\frac{y}{x} & x>0 \\
\arctan\frac{y}{x}+\pi & x<0, \ y\ge0 \\
\arctan\frac{y}{x}-\pi & x<0, \ y<0 \\
\frac{\pi}{2} & x=0, \ y>0 \\
-\frac{\pi}{2} & x=0, \ y<0 \\
\text{undefined} & x=0, \ y=0
\end{cases} \nonumber
\end{eqnarray}
\item Fourier transform of the non-relativistic Breit--Wigner distribution:
\begin{eqnarray}
\tilde{f}(t) & = & \int\text{d}\omega \ e^{-i\omega t}\frac{1}{\pi}\frac{\left(\frac{\Gamma}{2}\right)}{(\omega-M)^{2}+\left(\frac{\Gamma}{2}\right)^{2}} \nonumber \\
& = & e^{-iMt-\frac{\Gamma}{2}|t|} \ , \ \ \ t\ge0 \nonumber \\
& = & e^{-i\left(M-i\frac{\Gamma}{2}\right)t} \nonumber
\end{eqnarray}

\end{itemize}

 \label{chapter_appendixA}
\clearpage

\addcontentsline{toc}{chapter}{Appendix B - Used conventions}
\chapter*{Appendix B}

\medskip

\begin{itemize}
\item Natural units:
\begin{equation}
c = \hbar = 1 \nonumber
\end{equation}
\item Minkowski metric:
\begin{equation}
\text{d}s^{2} = \text{d}t^{2}-\text{d}\textbf{r}^{2} \ , \ \ \ \ \ (g_{\mu\nu}) = \diag(1,-1,-1,-1) \nonumber
\end{equation}
\item Normalization of states and commutator relations:
\begin{eqnarray}
|\textbf{p}\rangle & = & \sqrt{2E_{\textbf{p}}}a_{\textbf{p}}^{\dagger}|0\rangle \ , \nonumber \\
\langle \textbf{p}|\textbf{q}\rangle & = & 2E_{\textbf{p}}(2\pi)^{3}\delta^{(3)}(\textbf{p}-\textbf{q}) \ , \nonumber \\
\phi(x) & = & \int\frac{\text{d}^{3}p}{(2\pi)^{3}}\frac{1}{\sqrt{2E_{\textbf{p}}}}\Big(a_{\textbf{p}}^{\dagger}e^{ip\cdot x}+a_{\textbf{p}}e^{-ip\cdot x}\Big) \ , \nonumber \\
{[}a_{\textbf{p}},a_{\textbf{q}}^{\dagger}{]} & = & (2\pi)^{3}\delta^{(3)}(\textbf{p}-\textbf{q}) \nonumber
\end{eqnarray}
\item Spectral representation of Green's functions (with $\epsilon \to 0^{+}$):
\begin{eqnarray}
\text{QFT:} \ \ \ \ \ \ \Delta(p^{2}) & = & \frac{1}{\pi}\int_{0}^{\infty}\text{d}s^{2} \ \frac{\rho(s^{2})}{p^{2}-s^{2}+i\epsilon} \nonumber \\
& = & \frac{1}{\pi}\int_{0}^{\infty}\text{d}s \ \frac{2s\rho(s^{2})}{p^{2}-s^{2}+i\epsilon} \nonumber \\
& = & \frac{1}{\pi}\int_{0}^{\infty}\text{d}s \ \frac{d(s)}{p^{2}-s^{2}+i\epsilon} \ , \nonumber \\
\rho(s^{2}) & = & -\operatorname{Im}\Delta(s^{2}+i\epsilon) \ , \nonumber \\
d(s) & = & -2s\operatorname{Im}\Delta(s+i\epsilon) \nonumber \\
\nonumber \\
\text{QM:} \ \ \ \ \ \ G(E) & = & \frac{1}{\pi}\int_{-\infty}^{\infty}\text{d}\omega \ \frac{\rho(\omega)}{E-\omega+i\epsilon} \ , \nonumber \\
\rho(\omega) & = & -\operatorname{Im}G(\omega+i\epsilon) \nonumber
\end{eqnarray}

\end{itemize}

 \label{chapter_appendixB}
\clearpage

\addcontentsline{toc}{chapter}{Appendix C - Kinematics of two-body decays}
\chapter*{Appendix C}

\medskip

Consider a particle $S$ in its rest frame, decaying into two particles $A$ and $B$. Form kinematics we can find an expression for the three-momentum of the outgoing particles, namely $p_{SAB}\equiv |\textbf{p}_{SAB}|$:
\\
\begin{equation}
(p_{S}^{\mu}) = \begin{pmatrix}M_{S} \\ 0 \\ 0 \\ 0\end{pmatrix} = \begin{pmatrix}\sqrt{p_{SAB}^{2}+M_{A}^{2}} \\ p_{SAB} \\ 0 \\ 0\end{pmatrix}+\begin{pmatrix}\sqrt{p_{SAB}^{2}+M_{B}^{2}} \\ -p_{SAB} \\ 0 \\ 0\end{pmatrix} \ ,
\end{equation}
\begin{eqnarray}
M_{S}^{2} & = & M_{A}^{2}+M_{B}^{2}+2p_{SAB}^{2}+2\sqrt{(M_{A}^{2}+p_{SAB}^{2})(M_{B}^{2}+p_{SAB}^{2})} \ , \\
\nonumber \\
\Rightarrow \ \ \ p_{SAB} & = & \frac{1}{2M_{S}}\sqrt{M_{S}^{4}+(M_{A}^{2}-M_{B}^{2})^{2}-2(M_{A}^{2}-M_{B}^{2})M_{S}^{2}} \ .
\end{eqnarray}
\begin{figure}[h]
\centering
\includegraphics[scale=0.7]{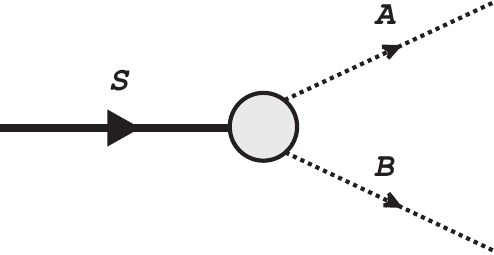}
\caption{Schematic decay process $S\rightarrow AB$.}
\label{figure_SABdecay}
\end{figure}
\\
And for $M_{A}=M_{B}$ this simplifies to
\begin{eqnarray}
p_{SAA} & = & \frac{1}{2M_{S}}\sqrt{M_{S}^{4}-4M_{A}^{2}M_{S}^{2}} \nonumber \\
& = & \sqrt{\frac{M_{S}^{2}}{4}-M_{A}^{2}} \ .
\end{eqnarray}

 \label{chapter_appendixC}
\clearpage

\addcontentsline{toc}{chapter}{Appendix D - Tree-level decay width after including a non-local interaction}
\chapter*{Appendix D}

\medskip

The tree-level decay width $\Gamma_{\text{tree}}(M)$, when including a non-local interaction term (\ref{equation_nonlocalL}) on the Lagrangian level, is modified by the regularization function $f_{\Lambda}(|\textbf{q}|=p_{S\phi\phi})$ just as a multiplication factor. We start in the same way as was done in the case without cutoff (see subsection \ref{subsection_decaywidth}):
\begin{eqnarray}
\hat{S} & = & \mathbbm{1}+ig\int\text{d}^{4}x \ \mathcal{T}\big{\{}\textbf{:}S(x)\int\text{d}^{4}y \ \phi(x+y/2)\phi(x-y/2)\Phi(y)\textbf{:}\big{\}} \nonumber \\
& = & \mathbbm{1}+\hat{S}^{(1)} \ ,
\end{eqnarray}
where $\hat{S}$ is the $\hat S$-Matrix, $\mathcal{T}$ is the time-ordering operator and the dots mark the normal ordering prescription. We need again to evaluate the matrix element
\begin{eqnarray}
\langle\text{final}|\hat{S}|\text{initial}\rangle & = & \langle\textbf{p}_{1}\textbf{p}_{2}|\hat{S}^{(1)}|\textbf{p}\rangle \nonumber \\
& = & \langle\textbf{p}_{1}\textbf{p}_{2}| \ ig\int\text{d}^{4}x \ \textbf{:} \ \xcancel{S^{(+)}\int\text{d}^{4}y \ \phi^{(+)}\phi^{(+)}\Phi(y)} \nonumber \\
&  & + \ \xcancel{S^{(+)}\int\text{d}^{4}y \ \phi^{(+)}\phi^{(-)}\Phi(y)}+\xcancel{S^{(+)}\int\text{d}^{4}y \ \phi^{(-)}\phi^{(+)}\Phi(y)} \nonumber \\
&  & + \ \xcancel{S^{(+)}\int\text{d}^{4}y \ \phi^{(-)}\phi^{(-)}\Phi(y)}+S^{(-)}\int\text{d}^{4}y \ \phi^{(+)}\phi^{(+)}\Phi(y) \nonumber \\
&  & + \ \xcancel{S^{(-)}\int\text{d}^{4}y \ \phi^{(+)}\phi^{(-)}\Phi(y)}+\xcancel{S^{(-)}\int\text{d}^{4}y \ \phi^{(-)}\phi^{(+)}\Phi(y)} \nonumber \\
&  & + \ \xcancel{S^{(-)}\int\text{d}^{4}y \ \phi^{(-)}\phi^{(-)}\Phi(y)} \ \textbf{:} \ |\textbf{p}\rangle \ .
\end{eqnarray}
Here, the crossed-out terms give no contribution because the creation and annihilation operators combine in such a way that their scalar product vanishes.
\begin{figure}[b]
\centering
\includegraphics[scale=0.56]{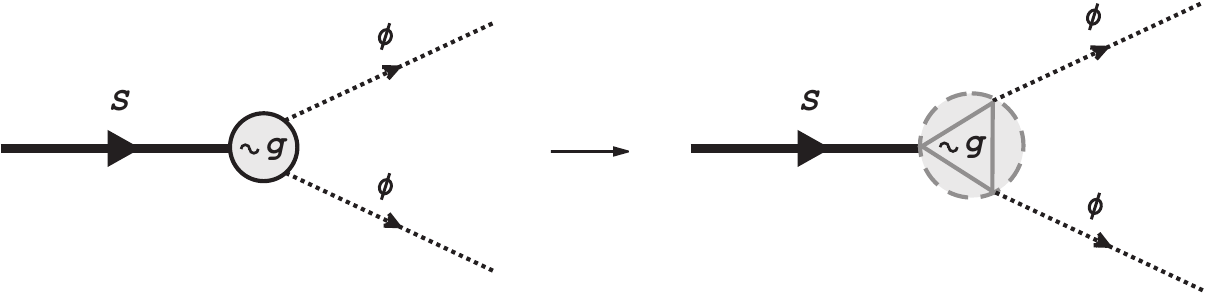}
\caption{Schematic decay process $S\rightarrow\phi\phi$ after including a non-local interaction. Every triangle corner in the new vertex marks a different point in space-time.}
\label{figure_Sphiphidecaynonlocal}
\end{figure}
Note that the superscript at the $S$- and $\phi$-fields denotes those parts of the field that contain a creation or annihilation operator, e.g.
\begin{eqnarray}
\phi^{(+)} & \equiv & \phi^{(+)}(x+y/2) \nonumber \\
& = &  \int\frac{\text{d}^{3}p}{(2\pi)^{3}}\frac{1}{\sqrt{2E_{\textbf{p}}}} \ b_{\textbf{p}}^{\dagger}e^{ip\cdot \left(x+\frac{y}{2}\right)} \ .
\end{eqnarray}
The remaining steps are straightforward:
\begin{eqnarray}
& = & ig\int\text{d}^{4}x\int\text{d}^{3}p_{1}^{\prime}\int\text{d}^{3}p_{2}^{\prime}\int\text{d}^{3}p^{\prime} \ \frac{\sqrt{2E_{\textbf{p}_{1}}2E_{\textbf{p}_{2}}2E_{\textbf{p}}}}{\sqrt{2E_{\textbf{p}_{1}^{\prime}}2E_{\textbf{p}_{2}^{\prime}}2E_{\textbf{p}^{\prime}}}(2\pi)^{9}} \nonumber \\
&  & \times \int\text{d}^{4}y \ e^{ip_{1}^{\prime}\cdot\left(x+\frac{y}{2}\right)}e^{ip_{2}^{\prime}\cdot\left(x-\frac{y}{2}\right)}e^{-ip^{\prime}\cdot x}\Phi(y)\langle0|b_{\textbf{p}_{2}}b_{\textbf{p}_{1}}b_{\textbf{p}_{1}^{\prime}}^{\dagger}b_{\textbf{p}_{2}^{\prime}}^{\dagger}a_{\textbf{p}^{\prime}}a_{\textbf{p}}^{\dagger}|0\rangle \nonumber \\
& = & ig\int\text{d}^{4}x\int\text{d}^{3}p_{1}^{\prime}\int\text{d}^{3}p_{2}^{\prime}\int\text{d}^{3}p^{\prime} \ \frac{\sqrt{2E_{\textbf{p}_{1}}2E_{\textbf{p}_{2}}2E_{\textbf{p}}}}{\sqrt{2E_{\textbf{p}_{1}^{\prime}}2E_{\textbf{p}_{2}^{\prime}}2E_{\textbf{p}^{\prime}}}}\int\text{d}^{4}y \ e^{ip_{1}^{\prime}\cdot\left(x+\frac{y}{2}\right)}e^{ip_{2}^{\prime}\cdot\left(x-\frac{y}{2}\right)}e^{-ip^{\prime}\cdot x}\Phi(y) \nonumber \\
&  & \times \ \Big\{\delta^{(3)}(\textbf{p}^{\prime}-\textbf{p})\delta^{(3)}(\textbf{p}_{1}-\textbf{p}_{2}^{\prime})\delta^{(3)}(\textbf{p}_{2}-\textbf{p}_{1}^{\prime})+\delta^{(3)}(\textbf{p}^{\prime}-\textbf{p})\delta^{(3)}(\textbf{p}_{1}-\textbf{p}_{1}^{\prime})\delta^{(3)}(\textbf{p}_{2}-\textbf{p}_{2}^{\prime})\Big\} \nonumber \\
& = & ig\int\text{d}^{4}x\int\text{d}^{4}y\left\{e^{ip_{2}\cdot\left(x+\frac{y}{2}\right)}e^{ip_{1}\cdot\left(x-\frac{y}{2}\right)}e^{-ip\cdot x}+e^{ip_{1}\cdot\left(x+\frac{y}{2}\right)}e^{ip_{2}\cdot\left(x-\frac{y}{2}\right)}e^{-ip\cdot x}\right\}\Phi(y) \nonumber \\
& = & ig\int\text{d}^{4}x \ e^{i(p_{1}+p_{2}-p)\cdot x}\int\text{d}^{4}y\left\{e^{iy\cdot\left(\frac{p_{2}-p_{1}}{2}\right)}+e^{iy\cdot\left(\frac{p_{1}-p_{2}}{2}\right)}\right\}\Phi(y) \ .
\end{eqnarray}
If the regularization function $f_{\Lambda}(q)$ is now simply the Fourier transform of $\Phi(y)$ and depends only on the magnitude of the three-momentum $\textbf{q}$, then the invariant amplitude $-i\mathcal{M}$ is changed only by the former:
\begin{eqnarray}
\langle\text{final}|\hat{S}|\text{initial}\rangle & = & 2igf_{\Lambda}(|\textbf{p}_{1}|)(2\pi)^{4}\delta^{(4)}(p_{1}+p_{2}-p) \nonumber \\
& \stackrel{!}{=} & -i\mathcal{M}(2\pi)^{4}\delta^{(4)}(p-p_{1}-p_{2}) \ , \\
\Rightarrow \ \ \ -i\mathcal{M} & = & 2igf_{\Lambda}(|\textbf{p}_{1}|) \ ,
\end{eqnarray}
by remembering that $\textbf{p}_{1}=-\textbf{p}_{2}$. The integral from Eq. (\ref{equation_Gammaint}) becomes
\begin{equation}
\Gamma_{\text{tree}}(M) = \frac{4\pi g^{2}}{(2\pi)^{2}M}\int_{0}^{\infty}\text{d}u \ \frac{Mu^{2}f_{\Lambda}^{2}(u)}{16(u^{2}+m^{2})\sqrt{\frac{M^{2}}{4}-m^{2}}} \ \delta\big(u-\sqrt{M^{2}/4-m^{2}}\big) \ ,
\end{equation}
where $u=|\textbf{p}_{1}|$. From this we obtain the real solution
\begin{equation}
\Gamma_{\text{tree}}(M) = \frac{g_{S\phi\phi}^{2} \ p_{S\phi\phi}}{8\pi M^{2}}f_{\Lambda}^{2}(p_{S\phi\phi}) \ \Theta(M-2m) \ ,
\end{equation}
with the modified coupling $g_{S\phi\phi}=\sqrt{2}g$ and the magnitude $|\textbf{p}_{1}|=|\textbf{p}_{2}|=p_{S\phi\phi}=\sqrt{M^{2}/4-m^{2}}$ of the two outgoing $\phi$-particles three-momenta.

 \label{chapter_appendixD}
\clearpage

\addcontentsline{toc}{chapter}{Appendix E - Spectral functions of the non-relativistic Lee model}
\chapter*{Appendix E}

\medskip

For the sake of completeness, we provide a compilation of selected plots of the continuous part of the spectral function for the non-relativistic Lee model. From this one can review the behaviour described during chapter 3.

\newpage

\begin{figure}
\begin{minipage}[hbt]{6cm}
\centering
\includegraphics[width=6.2cm]{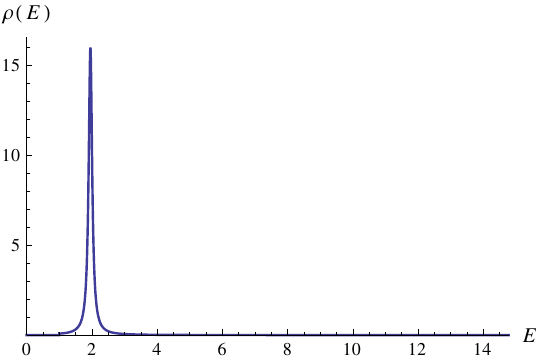}
\caption*{$g=0.5$}
\end{minipage}
\hfill
\begin{minipage}[hbt]{6cm}
\centering
\includegraphics[width=6.2cm]{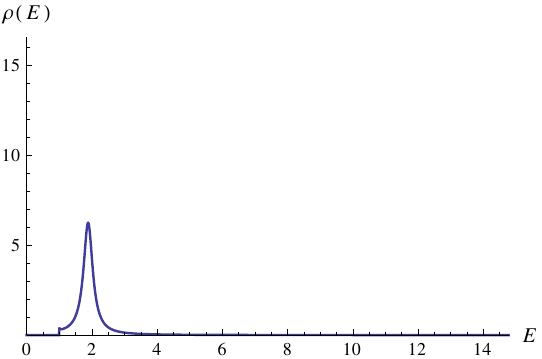}
\caption*{$g=0.8$}
\end{minipage}
\end{figure}
\begin{figure}
\begin{minipage}[hbt]{6cm}
\centering
\includegraphics[width=6.2cm]{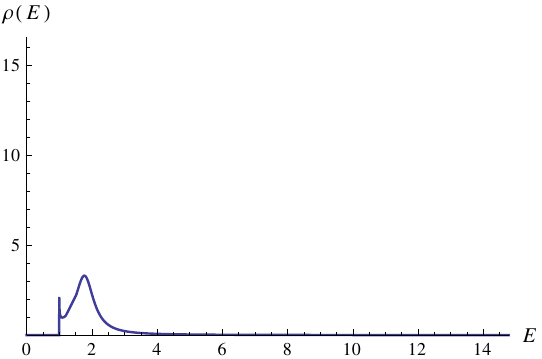}
\caption*{$g=1.1$}
\end{minipage}
\hfill
\begin{minipage}[hbt]{6cm}
\centering
\includegraphics[width=6.2cm]{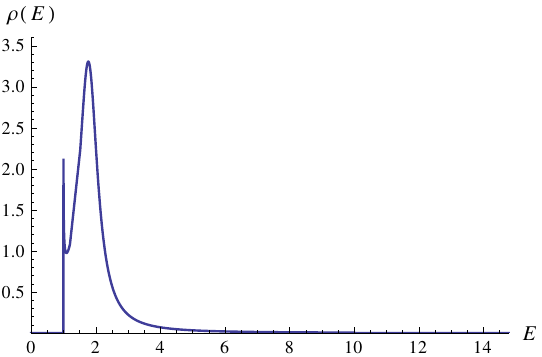}
\caption*{$g=1.1$}
\end{minipage}
\end{figure}
\begin{figure}
\begin{minipage}[hbt]{6cm}
\centering
\includegraphics[width=6.2cm]{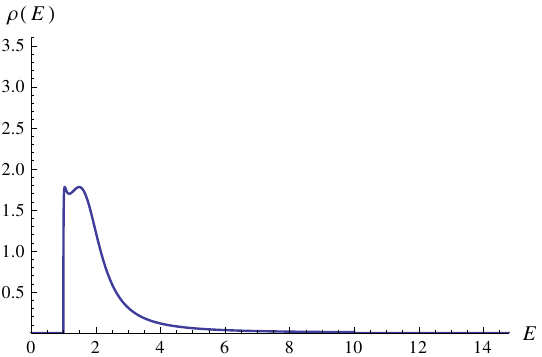}
\caption*{$g=1.5$}
\end{minipage}
\hfill
\begin{minipage}[hbt]{6cm}
\centering
\includegraphics[width=6.2cm]{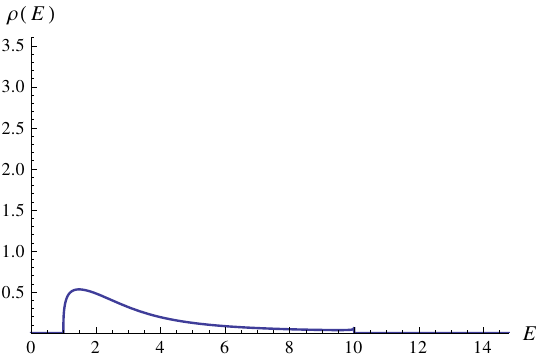}
\caption*{$g=2.4$}
\end{minipage}
\end{figure}
\begin{figure}
\begin{minipage}[hbt]{6cm}
\centering
\includegraphics[width=6.2cm]{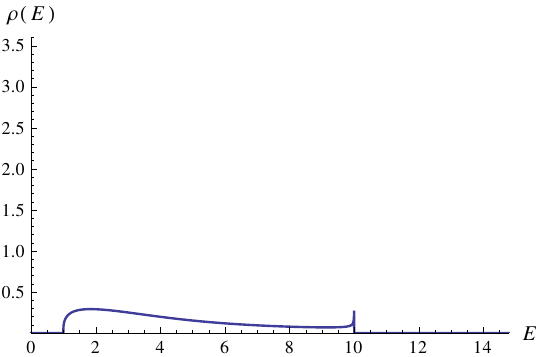}
\caption*{$g=3.1$}
\end{minipage}
\hfill
\begin{minipage}[hbt]{6cm}
\centering
\includegraphics[width=6.2cm]{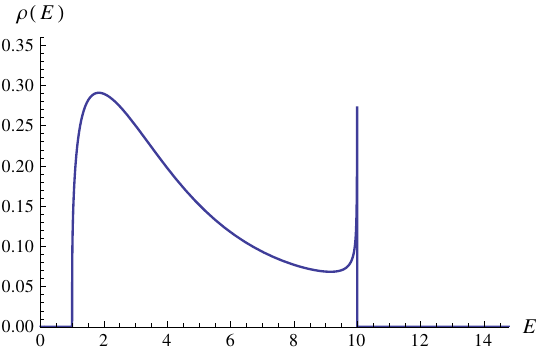}
\caption*{$g=3.1$}
\end{minipage}
\end{figure}
\begin{figure}
\begin{minipage}[hbt]{6cm}
\centering
\includegraphics[width=6.2cm]{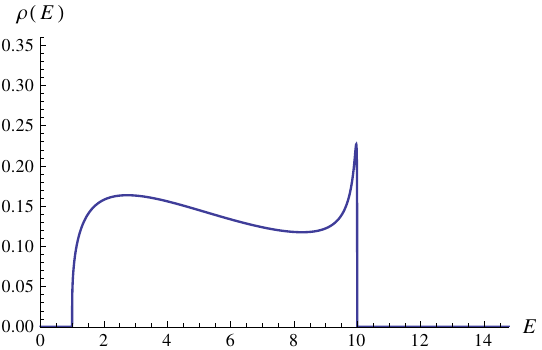}
\caption*{$g=4.2$}
\end{minipage}
\hfill
\begin{minipage}[hbt]{6cm}
\centering
\includegraphics[width=6.2cm]{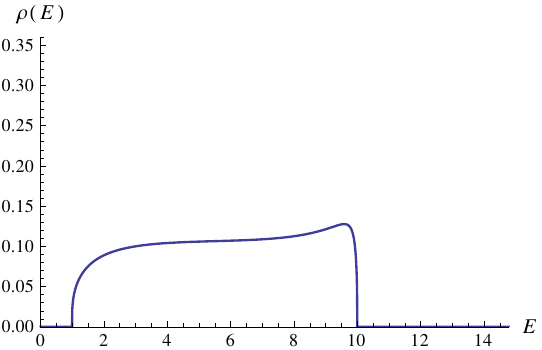}
\caption*{$g=5.6$}
\end{minipage}
\end{figure}
\begin{figure}
\begin{minipage}[hbt]{6cm}
\centering
\includegraphics[width=6.2cm]{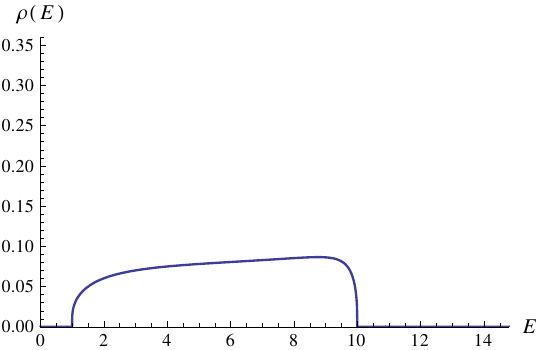}
\caption*{$g=6.8$}
\end{minipage}
\hfill
\begin{minipage}[hbt]{6cm}
\centering
\includegraphics[width=6.2cm]{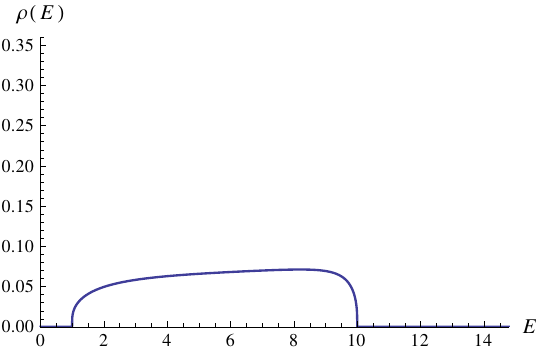}
\caption*{$g=7.5$}
\end{minipage}
\end{figure}
\begin{figure}
\begin{minipage}[hbt]{6cm}
\centering
\includegraphics[width=6.2cm]{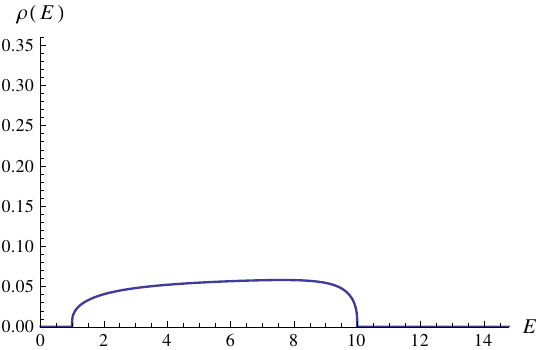}
\caption*{$g=8.3$}
\end{minipage}
\hfill
\begin{minipage}[hbt]{6cm}
\centering
\includegraphics[width=6.2cm]{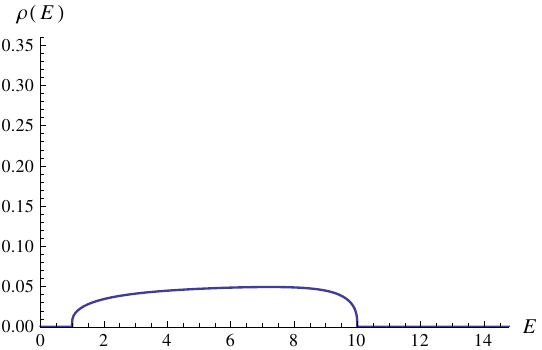}
\caption*{$g=9.0$}
\end{minipage}
\end{figure}
\begin{figure}
\begin{minipage}[hbt]{6cm}
\centering
\includegraphics[width=6.2cm]{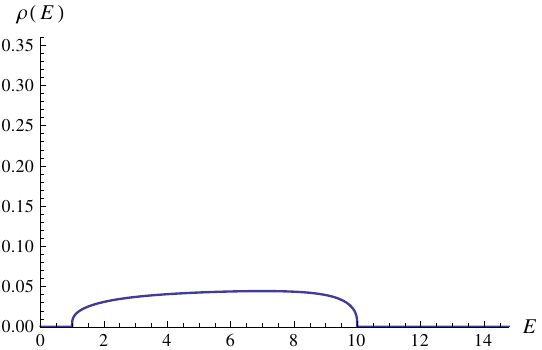}
\caption*{$g=9.5$}
\end{minipage}
\hfill
\begin{minipage}[hbt]{6cm}
\centering
\includegraphics[width=6.2cm]{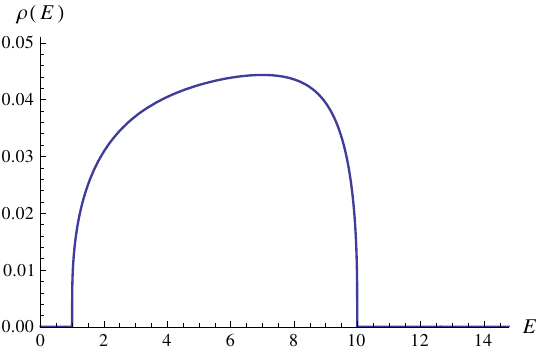}
\caption*{$g=9.5$}
\end{minipage}
\end{figure}

 \label{chapter_appendixE}
\clearpage

\addcontentsline{toc}{chapter}{Appendix F - Spectral functions of the $S\phi\phi$-model with sharp cutoff}
\chapter*{Appendix F}

\medskip

For the sake of completeness, we provide a compilation of selected plots of the continuous part of the spectral function for the $S\phi\phi$-model with sharp cutoff. From this one can review the behaviour described during chapter 4.

\newpage

\begin{figure}
\begin{minipage}[hbt]{6cm}
\centering
\includegraphics[width=6.2cm]{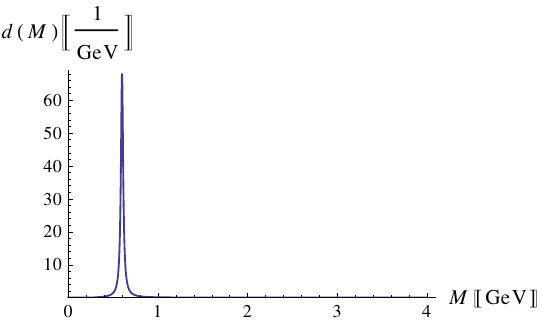}
\caption*{$g_{S\phi\phi}=1.0 \ \text{GeV}$}
\end{minipage}
\hfill
\begin{minipage}[hbt]{6cm}
\centering
\includegraphics[width=6.2cm]{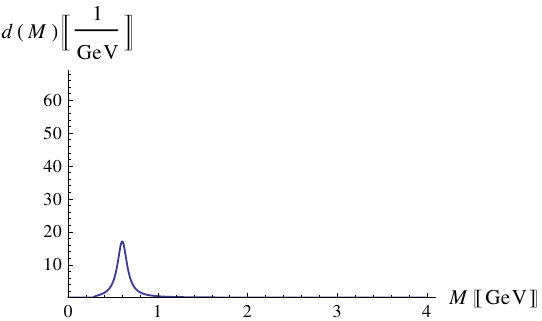}
\caption*{$g_{S\phi\phi}=2.0 \ \text{GeV}$}
\end{minipage}
\end{figure}
\begin{figure}
\begin{minipage}[hbt]{6cm}
\centering
\includegraphics[width=6.2cm]{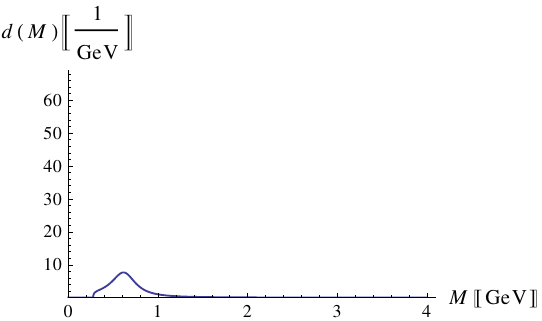}
\caption*{$g_{S\phi\phi}=3.0 \ \text{GeV}$}
\end{minipage}
\hfill
\begin{minipage}[hbt]{6cm}
\centering
\includegraphics[width=6.2cm]{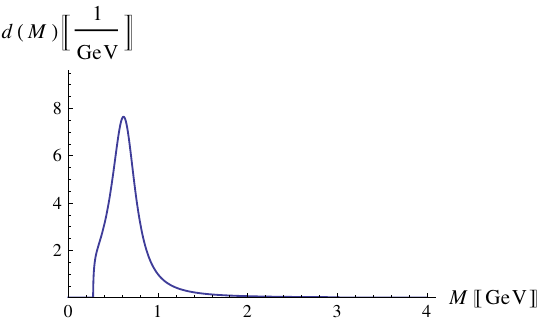}
\caption*{$g_{S\phi\phi}=3.0 \ \text{GeV}$}
\end{minipage}
\end{figure}
\begin{figure}
\begin{minipage}[hbt]{6cm}
\centering
\includegraphics[width=6.2cm]{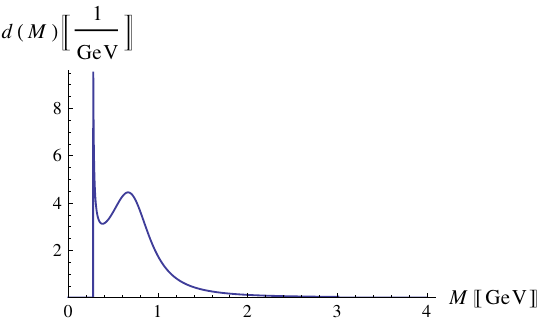}
\caption*{$g_{S\phi\phi}=4.0 \ \text{GeV}$}
\end{minipage}
\hfill
\begin{minipage}[hbt]{6cm}
\centering
\includegraphics[width=6.2cm]{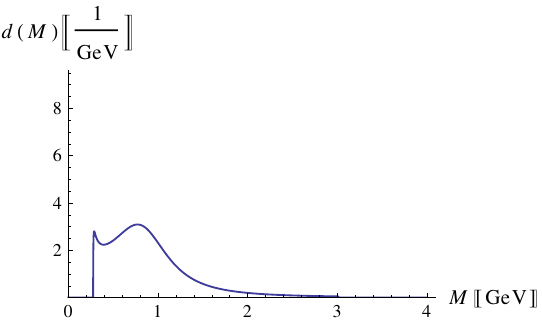}
\caption*{$g_{S\phi\phi}=5.0 \ \text{GeV}$}
\end{minipage}
\end{figure}
\begin{figure}
\begin{minipage}[hbt]{6cm}
\centering
\includegraphics[width=6.2cm]{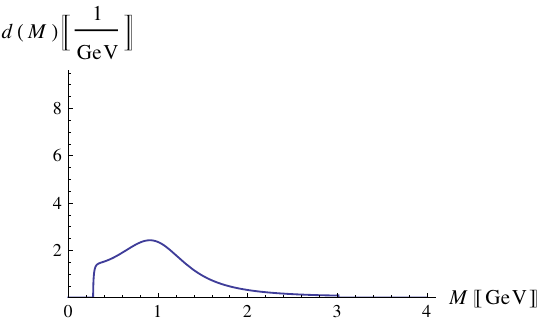}
\caption*{$g_{S\phi\phi}=6.0 \ \text{GeV}$}
\end{minipage}
\hfill
\begin{minipage}[hbt]{6cm}
\centering
\includegraphics[width=6.2cm]{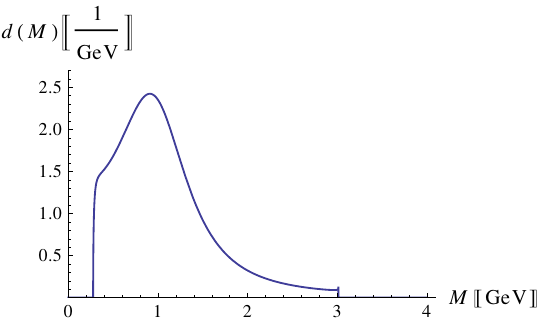}
\caption*{$g_{S\phi\phi}=6.0 \ \text{GeV}$}
\end{minipage}
\end{figure}
\begin{figure}
\begin{minipage}[hbt]{6cm}
\centering
\includegraphics[width=6.2cm]{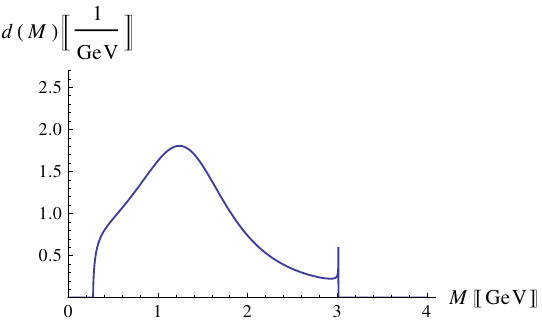}
\caption*{$g_{S\phi\phi}=8.0 \ \text{GeV}$}
\end{minipage}
\hfill
\begin{minipage}[hbt]{6cm}
\centering
\includegraphics[width=6.2cm]{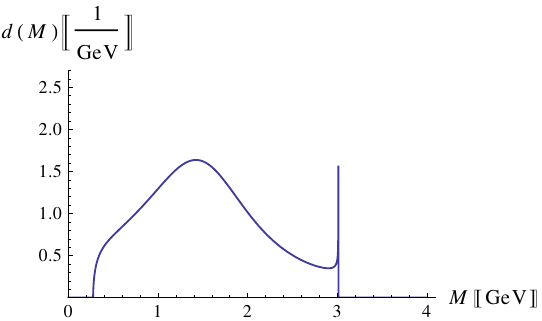}
\caption*{$g_{S\phi\phi}=9.0 \ \text{GeV}$}
\end{minipage}
\end{figure}
\begin{figure}
\begin{minipage}[hbt]{6cm}
\centering
\includegraphics[width=6.2cm]{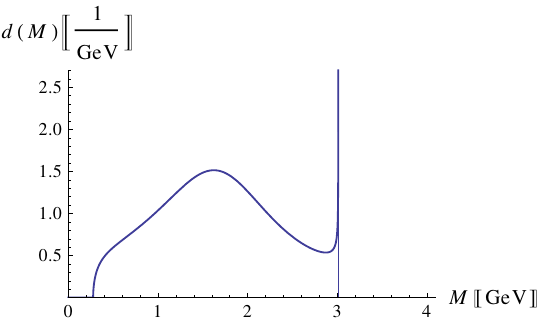}
\caption*{$g_{S\phi\phi}=10.0 \ \text{GeV}$}
\end{minipage}
\hfill
\begin{minipage}[hbt]{6cm}
\centering
\end{minipage}
\end{figure}

 \label{chapter_appendixF}
\clearpage

\addcontentsline{toc}{chapter}{Appendix G - Spectral functions of the $S\phi\phi$-model with smooth cutoff}
\chapter*{Appendix G}

\medskip

For the sake of completeness, we provide a compilation of selected plots of the continuous part of the spectral function for the $S\phi\phi$-model with smooth cutoff. From this one can review the behaviour described during chapter 4. The dark blue curves belong to the case of fixed $M_{\text{BW}}$, the red curves to the case of fixed $M_{0}$.

\newpage

\begin{figure}
\begin{minipage}[hbt]{6cm}
\centering
\includegraphics[width=6.2cm]{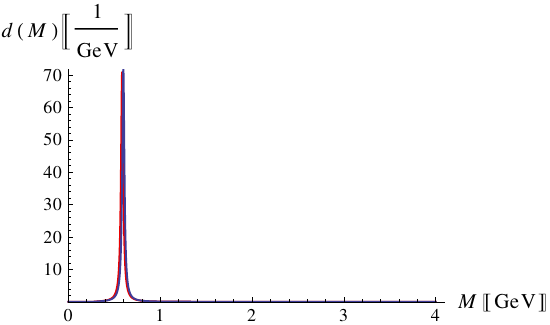}
\caption*{$g_{S\phi\phi}=1.0 \ \text{GeV}$}
\end{minipage}
\hfill
\begin{minipage}[hbt]{6cm}
\centering
\includegraphics[width=6.2cm]{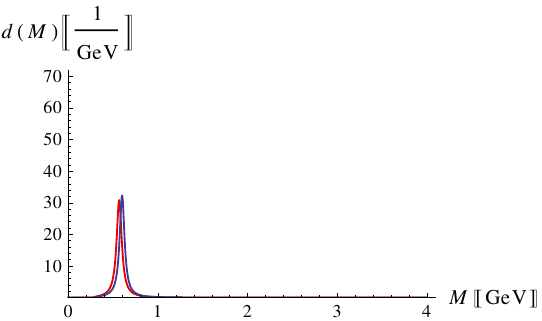}
\caption*{$g_{S\phi\phi}=1.5 \ \text{GeV}$}
\end{minipage}
\end{figure}
\begin{figure}
\begin{minipage}[hbt]{6cm}
\centering
\includegraphics[width=6.2cm]{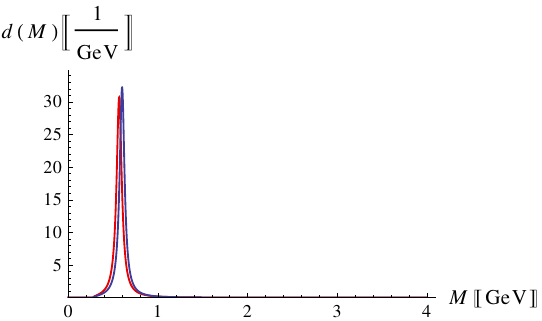}
\caption*{$g_{S\phi\phi}=1.5 \ \text{GeV}$}
\end{minipage}
\hfill
\begin{minipage}[hbt]{6cm}
\centering
\includegraphics[width=6.2cm]{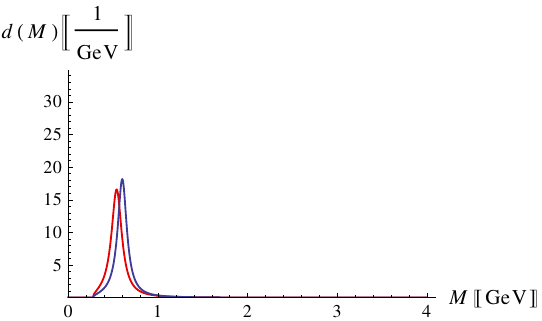}
\caption*{$g_{S\phi\phi}=2.0 \ \text{GeV}$}
\end{minipage}
\end{figure}
\begin{figure}
\begin{minipage}[hbt]{6cm}
\centering
\includegraphics[width=6.2cm]{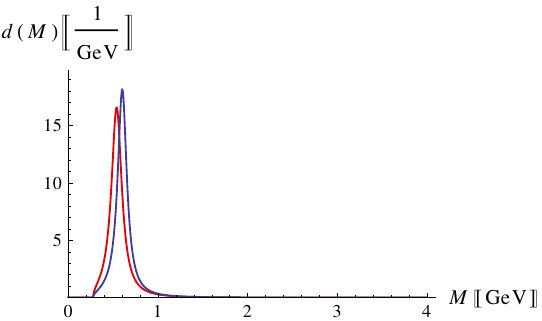}
\caption*{$g_{S\phi\phi}=2.0 \ \text{GeV}$}
\end{minipage}
\hfill
\begin{minipage}[hbt]{6cm}
\centering
\includegraphics[width=6.2cm]{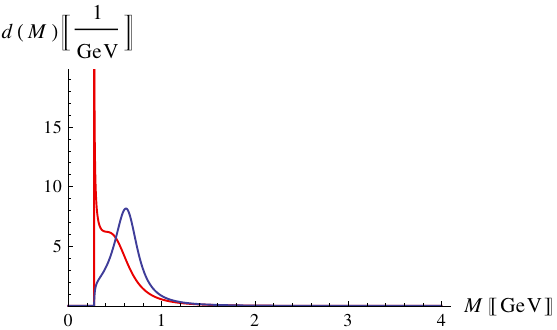}
\caption*{$g_{S\phi\phi}=3.0 \ \text{GeV}$}
\end{minipage}
\end{figure}
\begin{figure}
\begin{minipage}[hbt]{6cm}
\centering
\includegraphics[width=6.2cm]{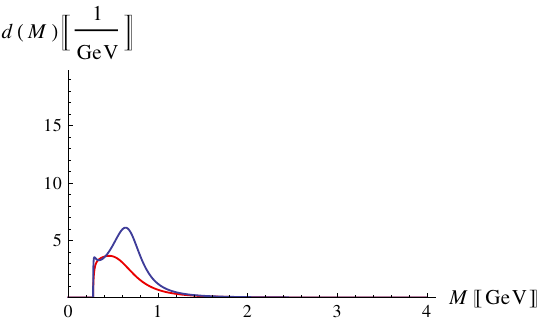}
\caption*{$g_{S\phi\phi}=3.5 \ \text{GeV}$}
\end{minipage}
\hfill
\begin{minipage}[hbt]{6cm}
\centering
\includegraphics[width=6.2cm]{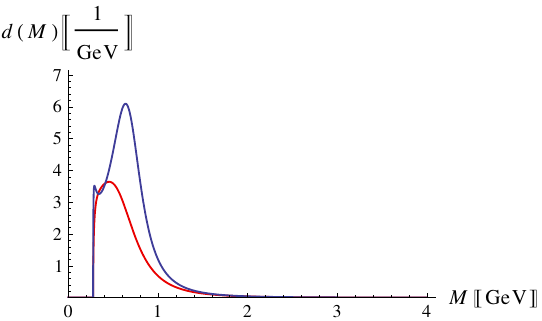}
\caption*{$g_{S\phi\phi}=3.5 \ \text{GeV}$}
\end{minipage}
\end{figure}
\begin{figure}
\begin{minipage}[hbt]{6cm}
\centering
\includegraphics[width=6.2cm]{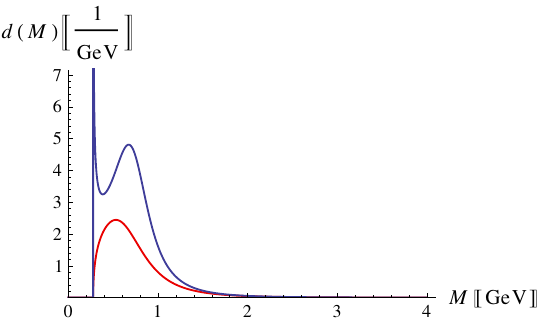}
\caption*{$g_{S\phi\phi}=4.0 \ \text{GeV}$}
\end{minipage}
\hfill
\begin{minipage}[hbt]{6cm}
\centering
\includegraphics[width=6.2cm]{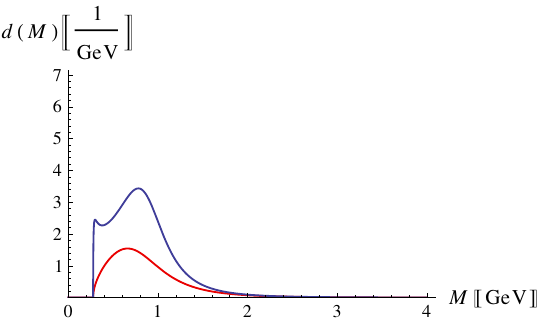}
\caption*{$g_{S\phi\phi}=5.0 \ \text{GeV}$}
\end{minipage}
\end{figure}
\begin{figure}
\begin{minipage}[hbt]{6cm}
\centering
\includegraphics[width=6.2cm]{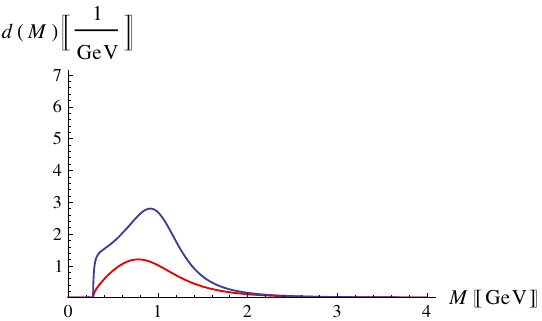}
\caption*{$g_{S\phi\phi}=6.0 \ \text{GeV}$}
\end{minipage}
\hfill
\begin{minipage}[hbt]{6cm}
\centering
\includegraphics[width=6.2cm]{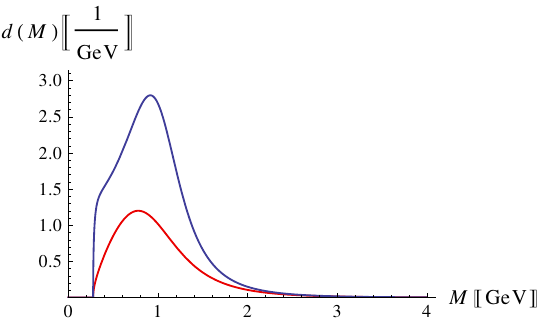}
\caption*{$g_{S\phi\phi}=6.0 \ \text{GeV}$}
\end{minipage}
\end{figure}
\begin{figure}
\begin{minipage}[hbt]{6cm}
\centering
\includegraphics[width=6.2cm]{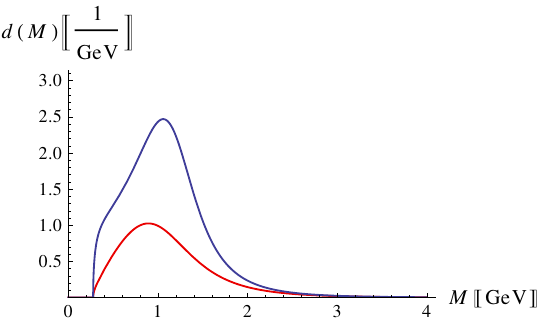}
\caption*{$g_{S\phi\phi}=7.0 \ \text{GeV}$}
\end{minipage}
\hfill
\begin{minipage}[hbt]{6cm}
\centering
\includegraphics[width=6.2cm]{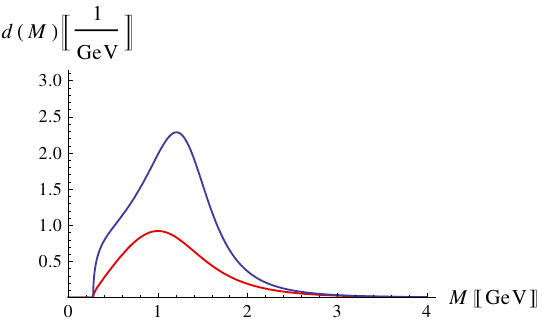}
\caption*{$g_{S\phi\phi}=8.0 \ \text{GeV}$}
\end{minipage}
\end{figure}

 \label{chapter_appendixG}
\clearpage

\thispagestyle{empty}
\

\newpage
\clearpage

\addcontentsline{toc}{chapter}{Bibliography}
\bibliographystyle{unsrt}
\bibliography{mylit}

\begin{thebibliography}{10}

\bibitem{yukawa}
H.~Yukawa.
\newblock On the {I}nteraction of {E}lementary {P}articles.
\newblock {\em Proc. Phys. Math. Soc. Japan}, 17, 1935.

\bibitem{gell-mann}
M.~Gell-Mann.
\newblock A schematic model of baryons and mesons.
\newblock {\em Phys. Lett.}, 8(3), 1964.

\bibitem{zweig}
G.~Zweig.
\newblock An {$SU(3)$} model for strong interaction symmetry and its breaking.
\newblock In D.~B. Lichtenberg and S.~P. Rosen, editors, {\em Developments In
  The Quark Theory Of Hadrons}, volume~1. Hadronic Press and CERN Geneva - TH.
  401 (REC.JAN. 64), 1964.

\bibitem{erlich}
J.~Erlich, E.~Katz, D.~T. Son, and M.~A. Stephanov.
\newblock {QCD} and a {H}olographic {M}odel of {H}adrons.
\newblock {\em Phys. Rev. Lett.}, 95, 2005.
\newblock J. Erlich, E. Katz, D. T. Son and M. A. Stephanov,
  \href{http://arxiv.org/abs/hep-ph/0501128}{arXiv:hep-ph/0501128}.

\bibitem{karch}
A.~Karch, E.~Katz, D.~T. Son, and M.~A. Stephanov.
\newblock Linear confinement and {A}d{S}/{QCD}.
\newblock {\em Phys. Rev. D}, 74, 2006.
\newblock A. Karch, E. Katz, D. T. Son and M. A. Stephanov,
  \href{http://arxiv.org/abs/hep-ph/0602229}{arXiv:hep-ph/0602229}.

\bibitem{weinberg}
S.~Weinberg.
\newblock Phenomenological {L}agrangians.
\newblock {\em Physica A}, 96, 1979.

\bibitem{scherer}
S.~Scherer.
\newblock Lecture notes: Chiral {D}ynamics, 2008.
\newblock
  \href{http://www.kph.uni-mainz.de/T/members/scherer/lecture/WS0708_SS08_chir%
al_dynamics.pdf}{http://www.kph.uni-mainz.de/T/members/scherer/lecture/WS0708$%
\_$SS08$\_$chiral$\_$dynamics.pdf} [17.09.2012].

\bibitem{mosel}
U.~Mosel.
\newblock {\em Fields, Symmetries, and Quarks}.
\newblock Springer-Verlag Berlin Heidelberg, 2010.

\bibitem{weise}
W.~Weise.
\newblock The {QCD} vacuum and its hadronic excitations.
\newblock W. Weise,
  \href{http://arxiv.org/abs/nucl-th/0504087}{arXiv:nucl-th/0504087}, 2005.

\bibitem{phdfrancesco}
F.~Giacosa.
\newblock {\em Glueball phenomenology within a nonlocal approach}.
\newblock PhD thesis, Eberhard Karls Universit{\"a}t, T{\"u}bingen, 2005.

\bibitem{phddenis}
D.~Parganlija.
\newblock {\em Quarkonium {P}henomenology in {V}acuum}.
\newblock PhD thesis, Johann Wolfgang Goethe-Universit{\"a}t, Frankfurt am
  Main, 2011.

\bibitem{machleidt}
R.~Machleidt and D.~R. Entem.
\newblock Chiral effective field theory and nuclear forces.
\newblock {\em Phys. Rep.}, 503, 2011.
\newblock R. Machleidt and D. R. Entem,
  \href{http://arxiv.org/abs/1105.2919}{arXiv:1105.2919 [nucl-th]}.

\bibitem{peskin}
M.~E. Peskin and D.~V. Schroeder.
\newblock {\em An Introduction to Quantum Field Theory}.
\newblock Perseus Books, 1995.

\bibitem{greiner}
J.~Reinhardt and W.~Greiner.
\newblock {\em Field Quantization}.
\newblock Springer-Verlag Berlin Heidelberg New York, 2008.

\bibitem{brown}
L.~S. Brown.
\newblock {\em Quantum Field Theory}.
\newblock Cambridge University Press, 1994.

\bibitem{kallen}
G.~K{\"a}ll{\'e}n.
\newblock On the {D}efinition of the {R}enormalization {C}onstants in {Q}uantum
  {E}lectrodynamics.
\newblock {\em Helv. Phys. Acta}, 25, 1952.

\bibitem{lehmann}
H.~Lehmann.
\newblock {\"U}ber {E}igenschaften von {A}usbreitungsfunktionen und
  {R}enormierungskonstanten quantisierter {F}elder.
\newblock {\em Nuovo Cim.}, 11(4), 1954.

\bibitem{giacosa}
F.~Giacosa.
\newblock Non-exponential {D}ecay in {Q}uantum {F}ield {T}heory and in
  {Q}uantum {M}echanics: {T}he {C}ase of {T}wo (or {M}ore) {D}ecay {C}hannels.
\newblock {\em Found. Phys.}, 42, 2012.
\newblock F. Giacosa, \href{http://arxiv.org/abs/1110.5923}{arXiv:1110.5923
  [nucl-th]}.

\bibitem{gross}
F.~Gross.
\newblock {\em Relativistic Quantum Mechanics and Field Theory}.
\newblock Wiley-Interscience, 1993.

\bibitem{beringer}
J.~Beringer et~al. (Particle Data~Group).
\newblock {R}eview of {P}article {P}hysics.
\newblock {\em Phys. Rev. D}, 86, 2012.
\newblock Particle Data Group,
  \href{http://pdg.lbl.gov/2012/reviews/rpp2012-rev-scalar-mesons.pdf}{http://%
pdg.lbl.gov/2012/reviews/rpp2012-rev-scalar-mesons.pdf}.

\bibitem{levy}
M.~Gell-Mann and M.~L{\'e}vy.
\newblock The {A}xial {V}ector {C}urrent in {B}eta {D}ecay.
\newblock {\em Nuovo Cim.}, 16(4), 1960.

\bibitem{schwinger}
J.~Schwinger.
\newblock A {T}heory of the {F}undamental {I}nteractions.
\newblock {\em Annals Phys.}, 2, 1957.

\bibitem{maiani}
L.~Maiani, F.~Piccinini, A.~D. Polosa, and V.~Riquer.
\newblock New {L}ook at {S}calar {M}esons.
\newblock {\em Phys. Rev. Lett.}, 93(21), 2004.
\newblock L. Maiani, F. Piccinini, A. D. Polosa and V. Riquer,
  \href{http://arxiv.org/abs/hep-ph/0407017}{arXiv:hep-ph/0407017}.

\bibitem{thooft}
G.~'t~Hooft, G.~Isidori, L.~Maiani, A.~D. Polosa, and V.~Riquer.
\newblock A theory of scalar mesons.
\newblock {\em Phys. Lett. B}, 662, 2008.
\newblock G. 't Hooft, G. Isidori, L. Maiani, A. D. Polosa and V. Riquer,
  \href{http://arxiv.org/abs/0801.2288}{arXiv:0801.2288 [hep-ph]}.

\bibitem{giacosaTetra}
F.~Giacosa.
\newblock Strong and electromagnetic decays of the light scalar mesons
  interpreted as tetraquark states.
\newblock {\em Phys. Rev. D}, 74, 2006.
\newblock F. Giacosa,
  \href{http://arxiv.org/abs/hep-ph/0605191}{arXiv:hep-ph/0605191}.

\bibitem{jaffe}
R.~J. Jaffe.
\newblock Multiquark hadrons. {I}. {P}henomenology of ${Q}^{2}\bar{Q}^{2}$
  mesons.
\newblock {\em Phys. Rev. D}, 15(1), 1977.

\bibitem{roos}
N.~A. T{\"o}rnqvist and M~Roos.
\newblock Confirmation of the {S}igma {M}eson.
\newblock {\em Phys. Rev. Lett.}, 76(10), 1996.
\newblock N. A. T{\"o}rnqvist and M Roos,
  \href{http://de.arxiv.org/abs/hep-ph/9511210}{arXiv:hep-ph/9511210v1}.

\bibitem{caprini}
I.~Caprini, G.~Colangelo, and H.~Leutwyler.
\newblock {M}ass and {W}idth of the {L}owest {R}esonance in {QCD}.
\newblock {\em Phys. Rev. Lett.}, 96, 2006.
\newblock I. Caprini, G. Colangelo and H. Leutwyler,
  \href{http://arxiv.org/abs/hep-ph/0512364}{arXiv:hep-ph/0512364}.

\bibitem{tornqvist}
N.~A. T{\"o}rnqvist.
\newblock Understanding the scalar meson $q\bar{q}$ nonet.
\newblock {\em Z. Phys. C}, 68, 1995.
\newblock N. A. T{\"o}rnqvist,
  \href{http://arxiv.org/abs/hep-ph/9504372}{arXiv:hep-ph/9504372}.

\bibitem{hoehler}
G.~H{\"o}hler.
\newblock {\"U}ber die {E}xponentialn{\"a}herung beim {T}eilchenzerfall.
\newblock {\em Zeits. f. Phys.}, 152, 1958.

\bibitem{pennington}
M.~Boglione and M.~R. Pennington.
\newblock Dynamical generation of scalar mesons.
\newblock {\em Phys. Rev. D}, 65, 2002.
\newblock M. Boglione and M. R. Pennington,
  \href{http://arxiv.org/abs/hep-ph/0203149}{arXiv:hep-ph/0203149}.

\bibitem{giacosaDynamical}
F.~Giacosa.
\newblock Dynamical generation and dynamical reconstruction.
\newblock {\em Phys. Rev. D}, 80, 2009.
\newblock F. Giacosa, \href{http://arxiv.org/abs/0903.4481}{arXiv:0903.4481
  [hep-ph]}.

\bibitem{peierls}
R.~E. Peierls.
\newblock Interpretation and properties of propagators.
\newblock In {\em Proceedings of the Glasgow Conference on Nuclear and Meson
  Physics}. Pergamon Press London, 1954.

\bibitem{levypoles}
M.~L{\'e}vy.
\newblock On the {D}escription of {U}nstable {P}articles in {Q}uantum {F}ield
  {T}heory.
\newblock {\em Nuovo Cim.}, 13(1), 1959.

\bibitem{aramaki}
S.~Aramaki and T.~Osawa.
\newblock On the existence of poles in the unphysical sheet.
\newblock {\em Prog. Theor. Phys.}, 29(3), 1963.

\bibitem{landshoff}
P.~V. Landshoff.
\newblock Poles and {T}hresholds and {U}nstable {P}articles.
\newblock {\em Nuovo Cim.}, 28, 1963.

\bibitem{bhattacharya}
T.~Bhattacharya and S.~Willenbrock.
\newblock Particles near threshold.
\newblock {\em Phys. Rev. D}, 47(9), 1993.

\bibitem{valencia1}
G.~Valencia and S.~Willenbrock.
\newblock Goldstone-boson equivalence theorem and the {H}iggs resonance.
\newblock {\em Phys. Rev. D}, 42(3), 1990.

\bibitem{willenbrock1}
S.~Willenbrock and G.~Valencia.
\newblock The {H}iggs resonance shape.
\newblock {\em Phys. Lett. B}, 247, 1990.

\bibitem{willenbrock2}
S.~Willenbrock and G.~Valencia.
\newblock On the definition of the ${Z}$-boson mass.
\newblock {\em Phys. Lett. B}, 259, 1991.

\bibitem{sirlin1}
A.~Sirlin.
\newblock Theoretical {C}onsiderations {C}oncerning the ${Z}^{0}$ {M}ass.
\newblock {\em Phys. Rev. D}, 67(16), 1991.

\bibitem{sirlin2}
A.~Sirlin.
\newblock Observations concerning mass renormalization in the electroweak
  theory.
\newblock {\em Phys. Lett. B}, 267, 1991.

\bibitem{morgan}
D.~Morgan and M.~R. Pennington.
\newblock New data on the ${K}\bar{K}$ threshold region and the nature of the
  $f_{0}({S}^{*})$.
\newblock {\em Phys. Rev. D}, 48(3), 1993.

\bibitem{escribano}
R.~Escribano, A.~Gallegos, J.~L.~Lucio M., G.~Moreno, and J.~Pestieau.
\newblock On the mass, width and coupling constants of the $f_{0}(980)$.
\newblock {\em Eur. Phys. J. C}, 28, 2003.
\newblock R. Escribano, A. Gallegos, J. L. Lucio M., G. Moreno and J. Pestieau,
  \href{http://de.arxiv.org/abs/hep-ph/0204338}{arXiv:hep-ph/0204338v3}.

\bibitem{koenigsberger}
K.~K{\"o}nigsberger.
\newblock {\em Analysis 1}.
\newblock Springer-Verlag Berlin, 2004.

\bibitem{weltner}
K.~Weltner.
\newblock {\em Mathematik f{\"u}r Physiker 1}.
\newblock Springer-Verlag Berlin Heidelberg, 2006.

\bibitem{pucker}
C.~B. Lang and N.~Pucker.
\newblock {\em Mathematische Methoden in der Physik}.
\newblock Spektrum Akademischer Verlag, 2005.

\bibitem{forster}
O.~Forster.
\newblock {\em Lectures on Riemann Surfaces}.
\newblock Springer-Verlag New York, 1984.

\bibitem{lepage}
W.~R. LePage.
\newblock {\em Complex Variables and the Laplace Transform for Engineers}.
\newblock Dover Publications, 1980.

\bibitem{nearing}
J.~Nearing.
\newblock {\em Mathematical Tools for Physics}.
\newblock Dover Publications, 2010.
\newblock
  \href{http://www.physics.miami.edu/~nearing/mathmethods/}{http://www.physics%
.miami.edu/$\sim$nearing/mathmethods/} [17.09.2012].

\bibitem{bronstein}
I.~N. Bronstein, K.~A. Semendjajew, G.~Musiol, and H.~M{\"u}hlig.
\newblock {\em Taschenbuch der Mathematik}.
\newblock Verlag Harri Deutsch, 2005.

\bibitem{fischerkaul}
H.~Fischer and H.~Kaul.
\newblock {\em Mathematik f{\"u}r Physiker}.
\newblock B. G. Teubner Verlag, 2005.

\bibitem{lee}
T.~D. Lee.
\newblock Some {S}pecial {E}xamples in {R}enormalizable {F}ield {T}heory.
\newblock {\em Phys. Rev.}, 95(5), 1954.

\bibitem{ghirardi}
L.~Fonda, G.~C. Ghirardi, and A.~Rimini.
\newblock Decay theory of unstable quantum systems.
\newblock {\em Rep. Prog. Phys.}, 41, 1978.

\bibitem{facchi}
P.~Facchi and S.~Pascazio.
\newblock Unstable systems and quantum {Z}eno phenomena in quantum field
  theory.
\newblock {\em Fundamental Aspects of Quantum Physics}, 17:222--246, 2003.
\newblock {\em Contribution to the Proceedings of the Japan-Italy Joint
  Workshop on Quantum Open Systems, Quantum Chaos and Quantum Measurement}, P.
  Facchi and S. Pascazio,
  \href{http://arxiv.org/abs/quant-ph/0202127}{arXiv:quant-ph/0202127}.

\bibitem{shankar}
R.~Shankar.
\newblock {\em Principles of Quantum Mechanics}.
\newblock Springer-Verlag New York, 1994.

\bibitem{feynmanhibbs}
R.~P. Feynman and A.~R. Hibbs.
\newblock {\em Quantum Mechanics and Path Integrals}.
\newblock McGraw-Hill, 1965.

\bibitem{moshinsky}
M.~Moshinsky, E.~Sadurn{\'\i}, and A.~del Campo.
\newblock Alternative {M}ethod for {D}etermining the {F}eynman {P}ropagator of
  a {N}on-{R}elativistic {Q}uantum {M}echanical {P}roblem.
\newblock {\em SIGMA}, 3, 2007.
\newblock {\em Contribution to the
  \href{http://www.emis.de/journals/SIGMA/symmetry2007.html}{Proceedings of the
  Seventh International Conference Symmetry in Nonlinear Mathematical
  Physics}}, M. Moshinsky, E. Sadurni and A. del Campo,
  \href{http://arxiv.org/abs/0711.3544}{arXiv:0711.3544 [quant-ph]}.

\bibitem{simons}
A.~Altland and B.~D. Simons.
\newblock {\em Condensed Matter Field Theory}.
\newblock Cambridge University Press, 2010.

\bibitem{higgspseudo}
G.~Passarino, C.~Sturm, and S.~Uccirati.
\newblock Higgs pseudo-observables, second {R}iemann sheet and all that.
\newblock {\em Nucl. Phys. B}, 824, 2010.
\newblock G. Passarino, C. Sturm and S. Uccirati,
  \href{http://arxiv.org/abs/1001.3360}{arXiv:1001.3360v1 [hep-ph]}.

\bibitem{sudarshan}
B.~Misra and E.~C.~G. Sudarshan.
\newblock The {Z}eno's paradox in quantum theory.
\newblock {\em J. Math. Phys.}, 18(4), 1977.

\bibitem{sakurai}
J.~J. Sakurai.
\newblock {\em Modern Quantum Mechanics}.
\newblock Prentice Hall, 1993.

\bibitem{itano}
W.~M. Itano, D.~J. Heinzen, J.~J. Bollinger, and D.~J. Wineland.
\newblock Quantum {Z}eno effect.
\newblock {\em Phys. Rev. A}, 41(5), 1990.

\bibitem{raizen}
S.~R. Wilkinson, C.~F. Bharucha, M.~C. Fischer, K.~W. Madison, P.~R. Morrow,
  Q.~Niu, B.~Sundaram, and M.~G. Raizen.
\newblock Experimental evidence for non-exponential decay in quantum
  tunnelling.
\newblock {\em Nature}, 387, 1997.

\bibitem{balzer}
Chr. Balzer, Th. Hannemann, D.~Rei{\ss}, Chr. Wunderlich, W.~Neuhauser, and
  P.~E. Toschek.
\newblock A relaxationless demonstration of the {Q}uantum {Z}eno paradox on an
  individual atom.
\newblock {\em Optics Communications}, 211, 2002.
\newblock Chr. Balzer, Th. Hannemann, D. Rei{\ss}, Chr. Wunderlich, W.
  Neuhauser and P. E. Toschek,
  \href{http://arxiv.org/abs/quant-ph/0406027}{arXiv:quant-ph/0406027}.

\bibitem{veltman}
M.~Veltman.
\newblock Unitarity and causality in a renormalizable field theory with
  unstable particles.
\newblock {\em Physica}, 29:186, 1963.

\bibitem{giacosaSpectral}
F.~Giacosa and G.~Pagliara.
\newblock Spectral functions of scalar mesons.
\newblock {\em Phys. Rev. C}, 76, 2007.
\newblock F. Giacosa and G. Pagliara,
  \href{http://arxiv.org/abs/0707.3594}{arXiv:0707.3594 [hep-ph]}.

\bibitem{giacosaHadrons}
G.~Pagliara and F.~Giacosa.
\newblock Non exponential decays of hadrons.
\newblock {\em Acta Phys. Polon. B Proc. Supp.}, 4, 2011.
\newblock G. Pagliara and F. Giacosa,
  \href{http://arxiv.org/abs/1108.2782}{arXiv:1108.2782 [hep-ph]}.

\bibitem{giacosaFermions}
F.~Giacosa and G.~Pagliara.
\newblock Spectral function of a scalar boson coupled to fermions.
\newblock F. Giacosa and G. Pagliara,
  \href{http://arxiv.org/abs/arXiv:1210.4192}{arXiv:1210.4192 [hep-ph]}, 2012.

\bibitem{steele}
G.~L.~Steele jr.
\newblock {\em Common LISP. The Language}.
\newblock Digital Press, 1990.
\newblock
  \href{http://www.cs.cmu.edu/Groups/AI/html/cltl/clm/clm.html}{http://www.cs.%
cmu.edu/Groups/AI/html/cltl/clm/clm.html} [17.09.2012].

\bibitem{kahan}
W.~Kahan.
\newblock Branch {C}uts for {C}omplex {E}lementary {F}unctions, or {M}uch {A}do
  {A}bout {N}othing's {S}ign {B}it.
\newblock {\em The State of the Art in Numerical Analysis}, pages 165--211,
  1987.
\newblock {\em Contribution to the Proceedings of the Joint IMA/SIAM Conference
  on the State of the Art in Numerical Analysis}.

\bibitem{achasov}
N.~N. Achasov and A.~V. Kiselev.
\newblock Propagators of light scalar mesons.
\newblock {\em Phys. Rev. D}, 70, 2004.
\newblock N. N. Achasov and A. V. Kiselev,
  \href{http://arxiv.org/abs/hep-ph/0405128}{arXiv:hep-ph/0405128}.

\bibitem{salam}
P.~T. Matthews and A.~Salam.
\newblock Relativistic {F}ield {T}heory of {U}nstable {P}articles.
\newblock {\em Phys. Rev.}, 112, 1958.

\bibitem{vanBeveren1}
E.~van Beveren and G.~Rupp.
\newblock First indications of the existence of a 38 {M}e{V} light scalar
  boson.
\newblock E. van Beveren and G. Rupp,
  \href{http://arxiv.org/abs/1102.1863}{arXiv:1102.1863 [hep-ph]}, 2011.

\bibitem{vanBeveren2}
E.~van Beveren and G.~Rupp.
\newblock Material evidence of a 38 {M}e{V} boson.
\newblock E. van Beveren and G. Rupp,
  \href{http://arxiv.org/abs/1202.1739}{arXiv:1202.1739 [hep-ph]}, 2012.

\bibitem{abraamyan}
Kh.~U. Abraamyan, A.~B. Anisimov, M.~I. Baznat, K.~K. Gudima, M.~A. Nazarenko,
  S.~G. Reznikov, and A.~S. Sorin.
\newblock Observation of the ${E}$(38)-boson.
\newblock Kh. U. Abraamyan, A. B. Anisimov, M. I. Baznat, K. K. Gudima, M. A.
  Nazarenko, S. G. Reznikov and A. S. Sorin,
  \href{http://arxiv.org/abs/1208.3829}{arXiv:1208.3829 [hep-ex]} Note: {\em
  The first version of the manuscript has been withdrawn for further
  verification and more detailed description of the experiment and data
  analysis. The second version is being prepared.}, 2012.

\bibitem{giacosaE38}
F.~Giacosa and T.~Wolkanowski.
\newblock Propagator poles and an emergent stable state below threshold:
  general discussion and the ${E}$(38) state.
\newblock {\em Mod. Phys. Lett. A}, 27(39), 2012.
\newblock F. Giacosa and T. Wolkanowski,
  \href{http://arxiv.org/abs/1209.2332}{arXiv:1209.2332 [hep-ph]}.

\bibitem{parganlijaBig}
D.~Parganlija, P.~Kovacs, G.~Wolf, F.~Giacosa, and D.~H. Rischke.
\newblock Meson vacuum phenomenology in a three-flavor linear sigma model with
  (axial-)vector mesons.
\newblock D. Parganlija, P. Kovacs, G. Wolf, F. Giacosa and D. H. Rischke,
  \href{http://arxiv.org/abs/1208.0585}{arXiv:1208.0585 [hep-ph]}, 2012.

\bibitem{rischke}
D.~Parganlija, P.~Kovacs, G.~Wolf, F.~Giacosa, and D.~H. Rischke.
\newblock Phenomenology of {A}xial-{V}ector {M}esons from an {E}xtended
  {L}inear {S}igma {M}odel.
\newblock D. Parganlija, P. Kovacs, G. Wolf, F. Giacosa and D. H. Rischke,
  \href{http://arxiv.org/abs/1208.2054}{arXiv:1208.2054 [hep-ph]}, 2012.

\bibitem{locher}
M.~P. Locher, V.~E. Markushin, and H.~Q. Zheng.
\newblock Structure of $f_{0}(980)$ from a coupled channel analysis of
  ${S}$-wave $\pi\pi$ scattering.
\newblock {\em Eur. Phys. J. C}, 4, 1998.
\newblock M. P. Locher, V. E. Markushin and H. Q. Zheng,
  \href{http://arxiv.org/abs/hep-ph/9705230}{arXiv:hep-ph/9705230}.

\end{thebibliography}
\clearpage

\thispagestyle{empty}
\

\newpage
\clearpage

\thispagestyle{empty}
\

\newpage
\clearpage

\thispagestyle{empty}
\section*{Acknowledgement}

I want to thank my supervisor Prof. Dr. Dirk H. Rischke who gave me the opportunity to do my master thesis in his group. I am grateful to him for carefully correcting the thesis and for his important remarks concerning my (peculiar) presentation of the results. Also, I want to thank all the members of the group for the friendly welcome.
\\
\\
Moreover, I would like to thank Dr. Francesco Giacosa for giving me such an interesting subject and for plenty of fruitful discussions about physics, philosophy and mathematics. I learned quite a lot about hadron physics over the last year. Francesco also gave an inspiring seminar on the foundations of quantum mechanics and held a lecture about decays in quantum field theory. He especially directed my attention to the issue regarding the putative novel scalar resonance $E(38)$, leading to my first scientific publication.
\\
\\
I also thank PD Dr. Dennis D. Dietrich for his useful comments on technical aspects of quantum field theory. Dennis gave an excellent introduction into physics beyond the Standard Model.
\\
\\
Vielen lieben Dank an Kira und Micha f\"ur ihre moralische Unterst\"utzung, aufmunternde
Art und f\"ur all die unz\"ahligen Gespr\"ache \"uber Physik, Frauen und was es sonst noch so gibt. Ein besonderer Dank geb\"uhrt auch Christian, der mir und meiner Familie die letzten Monate finanziell unter die Arme gegriffen hat. Ich entschuldige mich daf\"ur, dass ich das vergangene Jahr kaum Zeit und praktisch nichts anderes im Kopf hatte als diese nervigen Pole.
\\
\\
Ich danke dem Studentenwerk Frankfurt am Main f\"ur die z\"ugige Bearbeitung meiner BAf\"oG-Antr\"age; erst dadurch wurde es mir m\"oglich, \"uberhaupt ein Studium aufzunehmen und (fast in Regelstudienzeit) zu beenden.
\\
\\
Zu guter Letzt m\"ochte ich au{\ss}erordentlich meiner Frau Jessica danken, die mir ein ungeahntes Ma{\ss} an Verst\"andnis entgegenbrachte und mich in dunklen Zeiten nie alleine lie{\ss}, und unserer wundervollen Tochter Helene, die mir jeden Tag das sch\"onste Gl\"uck der Welt bescherte.

\clearpage

\thispagestyle{empty}
\

\newpage
\clearpage

\thispagestyle{empty}
\section*{Selbst\"andigkeitserkl\"arung}

Hiermit versichere ich, dass ich die vorliegende Masterarbeit selbst\"andig und nur mit den angegebenen Hilfsmitteln angefertigt habe und dass alle Stellen, die dem Wortlaut oder dem Sinne nach anderen Werken entnommen wurden, durch Angaben von Quellen als Entlehnung kenntlich gemacht sind. Diese Masterarbeit wurde zudem in gleicher oder \"ahnlicher Form in keinem anderen Studiengang als Pr\"ufungsleistung vorgelegt. 

\vskip 3cm

\hfill \hfill {\footnotesize Ort, Datum \ \ \ \ \ \ \ \ \ \ \ \ \ \ \ Unterschrift} \ \ \ \ \ \ \ \

\end{document}